%% file: section.tex
\documentclass[twoside]{article}

\usepackage{meta}
\usepackage{tufte-esque}
\usepackage{global}
\usepackage{local}

\begin{document}

\thispagestyle{empty}

{\sffamily
    \fontsize{20}{22}\selectfont%
    \par\noindent%
    \textcolor{darkgray}{%
        \allcaps{\longtitle}%
    }

    \vspace{1.3pc}

    \fontsize{18}{20}\selectfont%
    \par\noindent%
    \textcolor{black}{%
        \longauthors%
    }

    \vspace{-1.6\baselineskip}

    \par\noindent%
    \textcolor{darkgray}{%
        \hspace{0.6cm} \small{\contact}
    }

    \vspace{-2.5pc}

    \fontsize{14}{16}\selectfont%
    \par\noindent%
    \textcolor{darkgray}{%
        \begin{center}%
        \affiliations%
        \end{center}%
    }
}

\vspace{0.9\baselineskip}

{
    \begin{center}
    \begin{minipage}{0.92\textwidth}
    \noindent\rule{\textwidth}{0.6pt}%
    \justifying%
    \singlespacing%

    \vspace{-.5\baselineskip}

    {\noindent\ignorespaces%
        \input{0-abstract}
    }

    \noindent\rule{\textwidth}{0.6pt}
    \end{minipage}
    \end{center}
    \vfill
}

\pagebreak

\section{introduction}\label{sec:intro}
\input{1-intro}

\subsection{notational conventions}
\input{intro}

\section{background}\label{sec:bg}
\input{2-bg}

\section{the differential expression problem}\label{sec:de-objective}
\input{3-de-objective}

\section{differential expression robustness}\label{sec:de-robustness}
\input{4-de-influence}

\pagebreak
\section{experiments}\label{sec:exp}
\input{5-experiments}

\section{conclusions}\label{sec:conc}
\input{6-conclusions}

\section{acknowledgements}
\input{7-acknowledge}

\clearpage
{
    \small
    \singlespacing \sffamily
    \bibliographystyle{classybib}
    \renewcommand{\refname}{references}
    \bibliography{references}
}

\clearpage
\appendix
\renewcommand{\theequation}{A-\arabic{equation}}
\setcounter{figure}{0}
\renewcommand{\thefigure}{A-\arabic{figure}}
\input{_-appendix}

\end{document}

%% file: 0-abstract.tex
Analysis of differential gene expression plays a fundamental role in biology
toward illuminating the molecular mechanisms driving a difference between
groups (e.g., due to treatment or disease). While any analysis is run on
particular cells or samples, the intent is to generalize to future occurrences
of the treatment or disease. Implicitly, generalization is justified under the
assumption that present and future samples are independent and identically
distributed from the same population. Though this assumption is always false,
we might hope that any deviation from the assumption is small enough that A)
fundamental conclusions of the analysis still hold, and B) standard tools like
standard error, significance, and power still reflect generalizability.
Conversely, we might worry about these deviations, and reliance on standard
statistical tools, if conclusions could be substantively changed by dropping a
very small fraction of observations. While checking every small fraction is
computationally intractable, recent work develops an approximation to identify
when such an influential subset exists. Building on this work, we develop a
metric for dropping-data robustness of differential expression; namely, we cast
the analysis in a differentiable form suitable to the approximation, extend the
approximation to models whose hyperparameters depend on the full dataset, and
extend the notion of a data point from a single cell to a pseudobulk
observation. We then overcome the inherent non-differentiability of gene set
enrichment analysis to develop an additional approximation for the robustness
of top gene sets. We use our tool to assess the robustness of differential
expression for published single-cell RNA-seq data, and discover that thousands
of genes can have their results flipped by dropping <1\% of the data, including
hundreds of results that are sensitive to dropping a single cell (<0.07\%).
Surprisingly, we find that this non-robustness extends to high-level takeaways,
and that half or more of the top 10 gene sets can be changed by dropping <1–2\%
of cells---and two or more can be changed by dropping a single cell.

%% file: 1-intro.tex
Orchestration of \emph{gene expression} drives
differences between cell types within an organism, and
between cell states
in response to perturbation (such as disease or treatment with a drug).
Consequently, to understand the mechanism behind these differences,
\emph{differential expression} analysis---followed
by \emph{gene set enrichment} to detect a biologically meaningful signal among
differentially expressed genes---is
a fundamental and ubiquitous method in biology.

In the context of an experiment, researchers collect a finite number
of \textit{samples}
(for example, particular cells dissociated from particular tissue samples
collected from particular subjects within a study)
and use inferential statistics to make
fundamental statements about a broader \textit{population}
(the underlying molecular process behind a disease, external perturbation, or
phenotype).

For example, to better understand the etiology of a disease,
a typical differential expression analysis could entail
collecting tissue samples from a finite number of human subjects, some of whom
have been diagnosed with the disease and some of whom have not;
dissociating the tissue to quantify gene expression across many of its
constituent cells (via single-cell RNA-sequencing);
clustering and manually annotating cells to assign them to distinct cell types;
performing differential expression analysis (followed by gene set enrichment)
to detect meaningful differences between healthy and diseased cells of the same cell type;
and interpreting these results to
pose
hypotheses
about the underlying mechanism and effects of the disease.
In other words, the goal of such an experiment would be to
glean fundamental biological truths (or, at least, to generate hypotheses)
about the underlying disease process within that cell type---regardless of
\begin{itemize}
    \item the particular individuals chosen to represent healthy and diseased states,
    \item the particular tissue fractions that were sampled,
    \item the particular cells that were successfully sequenced,
    \item the particular subset of those cells that were classified as the relevant cell type, and survived QC,
    \item the particular transcripts in each cell that were sampled,
    \item or any anomalies or biases in the particular sequencing process (technology and batches)
        used to measure gene expression.
\end{itemize}
\vspace{\parskip}

Similarly, even prospective experiments in a cell line---where the observed
sample of cells is hypothetically drawn from
a purposely homogenous
population---are subject to
incidental differences
\begin{itemize}
    \item
        between cells (both intrinsic, like spontaneous mutations, and transient, like cell cycling) and
    \item
        between treatment conditions
        (like variable efficacy and off-target effects in
        genetic perturbation screens),
\end{itemize}
as well as
during
the measurement process itself
(detailed above).

While statistical tools like significance, power, and standard error are
essential for
quantifying the limitations of what a finite sample can say about the population
from which it was drawn,
all are predicated on the assumption that the data in hand are
an unbiased representation
of the target population.
As such, they cannot speak to the unmodeled idiosyncrasies within a particular RNA-seq
dataset---whether biological subpopulations or purely technical artifacts---that
may affect generalization to the \textit{desired} real-world
population.

To this end, recent work in the statistical robustness literature~\cite{ryan-amip}
develops a
tool to audit
generalizability based on
the extent to which the key takeaways of an analysis
are robust against dropping a small handful of observations from the dataset.
If key outcomes can be meaningfully changed by such a data perturbation,
then it may be unlikely that these outcomes
will generalize to future experiments
or are indicative of fundamental processes.
Further, such an analysis can point to interesting
structure within the data,
based on the particular data points that are highly influential.

In practice,
however,
such a metric is intractable to compute exactly
for even moderately sized datasets,
thanks to combinatorial explosion---for example,
${1000 \choose 10} >
10^{23}$ rounds of empirically rerunning the analysis
in order
to na\"ively identify the most influential 1\% of
$\ncells=1000$ observations.
Instead, the authors introduce a first-order approximation that is both
efficient and, they demonstrate,
sufficiently accurate to diagnose nonrobustness in published analyses
of real datasets.\footnote{
    Namely, basic and hierarchical linear regression of econometric data~\cite{ryan-amip}
}
Specifically, they use a first-order Taylor expansion
and automatic differentiation\footnote{
    Also known as \textit{autodiff}; encompassing various algorithms
    to evaluate derivatives of mathematical functions written as computer code~\cite{autodiff}.
    Using software that implements these techniques, we can write
    flexible and performant code for assessing dropping-data robustness
    (which hinges on differentiation)
    without working out each, potentially complex symbolic derivative.
}
to \textit{estimate}
the effect of dropping data points---enabling
a single (amortized) model fit and autodiff computation to yield
the approximate effect of excluding \textit{any} small combination of points.
Using this metric,
the authors identify published econometric analyses
where significant results with meaningful effect sizes are nonetheless
susceptible to having their effects erased, or even changed to a significant
result in the opposite direction, by dropping a small fraction of data points~\cite{ryan-amip}.

The ability to efficiently compute
such a
robustness metric for the key outcomes of differential expression---namely,
the \textit{sign}, \textit{magnitude}, and \textit{significance} of each treatment effect,
as well as \textit{higher-level patterns} in
biological functions enriched among
differentially expressed genes---would
provide a
relevant check on generalizability,
particularly for
inherently noisy single-cell RNA-seq data,
and through a distinct lens compared to existing tools for robustness.
To this end, we set out to develop a \emph{dropping-data robustness}
metric for differential expression
based on the minimal proportion of observations (i.e., cells, for single-cell RNA-seq data)
that can be excluded in order to reverse a finding.
Further, recognizing that these robustness results would be gene-specific,
whereas the outcomes of differential expression must be synthesized across genes
in order to form high-level takeaways,
we also set out to extend this data robustness metric to gene set enrichment
(a common downstream procedure to summarize the results of differential expression).

While the existing framework~\cite{ryan-amip} provides a means of estimating
dropping-data robustness for any Z-estimator\footnote{
    i.e., ``approximate zeros of data-dependent functions''~\cite{z-estimator}
}
(such as maximum likelihood estimation with a log-likelihood objective)
via a local first-order approximation,
we encounter several challenges in translating this approach
to generalized linear models for differential expression---including
data-dependent hyperparameters,
pathological failure to converge for a particular class of sparse genes,
rank-based corrections for multiple testing,
and test statistics with zero first-order derivative.

After reviewing
approaches to differential expression and robustness
({\cref{sec:bg}),
in \cref{sec:de-objective}, we cast the analysis and key gene-level outcomes of
differential expression in terms that are suitable for dropping-data robustness
by modifying a typical \texttt{DESeq2}~\cite{deseq}/\texttt{glmGamPoi}~\cite{ggp}-style analysis
(and verifying that the results of our modified analysis retain sufficient fidelity to the original).
In \cref{sec:de-robustness}, we review the existing approximation,
extend it to models with hyperparameters that depend on the full dataset,
and derive a means of computing dropping-data robustness
for both independent cell and pseudobulk approaches to differential expression
from single-cell RNA-seq.\footnote{
    While we focus on analysis of \textit{single-cell} data, our dropping-data robustness
    metric is, in theory, equally relevant to analysis of \textit{bulk} RNA-seq.
    However, our approximation to efficiently compute this metric (\cref{sec:de-robustness})
    works best when the number of observations
    is sufficiently large ($10^2$ or, ideally, $10^3$ or more).
    This large data setting is common for single-cell data
    (where cells are the unit of observation, whether or not they are treated as independent replicates),
    but less so for bulk data
    (where the number of replicates, technical or biological, tends to be much smaller).
    Alternately, for datasets with few observations
    (too few to trust the quality of the approximation),
    our dropping-data robustness may be exactly computable
    in reasonable time (via the jackknife).
}
Specifically, we derive estimators of the minimal number of cells that, if
dropped from the analysis, would flip the sign, meaningfully change the magnitude
(based a specified threshold),
and/or flip the significance
(based on standard or quasi-likelihood Wald testing) of a gene's treatment effect.

On the other hand, gene set enrichment---the canonical follow-up to differential expression,
to identify biologically meaningful patterns among differentially expressed genes---is
\textit{not} amenable to the existing robustness approximation because it is
based
on ranking and thresholding operations, both of which are
inherently non-differentiable.
Further, dropping observations affects differential expression results \textit{across} genes,
and therefore affects
the top enriched gene sets
(based on joint ranking of genes, followed by joint ranking of gene sets)
in
an intricate and combinatorial way.
Nonetheless, we develop a heuristic approach (\cref{sec:gsea-robustness}) to use
gene-level influence scores---an intermediate of our robustness metric---to
bound the dropping-data robustness of the top enriched gene sets
(based on hypergeometric testing),
a key high-level outcome of differential expression.

In sum, in order to make dropping-data robustness useful to biologists studying differential expression---a
foundational analysis in biology---here we
\begin{itemize}
    \item cast differential expression
        in terms that are suitable for dropping-data robustness;
    \item apply the
        dropping-data approximation to generalized linear models (GLMs),
        and extend it to models whose hyperparameters depend on the data;
    \item extend dropping-data robustness to multiple conceptions
        of a ``data point,''
        (e.g., a single measurement---corresponding to a single cell---or
         a pseudobulk observation comprising multiple cells);
    \item extend individual robustness results (per gene) into high-level
        robustness results (per pathway or gene set), despite inherent
        non-differentiability that precludes this analysis from being readily
        amenable to the
        existing framework;
    \item develop software (using Python and the autodifferentiation
                            library \texttt{jax}~\cite{jax})
          to quantify
          robustness for \texttt{DESeq2}/\texttt{glmGamPoi}-style
          differential expression analyses; and
    \item use these tools to analyze and interpret the robustness of differential expression
          results for published
          single-cell RNA-seq data.
\end{itemize}
\vspace{\parskip}

Namely, in \cref{sec:exp}, we demonstrate the accuracy and utility of our
dropping-data robustness metric by applying it to differential expression
(via Wald tests of negative binomial GLMs)
and gene set enrichment
(via hypergeometric tests)
for single-cell RNA-seq of healthy and diseased samples.\footnote{
    Specifically, goblet cells from healthy subjects and subjects with ulcerative colitis
}
Whereas exactly computing this metric would, na\"ively, take millennia,\footnote{
    For example, over 15,000,000 years to re-run differential expression analysis
    (assuming 1 minute per run) after dropping every subset of 5 cells
    from a dataset of 1000 cells
}
we
approximate the effect of excluding \textit{any} handful of cells,
for the key outcomes of differential expression across genes, in minutes.
As a result, we identify thousands of genes with meaningful nonrobustness---whose differential expression status,
with respect to statistical significance or effect size, can be flipped
by dropping a small handful of cells---and show that, for this particular dataset,
\textit{at least half}
of the top 10 gene sets
(enriched among upregulated or downregulated genes)
can be changed by dropping less than 1 or 2\% of cells (respectively);
\textit{four}
of the top 10 gene sets can be changed by dropping less than 0.5 or 0.3\%;
and \textit{two or three}
of the top 10 gene sets can be changed
by dropping a single cell (of >1000).

%% file: intro.tex
Throughout, we'll use the following notation---bold and capitalized for matrices (e.g. $\mat{M}$), and
bold and lowercase for vectors (e.g. $\vec{v}$), which will always be column vectors.
The $i\th$ row of $\mat{M}$ is $\vec{M}_i$,
the $j\th$ column is $\vec{M}^{(j)}$,
and the $(i,j)\th$ entry is $m_{i,j}$.
The $i\th$ entry of $\vec{v}$ is the scalar $v_i$.

Sometimes, we will explicitly define a vector or matrix's dimensions when
introducing it. For example, $\vec{v}\ofSize{J,1}$ is a length $J$ column
vector, and $\mat{M}\ofSize{J,K}$ is a $J \times K$ matrix.

A bolded number---such as $\vec{0}$ or $\vec{1}$---refers to a vector
(of the contextually appropriate size) whose entries are identical and equal to
that number---such as 0 or 1, respectively.

Encircled symbols denote component-wise analogs of their corresponding
operations; i.e., $\odot$ for the Hadamard product, and $\oplus$
for component-wise addition.

%% file: 2-bg.tex
\subsection{differential expression}

Since technology made it possible to measure
the expression level of a gene---first
for a few candidate genes
(by Northern blots in the `70s or quantitative PCR in the `80s);
then transcriptome-wide, across tens of thousands of genes
(by microarray in the `90s or bulk RNA-sequencing in the 2000s)~\cite{hist-seq,hist-seq-2};
and now at the precision of individual cells
(by single-cell RNA-sequencing, a.k.a. scRNA-seq, since the 2010s)~\cite{hist-seq-sc}---it
has been of interest to compare gene expression between groups.

All cells in an organism encode the same DNA sequence (to a first approximation);
therefore, differences between cells arise from differences in their orchestration
of gene \textit{expression} (transcription of DNA to RNA).
Similarly, cells maintain the same DNA sequence over time
(to a first approximation);
therefore, dynamic changes in response to a perturbation
(such as artificial perturbation with a drug or gene knockout,
or natural perturbation by disease)
arise from changes in expression.
The process of measuring the RNA content of a cell or biological sample
is generally destructive,
meaning that the same cell cannot be measured before \textit{and} after perturbation.
Often, such as when seeking to understand human disease, it is not even possible to
collect samples
from the same subject before and after perturbation.
So, scientists seeking biological insight
(into the molecular basis for differences between cell types in the same tissue,
or for the response to an external perturbation)
collect many samples from each group
(creating exchangeability through biological replicates and randomization design)
and use them to infer something fundamental about the population.
\emph{Differential expression} (DE) analysis---the
process of quantifying differences in levels of gene expression between
phenotypic
or other groups---is a fundamental analysis and workhorse of biology.

The overarching goal of differential expression is---for each gene---to:
\begin{itemize}
    \item test the null hypothesis that there is no difference in expression between groups, and
    \item make a point estimate of the effect size (often as a ``log-fold change'').
\end{itemize}
Then, after
assessing
each gene
(which generally number in the thousands or tens of thousands, depending on the organism),
the desired output is a reduced set and/or ranked list of \textit{differentially expressed}
genes prioritized for interpretation.
To find biologically meaningful patterns among these gene-level statistics,
they are often tested for \textit{gene set enrichment} of relevant gene sets
or pathways (assembled based on prior knowledge)---ultimately summarized as
a list of ``top'' gene sets.

The most common approaches to DE analysis of bulk or single-cell RNA-seq data are:
\begin{enumerate}
    \item t-tests or their nonparametric analog, Wilcoxon rank-sum tests; \label{it:ttest}
    \item generalized linear models (GLMs); and \label{it:glm}
    \item generalized linear mixed models, \label{it:lmm}
\end{enumerate}
where the latter two are then combined with a statistical test
(Wald, likelihood ratio, or score).

\ref{it:ttest} is simplest but often inadequate; it does not allow for covariate
structure, and either assumes Gaussian noise inappropriate for count data (t-test)
or sacrifices power by considering only ranks (Wilcoxon rank-sum).
Nonetheless---and perhaps thanks to their simplicity and non-customizability---
these are the default modes of DE analysis for popular single-cell software packages
(t-test for \texttt{scanpy.tl.rank\_genes\_groups}\footnote{
    \url{https://github.com/scverse/scanpy/blob/d26be443373549f26226de367f0213f153556915/scanpy/tools/_rank_genes_groups.py\#L541-L545}
} and
Wilcoxon rank-sum for \texttt{Seurat::FindMarkers}\footnote{
    \url{https://github.com/satijalab/seurat/blob/763259d05991d40721dee99c9919ec6d4491d15e/R/differential_expression.R\#L50}
}).
A survey of recent scRNA-seq publications involving differential expression~\cite{de-false-discoveries}
found that
\ref{it:ttest} was the most common approach
(mainly driven by Wilcoxon rank-sum).

For comparisons beyond the simplest of experimental designs, when
desiderata include
accounting for unwanted sources of variability
and
the power to detect effects beyond the most conspicuous,
the most common approach is \ref{it:glm}.
GLMs allow for the flexibility of exponential family distributions
to model the response (i.e., RNA transcript counts),
conditioned on interpretable linear predictors that determine
the natural parameter via a link function.
Models for gene expression are often parameterized as negative binomial family
(e.g., \texttt{DESeq2}~\cite{deseq}, \texttt{edgeR}~\cite{ql-edgeR})---whose
additional dispersion parameter accounts for the theoretical noise in the
measurement process and in biological variability, as well as the empirical
overdispersion observed in sequencing counts---though other forms are also used
(e.g., \texttt{MAST}~\cite{mast}, \texttt{limma-voom}~\cite{limma-voom};
 Gaussian with transformed observations).
Following Wilcoxon rank-sum, \texttt{DESeq2} was the most popular surveyed approach
to differential expression for scRNA-seq~\cite{de-false-discoveries},
despite originally being developed for bulk RNA-seq.

For single-cell RNA-seq data, which is the focus of our work, GLMs do pose a limitation
that may be addressed by \ref{it:lmm}.
Namely, cells collected from the same biological sample (e.g., cell culture, tissue, or subject)
are
inherently correlated
yet are na\"ively treated as independent samples,
yielding false power to detect population-level differences.
Because treatment is fully crossed with biological sample
(e.g., a subject is either healthy or diseased), the sample ID cannot be included as
a covariate in the GLM in order to regress out sample-level variability.
One solution is to use (generalized) linear mixed models (e.g., \texttt{NEBULA}),
which are multilevel models that can account for the hierarchical structure
of cells arising
from the same biological sample~\cite{de-glmm-nebula,de-glmm}.
However, in practice, this approach is impractical for the size of modern datasets
(e.g., >13 hours---versus minutes or less with various flavors of GLMs---to fit a
relatively small dataset of 1000 cells~\cite{de-false-discoveries}).

Alternately, a simple workflow to address the cell correlation problem
is to pool single-cell observations that arise from the same biological sample
(by summing RNA counts and collapsing covariates) to form a meta, \emph{pseudobulk} sample.\footnote{
    An early proposal for \textit{pooling} across cells is given by \cite{pseudobulk-first}---albeit
    in tandem with a more complex normalization scheme than is currently typical
    (e.g. \cite{de-false-discoveries}).
}
Then, a standard GLM can be fit to the reduced set of pseudobulk samples.
In a recent head-to-head, results for the pseudobulk analog of \ref{it:glm} were
equivalent to those for \ref{it:lmm}
(with respect to controlling the empirical false discovery rate)
while requiring a minuscule fraction of the compute power~\cite{de-false-discoveries}.
Confirming these findings, a commentary published in response
to some of the original proponents of the mixed model approach~\cite{de-glmm}
showed that pseudobulk methods were in fact superior
after
fixing limitations in
the original authors' methods
(for simulating data and benchmarking performance)~\cite{de-glmm-followup}.
In practice, ``independent cell'' GLMs (with the caveats noted above) and
pseudobulk GLMs (with the concomitant loss of single-cell resolution)
are both used.

\subsection{robustness}\label{sec:bg-robustness}

Robustness is a general concern in biology, given
\begin{enum-inline}
    \item the desire to use careful experimentation with limited samples to infer fundamental biological principles,
    \item the replication crisis~\cite{replication-crisis-reproducibility-project,replication-crisis-preclinical-cancer,replication-crisis-nature-survey,replication-crisis-cost,replication-crisis-generalizable-prs},
    and
    \item the growing complexity of data measurement, preprocessing, and analysis pipelines.
\end{enum-inline}
At the end of the day, scientists may wonder to what extent their inferences
generalizing from the data in hand
to make statements about the molecular underpinnings of a phenomenon
or the implications for the broader population-of-interest
are justified.

To this end, we propose a dropping-data robustness metric for the key outcomes
of differential expression analysis,
as a complement to the classical checks on robustness that are typically performed.
Namely,
we quantify the
\emph{minimal fraction of observations} (e.g., cells, subjects, or tissue samples)
that---if dropped from the analysis---would
materially change a key outcome of differential expression.
As a counterpart of this metric, we also quantify the \emph{maximal change to a key outcome}
(e.g., the sign, magnitude, or significance of a gene-level effect,
or the composition of the top gene sets enriched among differentially expressed genes)
that can be effected
by dropping no more than a given
fraction of observations.
These metrics, of a particular class of
data robustness, were first proposed
by \cite{ryan-amip} (where they were collectively termed
\ul{a}pproximate \ul{m}aximum \ul{i}nfluence \ul{p}erturbations, or AMIP);
we port, adapt, and extend them for the central conclusions
that can be drawn from inference on models of differential expression.

Our dropping-data metric is fundamentally distinct from,
and complementary to, existing checks on robustness and generalizability
that are commonly performed for differential expression analyses.

For example, many familiar metrics
revolve around the classical frequentist
concern of robustness to \emph{data sampling}---including
\textit{standard error}, \textit{confidence intervals}, \textit{significance levels}, and \textit{power}.
This umbrella (of robustness to data sampling) also encompasses methods to estimate
these quantities and other population-level statistics, including
resampling techniques
like the \textit{jackknife} (e.g., leave-one-out analysis),
the \textit{bootstrap} (resampling from the empirical distribution), and
\textit{random subsampling} (resampling without replacement).
These methods are designed to provide asymptotic coverage guarantees;
that is,
intervals that promise proper coverage
(i.e., have the specified probability of containing the true value,
 over repeated draws of new data from a fixed population)...so long as the
sample size $\ncells$ is infinite (or ``close enough'').
These inferential guarantees are valuable because---assuming we are willing to bet
that our data is ``large enough'' that asymptotics have kicked in---they
provide a calibrated means of reasoning about the underlying
population (the actual unit of interest), despite only observing a single, finite sample.
Essentially all (frequentist) analyses of differential expression involve
significance testing and/or confidence intervals, and some studies
have explicitly considered
the consistency of outcomes across random data resampling
(for differential expression~\cite{de-robustness-sc,de-robustness-bulk}
or downstream gene set enrichment~\cite{gsea-subsample}).
In contrast to these methods---which
assume that the data is sampled
precisely from the intended target population, and check robustness across future
samples from this population---we
focus on the dataset in hand, and examine robustness with respect to
dropping a handful of observations.
If key findings
are fragile to this small (and realistic) data perturbation,
then there may be reason to believe that the corresponding hypothetical population
may differ systematically from the real-world population of interest.
This is plausible for any type of data analysis, but it is particularly salient
for single-cell RNA-seq data, where
many axes of biological variation---sub-cell-type population structure,
spatial variation, cell cycling and other transient cell processes---both
orthogonal and correlated, along with technical effects,
co-exist~\cite{regev-axes-variation}.
Further, because these classical quantities are asymptotic measures of variability,
they necessarily vanish as the number of data points grows,
whereas our dropping-data metric does not.
For example, a dataset with very large $\ncells$ would have near-zero standard error, yet
may still give rise to empirical outcomes that can be reversed by dropping a
small \textit{fraction} of data points.

Other common checks on generalizability consider robustness to analysis decisions, such as
\emph{hyperparameter} setting and choice of \emph{model}, \emph{test}, and \emph{software}
(itself an agglomeration of model, test, and algorithm for inference and/or choosing hyperparameters).
For example, studies of differential expression have considered
consistency of results across software packages~\cite{de-false-discoveries,de-comparison-sc,de-reproducibility-sc,de-robustness-sc,de-comparison-bulk,de-reproducibility-bulk,de-robustness-bulk}
and consistency across model parameterizations or statistical tests within a given analysis~\cite{de-false-discoveries,de-robustness-sc}.
Other studies have examined the robustness of gene set enrichment results to analytical decisions,
like choice of threshold~\cite{gsea-threshold-microarrays}, metric~\cite{gsea-ranking-metrics},
or software~\cite{gsea-software}.
In contrast to these methods, which examine robustness of the analysis
with respect to a fixed dataset, here we
condition on the analysis and examine robustness
to perturbations of the data itself.
Both are useful, and complementary.
For examples, a result may be robust across hyperparameters, models, tests, and
implementations, yet still be brittle to dropping a small handful of data points
(and vice versa).

Other methods consider robustness to \emph{data collection}.
For example, differential expression analyses may attempt to regress out ``batch effects,''
and some studies have explicitly examined consistency across batches
(such as sequencing labs~\cite{de-reproducibility-bulk} or sequencing technologies~\cite{de-reproducibility-SMARTv10x},
for differential expression, or across studies~\cite{gsea-studies}, for gene set enrichment).
Whereas these methods provide a measure of robustness across \textit{known} sources of variation,
our dropping-data robustness metric can identify data points whose inclusion can substantively
change the outcomes of differential expression, even after accounting for known
structure within the data---potentially pointing to unpredicted axes of variation
(which may correspond to differences in biology and/or measurement).

Alternately, another line of research revolves around robustness to \emph{data corruption},
including \textit{gross error} or \textit{adversarial error}, as well as
\textit{outlier detection}.
Whereas these methods collectively consider arbitrarily adversarial
perturbations to the dataset (and are thus inherently model-specific),
our dropping-data approach is both model-agnostic and tailored
toward a more realistic perturbation for gene expression data.
For example, the former would be suited to detect data fabrication or manipulation
(a very useful task, but not relevant to the quotidian workflow for a biologist
analyzing their own data), whereas the latter---excluding a few cells from an
scRNA-seq dataset---is
a scenario that could reasonably arise when
collecting a new sample from, ostensibly, the same population.
So, our metric provides a more relevant check on generalizability
for (many) analyses of gene expression data.
Further, we focus on the implications of dropping data with respect to the key
outcomes of differential expression (and, in the process, identify the
particular data points whose exclusion would effect the worst-case change)
rather than generic outlier detection.

In contrast to these classical approaches to robustness, which are often
employed as checks on differential expression, the original
dropping-data robustness
paper finds
that
this form of data robustness
reflects the \textit{signal-to-noise ratio}~\cite{ryan-amip}---which
neither vanishes as the number of data points grows
nor can be fully accounted for by model misspecification.

Incidentally,
approximating this metric entails computing \emph{influence scores}
(based on an empirical ``influence function''),
which have a long history in the study of robustness and
are described (and related to leverage scores and consistency)
in the canonical textbook for GLMs~\cite{mccullagh-nelder-glm}.
For example, in the context of differential expression, \texttt{DESeq2}~\cite{deseq}
uses influence scores (a.k.a. ``Cook's distance'')\footnote{
    via \texttt{DESeq2::replaceOutliers}, a step in the default pipeline
    (\url{https://bioconductor.org/packages/release/bioc/vignettes/DESeq2/inst/doc/DESeq2.html\#approach-to-count-outliers})
} to identify and replace outlier samples in RNA-seq data.
Here, we leverage influence scores---which quantify the effect on an estimator of excluding a
single data point---as an approximation
toward estimating the worse-case dropping-data sensitivity of the key
statistical outcomes of an analysis.
Through this framework,
we improve on the utility and
interpretability of raw influence scores
by providing a natural and universal sense of scale
(\textit{minimal fraction} of observations to drop in order to
meaningfully change a key outcomes).

In practice, though in this work we compute approximate rather than exact dropping-data robustness
(and empirically provide a bound, by verifying predictions through an
additional model fit),
we find that many key results (meaningfully differentially expressed genes and gene sets)---which
have survived classical robustness procedures, like significance testing with
multiple-testing correction---are nonetheless susceptible to dropping a small
fraction of the data (i.e., a handful of cells).
In \cref{sec:conc}, we explore the implications of this finding and
how to interpret this type of nonrobustness when diagnosed.

%% file: 3-de-objective.tex
Recall that the existing framework~\cite{ryan-amip} provides a
means of computing the dropping-data robustness metric for any Z-estimator~\cite{z-estimator},
such as an estimator that optimizes a smooth objective.
Our first goal was to make differential expression analysis amenable to
this framework---addressing challenges including
 pathological convergence issues for a particular class of sparse genes,
data-dependent hyperparameters,
a test statistic that is not amenable to first-order methods, and
non-differentiable operations on gene-level results
(i.e., joint ranking and/or thresholding
for multiple testing correction and gene set enrichment).
In this section, we'll describe a typical
differential expression analysis
(\cref{sec:model,sec:inference,sec:testing,sec:gsea})
and explain (and justify) our modifications
(\cref{sec:mods,sec:de-analysis-for-sensitivity}).

The general setup is that we observe a cell-by-gene matrix of RNA molecule counts
(mainly messenger RNA transcripts),
$\mat{Y}\ofSize{\ncells,\ngenes} = [\ldots \, \vec{y}^{(\igene)} \, \ldots]$,\footnote{
    More precisely, we will use $\ngenes$ to refer to the number of genes
    included in the analysis, which may be smaller than the total number of
    genes measured.
    For example, genes are often excluded if they have very few nonzero
    observations (a common phenomenon in scRNA-seq data, for both biological
    and technical reasons).

    Beyond avoiding wasted computational effort to analyze these genes
    (which are exceedingly unlikely to contain sufficient signal
    to detect a difference between groups),
    this filtering step also increases sensitivity
    to detect differential expression
    (after correcting for the false discovery rate)
    by cutting down on the number of tests that are performed.~\cite{gene-filtering}
    This filtering step is kosher (statistically) so long as
    it is independent of the group delineation being tested.
    \label{fn:gene-filter}
}
and we additionally observe $\nbetas$ covariates,
$\mat{X}\ofSize{\ncells,\nbetas}$,\footnote{
    Or, more precisely, some number of covariates corresponding to
    $\nbetas$ independent regressors
    (since, e.g., a discrete covariate with $d$ categories would be encoded
     by $d-1$ regressors)
}
across cells.
One of these covariates corresponds to a group delineation---e.g., treated or untreated---and,
by regressing $\mat{Y}$ on $\mat{X}$, we hope to learn something about the effect of this
delineation on gene expression.
Specifically, the goal of DE---for each gene---is:
\begin{itemize}
    \item to determine whether some function of the
        estimated regression coefficients $\vec{\beta}$
        (often the function that picks out a single covariate-of-interest,
         such as the treatment\footnote{
            For simplicity, and because it's a common analysis, we'll frame
            differential expression as an exercise in looking for a difference between
            an unperturbed group of subjects or cells (\emph{control})
            and a perturbed group (\emph{treatment}).
            For single-cell sequencing, this treatment-versus-control comparison
            is generally performed \textit{within} a cell type
            (since it would otherwise be biased by changes in the proportion of different
             cell types---a potentially interesting, but separate, hypothesis to test).

            However, note that the same
            approach to differential expression---and our corresponding approach to dropping-data robustness---is
            equally suited to compare gene expression between
            other group delineations or phenotypes, such as cell types.
            Then, ``$\betaTreated$''
            captures the differences in gene expression \textit{across} cell types
            (rather than a typical ``treatment'' effect).
            \label{fn:treatment-effect}
        } effect $\betaTreated$)
        is significantly different from the null (usually a point mass at zero),
        and
    \item sometimes to also make a point estimate of that function (the \emph{effect size}).
\end{itemize}
\vspace{\parskip}

After regressing each $\vec{y}^{(\igene)}$ on $\mat{X}$ across genes
(which generally number in the thousands or tens of thousands---depending on the organism),
the desired output is a reduced set and/or ranked list of
\emph{``differentially expressed''} genes
prioritized for interpretation.
We formally outline this process in \cref{sec:model,sec:inference,sec:testing}.

Finally, a meta-analysis of individual gene results (\cref{sec:gsea})
is often performed to look for patterns in differential expression of
biologically meaningful gene sets or pathways.

\subsection{standard model} \label{sec:model}

We focus on the common DE modeling approach of overdispersed count GLMs,
as exemplified by \texttt{DESeq2}
and its recommended subroutine for single-cell data via \texttt{glmGamPoi}.
Namely, \texttt{DESeq2} posits a negative binomial\footnote{
    One way to understand the negative binomial as an \textit{overdispersed} count model
    (and therefore an attractive model for RNA-seq data)
    is to think of extending the Poisson with fixed rate---which has variance equal to the mean---to
    a Poisson with variable (gamma-distributed) rate---which has variance greater than the mean,
    according to a dispersion parameter that is determined
    by the gamma parameters. See \cref{app:gamma-poisson} for details.
}
GLM to model RNA transcript counts,
and \texttt{glmGamPoi} posits a ``quasi-likelihood''
variant of a negative binomial GLM.

Specifically, to regress RNA counts $\vec{y} \defeq \vec{y}^{(\igene)}$
(the $\igene\th$ column of $\mat{Y}$)
on the design $\mat{X}$ for a given gene $\igene$,
\begin{align}
    \vec{y} &\sim \distNamed{NB} \left( \vec{\mu}, \, \disp \right)
            && \text{for \texttt{DESeq2} } \nonumber
\shortintertext{or}
    \vec{y} &\sim \distNamed{NB}_{\QLdisp} \left( \vec{\mu}, \, \NBdisp \right)
            && \parbox[t]{6cm}{%
                \setstretch{1.2}
                for \texttt{glmGamPoi}, where ``$\distNamed{NB}_{\QLdisp}$''
                is not quite the negative binomial distribution, \\
                as elaborated upon in \cref{app:l-v-ql}
            } \nonumber
\shortintertext{with}
    \vec{\mu} &= \vec{\size} \, \odot \, \exp\left\{ \mat{X} \vec{\beta} \right\},
    \label{eq:mu-de}
\end{align}
where the coefficient vector $\vec{\beta}\ofSize{\nbetas,1}$ is the latent parameter of interest.
In particular, differential expression analysis seeks to estimate the coefficient
known as the \emph{treatment effect} (cf. \cref{fn:treatment-effect}),
\[
    \betaTreated \defeq \beta_{\ibeta} \;\;\;\mathrm{where}\;\;\;
    x_{\icell}^{(\ibeta)} = \1\left\{
        \mathrm{cell}\;\,\icell\;\,\mathrm{is}\;\,\mathrm{treated}
    \right\}.
\]

Both \texttt{DESeq2} and \texttt{glmGamPoi} first fit the gene dispersion
$\disp$ (or $\NBdisp$), and condition on it
when estimating
$\vec{\beta}$
(\cref{sec:inference}).

Cell size factors $\vec{\size}\ofSize{\ncells,1}$, which enter the model through
the negative binomial mean (\cref{eq:mu-de}),
are constants (estimated empirically by \texttt{DESeq2} or \texttt{glmGamPoi} up front)
to account for some notion of variation in exposure (e.g., library size or sequencing depth) across cells.
The default method for \texttt{glmGamPoi} is ``\texttt{normed\_sum},'' where
\begin{align}
    \total{\vec{y}} &\defeq \sum\overGenes \vec{Y}^{(\igene)} \nonumber
    && \mathcomment{total RNA count per cell}
\\
    \vec{\size} &\defeq \total{\vec{y}} \bigg/
                        \left( \prod\overCells \total{y}_{\icell} \right)^{1/\ncells} \nonumber
    && \mathcomment{size factor per cell}
\\
                &\hphantom{:}= \total{\vec{y}} \bigg/
                               \exp\left\{ \frac{1}{\ncells} \sum\overCells \log \total{y}_{\icell}
                                   \right\}
    && \mathcomment{\parbox{4.8cm}{ \singlespacing
        (as computed by \texttt{glmGamPoi}, \hphantom{(}for numerical stability)
        }} \label{eq:normed-sum}
\end{align}
---i.e., the total count per cell standardized by its geometric mean across cells.

Finally, both \texttt{DESeq2} and \texttt{glmGamPoi} incorporate light
regularization over the magnitude of the coefficients.
Specifically, both use L2 regularization, akin to placing
a normal prior over each coefficient,\footnote{
    Where the equivalence specifically is valid when MAP estimation is used to
    maximize the posterior
}
\[
    \beta_{\ibeta} \sim \normal(0, \; \sigma_{\ibeta}^2).~\footnotemark
\]\footnotetext{
    Note that, by default, this is a very wide prior
    that has little effect on the estimated coefficients
    (by design; in \texttt{DESeq2}, this default and recommended setting
     is denoted as \texttt{betaPrior=FALSE}).
    Specifically, $\sigma_{\ibeta}^2 = 10^6$ for \texttt{DESeq2},
    and $\sigma_{\ibeta}^2 = \ncells \times 10^{20}$ for \texttt{glmGamPoi}
    (where this form is explained in \cref{app:ggp-newton-prior}).
    \label{fn:beta-prior}
}

For brevity, and because each gene is fit by an independent GLM,
note that we omit gene index $\igene$ from gene-specific terms when
describing the model for a single gene.
Specifically,
\begin{itemize}
    \item counts $\vec{y}$, dispersion $\disp$, and coefficients $\vec{\beta}$
        are \textit{gene-specific}, whereas
    \item sizes $\vec{\size}$, covariates $\mat{X}$, and prior width $\vec{\sigma}^2$ are \textit{global}.
\end{itemize}

\subsubsection{pseudobulk}

The downside of the
model described above
is that cells are treated as independent samples,
whereas in reality the data is generally composed of many cells ($\ncells$) from a handful
of subjects ($\nsamples \Lt \ncells$).
Ideally we'd include subject as a covariate, but
inference on that GLM would be impossible---since
subject is totally crossed with the treatment effect,\footnote{
    i.e., each subject is either treated or not
}
so the linear model would not be full rank.
A common alternative
to the ``\emph{independent cell}'' model
(above)
is to form \emph{pseudobulk} samples from single-cell
measurements---by summing
the counts per gene across cells from each sample or subject---and
to fit these newly formed
data points as input to a GLM.\footnote{
    This can be done manually or, in \texttt{glmGamPoi}, with the routine
    \texttt{glmGamPoi::test\_de($\cdots$, pseudobulk\_by=$\;\cdot\,$)}.
}

Consider $\mat{\pseudobulkSelector}\ofSize{\nsamples,\ncells}$, an indicator matrix where
\[
    \pseudobulkSelector_{\isample,\icell} =
    \begin{cases}
        1 & \text{if the }\icell\th\text{ cell belongs to the }\isample\th\text{ sample}
\\
        0 & \text{otherwise.}
    \end{cases}
\]
Then, $\pseudobulk{\mat{Y}}\ofSize{\nsamples,\ngenes} \defeq \mat{\pseudobulkSelector}\mat{Y}$
is the pseudobulk analog of cell count observations $\mat{Y}\ofSize{\ncells,\ngenes}$.

The pseudobulk analog of the design matrix $\pseudobulk{\mat{X}}\ofSize{\nsamples,\nbetas}$
is formed by stacking one covariate vector per sample.
If all covariates for a sample's constituent cells are identical,
then that covariate vector (row of $\mat{X}$) is used.
Alternately---for covariates where this assumption does not hold---individual
cell covariates can be aggregated by other
$\R^{\ncells} \rightarrow \R$ operations,
such as averaging or summing~\cite{de-glmm,de-false-discoveries}.

The pseudobulk analog of size factors $\vec{\size}$ is
\begin{equation}
\begin{aligned} \label{eq:size-pseudobulk}
    \total{\pseudobulk{\vec{y}}} &\defeq \mat{\pseudobulkSelector} \total{\vec{y}}
\\
    \pseudobulk{\vec{\size}} &\defeq \total{\pseudobulk{\vec{y}}} \big/
                   \exp\left\{ \frac{1}{\nsamples} \sum\overSamples
                        \log \total{\pseudobulk{y}}_{\isample} \right\}.
\end{aligned}
\end{equation}

Substituting these parameters into \cref{eq:mu-de}
($\vec{y} \rightarrow \pseudobulk{\vec{y}}, \,
  \mat{X} \rightarrow \pseudobulk{\mat{X}}, \,
  \vec{\size} \rightarrow \pseudobulk{\vec{\size}}$)
yields the pseudobulk model.

\subsection{standard inference} \label{sec:inference}

\texttt{DESeq2} and \texttt{glmGamPoi} each implement a custom inference algorithm
(to estimate $\hat{\vec{\beta}}$)
that is motivated by minimizing deviance based on iteratively reweighted least squares.
Focusing on \texttt{glmGamPoi}, we verify---theoretically, through code inspection,
and empirically, through examination of intermediates during code execution---that
their implementation is functionally equivalent to performing Newton-Raphson
on the log-likelihood objective,\footnote{
    i.e., for
    a negative binomial with dispersion $\NBdisp$.
    This is not the model posited by \texttt{glmGamPoi}'s quasi-likelihood
    framework---which has no proper generative model---but nonetheless
    shares an equivalent objective for estimating $\vec{\beta}$
     (\cref{app:l-v-ql,app:ggp-newton}).
}
as expected~(\cref{app:ggp-newton}).

Having confirmed that their custom implementation optimizes a known objective---and so,
hypothetically, forms a valid Z-estimator---we are free to examine sensitivity
of the analysis as a whole
by focusing on the objective itself, rather than
their particular inference algorithm.\footnote{
    With the caveat that the optimization must not terminate before
    it has fully converged,
    as we will explore in \cref{sec:pseudocells}
}

\subsection{standard testing} \label{sec:testing}

Differentially expressed genes are specified as those that, at minimum,
satisfy a test of statistical significance for differential expression.
There are three classical
approaches to construct a (parametric) statistical
test of the null hypothesis
that there is no difference in expression between the treatment and control groups.
Namely,

\begin{enumerate}
    \item The \emph{likelihood ratio test} tests whether the log-likelihood of
          the fitted ``full'' model $\M$ (with all covariates)
          significantly\footnote{
              \label{fn:sig}
              i.e., \textit{statistically} significant(ly)
          } improves upon
          the fitted ``reduced'' model $\M\reduced$ (excluding $\betaTreated$ or, equivalently, fixing it to 0).
          Let $\LL(\vec{\beta}) \defeq \log \p(\vec{y}, \mat{X}; \vec{\beta}, \cdots)$
          be the log-likelihood function.
          The statistic is
          \[
              LR \defeq -2 \left[ \LL(\hat{\vec{\beta}}\reduced) - \LL(\hat{\vec{\beta}}) \right]
            \]
          where $\hat{\vec{\beta}}\reduced$ are the optimal coefficients
          within the restricted parameter space of model
          $\M\reduced$.
          The closer this statistic is to zero,
          the less evidence that it is advantageous
          (from a likelihood perspective)
          to fit a more complex model that incorporates treatment labels.
          \label{it:lr}
    \item The \emph{score test} tests the curvature of the log-probability density
          at the optimal restricted parameters
          to determine whether the incremental value of additional information
          would offer a significant\footref{fn:sig} improvement.
          The statistic is
          \[
            S \defeq
            \dd*{\LL(\vec{\beta})}{\vec{\beta}\T}
            \; \widehat{\Sigma}(\vec{\beta}) \;
            \dd*{\LL(\vec{\beta})}{\vec{\beta}}
            \; \Bigg|_{\vec{\beta} = \hat{\vec{\beta}}\reduced}
          \]
          ---i.e., the square score standardized by the estimated parameter covariance,
          evaluated at the fitted coefficients for the reduced model.
          Informally, when this statistic is small, the log-likelihood is near its
          optimum (has little curvature) even when optimized within the restricted parameter space,
          so adding treatment labels
          would be negligibly informative.
          \label{it:score}
    \item The \emph{Wald test} tests whether $\hatBetaTreated$
          differs significantly\footref{fn:sig} from the null hypothesized value (typically 0).
          The statistic is
          \[
              W \defeq \frac{ \hatBetaTreated - 0 }
                            { \se\left( \hatBetaTreated \right) }.
          \]
          When this statistic is small in magnitude, the treatment coefficient
          is close to the null value.
          \label{it:wald}
\end{enumerate}
\vspace{\parskip}

The hat over the central term in \ref{it:score} or the denominator in \ref{it:wald}
reflects the fact that the covariance or standard error, respectively, is
itself an estimate, with different approaches for estimation under different assumptions.
Namely, the Fisher estimator derives from the assumption that the model is
well-specified---and its behavior is not guaranteed outside of this regime---whereas
the ``robust'' sandwich estimator holds regardless of model misspecification
(\cref{app:cov-estimators}).

Each statistic has an associated sampling distribution (under the null hypothesis)
that can be used to construct a confidence interval and to test significance.
Namely, $\sign\left[\hatBetaTreated\right] \, \sqrt{LR}$,
        $\,\sign\left[\hatBetaTreated\right] \, \sqrt{S}$, and $W$
are all asymptotically $z$-distributed\footnote{
    a.k.a. standard-normal; $z \sim \normal(0, 1)$
}
(equivalently, $LR$, $S$, and $W^2$ all follow a $\chi^2_1$ distribution)
for a single coefficient-of-interest.\footnote{
    i.e., when the difference in degrees of freedom between $\M$ and
    $\M\reduced$, $\df$, is 1
}
Asymptotically in $\ncells$, all three tests are equivalent.

By default, \texttt{DESeq2} performs a Wald test (but recommends the likelihood ratio
approach for single-cell data)\footnote{
    Based on the approach of \texttt{glmGamPoi}, which (under the hood) is
        their recommended engine for single-cell data.
    They also use a normally-distributed Wald null statistic by default,
    but suggest
    an alternative $\distNamed{t}$-distributed null
    with heavier tails to cut down on the number of significant genes~\cite{deseq-manual}.
}~\cite{deseq-manual} whereas
\texttt{glmGamPoi} performs a quasi-likelihood analog of the likelihood ratio test.
Specifically, they compute
\[
    LR' \defeq \frac{ LR / \df }{\hat{\QLdisp}}
\]
where $\hat{\vec{\beta}}, \; \hat{\vec{\beta}}\reduced$ (to compute $LR$)
were fit given $\NBdisp$ rather than $\disp$ (\cref{eq:disp}),
and $\df$ (the difference in degrees of freedom between $\M$ and $\M\reduced$)
is generally 1.\footnote{
    i.e., for a two-group comparison with a single treatment coefficient
}
This modified statistic is then assumed to follow an
F-distribution\footnote{
    $\distNamed{F}_{a,b} \defeq \dfrac{\chi^2_a/a}{\chi^2_b/b}$
}
with
$(\df, \df_{\QLdisp})$ degrees of freedom, where the latter is
based on an empirical Bayes prior for $\QLdisp$.\footnote{
    This line of reasoning traces back to the differential expression library
    \texttt{edgeR}~\cite{ql-edgeR}, which in turn cites Tjur (1998)~\cite{ql-tjur}.
    Formally, $LR'$ is F-distributed for Gaussian observations---which
    sparse transcript counts certainly are not (and are not modeled as).
    Tjur's line of informal reasoning:
    ``common sense suggests that it is better to perform this correction
    for randomness [of the dispersion estimate]...than not to perform
    any correction at all...''~\cite{ql-tjur}

    For RNA-seq data, \texttt{edgeR} presents simulation results
    (p-value coverage, etc) to justify the quasi-likelihood F-test analog.
    However, these simulations are based on parameters corresponding to bulk data,
    and additionally use rejection sampling to ensure simulated genes have means > 1~\cite{ql-edgeR}.
    Neither \texttt{glmGamPoi} nor \texttt{edgeR} explore the extent to which
    the F-distribution approximation is justifiable for highly sparse single-cell data.

    Further, \texttt{edgeR} estimates a single parameter $\NBdisp$ for the entire
    dataset, whereas \texttt{glmGamPoi} estimates $\ngenes$ times as many parameters,
    i.e., $\NBdisp$ per gene. (Both also estimate $\QLdisp$ per gene.)
}
(Under the same logic,
$S/\hat{\QLdisp}$ and $W^2/\hat{\QLdisp}$
have equal claim to being F-distributed and could serve as quasi-likelihood analogs
of their respective tests.)

Finally, since differential expression analyses can involve tens of thousands of
tests (across genes), a multiple-testing correction is generally applied to p-values
in order to control the false discovery rate. \texttt{DESeq2} and \texttt{glmGamPoi}
use a
Benjamini-Hochberg (BH) procedure~\cite{bh} where p-values are
ascendingly ranked (rank=$\irank$) and inflated by $\ngenes/\irank$
before being subjected to a chosen level.

\subsection{standard downstream analysis of gene set enrichment} \label{sec:gsea}

The ultimate outcome for DE
analysis is
often not a table of significance testing for tens of thousands of genes,
but rather a functional meta-analysis of gene-level results
to identify biologically meaningful patterns in differential expression.
Specifically,
\emph{gene set enrichment analysis} (GSEA)
seeks to determine which biologically meaningful gene
sets---predetermined groupings based on prior knowledge that correspond to,
say, known biological pathways or shared biological functions---are
overrepresented among genes
whose expression differs between groups.
Often, the top
ten gene sets (based on p-values of a downstream test for enrichment)
are then reported and prioritized for interpretation.

Approaches to GSEA generally fall into two buckets:
\textit{threshold}-based
(where genes are thresholded by some criterion,
such as a maximal p-value and/or minimal effect size)
or
\textit{rank}-based
(where genes are ranked by some criterion, such as
p-value multiplied by the sign of the effect).
Both construct an enrichment test around subsetting or ranking genes (with
respect to a reference dictionary of gene sets),
then use the p-value of that test to rank differentially expressed gene sets
(where the number of differentially expressed gene sets is hypothetically much
smaller than the number of differentially expressed genes).

Neither approach is amenable to our dropping-data sensitivity approximation---as
neither operation (thresholding or ranking) is differentiable.
Further, unlike gene-level outcomes, both approaches require
consolidating the impact of dropping data \textit{across} genes.
Nonetheless,
we develop a heuristic procedure
using gene-level influence scores
to approximate the sensitivity of the top-ranked gene sets
(to dropping a small handful of cells)---and
show that, in practice, this procedure yields meaningful bounds on the
robustness of this high-level outcome of differential expression.

We focus on the simplest, and an extremely common, method for functional
enrichment analysis: the hypergeometric test
(described in brief here, or see \cref{app:gsea} for more details).
First, gene p-values from differential expression testing (\cref{sec:testing})
are ranked and BH-corrected, and a subset of significant genes
(the ``targets'') are selected based on a cutoff
at the desired significance level.\footnote{
    Potentially segmented into two target sets,
    upregulated or downregulated, based on the sign of the treatment effect
    ~\cite{gsea-up-down}
}
These targets comprise a subset of the greater ``gene universe'': the set of
all genes that were tested for differential expression.
A predetermined collection of gene sets, grouped by common biological function
based on prior knowledge, is chosen.
For each gene set, a hypergeometric test is run to determine whether
differentially expressed targets are overrepresented, versus
what would be expected from the gene universe.
Finally, across all gene sets, hypergeometric p-values are corrected
for multiple testing and ranked,
and the biological descriptions of the top-ranked gene sets are reported.

\subsection{modifications} \label{sec:mods}

The above model, inference, and testing framework presents multiple
roadblocks
to
automatic robustness analysis.
In this work, we make several modifications to surmount these obstacles
while balancing fidelity to a standard differential expression procedure.

\subsubsection{pseudocell prior}\label{sec:pseudocells}

We observed that optimization by \texttt{glmGamPoi} (\texttt{DESeq2}'s recommended
engine for single-cell data) fails to converge---based
on the magnitude of the gradient and the condition number of the Hessian
at
the coefficients estimated by \texttt{glmGamPoi},
$\hat{\vec{\beta}}^{\texttt{ggp}}$---for
a subset of sparse genes.
Specifically, we observed pathological failure to converge
for genes where one group (treatment or control) had all zero counts---a
scenario that is fairly frequent for naturally sparse single-cell measurements, for
both biological and technical reasons.
We'll refer to these as \emph{zero-group genes}.

Local sensitivity analysis is contingent on the objective-of-interest being fully optimized.
Influences are computed by approximating small perturbations around the optimum,
so if the starting point does not in fact optimize the objective,
the effect of these perturbations is effectively swamped by noise.

We surmise that this pathology occurs because the
\texttt{DESeq2} and \texttt{glmGamPoi} objectives are
not well-defined for genes where no nonzero counts are observed in one group.

One way to think about this phenomenon is that the treatment effect $\betaTreated$
is effectively a log-ratio
between the treatment and control groups.
This quantity is ill-posed if the numerator or denominator is zero;
that is, it tends toward positive or negative infinity
(though forced to take on an arbitrary finite value by the optimization procedure
 and its termination rules),
and it does not vary smoothly with the level of expression in the other group
(in contrast with the behavior of the log-ratio when counts in one group
are small but not entirely zero).

Another way to think about this pathology is that the Hessian of the log-likelihood
is proportional to the mean
(\cref{eq:hessian}),
so when
mean estimates $\mu_{\icell}$ are vanishingly small for all cells $\icell$
that span a particular direction in the regressor space---e.g.,
all treatment observations or all control observations---the
inverse of the Hessian is ill-defined.
This is particularly problematic because the inverse Hessian (or inverse Fisher information)
is essential to rescale the gradient during Newton-Raphson optimization (\cref{app:ggp-newton}),
as well as to estimate the standard error (\cref{app:cov-estimators})---and,
if the Hessian is singular, the objective does not have a unique solution
and we cannot apply the implicit function theorem to form a sensitivity
approximation (\cref{sec:amip,app:dbeta-dw}).

The \texttt{DESeq2} library previously sought to address this problem by
\begin{enumerate}
    \item posing a zero-centered prior over $\vec{\beta}$,\footnote{
        Specifically, when \texttt{betaPrior=TRUE}, enforcing
        a stronger-than-default\cref{fn:beta-prior} prior
        over $\{ \beta_{\ibeta} \,:\, 0 < \ibeta \leq \nbetas \}$;
        i.e., all coefficients except the intercept
        }
        to regularize estimates that would otherwise trend toward
        infinity,\footnote{
            Possibly a heavy-tailed prior, to prevent over-regularizing
            large effects~\cite{deseq-heavy-tail}
} and \label{it:reg-ridge}
    \item imposing a minimum on each $\mu_{\icell}$
        ($10^{\minus6}$ by default,
        although it is recommended to eliminate this limit
        ``\texttt{minmu}'' for single-cell data)~\cite{deseq-manual}.\footnote{
            Presumably this cutoff stabilizes inference without much consequence
            for what is essentially an edge case in bulk RNA-seq
            (the measurement for which \texttt{DESeq2} was originally developed).
            On the other hand, very small means are \textit{de rigueur}
            for sparse scRNA-seq data, and the authors presumably recognized
            that enforcing this threshold in this context
            would meaningfully warp results.
        }\label{it:reg-minmu}
\end{enumerate}
The \texttt{glmGamPoi} library retains the prior over coefficients
(albeit so wide as to be meaningless) and eliminates the \texttt{minmu} threshold.
However, we observe that neither library's strategy is
sufficient (to ensure convergence and correct the flaws noted above).
\ref{it:reg-ridge} places a prior over $\vec{\beta}$, when theoretically
we would in fact like to place a prior over $\mu$
(to prevent it from going to zero---which we are not able to control
by regularizing the magnitude of $\vec{\beta}$, since a group's $\mu$
can still approach zero even when the coefficients are far from zero\footnote{
    e.g., trivially, for the zero-level group in the model
    $\beta_0 + \beta_1 \texttt{is\_treated}$ when observed counts in that group
    are sparse---as \texttt{betaPrior=TRUE} does not affect regularization of
    the intercept $\beta_0$
}).
\ref{it:reg-minmu}
nonsensically distorts the results when enforced
(by hard-thresholding small values of $\mu_{\icell}$, which are much more common in single-cell
 than in bulk data), and is rightly not recommended for scRNA-seq~\cite{deseq-manual}.

To address the pathological lack of convergence for zero-group genes,
we propose placing an intuitive pseudocount prior over $\mu$---analogous
to the
prior $\distNamed{Beta}\left(H, T\right)$
for the Bernoulli, which effectively ``seeds'' coin flip data with
$H$ heads and $T$ tails.
By choosing these pseudocounts, the modeler can intuitively express their belief
about the bias of the coin ($H/T$) and the strength of the prior ($H+T$).
However, whereas the beta is conjugate to the Bernoulli, there is no equivalent
conjugate prior for the mean of a negative binomial.

Instead, we effectively impose a pseudocount prior over $\mu_{\text{treatment}}$
and $\mu_{\text{control}}$---$\mu_{\icell}$ marginalized over each cell $\icell$
that belongs to the treatment or control group, respectively---by
incorporating two \emph{pseudocells} as additional data points, one per group.
Each pseudocell has the same observation $y_{\text{pseudo}}$ per gene,
and size factor $\size_{\text{pseudo}}=1$.\footnote{
    The geometric mean of all cell size factors $\vec{\size}$, by definition, when
    calculated via \texttt{normed\_sum} (\cref{eq:normed-sum})
}
If additional covariates are included in the regression, we take their median
values, $\median\limits_{\icell} x^{(\ibeta)}_{\icell}$, in order to construct
$\vec{x}_{\text{pseudo:treatment}}$ and $\vec{x}_{\text{pseudo:control}}$
(the covariate vectors
for, respectively,
the pseudocell assigned to the treatment group and
the pseudocell assigned to the control).

\begin{figure}[htbp!]
    \centering
    \includegraphics[width=\linewidth]{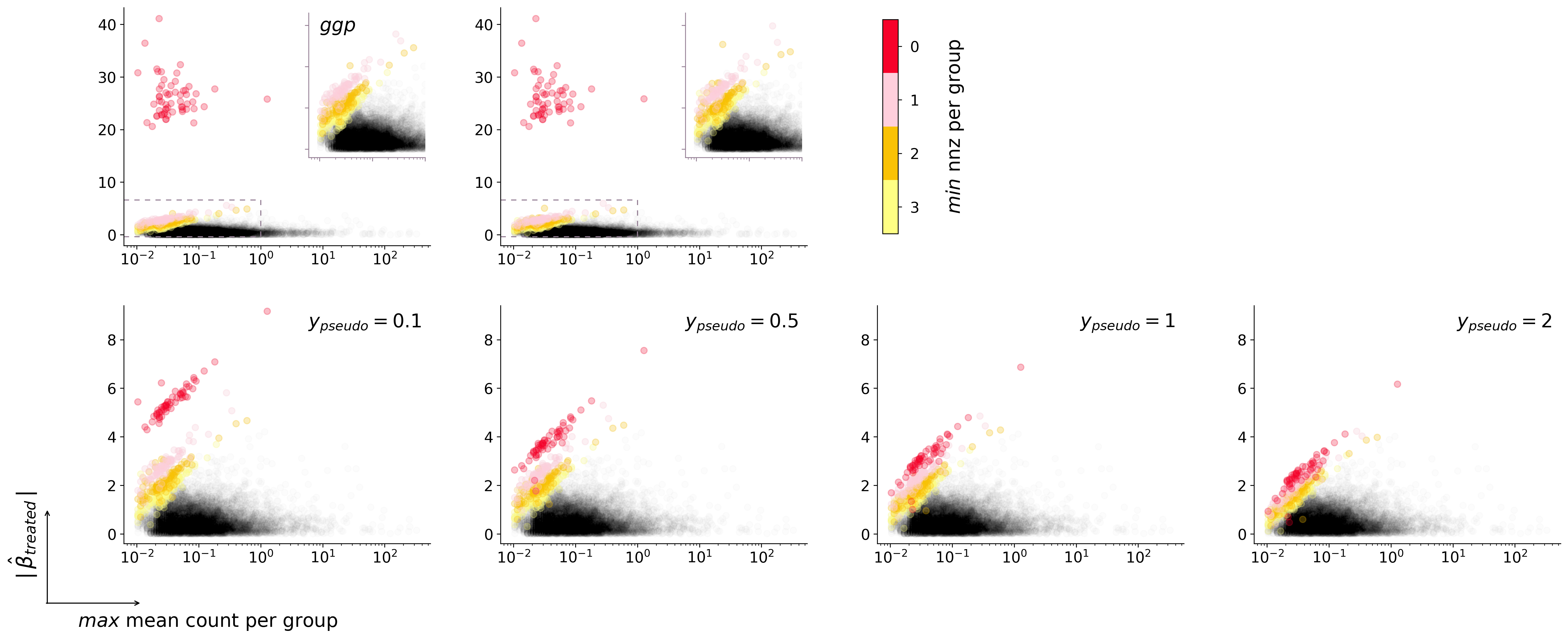}%
    \caption[foo bar]{
        \textbf{Pseudocell prior makes estimated treatment effects sensical for zero-group genes.}
        Each plot shows the relationship between the number of observed transcript counts
        per group
        and the magnitude of the estimated treatment effect (\textit{y-axis})
        across genes (\textit{points})
        for a representative scRNA-seq dataset (\cref{app:data}).
        To focus on genes that could realistically show up as meaningful within a
        differential expression analysis, we plot the $\ngenes=9485$ genes where
        at least one group contains 10 or more nonzero counts.
        \captionbr
        Specifically, on the x-axis genes are plotted according to the mean
        observed count among cells in whichever group (treatment or control)
        has the largest empirical mean.
        Genes are colored if they have very few nonzero counts ($\leq 3$)
        in the group with the smallest number of nonzero counts (\textit{nnz}).
        In other words, zero-group genes are red, and other highly sparse but
        non-zero-group genes are pink, orange, or yellow.
        Hypothetically, we expect that genes with many large counts in one
        group (\textit{farther right along x-axis}) and
        very few nonzero counts in the other group (\textit{colored})
        will have larger inferred treatment effects
        (i.e., \textit{fall higher on the y-axis}).
        \captionbr
        This relationship is plotted for various estimates $\hatBetaTreated$
        under different modeling assumptions:
        \captionbr
            Top row, treatment effects estimated with no pseudocell prior---either
            directly from \texttt{glmGamPoi} (\textit{ggp; left}) or after refitting
            with our modified model (\cref{sec:we-dont-do-ql,sec:our-inference}; \textit{right}).
            Note the differing y-axis scale between the two rows; estimates for
            zero-group genes (\textit{red}) without a pseudocell prior are an order
            of magnitude greater than estimates for any other gene.
            Inset (\textit{upper right} of each plot) zooms into the region
            outlined by a dotted line, where axis ticks are on the same scale
            as the bottom row.
            Treatment effect estimates for zero-group genes do \textit{not}
            scale with the average number of counts in the other group
            (i.e., red points have no strong x-y correlation, unlike other colored points).
        \captionbr
            Bottom row, treatment effect estimates when pseudocells are incorporated.
            The size of the observation $\ypseudo$ assigned to each pseudocell increases from left to right.
            When a pseudocell prior is enforced, treatment effect estimates for zero-group genes
            are
            \begin{enum-inline}
                \item on the same scale as other sparse genes, and
                \item strongly positively correlated with the
                mean observed count
                in the other group, as other highly sparse genes are.
            \end{enum-inline}
    } \label{fig:sparsity-v-lfc-pseudo}
\end{figure}

\begin{figure}[htbp!]
    \centering
    \includegraphics[width=\linewidth]{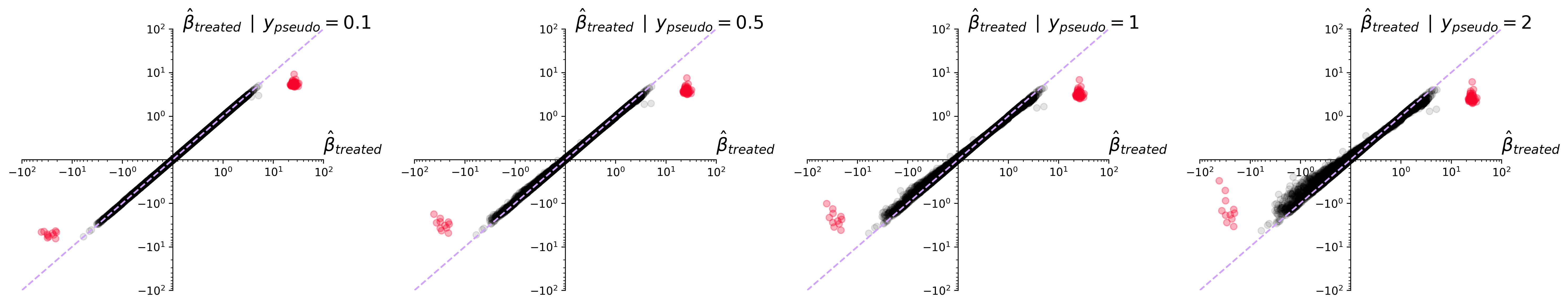}%
    \caption{
        \textbf{Changes to estimated effect size under pseudocell prior of varying strength.}
        The estimated treatment effect $\hatBetaTreated$ across genes (\textit{points})
        under a model with no pseudocell prior (\textit{x-axis}) or when a
        pseudocell prior is enforced (\textit{y-axis}).
        Zero-group genes are highlighted in red.
        The size of the observation per pseudocell, $\ypseudo$, increases across plots from left to right.
        The data used to fit $\hat{\vec{\beta}}$ is the same as in \cref{fig:sparsity-v-lfc-pseudo}.
        See \cref{supp-fig:betas-pseudo} to compare all coefficients.
    } \label{fig:betas-pseudo}
\end{figure}

\begin{figure}[t!]
    \centering
    \includegraphics[width=\linewidth]{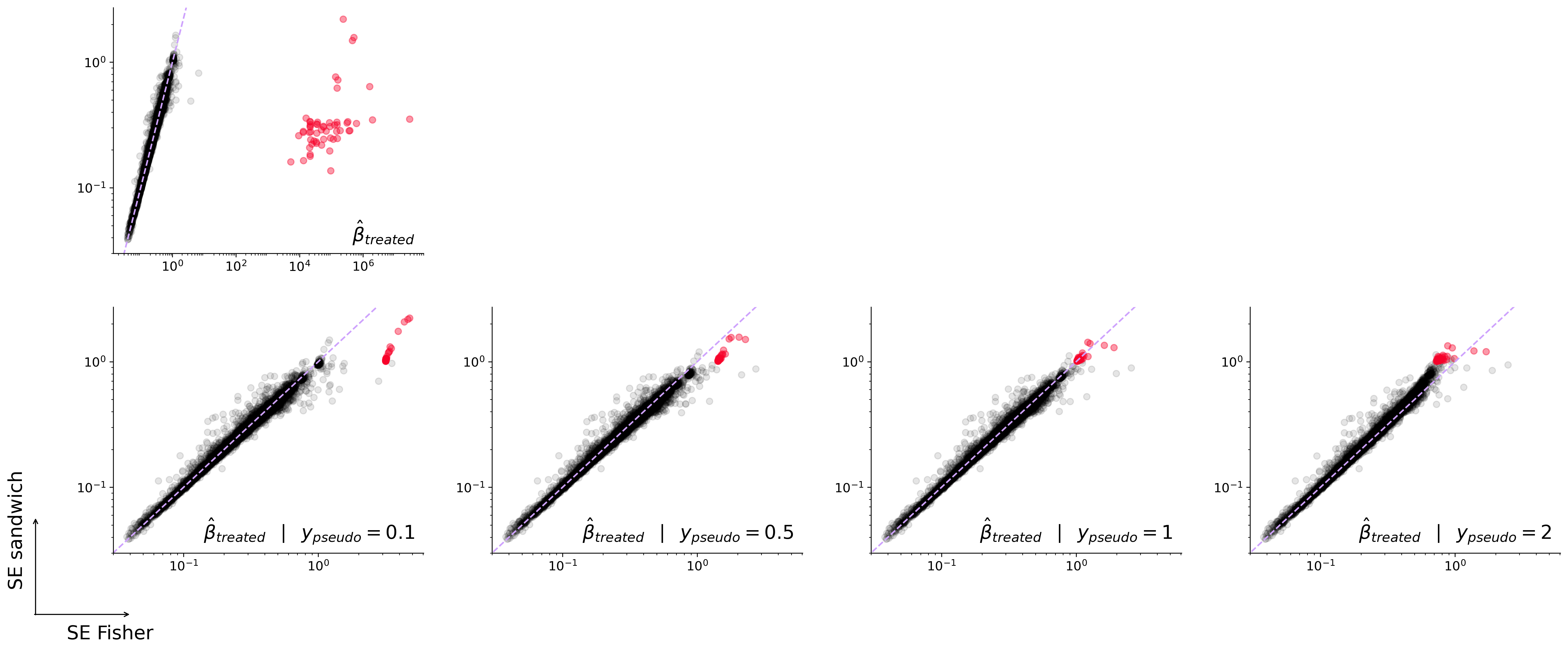}%
    \caption{
        \textbf{Relationship between standard error estimators under pseudocell prior of varying strength.}
        The relationship between the Fisher standard error (\textit{x-axis}) and
        sandwich standard error (\textit{y-axis}) across genes (\textit{points}) when
        coefficients are estimated under a variety of model likelihoods.
        The data used to fit $\hat{\vec{\beta}}$ is the same as in \cref{fig:sparsity-v-lfc-pseudo}.
        Zero-group genes are highlighted in red.
        \captionbr
        Top row, no pseudocell prior.
        The correlation between standard error estimators is 0.54.
        \captionbr
        Bottom row, pseudocell prior where strength (size of the pseudocell observation $\ypseudo$)
        increases from left to right.
        With a pseudocell prior of at least 0.5,
        the correlation between standard errors rises to $\apprx$0.97--0.98
        (or 0.84 at $\ypseudo=0.1$).
    } \label{fig:fisher-sandwich-pseudo}
\end{figure}

\begin{figure}[htb!]
    \centering
    \subcaptionbox{\label{fig:convergence-alpha}}{\includegraphics[width=\linewidth]{%
        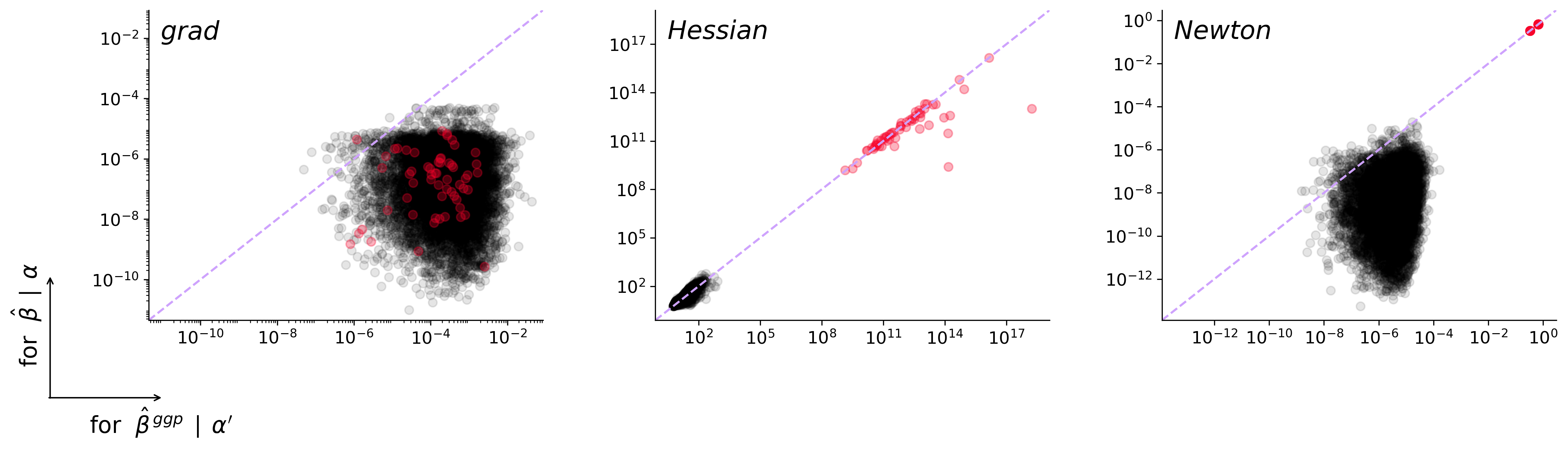}}%
    \hfill%
    \subcaptionbox{\label{fig:convergence-pseudo}}{\includegraphics[width=\linewidth]{%
        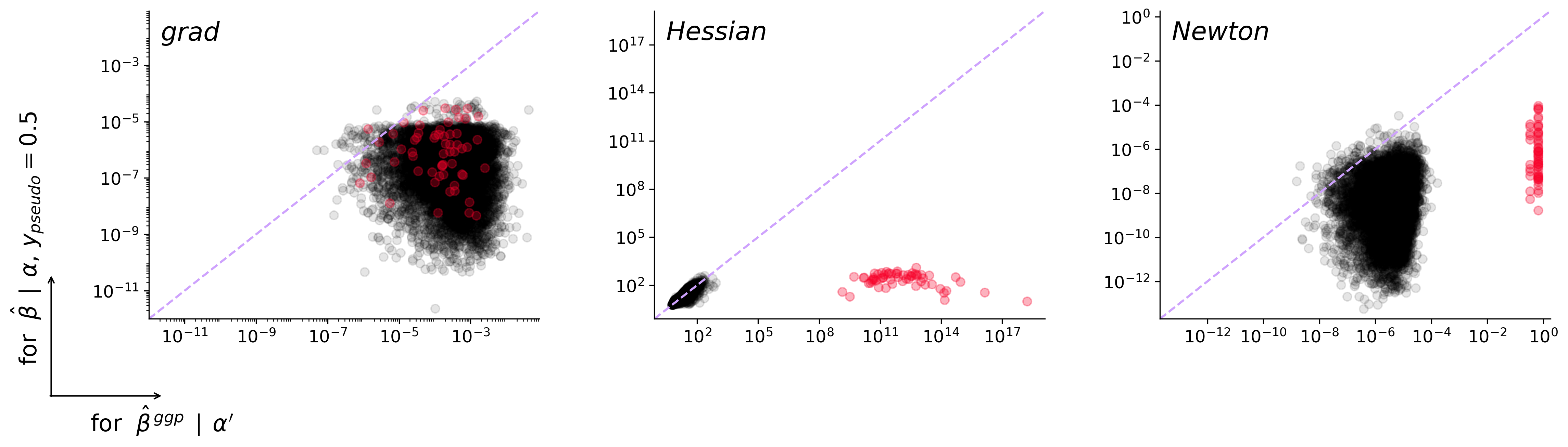}}%
    \caption{
        \textbf{Metrics of convergence for fitted coefficients under modified model.}
        Plots show summary statistics of convergence
        (1-d functions of the gradient, Hessian, and Newton step) per gene (\textit{point})
        for a representative scRNA-seq dataset (\cref{app:data},
        where genes are filtered as described in \cref{fig:sparsity-v-lfc-pseudo}).
        For each metric, smaller values indicate better convergence.
        Zero-group genes are highlighted in red.
        \captionbr
        The resulting statistics after refitting (\textit{y-axis}) are plotted against
        the original statistics for the coefficients output by \texttt{glmGamPoi}
        (\textit{ggp; x-axis}),
        which serve as the initialization for our own inference.
        The gradient and Newton are summarized by
        the mean absolute value of each respective $\nbetas$-vector.
        The Hessian is summarized by its condition number, where a large value
        is indicative of a problem that is ill-conditioned
        (i.e., a Hessian matrix that is close to singular).
    }
    \label{fig:convergence}
\end{figure}

We empirically experiment with the size of the pseudocount and find that
$y_{\text{pseudo}}=0.5$ has the salutary properties that we seek; i.e.,
restores expected behavior, and fixes the convergence problem, for zero-group genes
while leaving other genes' results intact
(\cref{fig:sparsity-v-lfc-pseudo,fig:betas-pseudo,fig:fisher-sandwich-pseudo,fig:convergence}).

First, this prior brings effect size estimates
for zero-group
genes in line with estimates for genes that have highly sparse---but not
entirely zero---counts per group
(observe that zero-group genes, highlighted in red, are initially an
order-of-magnitude greater than those for any other gene, but are restored to
similar magnitude
when a pseudocell prior is enforced; \cref{fig:sparsity-v-lfc-pseudo,fig:betas-pseudo}).
Further, $\hatBetaTreated$ for zero-group genes then scales as expected with
the size of the counts in the other (more plentifully observed)
group, whereas it previously bore no evident relationship
(\cref{fig:sparsity-v-lfc-pseudo}).
On the other hand,
for all other genes---including those that are highly sparse in one group
(few and small counts, but not entirely zero)---we observe that the
prior at $y_{\text{pseudo}}=0.5$ is effectively diluted by the observations,
and treatment effect estimates remain unaffected
(black points in \cref{fig:betas-pseudo} remain on the one-to-one line
comparing estimates before and after).

The pseudocell prior also fixes the Fisher standard error
(which is otherwise vastly overestimated)
and the sandwich standard error
(which is otherwise underestimated)
for zero-group genes---restoring the approximate fidelity between
the two estimators (i.e., restoring red points to the one-to-one line between
estimators; \cref{fig:fisher-sandwich-pseudo} \textit{bottom row}).
Otherwise, sans prior,
Fisher standard errors for zero-group genes
are systematically $\apprx$four to six orders-of-magnitude larger than
those for any other gene, and
similarly eclipse
their sandwich counterparts
(\cref{fig:fisher-sandwich-pseudo} \textit{top row}).
As a result, the pseudocell prior restores correlation between
Wald Fisher and Wald sandwich p-values for zero-group genes---as
well as between Wald Fisher and likelihood ratio test p-values
(\cref{supp-fig:pvals-pseudo}),
which are asymptotically equivalent.

Finally, enforcing this prior fixes the convergence problem for zero-group genes
(based on metrics of the gradient and Hessian; \cref{fig:convergence}).
In particular, the Newton step for the log-likelihood objective---which ought
to be vanishingly small
at the maximum likelihood parameter estimate---is
concerningly large (average magnitude $\apprx$1 across coefficients)
for all zero-group genes under the original model.
On the other hand,
under our modified model where a pseudocell prior is enforced,
this metric of the Newton shrinks to $10^{\minus4}$--$10^{\minus9}$
across zero-group genes
(red points in \cref{fig:convergence-pseudo}),
indicating that Newton-Raphson has converged.
This fix, we show, is the result of the Hessian for zero-group genes
moving from clearly ill-conditioned (very large condition number)

This fix is \textit{not} due to intrinsic differences with our optimization algorithm,
or to choosing a different
dispersion (as we propose in \cref{sec:we-dont-do-ql}),
which together decrease the size of the gradient across genes---as
well as the Newton step for non-zero-group genes
(which is already reasonably small)---but do not meaningfully impact the Newton
step for zero-group genes
(i.e., red points retain a large Newton metric whereas
black points are decreased; \cref{fig:convergence-alpha}).

\vspace{\baselineskip}

While we include a pseudocell prior by necessity
(because optimization convergence is essential to the sensitivity approximation),
we recommend this approach more generally when using GLMs to infer
differential expression for scRNA-seq data---particularly if effect sizes are
reported and/or used to threshold or rank genes.
Without this pseudocell prior, estimates of the effect size and its standard
error for zero-group genes (which are common in scRNA-seq data) are
untrustworthy, and
dictated more by
the quirks of the particular optimization algorithm than the
data itself.\footnote{
    Note that our approach is distinct from a pseudo-count prior,
    such as the ``\texttt{prior.count}'' option in \texttt{edgeR},
    which suggests adding a small \textit{count} of \textit{varying} size to each \textit{observation}
    (to avoid logging 0)~\cite{ql-edgeR}.
    This option has similar drawbacks to \texttt{DESeq}'s ``\texttt{minmu}''
    parameter (i.e., distortion), and also is not automatically diluted
    in the presence of copious data, as a prior effect should be~\cite{deseq-heavy-tail}.
    In contrast, we propose adding a pseudo-\textit{observation} (with
    \textit{fixed} count) to each \textit{group}.
}

\subsubsection{Wald testing with standard likelihood}\label{sec:we-dont-do-ql}

Local sensitivity analysis seeks to quantify how small perturbations
to parameters perturb some statistic-of-interest,
based on a Taylor expansion approximation (as we will develop in \cref{sec:de-robustness}).
The likelihood ratio test (or its quasi-likelihood analog) is not readily amenable
to a first-order sensitivity approximation since, by definition, the
log-likelihood terms in the $LR$ statistic have zero gradient at their
respective optima (\cref{app:lrt-sensitivity}).
We leave it to future work to develop an efficient second-order method for
sensitivity analysis of statistics like these.

Instead, we focus on sensitivity analysis of the Wald test
(which is asymptotically equivalent to the likelihood ratio test, requires only
one model fit rather than two or more,
and---we show empirically---has equivalent coverage for scRNA-seq data).
We provide sensitivity analysis of the Wald test for both Fisher and sandwich
standard error estimators. While \texttt{DESeq2} implements only the Fisher
estimator for its Wald test, we recommend the sandwich estimator for its
theoretical robustness to misspecification and empirical coverage
properties (\cref{app:cov-estimators}).

For simplicity, and clearer statistical justification,
we also focus on standard GLMs and statistical testing
rather than their quasi-likelihood analogs.
However, note that sensitivity to the quasi-likelihood Wald test could readily
be calculated by conditioning on $\QLdisp$ (see \cref{app:ql-stats-of-interest}),
or, with more work, by propagating sensitivities through dispersion estimation.

In practice, we use \texttt{glmGamPoi} to estimate $\QLdisp$ and $\NBdisp$,
then compute $\disp$ based on~\cref{eq:disp} (and condition on this dispersion
when fitting the coefficients and estimating robustness).
See \cref{supp-fig:betas-alpha} for the effect on estimated coefficients.

\newlength{\vennHeight}
\setlength{\vennHeight}{3.5cm}
While direct approximation of \texttt{glmGamPoi}'s analysis is not our goal
(given, e.g., changes to the model and inference to ensure convergence,
changes to the statistical test, and our use of likelihood rather than quasi-likelihood),
we nonetheless observe that significance results remain largely concordant.\footnote{
    For example, for a sample scRNA-seq dataset (\cref{app:data}),
    p-values are >98\% correlated across genes,
    and significant genes (BH-corrected $p < 0.01$) are highly overlapping:

    \includegraphics[height=\vennHeight,clip,trim={0 0.5cm 0 2cm}]{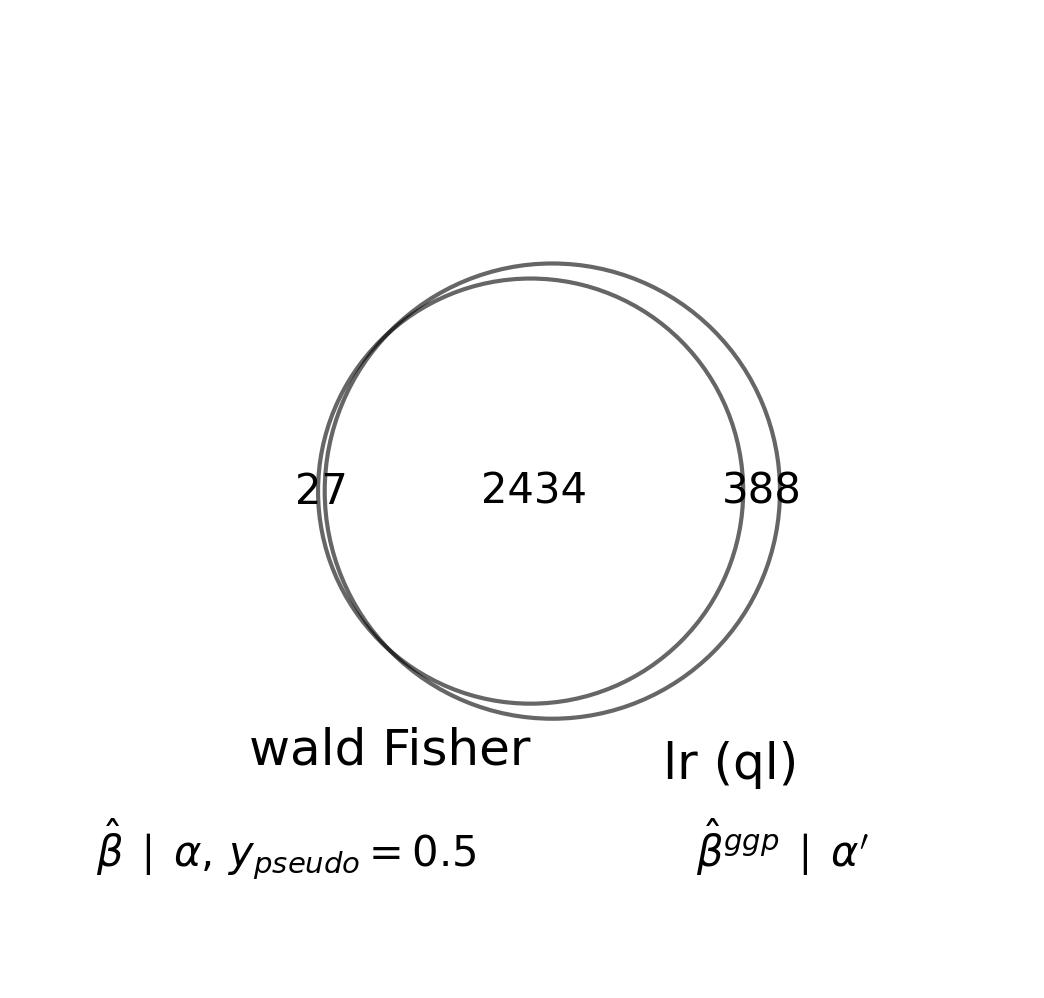}
    \includegraphics[height=\vennHeight,clip,trim={0 0.5cm 0 2cm}]{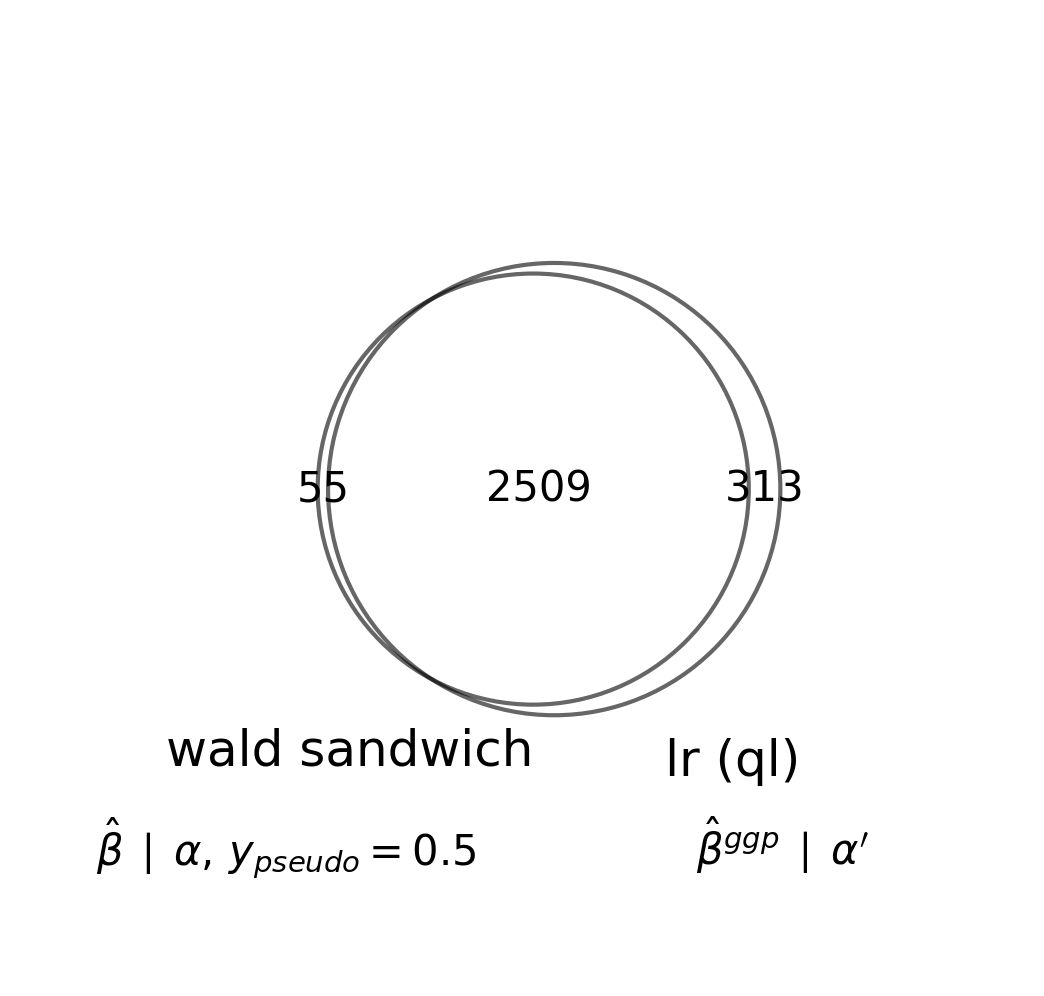}
}

\subsubsection{generic maximum-likelihood inference with autodiff}\label{sec:our-inference}

To ensure convergence, and to perform inference on our
tweaked version of the model
(i.e., with pseudocells and dispersion $\disp$ rather than $\NBdisp$), we
estimate $\vec{\beta}$
by minimizing the negative
log-likelihood objective
using ready-made optimizers and automatic differentiation
(first by BFGS---via \texttt{jax.scipy.optimize}---and then
 by the second-order method Newton conjugate-gradient trust region---via
 \texttt{scipy.optimize}---for genes where the first-order method fails\footnote{
     i.e., the \texttt{jax} routine with default settings for convergence
     ($L^{\infty}$ norm of the gradient $< 10^{\minus5}$
     and up to $200 \, \nbetas$ iterations;
     $\nbetas=3$ for the examples plotted in this section)
     returns \texttt{NaN}
}).
We initialize our inference with the coefficients output by \texttt{glmGamPoi},
$\hat{\vec{\beta}}^{\texttt{ggp}}$.

Together, both optimizers readily generalize to changes in the model
(thanks to autodiff via \texttt{jax} to compute the gradient and Hessian)
and efficiently fit the tens of thousands of objectives (genes)
necessary for each analysis
(thanks to \texttt{jax}'s built-in GPU acceleration, and
CPU parallelization of \texttt{scipy.optimize} via \texttt{multiprocessing}).

\subsubsection{unaccounted-for sources of sensitivity}~\label{sec:not-accounted-for}

\vspace{-1.25\baselineskip}

For completeness, these are the
factors that we don't incorporate into sensitivity calculations:
\begin{itemize}
    \item Negative binomial \emph{dispersion} (and/or quasi-likelihood dispersion).
        We condition on the dispersion when estimating robustness,
        as \texttt{DESeq2} and \texttt{glmGamPoi} do when fitting $\vec{\beta}$.
        Hypothetically, sensitivities could be propagated through the dispersion estimation step
        as well, though work would be needed to transform this iterative, heuristic step into
        a form that is amenable to sensitivity (such as a well-formed optimization objective).

    \item The \emph{prior coefficient width} $\sigma$.
        By default in \texttt{DESeq2} and \texttt{glmGamPoi}, $\sigma$ is fixed
        to a large constant that is not data-dependent.
        However, if it is set empirically (\texttt{betaPrior=TRUE} in \texttt{DESeq2}),
        then sensitivities could hypothetically be propagated through the prior
        estimation step, as above.

    \item More complex algorithms for estimating \emph{size factors} $\vec{\size}$.
        Throughout our experiments, we estimate $\vec{\size}$ via \texttt{glmGamPoi}'s
        default method, \texttt{normed\_sum}, and do propagate sensitivities through this step.
        However, if a more complex algorithm is used to estimate size factors,
        we provide an interface to either condition on $\vec{\size}$ as a fixed
        hyperparameter\footnote{
            Especially if dropping a small subset of data points is
            unlikely to meaningfully impact the estimated size factors
        } \textit{or} to automatically compute sensitivity for an alternate
        parameterization of $\vec{\size}$ (so long as it can be written as a
        smooth function of data weights\footnote{
            e.g., analytically, or through the estimating equation of an
            optimization problem
        }).
\end{itemize}

\subsection{a differential expression analysis amenable to sensitivity} \label{sec:de-analysis-for-sensitivity}

Drawing together what we've described in this section,
we now formally outline a differential expression procedure
for which we can automatically compute sensitivity
(with respect to dropping observations).

Specifically, we are interested in the estimator $\hat{\vec{\beta}}$ for each
gene: a Z-estimator whose estimating equations are given by the gradient of
the log-likelihood, since this function goes to zero at the optimal solution.
Namely, the Z-estimator will be the solution $\hat{\vec{\beta}}$ such that
\begin{equation} \label{eq:estimating}
    \nabla \LL\ofBeta
    \defeq \dd*{}{\vec{\beta}} \, \LL(\vec{\beta}, \cdots)
    \defeq \estimating_0\ofBeta +
           \sum\overCells \estimating_{\icell}\ofBeta
    = \vec{0}\ofSize{\nbetas,1}
\end{equation}
for estimating equations $\estimating_{\icell} \defeq \nabla \ll\left( \vec{\beta}; \vec{x}_{\icell}, y_{\icell} \right)$---the
gradient of the log-likelihood with respect to the $\icell\th$ data point---and
optional regularization $\estimating_0$.
In particular, for a differential expression GLM,
\begin{align}
    \estimating_{\icell}\ofBeta &=
        \frac{ y_{\icell} - \size_{\icell} \exp\left\{\xbn\right\} }
             { 1 + \disp \, \size_{\icell} \exp\left\{\xbn\right\} } \, \vec{x}_{\icell}
        \label{eq:G-de}
        && \mathcomment{gradient of the GLM}
    \shortintertext{and}
    \estimating_0\ofBeta &=
        - \frac{\vec{\beta}}{\vec{\sigma}^2}.
        \nonumber
        && \mathcomment{gradient of the coefficient prior}
\end{align}
See \cref{app:z-estimators} for more details, starting with ordinary least squares
and working up to generalized linear models.

This estimator is fit to the augmented dataset
\[ \big\{ \,
    (\vec{x}_1, y_1), \, \ldots \, ,
    (\vec{x}_{\ncells}, y_{\ncells}),\,
    (\vec{x}_{\text{pseudo:treatment}}, y_{\text{pseudo}}),\,
    (\vec{x}_{\text{pseudo:control}}, y_{\text{pseudo}})
  \, \big\}.
\]

Alternately, for the pseudobulk model, replace
cell observations with
pseudobulk observations (aggregating
cells within a sample)
in \cref{eq:estimating,eq:G-de}.\footnote{
    i.e.,
    $
     \estimating_{\icell} \rightarrow \estimating_{\isample}, \,
     \vec{x}_{\icell} \rightarrow \pseudobulk{\vec{x}}_{\isample}, \,
     y_{\icell} \rightarrow \pseudobulk{y}_{\isample}, \,
     \size_{\icell} \rightarrow \pseudobulk{\size}_{\isample}, \,
     \ncells \rightarrow \nsamples
    $
}

The key statistical outcomes of this analysis revolve around the inferred $\hatBetaTreated$
(also known as the log-fold change; \emph{LFC}) for each gene; namely,

\begin{enumerate}
    \item the \emph{sign} of the treatment effect, $\sign\left[\hatBetaTreated\right]$---a
        minimal outcome that signifies whether ``treatment'' is associated with
        increased or decreased expression of a gene;
        \label{it:outcome-sign}

    \item the \emph{magnitude} of the effect, $|\hatBetaTreated|$,\footnote{
            Note that the coefficients $\vec{\beta}$ are modeled in natural log
            space (the canonical link), but
            often reported in base two
            (for interpretation as log twofold-change).
            In other words, the effect size is reported as
            $\log_2( \exp \hatBetaTreated ) = \hatBetaTreated  \times \log_2(\e)$.
        }
        which quantifies the difference in expression between groups
        and may be used to rank or threshold genes (by a minimal ``meaningful''
        effect size) for downstream analysis;
        \label{it:outcome-magnitude}

    \item the bounds of the \emph{confidence interval} containing the estimated effect,
        $\hatBetaTreated \,\pm\, \CIwidth$, where $\CIwidth$ is the one-sided width
        of the confidence interval at the chosen significance level.
        The treatment effect is ruled significant (i.e., that gene is differentially
        expressed between groups) $\iff$ the interval does not contain zero.
        \captionbr
        Recall that this level is typically used to threshold differential
        expression p-values that have first been corrected for multiple testing
        (across genes); \texttt{DESeq2} and \texttt{glmGamPoi} use the BH
        step-up procedure~\cite{bh}.
        Because the BH procedure involves ranking, it is \textit{not} differentiable and
        therefore not readily amenable to sensitivity analysis.
        As a proxy,
        we instead construct confidence intervals under the empirical cutoff
        (on raw p-values)
        corresponding to the desired significance level for BH-corrected
        p-values. This effectively entails conditioning on the number of genes
        $\nrank$ ruled significant (under BH correction) at the desired level,
        then multiplying that level by $\nrank / \ngenes$.
        \label{it:outcome-ci}
\end{enumerate}

Finally, a key high-level outcome is
\begin{enumerate}[resume]
    \item a list of the \emph{top 10 gene sets}
    based on a hypergeometric enrichment test of differentially expressed genes.
        \label{it:outcome-gene-sets}
\end{enumerate}

%% file: 4-de-influence.tex
In this section, we'll lay out our approach to measuring dropping-data robustness of
the procedure outlined above---based on
approximating how these key statistical outcomes of differential expression
can be maximally perturbed by dropping a handful of data points.

At a high level, our approach is to introduce a vector of data weights $\vec{w}$,
comprised of weights $w \in [0, 1]$ that modulate the contribution of each observation
(cell or sample).
These weights serve as a continuous, and therefore differentiable, proxy for
the binary inclusion/exclusion of data points.
By rewriting the optimization problem, and its key statistical outputs, as
a function of data weights, we can approximate how these outputs would
change if a fraction of data points (cells) were not observed.
In other words, we seek to quantify robustness
by identifying the most influential cells
for each statistic-of-interest in differential expression,
and approximating how each statistic would change
under a data perturbation where
a small number
of those influential cells were dropped.

In this section, we begin by formally outlining the key
statistical outcomes $\stat$ of differential expression.
Then, after reviewing the original framework for
\ul{a}pproximate \ul{m}aximum \ul{i}nfluence \ul{p}erturbations (AMIP)~\cite{ryan-amip},
we derive how to compute
$\ddd{}{\vec{w}} \stat$---approximating how small perturbations to
the inclusion of cells in a \texttt{DESeq2}/\texttt{glmGamPoi}-esque analysis
would perturb these key gene-level outcomes.
We explain how to transform this quantity (cell influences per gene statistic)
into a useful metric of dropping-data robustness for differential expression.
Finally, we extend these gene-level results
(across thousands or tens of thousands of genes)
to derive insight into the robustness of a high-level outcome of differential
expression; namely, the top gene sets enriched among differentially expressed
genes.

While the gene-level outcomes of differential expression are amenable
to direct sensitivity approximation, with minor modifications (\cref{sec:mods}),
the latter is particularly challenging.
First, it requires a thresholding operation (into significant and nonsignificant genes),
which is inherently discrete and therefore \textit{not differentiable}.
Second, this thresholding operation is performed on p-values that
have been ranked (in order to apply a multiple-testing correction).
Corrected p-values are therefore smooth only within a given rank,
whereas a jump in ranking is discrete and therefore \textit{not differentiable}.
Moreover, we are concerned not just with a single gene, but with
a joint ranking problem (where dropping a given handful of cells will affect
the differential expression test for each gene,
leading to combinatorial complexity in computing the effect on rankings \textit{across} genes).
Finally, the top gene sets are based on hypergeometric testing of
overlap among discrete sets, followed by another ranking operation
(of hypergeometric p-values),
both of which are inherently \textit{not differentiable}.

After showing how to directly approximate the dropping-data robustness
of the key gene-level outcomes of differential expression
(\cref{sec:amip,sec:stats-of-interest,sec:robustness-independent-cells,sec:robustness-pseudobulk}),
we describe a procedure to
cluster and score cell influence vectors (across genes)
in order to bound the dropping-data robustness of the top gene sets
(\cref{sec:gsea-robustness}).

\subsection{statistics-of-interest} \label{sec:stats-of-interest}

Recall that the key outcomes of differential expression revolve
around the \emph{sign}, \emph{size}, and \emph{significance}
of the treatment effect, $\hatBetaTreated$, for each gene.\footnote{
    A fourth key outcome, the top gene sets enriched among differentially
    expressed genes, will be addressed in \cref{sec:gsea-robustness}
}
For each outcome-of-interest, we will now formally define a 1-d statistic $\stat$.
The desiderata for each $\stat$ is
that it will be defined as a function of data weights $\vec{w}$---either
explicitly or implicitly, through the dependence $\vec{\beta} = \hat{\beta}(\vec{w})$---and
that a change in its sign will correspond to a ``meaningful'' change in the
corresponding differential expression outcome
(such as a change in the direction of a treatment effect,
or a statistically significant finding becoming nonsignificant,
or vice versa).

We begin by defining two statistics that will serve as building blocks for the others:
\begin{align*}
    \stat_{\text{LFC}}^+\ofBeta &=
        \sign \left[ \vec{\contrast}\T \, \hat{\beta}(\vec{1}) \right] \times
         \vec{\contrast}\T \, \vec{\beta}
    && \mathcomment{unsigned treatment effect}
\shortintertext{and}
    \stat_W^+\ofBetaW &=
        \sign \left[ \vec{\contrast}\T \, \hat{\beta}(\vec{1}) \right] \times
        \frac{\vec{\contrast}\T \, \vec{\beta}}
             {\sqrt{\vec{c}\T \cdot \widehat{\Sigma}\ofBetaW \cdot \vec{\contrast}}}
    && \mathcomment{unsigned Wald statistic}
\end{align*}
where
$\widehat{\Sigma}$ is an estimator of $\cov\big[ \vec{\beta} \big]$
(namely, either the Fisher or robust sandwich estimator; \cref{app:cov-estimators}).
For mathematical convenience, the treatment effect $\hatBetaTreated$ is computed as
$\vec{\contrast}\T \, \hat{\vec{\beta}}$,
where $\vec{\contrast}$ is the \emph{contrast} vector that picks out the
coefficient-of-interest.

Multiplication by the sign of the original effect ensures that each is
positive at $\vec{w}=\vec{1}$;
this allows for direct influence comparisons across genes
(regardless of the direction of each treatment effect)
and unifies the definitions of the statistics below.

Specifically, we design each $\stat$ to correspond to a decision function with
a boundary at zero, where the outcome for the original analysis,
$\stat(\vec{1})$, is negative and the subscript becomes true $\iff$ $\stat > 0$.
So, for example,
$\stat_{\text{erase significance}}$ is negative for any gene that is originally significant
(i.e., where the lower bound of the positive effect is above zero).
Then, we can erode the signal for these genes
by increasing this statistic
(i.e., pushing the lower bound below zero).

The key statistics for differential expression
(corresponding to
the outcomes
outlined in \ref{it:outcome-sign}$\,$--$\,$\ref{it:outcome-ci}; \cref{sec:de-analysis-for-sensitivity})\footnote{
    Ignoring---for now---the final, inter-gene outcome
    (\ref{it:outcome-gene-sets}; \cref{sec:de-analysis-for-sensitivity}),
    which will be addressed in \cref{sec:gsea-robustness}
}
are:
\begin{align*}
    \stat_{\text{flip sign}}\ofBeta &= -\stat_{\text{LFC}}^+\ofBeta
    \\
    \stat_{\text{shrink below threshold}}\ofBeta \,&=\;
       \LFCthreshold - \stat_{\text{LFC}}^+\ofBeta
    \\
    \stat_{\text{increase above threshold}}\ofBeta \,&=\;
       \stat_{\text{LFC}}^+\ofBeta - \LFCthreshold
    \\
    \stat_{\text{erase significance}}\ofBetaW \,&=\;
        -\left[ \stat_W^+\ofBetaW - \CIwidth \right]
    && \mathcomment{-CI lower bound}
    \\
    \stat_{\text{bestow significance}}\ofBetaW \,&=\;
        +\left[ \stat_W^+\ofBetaW - \CIwidth \right]
    && \mathcomment{\hphantom{-}CI lower bound}
    \\
    \stat_{\text{flip sign w/ significance}}\ofBetaW \,&=\;
        -\left[ \stat_W^+\ofBetaW + \CIwidth \right]
    && \mathcomment{-CI upper bound}
\end{align*}
where
\begin{itemize}
    \item $\tau$ is a chosen effect size threshold
        (minimal log-fold change with a ``meaningful'' magnitude),\footnote{
            Note that since $\hat{\vec{\beta}}$ is fit in natural log space,
            while the effect size threshold is often set in $\log_2$ (for interpretation as log twofold-change),
            the threshold-of-interest would be $\tau = \tau' / \log_2 e$ (for $\tau'$ in $\log_2$ space).
        } as determined by the scientist for a given experiment,
    and
    \item $\CIwidth$ is defined as the one-sided width of a confidence interval (CI)---so,
        for a 95\% CI (significance level 0.05),
        $
            \CIwidth \defeq F\inv\left(1-\frac{1-0.95}{2}\right) \approx 1.96,
        $
        where $F\inv$ is the inverse CDF
        of the null distribution (and the approximation
        corresponds to a standard Gaussian null).
        Recall (\ref{it:outcome-ci}; \cref{sec:de-analysis-for-sensitivity}) that we
        approximate a significance level for BH-corrected p-values
        by correcting the effective level (for raw p-values) based on the
        number of genes ruled significant in the original analysis.
\end{itemize}
\vspace{\parskip}

See \cref{app:ql-stats-of-interest} for the quasi-likelihood counterpart of the
statistics involving significance.

For convenience, we may write each statistic as $\stat\ofW$ to emphasize
its implicit dependence on data weights $\vec{w}$
through the estimated coefficients $\hat{\vec{\beta}}$.

Having defined six statistics-of-interest,\footnote{
    Nine if considering signicance
    under both Fisher \text{and} sandwich Wald testing
} we might na\"ively expect that six influence computations
would be required to compute $\ddd{}{\vec{w}}\stat$ and estimate dropping-data
robustness of each statistic (per gene).
However, all of these statistics-of-interest ultimately
revolve around
just two decision functions,\footnote{
    Three if considering signicance
    under both Fisher \text{and} sandwich Wald testing
} with varying decision boundaries (corresponding to different outcomes-of-interest).
We leverage this fact
by laying out our statistics-of-interest modularly
(as affine functions of $\stat_{\text{LFC}}^+$ or $\stat_W^+$),
such that we can perform a single influence computation of each building block
($\ddd*{}{\vec{w}}\stat_{\text{LFC}}^+$ and $\ddd*{}{\vec{w}}\stat_W^+$)
and construct all influences-of-interest by cheap arithmetic,
including varying the confidence level and null distribution.

\subsection{approximate maximum influence perturbations} \label{sec:amip}

We seek to estimate how well these key outcomes of DE
hold up under a small data perturbation; namely, dropping the most influential
handful of observations.
Here, we'll outline the dropping-data approach in general terms,
before delving into our particular application and adaptation of this approach
to models for differential expression.

The foundational
dropping-data robustness study~\cite{ryan-amip} showed that we can approximate
how
a 1-d statistic-of-interest $\stat$ would change, if a fraction of
data points were dropped from an analysis, by
\begin{itemize}
    \item introducing data weights $\vec{w}$ to form a weighted analog of the
        original Z-estimator $\hat{\theta}$,
    \item fitting $\hat{\theta}$ to
        the original dataset ($\vec{w}=\vec{1}$),
    \item computing a vector of partial derivatives $\ddd*{}{\vec{w}}\stat$
        (analogous to influence scores), and
    \item linearly extrapolating to form an approximation of $\stat$
        when $w_{\icell} = 0$ (i.e., dropping the $\icell\th$ observation)
        for any handful of data points.
\end{itemize}
Finally, the authors introduced, as a measure of data robustness,
the (approximate) minimal fraction of data points required to enact a ``meaningful'' change in
$\stat$. If a key statistic can be meaningfully perturbed by dropping a trivial
fraction of observations (where ``meaningful'' and ``trivial'' are dataset-, and
perhaps researcher-, dependent),
then it is not---through this particular lens---robust.

Specifically, to briefly review \cite{ryan-amip}, let the original analysis
entail fitting parameters $\vec{\theta}$
via the Z-estimator whose
estimating equations are given by
\begin{equation}
    \sum\overCells \estimating_{\icell}( \vec{\theta} ) = \vec{0}.
\end{equation}
Then, under the \textit{weighted} dataset with data weights $\vec{w}$,
let the corresponding parameter estimate be given by
$\hat{\theta}(\vec{w})$---i.e.,
the solution to the
\textit{weighted} Z-estimator
given by
\begin{equation} \label{eq:amip-w-estimating}
    \sum\overCells w_{\icell} \estimating_{\icell} \ofThetaHatW* = \vec{0}.
\end{equation}
In other words, if $\estimating_{\icell}$ is the gradient of the $\icell\th$
data point, the weight $w_{\icell}$ acts to modulate the contribution of that
data point to the objective.\footnote{
    Note that this definition assumes that the gradient itself
    depends on the weights only via a functional dependence on
    $\vec{\theta}=\hat{\theta}(\vec{w})$.
    Later, we'll extend the original dropping-data robustness framework~\cite{ryan-amip}
    to relax this assumption.
}
When $\vec{w} = \vec{1}$, we recover the original analysis.
The only restriction on $\estimating_{\icell}$---for the sake of the following
paragraphs---is that it must be
smooth (twice continuously differentiable).

A linear approximation of how perturbing $\vec{w}$
(by zeroing a small fraction of its entries---akin to dropping those data points)
will perturb the statistic-of-interest $\stat$ is given by the first-order Taylor expansion
\begin{align} \label{eq:taylor-expansion}
    \stat(\vec{w}) \approx
    \stat(\vec{1}) +
        \sum\overCells (w_{\icell} - 1) \, \influence_{\icell}
    \eqdef \hat{\stat}(\vec{w})
\end{align}
where $\influence_{\icell}$ is the \emph{influence}
of the $\icell\th$ data point
\begin{align} \label{eq:influence}
    \influence_{\icell} \defeq \dd*{\stat\ofThetaHatW}{w_{\icell}} \bigg|_{\vec{w}=\vec{1}}.
\end{align}

To calculate $\influence_{\icell}$,
expand the partial derivative via the chain rule:
\begin{equation} \label{eq:influence-calc}
    \dd*{\stat\ofThetaHatW}{w_{\icell}} \bigg|_{\vec{w}} =
    \underbrace{
        \strut
        \dd*{\stat\ofThetaW}{\vec{\theta}\T}
            \bigg|_{\hat{\theta}(\vec{w}), \vec{w}}
    }_{\circled{1}}
    \, \cdot \,
    \underbrace{
        \strut
        \dd*{\hat{\theta}(\vec{w})}{w_{\icell}} \bigg|_{\vec{w}\vphantom{\hat{\theta}(\vec{w})}}
    }_{\circled{2}}
     +
    \underbrace{
        \strut
        \dd*{\stat\ofThetaW}{w_{\icell}}
            \bigg|_{\hat{\theta}(\vec{w}), \vec{w}}
    }_{\circled{3}}.
\end{equation}

\circled{1} and \circled{3} are readily computed using automatic differentiation,\footnote{
    Or, substituting manual derivatives as desired
}
evaluated at the original maximum likelihood estimate
$\hat{\theta}(\vec{1})=\hat{\vec{\theta}}$.

\circled{2} can be computed by considering that we've implicitly
defined the estimator $\hat{\theta}$ as a function of the weights
through the weighted estimating equation that it solves (\cref{eq:amip-w-estimating}).
Then, by the implicit function theorem and some algebra~\cite{influence-derivative,ryan-amip} (\cref{app:dbeta-dw}),
\begin{equation}
\begin{aligned}\label{eq:amip-dtheta-dw}
    \dd{\hat{\theta}(\vec{w})}{\vec{w}\T} \bigg|_{\vec{w}}
    =
    -&\left(
        \sum\overCells
        w_{\icell} \,
        \dd*{\estimating_{\icell}\ofTheta}{\vec{\theta}\T }
            \bigg|_{\hat{\theta}(\vec{w}), \vec{w}}
     \right)\inv
     \cdot \;
         \bigg[
            \estimating_1\ofThetaHatW*,
            \,\ldots\, ,
            \estimating_{\ncells}\ofThetaHatW*
        \bigg]
\end{aligned}
\end{equation}
where the partial derivatives are again computable by autodiff.

Importantly, we can use \cref{eq:taylor-expansion}
to approximate the effect of dropping \textit{any} fraction of data points
(by manipulating $\vec{w}$),
at the cost of a single amortized
computation.
Specifically, beyond the original model fit
(at $\vec{w}=\vec{1}$; i.e., the analysis that the researcher will
have conducted anyway), the computational burden is
a one-off round of autodiff calculations to compute
$\vec{\influence} \defeq [\influence_1, \ldots, \influence_{\ncells} ]$
(\cref{eq:influence,eq:influence-calc,eq:amip-dtheta-dw}).
Once this (reasonably cheap) cost is paid up front, we can approximate the
effect of dropping any
subset of data points for free.\footnote{
    Of course, the fidelity of the approximation will be best for small subsets
    ($\vec{w}$ ``near'' $\vec{1}$), since it is based on a Taylor expansion at the
    original data weights.
}

A key contribution of dropping-data robustness~\cite{ryan-amip}
is transforming these influences per data point into
a useful metric of robustness.
First, explicitly defining each statistic $\phi$
based on a decision boundary that corresponds to a
``meaningful'' change---e.g., flipping the sign or erasing the significance of
an effect---introduces a natural, common scale across influences.
Without loss of generality, assume that $\stat$ is defined as a statistic
that is negative when $\vec{w}=\vec{1}$, and becomes positive $\iff$
the
corresponding change is effected.
Then, by \cref{eq:taylor-expansion}, the
observations whose removal would maximally impact $\stat$
in the direction of the decision boundary
are those with the most negative influence scores.
Let $\pi$ be a permutation of the influence scores such that
$\pi(\vec{\influence}) \defeq \left[ \influence_{(1)}, \influence_{(2)}, \ldots, \influence_{(\ncells)} \right]$
and
$\influence_{(1)} \leq \influence_{(2)} \leq \cdots \leq \influence_{(\ncells)}$.
Then, the most influential $\ninfluence$ observations are defined by the subscripts of the first
$\ninfluence$ entries of $\pi(\vec{\influence})$, and the approximate
change induced by
dropping this data subset---corresponding to the weight vector $\wdrop$
where the entries
$w_{(1)}, w_{(2)}, \ldots, w_{(\ninfluence)}$ are zeroed
and the rest are 1---is
$\sum\limits_{\iinfluence=1}^{\ninfluence} \influence_{(\iinfluence)}$.

The dropping-data
robustness metric is the minimal portion of observations
that, when removed, we predict will be sufficient
to enact the change-of-interest defined by $\stat$
(e.g., ``flip sign'' or ``erase significance''; \cref{sec:stats-of-interest});
namely,
\begin{equation} \label{eq:amip-metric}
    \inf \bigg\{ \ninfluence / \ncells \, : \,
        \sum\limits_{\iinfluence=1}^{\ninfluence}
        \influence_{(\iinfluence)} < \underbrace{
                                        \strut
                                        \stat\ofThetaHatW[\vec{1}]
                                    }_{< \, 0} \bigg\}
    \iff
    \inf \bigg\{ \ninfluence / \ncells \, : \,
        \hat{\stat}\ofThetaHatW[\wdrop] > 0 \bigg\}.
\end{equation}
In other words, we iterate over $\pi(\vec{\influence})$ from left to right,
including data points in the minimal set to drop until the cumulative sum of
their influence scores surpasses the original statistic.

If we predict that no set of observations will be sufficient,
we consider the minimal portion to be \texttt{NaN}
(or $\infty$)---meaning that $\stat$ is fully robust
against small data perturbations (at a first-order approximation,
formed locally around $\vec{w}=\vec{1}$).
In other words, the statistical outcome represented by $\stat$
(such as whether the treatment effect is greater than a minimal threshold,
or whether it is significant) is estimated to be stable
(i.e., consistent with the original analysis)
when any small fraction of observations are dropped.
In reality $\stat$ may be perturbed by dropping many data
points---so many such that the local approximation is no longer valid---but
also so many such that we are not concerned about $\stat$'s dropping-data robustness.

\subsection{independent cells} \label{sec:robustness-independent-cells}

We now return
to the differential expression problem at hand.
To quantify the dropping-data robustness of
our key statistics-of-interest (\cref{sec:stats-of-interest}),
we'll now apply and adapt the original robustness
procedure~\cite{ryan-amip} for
differential expression analysis of scRNA-seq data
where cells are treated as independent observations.\footnote{
    Or, equivalently, for
    analysis of bulk data where \textit{samples} are the independent unit of observation.

    Because the Taylor approximation with respect to data weights deteriorates at small $\ncells$,
    this approach should only be applied to datasets with a sufficient number of observations
    (at least $10^2$ and, ideally, $10^3$ or more).
    While this number of replicates is less common for bulk data,
    the number of cells in a single-cell experiment routinely surpasses it
    (extending to $10^6$ or, recently, $10^7$)~\cite{nxn-ncells}.
    Throughout this section and beyond, we will refer to data points
    as \textit{cells} and assume the context of scRNA-seq.

    In fact, for datasets
    of moderate but insufficient size to trust the
    quality of the approximation,
    dropping-data robustness may reasonably be computed exactly
    (by empirically dropping all subsets of a given size),
    if
    the regime of insurmountable combinatorics has not yet kicked in.
    Alternately,
    in the regime of
    small data (say, $\ncells < 100$),
    dropping-data robustness may not be a sensical axis of robustness
    to care about, as we could reasonably anticipate \textit{a priori} that any
    statistical outcome from such a small dataset would be strongly dependent
    on each observation.
}

Recall that the solution to the differential expression objective is given
by a Z-estimator (\cref{eq:estimating})---i.e., the root of a data-dependent
equation---since the GLM log-likelihood objective is maximized where its
gradient is zero.
This estimator yields the fitted model parameters
(generically called $\hat{\vec{\theta}}$ in \cref{sec:amip});
namely, the GLM coefficients $\hat{\vec{\beta}}$.

To smoothly modulate the contribution of each cell to the estimator,
we introduce a new parameter $\vec{w}\ofSize{\ncells,1}$ of data weights,
where each $w_{\icell} \in [0,1]$ is applied to each cell observation $(\vec{x}_{\icell}, y_{\icell})$.
The optimal coefficients under this \textit{weighted}
dataset are $\hat{\vec{\beta}}=\hat{\beta}(\vec{w})$---namely,
the solution to the \textit{weighted} Z-estimator
\begin{equation} \label{eq:w-estimating}
    \estimating_0 \ofBetaHatW* +
    \sum\overCells
        w_{\icell} \, \estimating_{\icell} \ofBetaHatW
    = \vec{0}
\end{equation}
(cf. the unweighted estimator, \cref{eq:estimating}).

Recall that the $\estimating_0$ term---absent in the original dropping-data robustness
setup
(\cref{eq:amip-w-estimating})---captures regularization (in the form of a prior
over coefficients $\vec{\beta}$), and so only depends on data weights through
its dependence on $\hat{\beta}(\vec{w})$.
The remaining terms, $\estimating_{\icell}$,
capture the gradient with respect to each cell,
$\nabla \ll(\vec{\beta}; \vec{x}_{\icell}, y_{\icell})$.

A small notational---but key functional---difference from \cite{ryan-amip}
is the dependence of each $\estimating_{\icell}$
on both $\hat{\beta}(\vec{w})$ \textit{and} \vec{w} (cf. \cref{eq:amip-w-estimating}).
Whereas the original setup implicitly assumed
that the gradient $\estimating_{\icell}$ was dependent on data weights only
through its functional dependence on $\hat{\vec{\beta}}$,
the \texttt{DESeq2}/\texttt{glmGamPoi}
log-likelihood involves
hyperparameters that are
set empirically based on the full dataset. This introduces a data weight
dependency that requires $\estimating_{\icell}$ to be redefined as
an explicit function of $\vec{w}$.\footnote{
    Alternately, this dependence can be ignored---by conditioning on such
    hyperparameters---for expediency, with some concomitant loss of quality of
    the sensitivity approximation.
    This is our approach for the negative binomial dispersion $\disp$, which is
    fit by a somewhat heuristic procedure that would be nontrivial to translate
    into a differentiable
    dependence on $\vec{w}$.
    On the other hand, for cell size factors $\vec{\size}$,
    we do incorporate
    their data weight dependence
    in our experiments.
}

Specifically, recall that size factors $\vec{\size}$ are computed as a function
of all observed counts $\mat{Y}$
(to account for some trend in library size across cells).
As a consequence, size factors are a function of cell weights,
$\vec{\size} \defeq \sizeFn\left( \vec{w} \right)$.
For example, assuming sizes are computed by the default \texttt{normed\_sum} method,
\begin{equation} \label{eq:normed-sum-w}
    \sizeFn\left(\vec{w}\right) \defeq
        \total{\vec{y}} \big/
        \exp\left\{ \frac{1}{\sum\limits_{\icell} w_{\icell}}
                             \sum\limits_{\icell} w_{\icell} \log \total{y}_{\icell} \right\}
\end{equation}
---i.e., the weighted analog of \cref{eq:normed-sum}.
The terms of the weighted estimating equations can then be rewritten as
\begin{equation} \label{eq:Gn-w}
    \estimating_{\icell} \ofBetaHatW
    \defeq
      \frac{ y_{\icell} - \sizeFn\left( \vec{w} \right)_{\icell} \exp\left\{ \xbnHatW \right\} }
           { 1 + \disp \, \sizeFn\left( \vec{w} \right)_{\icell} \exp\left\{ \xbnHatW \right\} } \, \vec{x}_{\icell}.
\end{equation}

In \cref{sec:stats-of-interest}, we defined several key outcomes
$\stat$ for differential expression
(revolving around the sign, magnitude, and significance of the treatment effect).
The
dropping-data
framework~\cite{ryan-amip} provides a
tractable approximation of how dropping a small fraction of cells
(i.e., perturbing $\vec{w}$)
will perturb each outcome-of-interest $\stat$
by linearizing $\ddd{}{\vec{w}} \stat$
with a Taylor expansion (\cref{eq:taylor-expansion})
and
using it
to efficiently and automatically
calculate cell influence scores $\influence_{\icell}$
(\cref{eq:influence-calc,eq:amip-dtheta-dw}).

Now that $\estimating_{\icell}$ depends explicitly on $\vec{w}$,
we must modify the computation of
\[
    \influence_{\icell} = \dd*{}{w_{\icell}} \stat \bigg|_{\vec{w}=\vec{1}}.
\]
Specifically,
term \circled{2} in \cref{eq:influence-calc}
(the chain rule expansion of $\influence_{\icell}$)
becomes
\begin{equation}
\begin{aligned}\label{eq:dbeta-dw}
    \dd*{\hat{\beta}(\vec{w})}{w_{\icell}} \bigg|_{\vec{w}}
    =
    -&\left(
        \dd*{\estimating_0\ofBeta}{\vec{\beta}\T}
            \bigg|_{\hat{\beta}(\vec{w})}
        + \;
        \sum\overCells
        w_{\icell} \,
        \dd*{\estimating_{\icell}\ofBetaHatW}{\vec{\beta}\T }
            \bigg|_{\hat{\beta}(\vec{w}), \vec{w}}
     \right)\inv
     \\
     \cdot \; &\Bigg(
        \underbrace{
            \strut
            \sum\overCells
            w_{\icell} \dd*{\estimating_{\icell}\ofBetaW}{\vec{w}\T}
            \bigg|_{\hat{\beta}(\vec{w}), \vec{w}}
        }_{\star}
        + \;
         \bigg[
            \estimating_1\ofBetaHatW,
            \,\ldots\, ,
            \estimating_{\ncells}\ofBetaHatW
        \bigg]
     \Bigg)
\end{aligned}
\end{equation}
where $\;\star\;$ is a new term that disappears only when $\estimating_{\icell}$
is not an explicit
function of
$\vec{w}$,\footnote{
    In other words, when $\estimating_{\icell}$ depends on $\vec{w}$ only through
    the parameter estimate $\hat{\vec{\beta}}$
} as is assumed throughout \cite{ryan-amip}
(recovering \cref{eq:amip-dtheta-dw}).
For details,
see \cref{app:dbeta-dw}.

Through this modified dropping-data
approximation, we can compute
the combinatorial effect on $\stat$
(a meaningful function of a gene's $\betaTreated$)
of removing any subset of cells.
Specifically, to automatically approximate dropping cells, we
\begin{itemize}
    \item Fit the \emph{original differential expression analysis},
        which entails fitting $\ngenes$ GLMs,
        to compute $\hat{\vec{\beta}} = \hat{\beta}(\vec{1})$ for each gene.
        For a sample scRNA-seq dataset of
        $\ncells=1440$ cells and $\ngenes=10,502$ genes\footnote{
            Specifically, $15,516$ total genes measured with at least one nonzero observation
            (all of which are fit with \texttt{glmGamPoi}, which estimates
            dispersions and size factors based on trends across genes),
            and $10,502$ genes selected for
            further analysis based on a criterion of 10 or more nonzero observations.
            For a note on this choice, see
            \cref{fn:gene-filter-exp}.
        } (\cref{app:data}),
        this costs \textbf{$\apprx$2.5 minutes} at an amortized cost of
        \textbf{<0.015 seconds per gene}.
        Specifically, the computation is broken down into
        \textbf{48 seconds} to run \texttt{glmGamPoi::glm\_gp}
        (to estimate $\disp$ and fit an initial estimate $\hat{\vec{\beta}}^{\texttt{ggp}}$ per gene),
        and \textbf{106 seconds} to fit $\hat{\vec{\beta}}$ across genes
        under our modified objective (\cref{sec:de-analysis-for-sensitivity}).\footnote{
            More specifically, the first stage of fitting uses \texttt{jax} to take
            advantage of parallelization on a GPU
            (GeForce RTX 2080 SUPER with 8GB RAM);
            this takes \textbf{90 seconds}.
            The second stage uses second-order optimization and CPU parallelization
            (Intel Xeon W-2295 with 32 cores and a generous 250GB RAM)
            to fit any genes where first-order optimization failed;
            this takes \textbf{16 seconds}
            (across the 480 optimizations that had to be repeated for this dataset).
        }

    \item Use autodiff to \emph{compute cell influences} $\vec{\influence}$ for each gene.
        The final output is a cell-by-gene influence matrix
         $\mat{\influence}\ofSize{\ncells, \ngenes}$.
        With GPU acceleration, this costs \textbf{$\leq$4 seconds} total.\footnote{
            Specifically,
            \textbf{1 second} to compute influences for $\stat$ based on the treatment effect,
            \textbf{3 seconds} to compute influences for $\stat$ based on the Wald Fisher statistic,
            or
            \textbf{4 seconds} to compute influences for $\stat$ based on the Wald sandwich statistic
        }
        While influences are specific to the chosen statistic $\stat$,
        we can compute $\mat{\influence}$ for \textit{all} key statistics-of-interest
        by performing this step twice---to compute influences for
        building blocks $\stat_{\text{LFC}}^+$ and $\stat_W^+$---and
        constructing all influence matrices-of-interest by simple arithmetic
        (as linear functions of
         $\ddd{}{\vec{w}}\stat_{\text{LFC}}^+$ and $\ddd{}{\vec{w}}\stat_{W}^+$;
        \cref{sec:stats-of-interest}).
\end{itemize}
For this one-time cost of \textbf{2 minutes and 39 seconds} total (\textbf{0.015 seconds per gene}),
we can approximate the effect of removing \textit{any} subset of cells
on \textit{any} key gene-level outcome-of-interest for differential expression
(\cref{sec:stats-of-interest}).

In contrast, recall that the exact computation that our procedure approximates
would require fitting
$\ngenes \times \ncells$ GLMs to determine
the effect of dropping each cell on each gene outcome---and
$\ngenes \times {\ncells \choose \ninfluence}$ GLMs
to exactly determine
the effect per gene of dropping any $\ninfluence$ cells.
For the sample dataset described above, these exact computations would
require almost \textbf{51 hours} (to measure the effect of dropping each cell)
and, e.g., $\mathbf{>2.5 \times 10^8\ years}$ (to measure the effect of dropping each cell
subset of size $\ninfluence=5$).\footnote{
    These time estimates are calculated based on
    \textbf{154 seconds} for the original analysis;
    \textbf{127 seconds} to refit after dropping each cell, one at a time
    (across 1440 cells); and
    \textbf{156 seconds} to refit after dropping each cell subset
    of size $\ninfluence=5$ (across $>5.1 \times 10^{13}$ possible subsets).
    Times are estimated based on dropping 100 random draws (of size 1 or 5) without replacement.

    Note that the time to refit increases as $\ninfluence$ increases and the
    weights move farther from where the optimization was initialized, at
    $\hat{\beta}(\vec{1})$---in addition to the combinatorial explosion in the number of possible subsets.
    For example, it takes \textbf{189 seconds} (cf. 156 seconds at $\ninfluence=5$)
    to refit after dropping each cell subset of size $\ninfluence=10$---and
    $\mathbf{>6.1 \times 10^{19}\ years}$ to fit all >$10^{25}$ possible subsets.
}

The key takeaway is that this procedure allows us to efficiently
quantify the dropping-data robustness of each gene-level outcome
by estimating the minimal fraction of cells that, when dropped, would
effect a meaningful change to each outcome, for each gene
(\cref{eq:amip-metric}).
For example, we can predict the minimal portion of cells we'd need to remove
in order to flip the sign, or erase the significance, of each gene's treatment effect.
Later, we'll extend these results to identify cell subsets that, when removed,
effect biologically meaningful changes to results \textit{across} genes
(\cref{sec:gsea-robustness}).

\subsection{pseudobulk} \label{sec:robustness-pseudobulk}

For the pseudobulk approach to scRNA-seq, there are two different weighted
estimators of interest: one to approximate
generalizability and robustness with respect to \textit{samples}
(e.g., particular tissue samples or subjects),
and one to approximate with respect to \textit{cells}.
In other words, the former can identify if a small number of samples
are driving differential expression results---perhaps due to
unmodeled biological variability (like background genotype)
or technical variability (like tissue preparation)
among \textit{samples}.
On the other hand, the latter can identify if a small number of cells
are driving differential expression results---perhaps due to
unmodeled biological variability (like cell cycling, or sub-types within a cell type)
or technical variability (like doublets)
among \textit{cells}.

For robustness with respect to \emph{dropping samples}
(i.e., the level at which single-cell measurements are aggregated),
the approach to calculating the sample influence matrix
$\mat{\influence}\ofSize{\nsamples, \ngenes}$ is identical
to the logic of \cref{sec:robustness-independent-cells}---but
replacing cell indices with sample indices
($\icell \rightarrow \isample, \, \ncells \rightarrow \nsamples$)
and cell observations with pseudobulk observations
($\vec{y} \rightarrow \pseudobulk{\vec{y}}, \,
  \mat{X} \rightarrow \pseudobulk{\mat{X}}, \,
  \vec{\size} \rightarrow \pseudobulk{\vec{\size}}, \,
  \vec{w} \rightarrow \pseudobulk{\vec{w}}
  $).
In other words,
the Z-estimator for the \textit{weighted} dataset is
\[
    \estimating_0 \ofBetaHatWPseudo* +
    \sum\overSamples
        \pseudobulk{w}_{\isample} \, \estimating_{\isample} \ofBetaHatWPseudo
    = \vec{0}
\]
(cf. the weighted estimator for independent cells, \cref{eq:w-estimating});
the terms of the estimating equation are
\[
    \estimating_{\isample} \ofBetaHatWPseudo
    \defeq
      \frac{ \pseudobulk{y}_{\isample} - \pseudobulk{\sizeFn}\left( \pseudobulk{\vec{w}} \right)_{\isample} \exp\left\{ \xbnpseudoHatW \right\} }
           { 1 + \disp \, \pseudobulk{\sizeFn}\left( \pseudobulk{\vec{w}} \right)_{\isample} \exp\left\{ \xbnpseudoHatW \right\} } \, \pseudobulk{\vec{x}}_{\isample}
\]
where $\pseudobulk{\vec{x}}_{\isample}$ are the consensus covariates across all cells in sample $\isample$
(cf. those for independent cells, \cref{eq:Gn-w}),
with pseudobulk sizes computed (as per \texttt{normed\_sum}) by
\[
    \pseudobulk{\sizeFn}\left( \pseudobulk{\vec{w}} \right) \defeq
        \total{\pseudobulk{\vec{y}}} \big/
        \exp\left\{ \frac{1}{\sum\limits_{\isample} \pseudobulk{w}_{\isample}}
                             \sum\limits_{\isample} \pseudobulk{w}_{\isample} \log \total{\pseudobulk{y}}_{\isample}
            \right\}
\]
(cf. \cref{eq:normed-sum-w});
and the data weight vector $\pseudobulk{\vec{w}}$ is
$\nsamples$-dimensional rather than $\ncells$-dimensional.

For robustness with respect to \emph{dropping cells},
the weight vector $\vec{w}$ remains $\ncells$-dimensional
(as per the individual cell model),
but the effect of data weights is more complex.
First, since the gradient no longer factorizes over cell weights,
the Z-estimator is the solution $\hat{\vec{\beta}}$ to the weighted estimating equation
\begin{equation}
    \estimating_0 \ofBetaHatW* +
    \sum\overSamples
        \estimating_{\isample} \ofBetaHatW
    = \vec{0}
    \label{eq:w-estimating-pseudobulk}
\end{equation}
(cf. \cref{eq:w-estimating}).
Then, \circled{2} from \cref{eq:influence-calc}
(the chain rule expansion of $\influence_{\icell}$)
is instead computed as
\begin{align*}
    \dd*{\hat{\beta}(\vec{w})}{w_{\icell}} \bigg|_{\vec{w}}
    =
    -&\left(
        \dd*{\estimating_0\ofBeta}{\vec{\beta}\T}
            \bigg|_{\hat{\beta}(\vec{w})}
        + \;
        \sum\overSamples
        \dd*{\estimating_{\isample}\ofBetaHatW}{\vec{\beta}\T }
            \bigg|_{\hat{\beta}(\vec{w}), \vec{w}}
     \right)\inv
\end{align*}
(cf. the formula for individual cells, \cref{eq:dbeta-dw}).

Finally, by necessity, sensitivity calculations \textit{must} incorporate
size factor estimation
(i.e., $\estimating_{\isample}$ must depend directly on $\vec{w}$,
 so $\;\star\;$ must be included in \cref{eq:dbeta-dw}).
Otherwise, by conditioning on a fixed $\vec{\size}$,
we'd diminish counts from a pseudobulk sample (by dropping cells)
without changing the overall size (``exposure'') of the sample.
The weighted analog of \cref{eq:size-pseudobulk} is
\begin{align}
    \total{\pseudobulk{\vec{y}}} &\defeq \mat{\pseudobulkSelector} \cdot (\vec{w} \odot \total{\vec{y}})
    \nonumber
\\
    \pseudobulk{\vec{\size}} &\defeq
        \total{\pseudobulk{\vec{y}}} \big/
        \exp\left\{ \frac{1}{\nsamples} \sum\limits_{\isample} \log \total{\pseudobulk{y}}_{\isample} \right\}
    \label{eq:size-pseudobulk-w}
\end{align}
---where cell weights $\vec{w}$ modulate the contribution of each cell's total
count to its corresponding pseudobulk sample (allocated via $\mat{Z}$),
and sample sizes are computed based on the resulting pseudobulk total counts.
Then, sample sizes are implicitly defined as a function of cell weights,
$\pseudobulk{\vec{\size}}=\pseudobulk{\size}(\vec{w})$.

Note that \cref{eq:size-pseudobulk-w}
assumes we never fully drop all cells in a sample;
i.e., the geometric mean is taken across all $\nsamples$ samples.
Because it is not straightforward to relax this assumption and retain differentiability,
and because our experiments in this work do not involve this model,
we do not address this assumption here.
For now, we recommend
separately analyzing dropping-data robustness with respect to \textit{samples} and
dropping-data robustness with respect to \textit{cells}, where all samples
retain at least one cell
(enforced as a constraint when enumerating the most influential cells).

With these cell weight dependencies in mind,
the terms of the estimating equation (\cref{eq:w-estimating-pseudobulk}) are
\[
    \estimating_{\isample}\ofBetaHatW
    \defeq
      \frac{ \pseudobulk{y}_{\isample} - \pseudobulk{\size}(\vec{w})_{\isample} \exp\left\{\xbnpseudoHatW\right\} }
           { 1 + \disp \, \pseudobulk{\size}(\vec{w})_{\isample} \exp\left\{\xbnpseudoHatW\right\} } \, \pseudobulk{\vec{x}}_{\isample}
\]
where $\pseudobulk{y}_{\isample}$ is the RNA count of the $\isample\th$
pseudobulk sample, for a given gene (column vector $\pseudobulk{\vec{y}}^{(\igene)}$)
of the weighted observed count matrix
$\pseudobulk{\mat{Y}} \defeq (\vec{w}\T \odot \mat{\pseudobulkSelector}) \, \mat{Y}$.

Thanks to autodiff, cell influence scores $\mat{\influence}\ofSize{\ncells, \ngenes}$
for the pseudobulk model still automatically shake out from
an analogous procedure
to that
described above
(\cref{eq:influence-calc} \& \cref{sec:robustness-independent-cells}).

\subsection{gene set enrichment robustness}\label{sec:gsea-robustness}

We've now
outlined a procedure to identify the most influential set of cells
(and quantify the effect of its removal) for each key gene-level outcome of
differential expression. This yields a measure of robustness, and an associated
set of influential cells, for each of thousands or tens of thousands of genes.

While this
allows us to scrutinize
individual differential expression results
at high resolution, what is lacking is the ability to \textit{zoom out} and characterize
robustness of the differential expression experiment as a whole---and
to identify a single set of influential cells whose removal would meaningfully
perturb the biological takeaway from the entire experiment.
To this end, we develop a procedure to approximate the robustness of a
key high-level outcome of differential expression---namely, the top 10
gene sets from a hypergeometric test for
enrichment
(of differentially expressed genes, across a defined collection of functionally related gene sets;
\cref{sec:gsea,app:gsea}).
This is particularly challenging because of the
inherently discrete, and therefore non-differentiable, nature of GSEA
(due to ranking and thresholding, and comparison of discrete sets),
as well as the combinatorial challenge of considering the impact of
dropping observations \textit{across} genes.

In order to perturb the composition of the top gene sets,
we need to perturb which genes are
selected as differentially expressed.\footnote{
    Some analyses further filter ``differentially expressed'' genes based on
    the magnitude of their estimated effect size, in additional to their
    significance.
    Here we lay out a dropping-data procedure for GSEA among genes
    filtered based on significance only, but note that future work could
    develop a procedure for GSEA based on both significance and minimal
    effect size filters by
    incorporating influence scores for log-fold change
    (functions of $\stat_{\text{LFC}}^+$)
    in addition to significance (functions of $\stat_W^+$)
    when clustering and scoring cells to drop.
    Alternately, the statistical test could be constructed against a null
    hypothesis of a minimal ``meaningful'' magnitude, rather than zero,
    and the resulting influences $\stat_W^+$ could be used to estimate
    dropping-data robustness of GSEA as described.
}
In other words, we seek to identify cells that, when dropped,
would \textit{demote} the top-ranking gene sets---by erasing
the significance of differentially expressed
genes that overlap with the top 10 gene sets---and
\textit{promote} lower-ranked gene sets---by bestowing significance upon
nonsignificant genes; preferably those that overlap with lower-ranked gene sets,
but not the top 10.

To this end, we develop a series of heuristics to
\begin{itemize}
    \item \emph{cluster} cell influence vectors
        to find groups of cells that, when dropped, act synergistically
        across genes-of-interest---especially genes that appear nonrobust in
        the direction-of-interest (toward erasing or bestowing significance), and
    \item \emph{score} cell clusters, based on their predicted effect
        on relevant genes and gene sets.
\end{itemize}
\vspace{\parskip}

These heuristics serve to greatly funnel the combinatorial number of possible
cell subsets into a small handful of influential subsets to verify
(by actually dropping cells and rerunning the analysis)
and---ultimately---to bound
the dropping-data robustness of the overall
biological conclusions drawn from differential expression.\footnote{
    In other words, we bound robustness by upper-bounding
    the size of the minimal influential cell subset that has
    the intended disruptive effect on the top gene sets
    (i.e., by identifying and validating that dropping a particular set of
    $K$
    cells changes the composition of the
    top-ranked gene sets as intended---while leaving open the possibility that
    a smaller set of cells with a similar effect may exist).
    \captionbr
    As a counterpart, we also bound robustness by lower-bounding
    the maximal disruption to the top gene sets that can be
    effected by dropping a set of cells of fixed size $K$
    (while leaving open the possibility that another set of $K$ cells
    with a more disruptive effect on top gene sets may exist).
}

Throughout the remainder of \cref{sec:gsea-robustness}, we will make
pronouncements about heuristic settings that are ``better'' or ``worse''
than other settings.
Here, we
explicitly define the comparator for these pronouncements.
Let $\genesetPerturbationAtK_{\text{method}}$ be the maximal number of top 10
gene sets that are perturbed (i.e., displaced from the top 10, and replaced by
originally lower-ranked gene sets) when a cell cluster of size $K$,
among those identified
through algorithmic choice
``$\text{method}$,''\footnote{
    Where, concretely, the particular
    methodological decision
    is evaluated
    while holding the rest of
    our algorithm
    (laid out in \cref{sec:gsea-clustering,sec:gsea-scoring,sec:gsea-algo})
    constant
}
is dropped.
If
$\genesetPerturbationAtK_{\text{method A}} \geq \genesetPerturbationAtK_{\text{method B}}$
for all $K$ tested
(across GSEA of both downregulated and upregulated genes),
then method A is \textit{strictly superior} to method B
(or, B is \textit{strictly worse} than A)---unless this quantity is
precisely equal across all $K$, in which case A and B are \textit{equivalent}.
Otherwise, if
$\genesetPerturbationAtK_{\text{method A}} \geq \genesetPerturbationAtK_{\text{method B}}$
for the \textit{majority} of settings of $K$ tested
(across GSEA of both downregulated and upregulated genes),
then method A is \textit{superior} to method B (or, B is \textit{worse}).
In other words, \emph{superiority} connotes a method that generally
leads to a tighter empirical bound
on the maximal disruption to the top 10 gene sets
by dropping a given number of cells.
Throughout this section,
we evaluate superiority with respect to the dataset described in \cref{app:data};
future work should evaluate a wider variety of RNA-seq datasets in order to
verify how well these settings generalize
(or, to define a procedure to estimate good heuristic settings,
leading to tighter empirical bounds,
based on characteristics of the data).

\subsubsection{filtering genes}\label{sec:gsea-filtering-genes}

In order to estimate the dropping-data robustness of GSEA,
we first
identify the genes of highest interest:
\begin{itemize}
    \item $\geneset{top 10}$, the set of all genes (tested
        for differential expression) that overlap with the top 10 pathways
        (from the original gene set enrichment analysis), and
    \item $\geneset{top 11--\ngenesets}$, the set of all genes
        that overlap with the top 11 to $\ngenesets$ pathways but not the top 10.
\end{itemize}
\vspace{\parskip}

Here, $\ngenesets$ is chosen to balance \textit{focus} on a smaller number
of target gene sets (such that progress toward
promoting genes into the differentially expressed set
is not diluted over too many disparate genes, or too many disparate gene sets)
with \textit{inclusivity} of potential targets
(such that gene sets that include many genes on the cusp of
significance, with respect to dropping-data sensitivity, are not excluded).
We find that $\ngenesets=30$ works well.

(Specifically,
we find that $\ngenesets=20$ is worse,
and $\ngenesets=40$ is strictly worse, than $\ngenesets=30$.
This hard filter at $\ngenesets=30$ is also strictly superior to
a soft-filter approach to gene feature selection---i.e.,
applying a very lax filter up front
($\ngenesets=100$)
and, later, applying
decaying weights
based on gene set rank to prioritize
those that are closer to the top 10.\footnote{
    Specifically, we experiment with setting $\geneWeights_{\igene}$ in
    \cref{eq:greedy-cluster}, for genes $\igene$ targeted for upranking,
    based on gene set weights $\genesetWeights_{\igeneset}$ (summed over all
    relevant gene sets $\igeneset=11,\ldots,\ngenesets$ that contain gene $\igene$).
    We try
    calculating gene set weights
    based on linear decay
    ($\genesetWeights_{\igeneset} = 1 - \frac{\igeneset - 11}{\ngenesets - 11}$),
    power-law decay
    ($\genesetWeights_{\igeneset} = \left[ 1 - \frac{\igeneset - 11}{\ngenesets - 11} \right]^2$),
    or exponential decay
    ($\genesetWeights_{\igeneset} = \exp\left[ - \frac{\igeneset - 11}{\ngenesets - 11} \right]$).
    Alternately, when $\genesetWeights_{\igeneset} = 1$ for all $\igeneset$, we
    recover our standard method (described below; \cref{eq:greedy-cluster}).
})

In order to cluster cells, we further filter these genes-of-interest to
identify those that we seek to \textit{demote from} or \textit{promote to} the
differentially expressed target set
(i.e., genes with significant treatment effects in the relevant direction,
since we separately assess enrichment for genes that are upregulated versus
downregulated among treated cells)---with a reasonable chance of success after
dropping $K$ cells. Namely, we identify
\begin{itemize}
    \item $\geneset{demote}^K$, the subset of genes in
        $\geneset{top 10}$ that are already targets---i.e., already significant,
        with treatment effects in the relevant direction---but on the verge
        of being knocked out---i.e., (we estimate) require dropping
        $\leq \lceil K/2 \rceil$ cells to erase significance; and
    \item $\geneset{promote}^K$, the subset of genes in
        $\geneset{top 11--\ngenesets}$ that are not targets---either
        because they have treatment effects in the relevant direction but
        are not significant, \textit{or} because they have treatment effects
        in the opposite direction---but on the verge of being knocked in---i.e.,
        require dropping $\leq \lceil K/2 \rceil$ cells to
        bestow significance \textit{or} flip the sign of the effect with
        significance, respectively.
\end{itemize}
These sets of genes are decorated with a $K$ to emphasize that the
gene features selected for clustering vary with the intended number of cells to drop.

As with $\ngenesets$ above, the threshold $\lceil K/2 \rceil$
(for minimal cells to drop in order to effect the change-of-interest)
is chosen to balance \textit{focus}
(on a smaller number of the most relevant genes)
with \textit{inclusivity}
(of potentially impactful genes that can be knocked in or out of the target set
within the given cell ``budget'').
We experiment with this cutoff and find it provides good results
(i.e., leads to discovery of clusters with maximal observed disruption of gene
sets) across a range of $K$s
(superior to,
e.g., not enforcing a threshold, setting the threshold to $K$,
or setting the threshold to a low fixed value like two or three).
In addition to improving results, this filtering step speeds clustering by
reducing the number of gene features by one to two
orders of magnitude.

\subsubsection{clustering cells}\label{sec:gsea-clustering}

Having selected gene features for clustering, we
filter the influence matrix $\mat{\influence}$
(with respect to the unsigned Wald statistic $\stat_W^+$; \cref{sec:stats-of-interest})
to these gene columns and
cluster the rows to find sets of $K$ cells that,
when dropped, act synergistically across genes.
Specifically, starting with each cell as a seed, we iteratively and greedily
add cells to the cluster based on heuristics (described below) intended to
prioritize cells that, together, will maximally disrupt top gene
sets.\footnote{
    We find empirically that this seed-based approach (to greedily choose cells
    $2,\ldots,K$, starting with each cell as a seed) is strictly superior to
    fully greedy selection (to identify a single cluster of cells $1,\ldots,K$).
}

Let $\mathbb{\ncells}$ be the set of all cells and let
$\mathbb{K}$ be the set of all cells in the cluster so far.
Then, the next cell we'd add to the cluster is
\begin{equation}
    \argmax\limits_{\icell \,\in\, \mathbb{\ncells} \setminus \mathbb{K}}
    \; \sum\limits_{\substack{
            \setlength{\jot}{-0.95\baselineskip}
            \begin{aligned}
                \igene \,\in\, &\geneset{promote}^K \,\cup \\
                               &\geneset{demote}^K
            \end{aligned}
        }} \,
        \influence^{(\igene)}_{\icell}
        \,\times\,
        \signPattern_{\igene}
        \,\times\,
        \geneWeights_{\igene}
        \,\times\,
        \isRelevant_{\igene}
    \label{eq:greedy-cluster}
\end{equation}
for
\begin{itemize}
    \item \emph{sign pattern} $\vec{\signPattern}$,
        a vector with entries $\signPattern_{\igene} = \pm 1$
        such that multiplication with
        $\mat{\influence}$
        ensures that more positive influences correspond to cells
        whose removal would push the corresponding genes in the
        most disruptive direction (with respect to top gene sets).
        Recall that we have already classified genes based on changes
        that would be maximally impactful to top gene sets from GSEA (\cref{sec:gsea-clustering}):
        those we are targeting to \textit{erase significance} (all genes in $\geneset{demote}^K$),
        those we are targeting to \textit{bestow significance} (some genes in $\geneset{promote}^K$),
        and those we are targeting to \textit{flip sign with significance} (the remaining genes in $\geneset{promote}^K$).
        Then, since the most influential cells are those with the most negative
        influence scores (\cref{sec:amip}),
        and based on the affine transformations of $\stat_W^+$ to calculate the
        statistics-of-interest (\cref{sec:stats-of-interest}),\footnote{
            Such that $\nicefrac{\d\stat}{\d\stat_W^+}$ is
            -1 for $\stat_{\text{erase significance}}$,
            +1 for $\stat_{\text{bestow significance}}$, and
            -1 for $\stat_{\text{flip sign w/ significance}}$
        }
        the sign pattern vector is comprised of entries
        \[
            \signPattern_{\igene} =
            \begin{cases}
                +1 & \text{if the }\igene\th\text{ gene is being targeted to \textit{erase} significance} \\
                -1 & \text{if the }\igene\th\text{ gene is being targeted to \textit{bestow} significance} \\
                +1 & \text{if the }\igene\th\text{ gene is being targeted to \textit{flip sign with} significance.}
            \end{cases}
        \]

    \item \emph{gene weights} $\vec{\geneWeights}$, a positive vector weighting
        genes by their impact on top gene sets.
        Namely, $\geneWeights_{\igene}$ is the number of top 10 gene sets that include gene $\igene$
        (if $\igene$ is targeted to be knocked \textit{out} of the differentially expressed set)
        or else the number of top 11--$\ngenesets$ gene sets that include $\igene$
        (if $\igene$ is targeted to be knocked \textit{in}).
        After observing that cells whose exclusion has an outsized
        impact on top gene sets often
        disrupt
        multiple gene sets with similar biological function,
        we decided to intentionally incorporate this into the clustering process.

    \item \emph{gene selector} $\vec{\isRelevant}$, which acts to iteratively
        eliminate genes as features once we estimate that ``success'' has been
        achieved by dropping cells in the cluster so far.
        Namely, for each round of clustering, we compute
        \[
            \isRelevant_{\igene} =
            \begin{cases}
                0 & \text{if }\sum\limits_{\icell \,\in\, \mathbb{K}} \influence_{\icell}^{(\igene)} \geq -\stat(\vec{1})
                \\
                1 & \text{otherwise}
            \end{cases}
        \]
    where $\stat\ofW$ is either $\stat_{\text{erase significance}}$,
    $\stat_{\text{bestow significance}}$, or $\stat_{\text{flip sign w/ significance}}$
    (\cref{sec:stats-of-interest}),
    depending on which change is being targeted (\cref{sec:gsea-filtering-genes}).
\end{itemize}
\vspace{\parskip}

We find through ablation that our greedy objective is superior to
eliminating any individual element (like weighting by pathways via $\vec{\geneWeights}$,
or iteratively selecting
features as success is achieved via $\vec{\isRelevant}$).
We also experiment with variations---like weighting by overlapping (summed)
gene set completeness rather than the overlapping number of gene sets,
stricter thresholds for filtering features,
and recomputing the sign vector over the course of clustering rather than
designating a fixed
pattern---and find that the method we describe here is superior
(with a couple notable exceptions at particular $K$; \cref{app:gsea-alt}).

\subsubsection{scoring \& verifying clusters}\label{sec:gsea-scoring}

The output of the step above is $\ncells$ soft clusters\footnote{
    i.e., comprised of overlapping cells
}
whose impact we then score in order to prioritize a few clusters for verification
(by actually dropping cells).

First, using influences as a linear approximation,\footnote{
    To compute the predicted Wald statistic for each gene after
    dropping a given set of cells, then using these predicted statistics to
    compute, rank, and BH-correct predicted p-values
}
for each cluster we predict which genes would be ruled differentially expressed---i.e.,
significant, with a treatment effect in the specified direction---after
dropping those $K$ cells. Call these predicted targets (after dropping cells)
$\geneset{DE}^w$ and the original targets $\geneset{DE}$.
We then compute a score per cluster of
$\left( \ref{it:score-combined},\: \minus\,\ref{it:score-top10},\: \ref{it:score-top11+} \right)$,
where
\begin{enumerate}
    \item is the combined productive change in targets compared to the original analysis,
        \\$
            \max \left( 0, \;
                | \geneset{top 10} \cap \geneset{DE} | -
                | \geneset{top 10} \cap \geneset{DE}^w |
                \right)
            \; + \;
            \max \left( 0, \;
                | \geneset{top 11--\ngenesets} \cap \geneset{DE}^w | -
                | \geneset{top 11--\ngenesets} \cap \geneset{DE} |
                \right);
        $
        \label{it:score-combined}

    \item is the number of predicted gene targets that overlap with the original top 10 gene sets,
        \\$
            | \geneset{top 10} \cap \geneset{DE}^w |;\text{ and}
        $
        \label{it:score-top10}

    \item is the number of predicted gene targets that overlap with the original top 11--$\ngenesets$ gene sets,
        \\$
            | \geneset{top 11--\ngenesets} \cap \geneset{DE}^w |.
        $
        \label{it:score-top11+}
\end{enumerate}
In other words, we seek to minimize the overlap of targets (after dropping cells)
with the original top 10 gene sets and to maximize the overlap with gene sets
that are originally ranked below, but not too far beyond, the top 10.
We rank scores according to the first component ($\ref{it:score-combined}$),
using successive components to break ties.

Next, we further characterize
the highest-scoring clusters
by directly predicting the effect on the top 10 gene sets.
Whereas the above step (to compute $\ref{it:score-combined},\: \ref{it:score-top10},\: \ref{it:score-top11+}\,$)
is computable in seconds, since it only requires arithmetic
on precomputed
influences followed by a single ranking of genes,
this step is a bit more expensive since it requires computing many exact (hypergeometric) tests.
For this reason, we only score gene sets for the top 50 cell clusters (based on the criteria above).
For those clusters, we compute
\begin{enumerate}[resume]
    \item the number of gene sets in the top 10 that are predicted to be perturbed
        (i.e., displaced and replaced by new gene sets)
        \label{it:score-genesets}
\end{enumerate}
where, for efficiency, we only run hypergeometric tests (on estimated targets)
for the top 100 gene sets from the original analysis
(versus, e.g., $\apprx$8000 gene sets in the
\texttt{GO:BP} collection of gene sets related to biological
processes~\cite{go-2000,go-2020}).

Finally, based on overall scores
$\left( \ref{it:score-genesets},\: \ref{it:score-combined},\: \minus\,\ref{it:score-top10},\: \ref{it:score-top11+} \right)$,\footnote{
    Again ranked according to the first component,
    using successive components to break ties
}
we test the top clusters
to verify their effect\footnote{
    By refitting $\hat{\vec{\beta}}=\hat{\beta}(\vec{w})$
    (where $\vec{w}$ corresponds to dropping the cells-of-interest),
    selecting differentially expressed targets based on BH-corrected Wald tests,
    and running hypergeometric enrichment tests to identify the actual
    top gene sets when those cells are dropped
} and
to estimate (or, more precisely, to bound)
the maximal impact on the top 10 gene sets by dropping $K$ cells.
This is the most expensive step in this process,
where the time is dominated by refitting $\ngenes$
sets of coefficients.
For a dataset with $\ncells \approx 10^3$ and $\ngenes \approx 10^4$,
refitting $\hat{\vec{\beta}}$ across all $\ngenes$ GLMs takes
about a minute.
So, testing the 10 top-scoring clusters across
six settings of $K$ (to cover the grid we recommend in \cref{sec:gsea-algo})
would require about an hour.
We make two concessions to cut
this time down (to $\apprx$10 minutes total)
while retaining equivalent results.

First, rather than testing 10+ clusters, we test the top
\textit{two} (for clusters of size $K < 1\% \times \ncells$)
or top \textit{six} (for clusters of size $K \geq 1\% \times \ncells$).
We
test more clusters (by actually dropping cells) at the highest
settings of $K$ because accuracy of the dropping-data robustness approximation
(upon which scores are based)
degrades with increasing $K$
(since the corresponding data weight is farther from where the
approximation was formed, at $\vec{w}=\vec{1}$, and there are more opportunities
for the linearity assumption across data points to be violated).
Indeed, we observe empirically that this gene-level prediction is borne out by
cross-gene-set-level results (in other words, that the cluster with the largest
``actual'' perturbation to the top 10 gene sets tends to be lower-ranked,
based on prediction-based scores, for clusters at larger $K$).
While this schematic (two for $K < 1\% \times \ncells$; else six)
proved effective for our particular dataset---leading to the discovery
of equivalent disruption to the top gene sets as testing the top 15 clusters,
across $K$ and across up- and down-regulated genes---we note that future work
could focus on tuning this and other heuristic choices for future datasets.
For example, the accuracy of gene-level approximations
when a given fraction of influential cells are dropped
(as we later plot in \cref{fig:pred-v-actual})
could be used to set the number of top-scoring clusters to test at each size $K$.

Second,
when verifying clusters,
we save time by only refitting
$\hat{\vec{\beta}}$ for genes up to the maximal rank of the gene whose
significance status is predicted to be affected by dropping cells.
Very roughly, this shortcut halves the time to re-fit coefficients
and re-rank genes after dropping a cluster of cells,
which scales roughly linearly with $\ngenes$.

Finally, for the cluster with the maximal estimated impact,
we fit $\hat{\vec{\beta}}$ across the remaining genes to
verify
the effect of dropping those cells on GSEA.\footnote{
    Anecdotally, across all runs to date, this step has yet
    to contradict the estimated number of gene sets
    disrupted based on fitting only the described subset of genes.
}
The outcome serves to lower-bound the
disruption to the top 10 gene sets
by dropping $K$ cells or, equivalently, to upper-bound the minimal number of
cells to drop in order to
effect the observed disruption to the top 10 gene sets.

We confirm empirically
that this accelerated process
(scoring with gene sets for the top 50 clusters, and verifying two to six)
yields clusters that are as influential as---yet much less time-consuming
than---scoring with gene sets for all $\ncells$ clusters and verifying the top 50.

\subsubsection{high-level algorithm} \label{sec:gsea-algo}

Putting these steps together,
we first identify the genes that are most relevant
to the composition of the top 10
gene sets, $\geneset{top 10}$ and $\geneset{top 11--\ngenesets}$.
Then, across a grid of
settings of $K$
(e.g., $K \in \{\, \lfloor 2\% \times \ncells \rfloor,
                   \lfloor 1\% \times \ncells \rfloor,
                   \lfloor 0.5\% \times \ncells \rfloor,
                   \ldots, 1 \,\}$),
we further filter gene features
(to those whose differential expression status can reasonably be flipped in the
desired direction by dropping $\Lt K$ cells)
and apply greedy iterative clustering to generate $\ncells$ cell clusters of size $K$.
We score these clusters based on their predicted impact and, ultimately,
verify a few to find the cluster of a given size with the maximal impact on the top 10 gene sets.

While our approximate algorithm does not have guaranteed error bounds,
we show that---in practice---it is sufficient to identify meaningful
dropping-data sensitivity for gene set enrichment analysis of real scRNA-seq
data (\cref{sec:exp-gsea}).

%% file: 5-experiments.tex
\newlength{\volcanoWidth}
\setlength{\volcanoWidth}{0.47\textwidth}
\newlength{\goHeight}
\setlength{\goHeight}{2.3in}
\newlength{\vennHeightTaller}
\setlength{\vennHeightTaller}{4cm}
\newcommand{\standout}{\textbf}
\newcommand{\makeSubcaption}[1]{
    $\;$ drop \textit{#1}
}

As demonstration, we'll start by presenting results
from a single-cell RNA-seq study of ulcerative colitis (UC)~\cite{data-uc}.
In this dataset, \emph{treatment} is the natural biological ``perturbation'' of
disease---i.e., cells from subjects with UC.
Specifically, we examine differential expression
within goblet cells (based on the original authors'
annotations)
to compare cells that are ``healthy'' versus ``inflamed,''\footnote{
    Ignoring the third health status ``non-inflamed''
} where $\ncells=1440$ cells and $\ngenes=10,502$ genes with at least
10 nonzero counts (reduced from 20,028 total genes measured).\footnote{
    This (fairly non-stringent) threshold eliminates irrelevant genes that are
    not meaningfully expressed under either condition,
    and would be very unlikely to contain sufficient signal to detect
    a difference between groups.
    This sort of filtering step is common when analyzing
    RNA-seq data;
    see \cref{fn:gene-filter}.
    \label{fn:gene-filter-exp}
}
Cells are sampled from 12 healthy subjects and 14 subjects with UC.
We examine differential expression within this cell type for the model
where
$$
    \mu_{\icell} = \size_{\icell} \, \exp\left\{
        \beta_0 + \beta_1 \, \texttt{nUMI\_scaled}_{\icell} + \beta_2 \, \texttt{Health}_{\icell}
    \right\}.\footnote{
    Equivalently, \texttt{y $\sim$ nUMI\_scaled + Health}
}
$$
The covariate $\texttt{nUMI\_scaled}_{\icell}$ is the total number of transcripts
(UMIs) for cell $\icell$, standardized across cells
(i.e., centered and scaled to unit variance) as per \texttt{DESeq2}'s advice
for variables with a large range.
The covariate $\texttt{Health}_{\icell}$ encodes the health status of cell $\icell$;
zero if the cell is sampled from healthy tissue, and one if it is sampled from
inflamed UC tissue.
After fitting this model
to estimate $\betaTreated$ ($= \beta_2$)
for each gene,
and determining the overall set of genes that are significantly
differentially expressed in goblet cells,
we compute influence scores and measure the robustness of these results.

In this section,
we show that dropping-data robustness yields insight that is distinct from that
revealed by classical tools for robustness that are already employed for
differential expression (\cref{sec:exp-other-metrics}).
Using our efficient approximation,
we report widespread dropping-data sensitivity for gene statistics
related to
treatment effect size and significance for the UC dataset (\cref{sec:exp-uc-genes})
and show that this approximation is accurate within the regimes we care about (\cref{sec:exp-accuracy}).
Further, we find that dropping-data sensitivity extends to high-level takeaways
from differential expression, in that a meaningful portion of the top gene sets
from GSEA
can be disrupted by dropping a handful of cells (\cref{sec:exp-gsea}).
We close by delving into how to interpret
dropping-data sensitive results
like these
in the context of differential expression (\cref{sec:exp-interpretation}).

\subsection{differential expression results are comparable across tests} \label{sec:exp-tests}

First, we find that results across significance tests---likelihood ratio (\emph{lr}),
Wald with Fisher estimator (\emph{Wald Fisher}), and
Wald with sandwich estimator (\emph{Wald sandwich})---are
largely equivalent.\footnote{
    P-values are >98--99\% correlated across genes (on linear or log scales, BH-corrected or not),
    and significant genes (BH-corrected $p < 0.01$) are highly overlapping:

    \includegraphics[height=\vennHeight,clip,trim={0 0.5cm 0 2cm}]{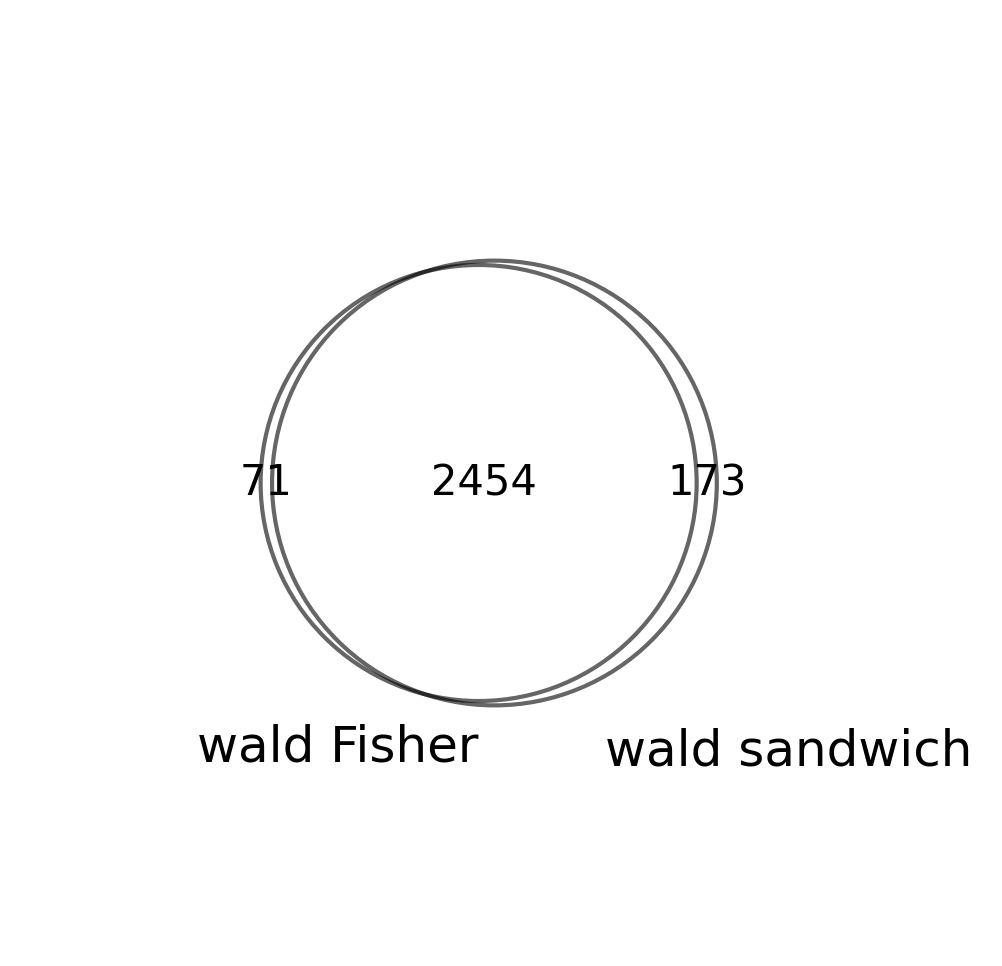}
    \includegraphics[height=\vennHeight,clip,trim={0 0.5cm 0 2cm}]{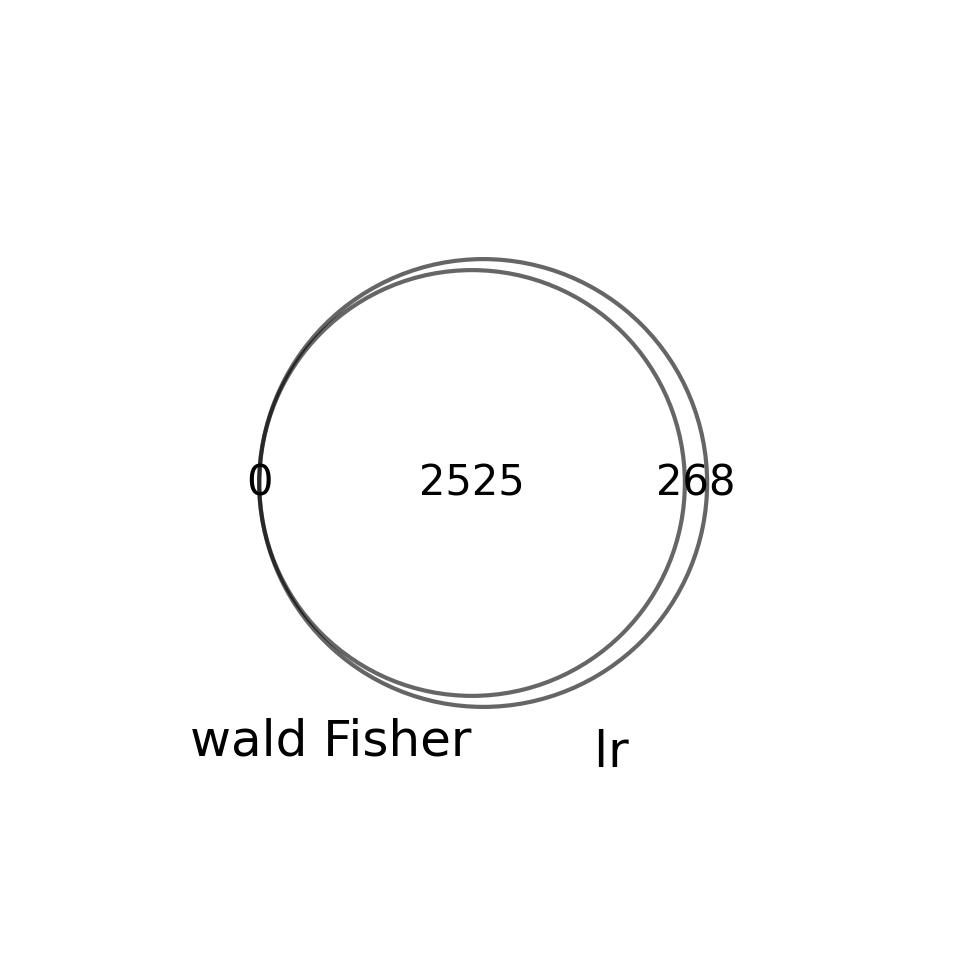}
    \includegraphics[height=\vennHeight,clip,trim={0 0.5cm 0 2cm}]{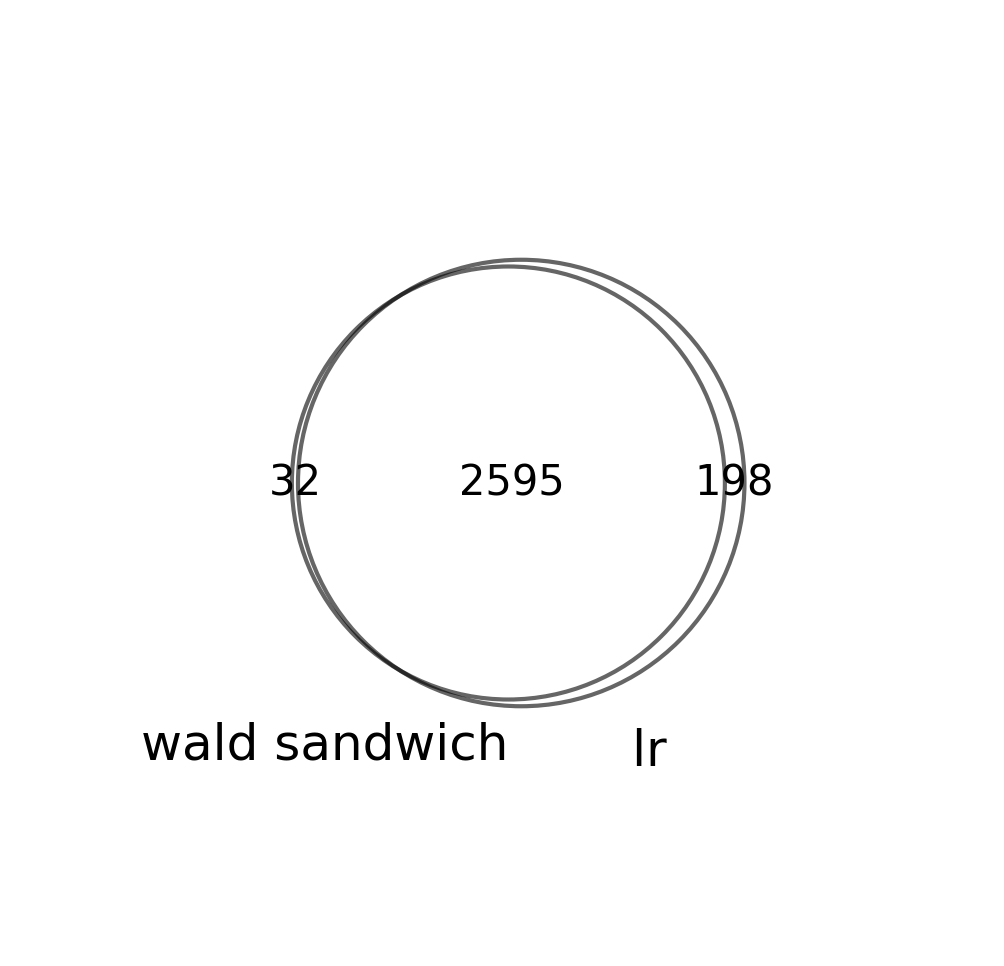}
}
This observation suggests that---at least for this dataset---we are nearing
the asymptotic equivalency between tests (\cref{sec:testing}).
So, though we will only directly approximate sensitivity of the Wald tests,
these results
are comparable to results under the LR test.\footnote{
    Recall that LR is the only test for \texttt{glmGamPoi}, and is recommended
    for single-cell data by \texttt{DESeq2}, whereas Wald Fisher is the default
    test for \texttt{DESeq2}.
    On the other hand, among Wald tests, the Wald sandwich is the statistically
    preferable
    choice (since it does not rely on the model being well-specified,
    which we know \textit{a priori} to be false;
    \cref{sec:we-dont-do-ql,app:cov-estimators}).
}

\subsection{dropping-data robustness reveals trends that are distinct from standard metrics} \label{sec:exp-other-metrics}

We hypothesize that dropping-data sensitivity will reveal patterns
in the robustness of differential expression outcomes that are
distinct from tools
typically used to rank and threshold differentially expressed genes;
namely, effect size and significance
(via standard error as well as multiple-testing correction).

\begin{figure}[hp!]
    \centering
    \includegraphics[width=\volcanoWidth]{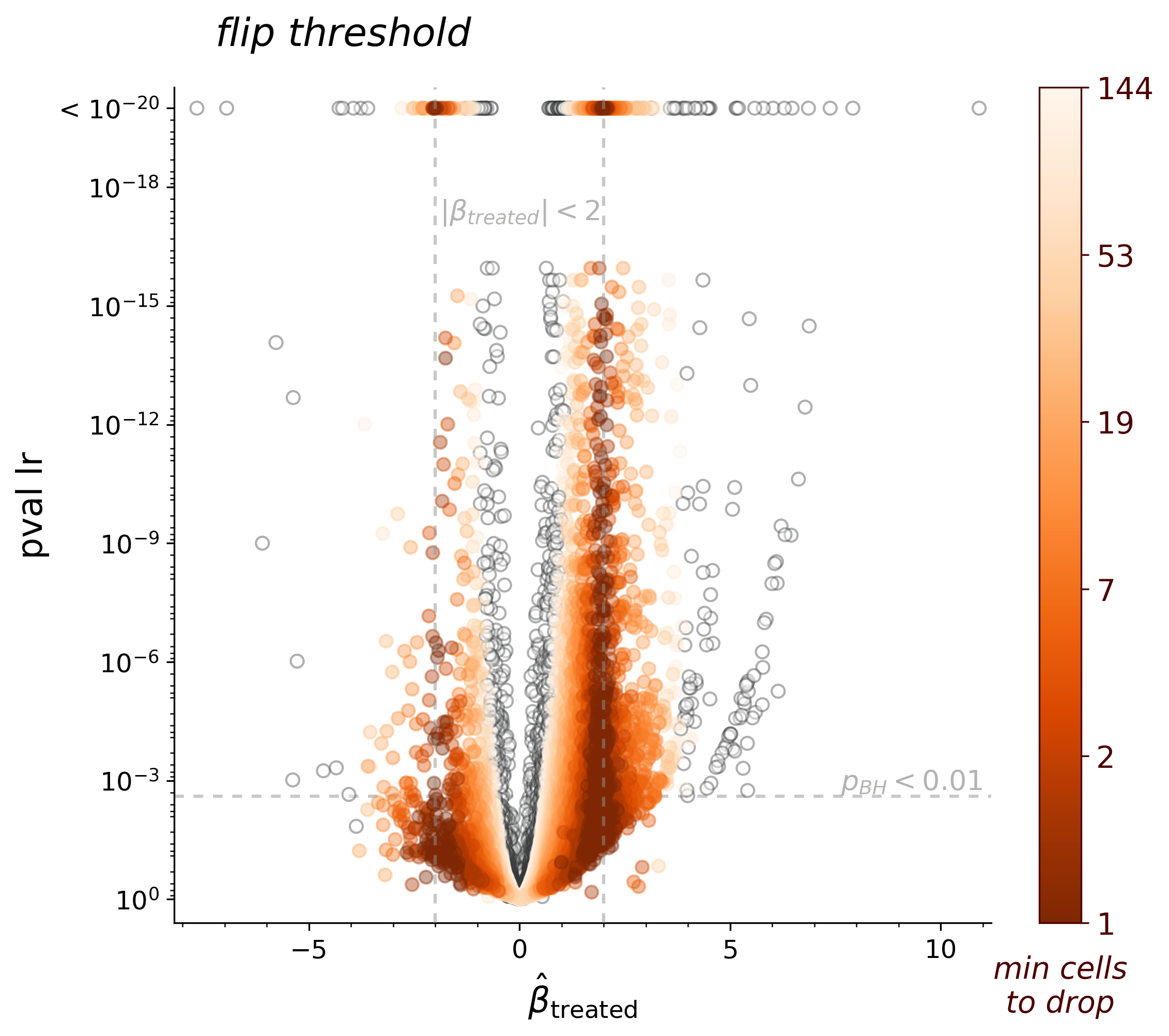}%
    \hfill%
    \includegraphics[width=\volcanoWidth]{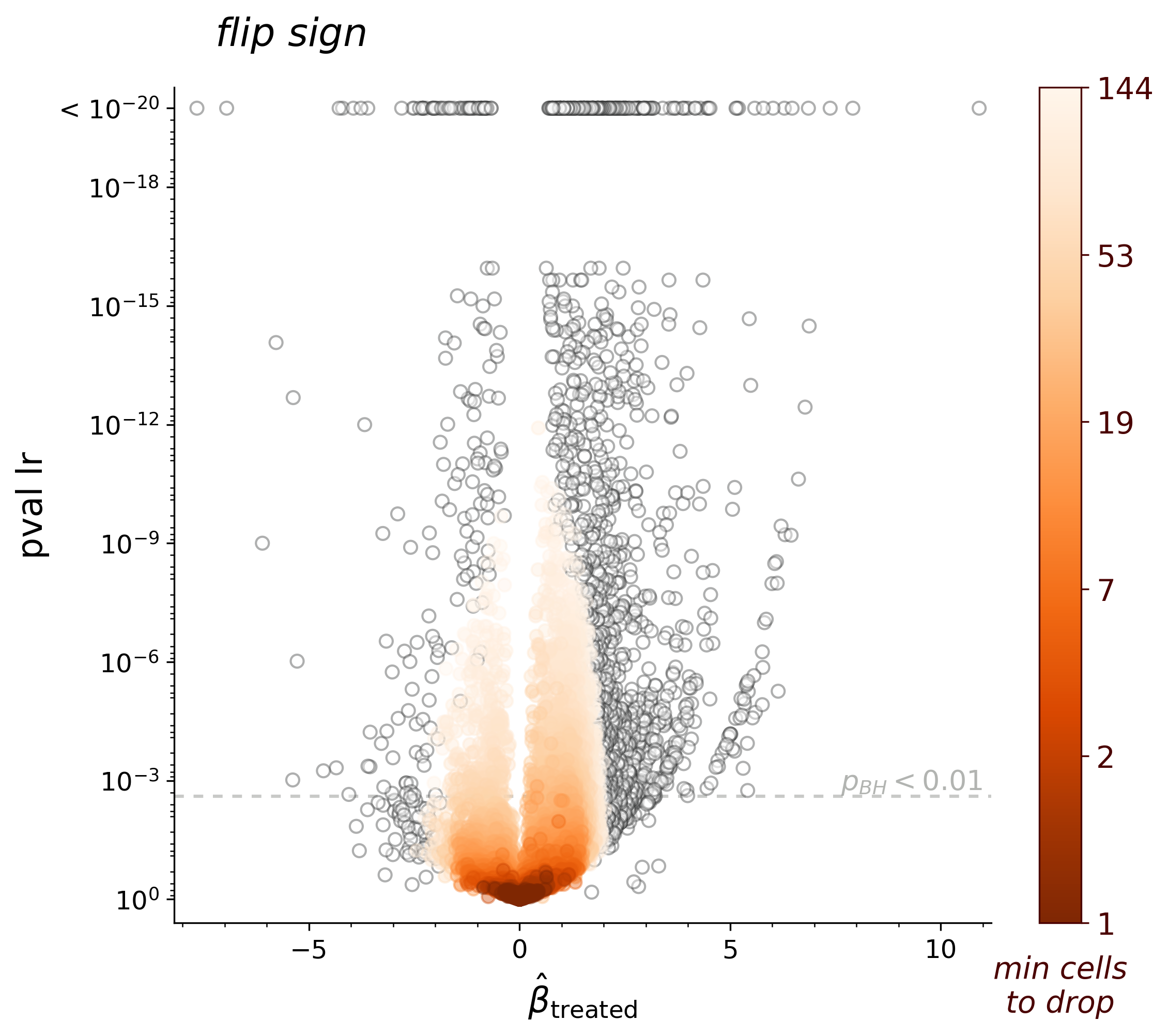}%
    \vspace{\vspaceBwRows}
    \includegraphics[width=\volcanoWidth]{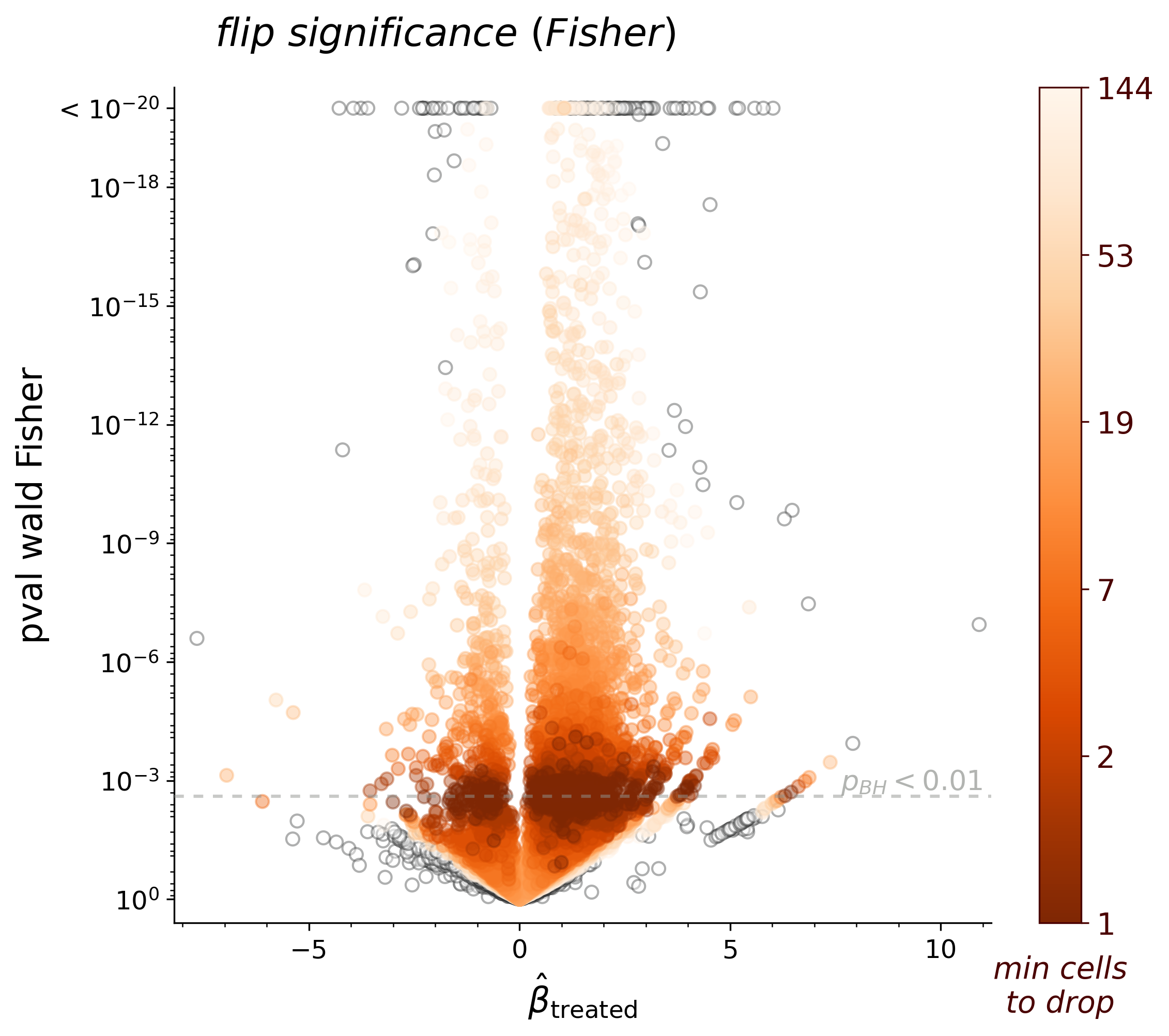}%
    \hfill%
    \includegraphics[width=\volcanoWidth]{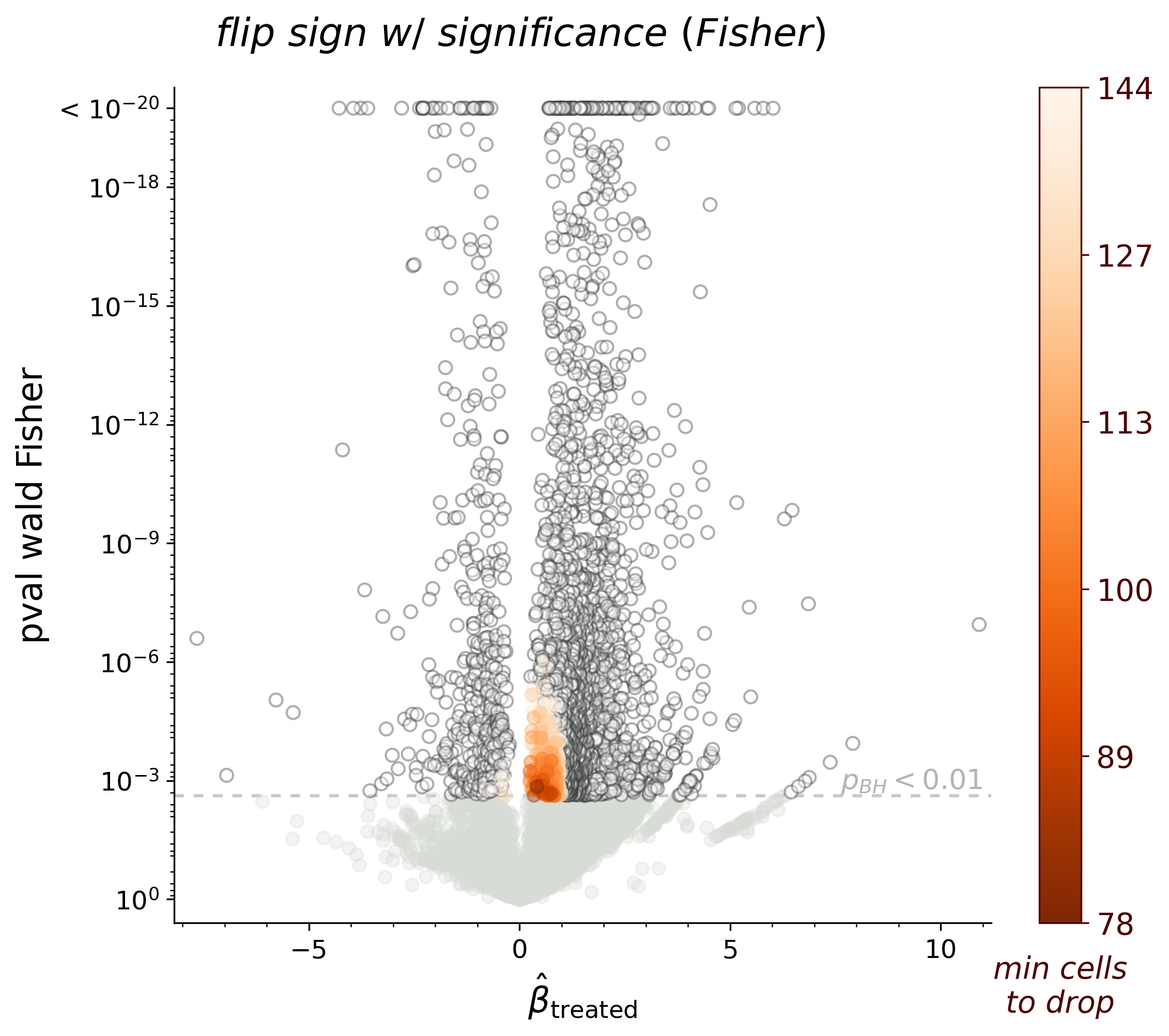}%
    \vspace{\vspaceBwRows}
    \includegraphics[width=\volcanoWidth]{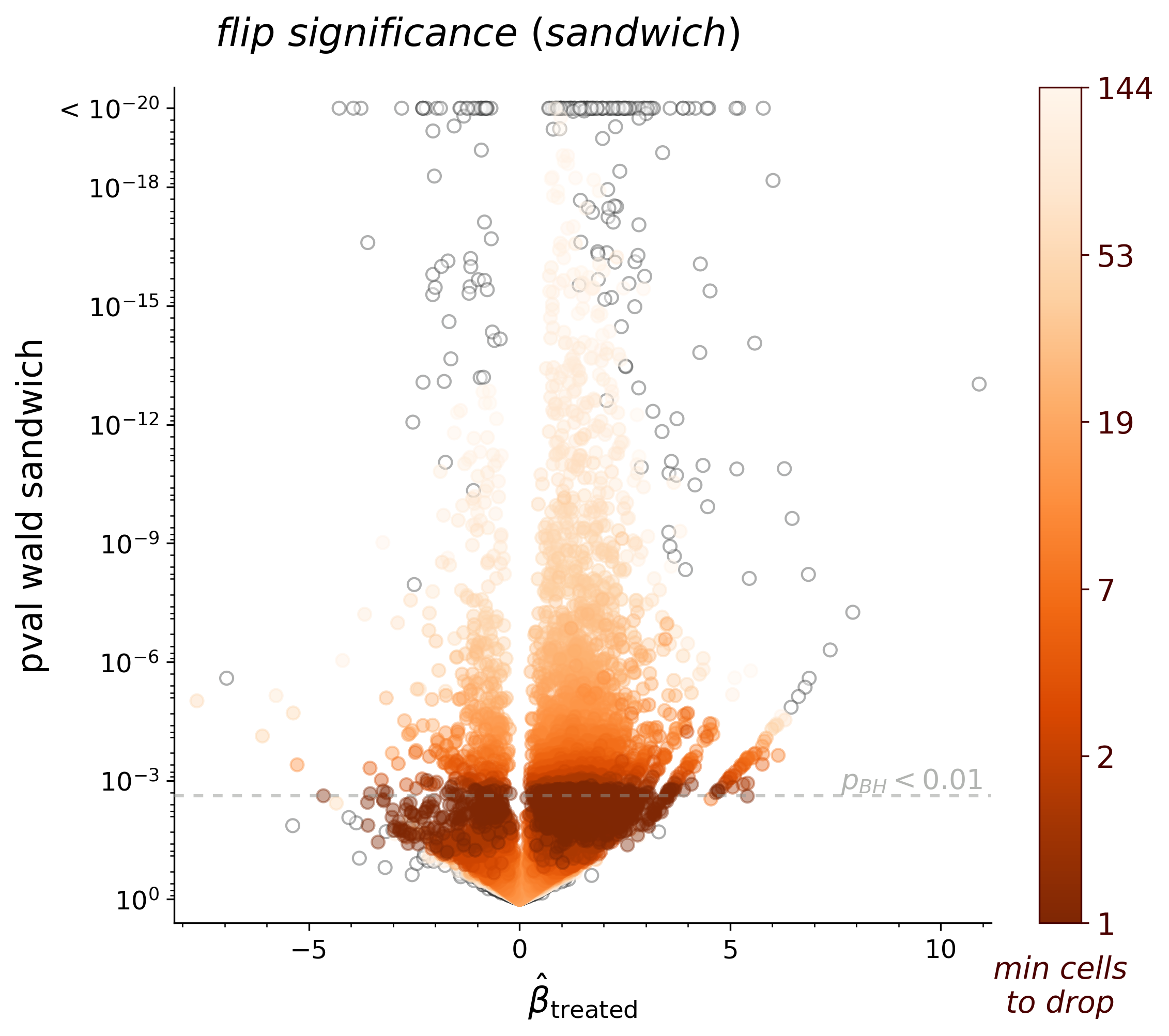}%
    \hfill%
    \includegraphics[width=\volcanoWidth]{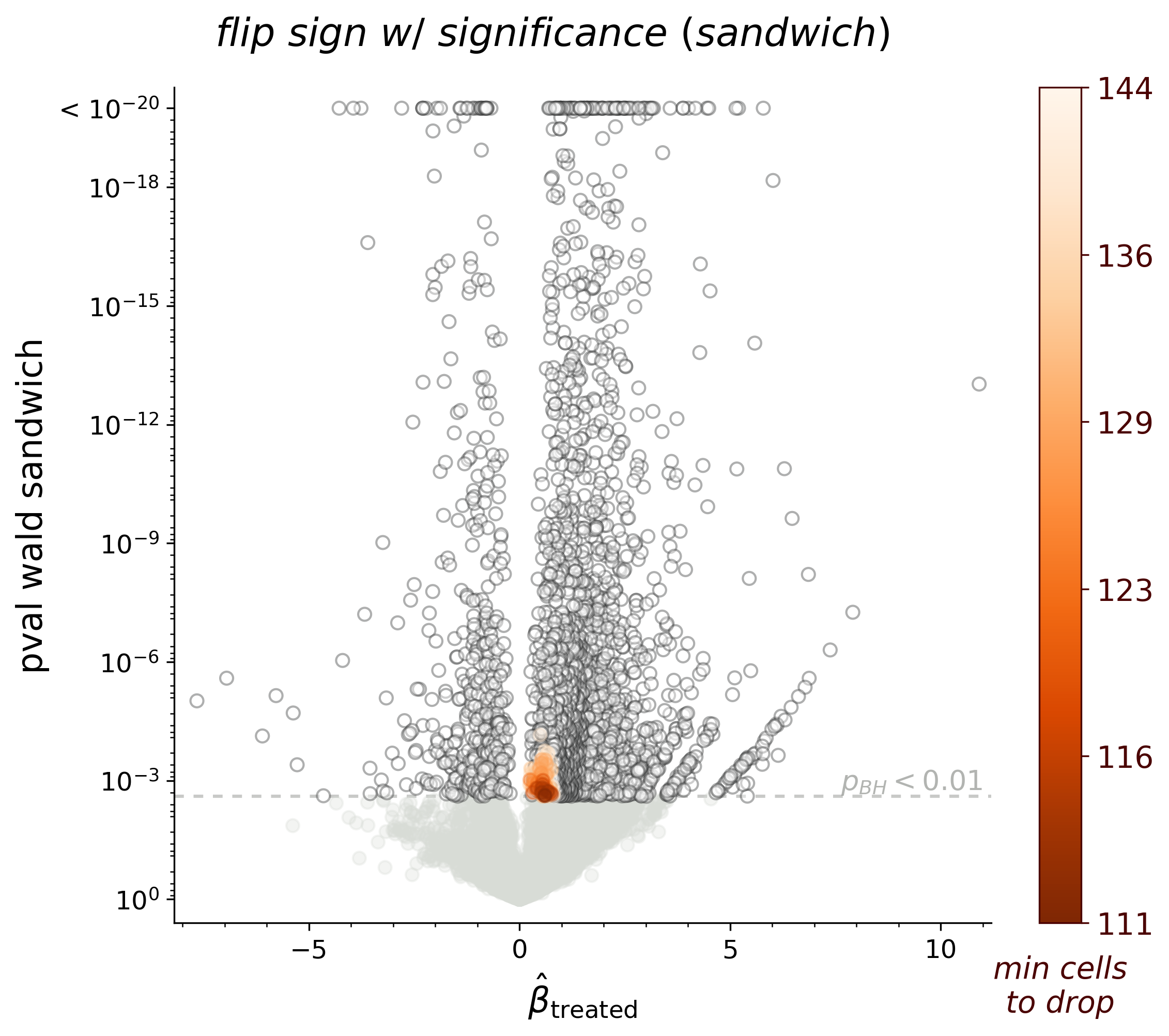}%
    \caption{
        \textbf{Minimal number of cells to drop to enact the change-of-interest, across genes.}
        Volcano plots of effect size (on a $\log_2$ scale) versus p-value
        (for the test indicated on the y-axis), for the dataset ($\ncells=1440$ cells)
        described in \cref{app:data}.
        Genes (\textit{points}) are colored by the size of the minimal cell
        subset---up to 10\% of cells---that, when dropped, are predicted to
        effect the change-of-interest (\textit{title}).
    } \label{fig:min-to-drop}
\end{figure}

To this end, we compute sensitivities and estimate the minimal number
of cells to drop in order to flip various outcomes-of-interest for differential
expression related to the sign, magnitude, and significance of the treatment effect
(\cref{fig:min-to-drop}).
Across outcomes, we
find that our measure of
robustness (\textit{point color}) is correlated with, but distinct from, the
p-value (\textit{y-axis}) and the magnitude of the effect size (\textit{x-axis}).

Specifically, for outcomes revolving around the \emph{size of the treatment effect}
(top row of \cref{fig:min-to-drop}),
point color radiates outward (along the x-axis) from the decision boundary of zero
(for \textit{``flip sign''}) or $\pm 2$ (for \textit{``flip threshold''}),
meaning that genes nearer the decision boundary are, unsurprisingly, more likely
to be susceptible to dropping a small number of data points.
For both outcomes, genes with effect sizes that are four-fold larger or smaller
than the decision boundary (i.e., two or more ticks away along the x-axis)
are fully robust against dropping-data perturbations (up to 10\% of cells).
Similarly, among genes at a given effect size---most evidently for \textit{``flip sign''}---those
with smaller p-values (higher along y-axis) are more likely to be robust.

However, dropping-data robustness is \textit{not} fully predictable from
effect size and significance (and certainly not from either alone); see
neighboring points on both plots with visible
differences in point color.
This observation is made especially clear
by comparing \cref{fig:min-to-drop} with \cref{supp-fig:min-to-drop-reversed},
where the same data is plotted in reverse order in order to reveal the
extremities in color (robustness) of overlapping points.

For outcomes revolving around \emph{significance}
(last two rows of \cref{fig:min-to-drop}),
genes sensitive to dropping data are
similarly concentrated around the decision boundary (\textit{horizontal dotted line}),
but the discrepancy in information revealed by traditional robustness
metrics versus dropping-data robustness is even more striking
(cf.~\cref{supp-fig:min-to-drop-reversed}).
In other words,
dropping-data robustness is conspicuously
not monotonic with respect to p-value; see, for example, dark red points
(genes whose significance can be flipped by dropping a single cell)
that sporadically crop up beyond the p-value cutoff for
\textit{``flip significance (Fisher).''}
Notably, this cutoff already reflects an additional check on robustness via
multiple testing correction.
These nonrobust results also span nearly the full gamut of effect sizes
(dark red points ranging up to five or six ticks in either direction
along the x-axis,
representing genes with more than $2^5=32$-fold difference in
expression between treatment groups---and whose significance is estimated
to be flipped by dropping no more than a couple cells).

\continueCaption{%
        Specifically, \textit{``flip significance''} is composed of
        $\stat_{\text{erase significance}}$ (for genes that are originally significant at $p_{BH}=0.01$)
        and $\stat_{\text{bestow significance}}$ (for genes that are not originally significant),
        and \textit{``flip threshold''} is composed of
        $\stat_{\text{shrink below theshold}}$
        (for genes whose log two-fold change is originally above the minimal threshold $\,\tau=2$)
        and $\stat_{\text{increase above threshold}}$
        (for genes whose log-fold change is originally below the minimal threshold).
        \captionbr
        Hollow points represent genes where results are robust up to dropping 10\% of cells,
        and solid gray points represent genes that are not germane to the change-of-interest.
        Dashed horizontal lines represent significance cutoffs for raw p-values
        corresponding to level 0.01 for BH-corrected p-values.
        \captionbr
        Note that genes are plotted from most to least robust, in order to highlight
        those with the most concerning dropping-data sensitivity. In
        \cref{supp-fig:min-to-drop-reversed}, we plot the same data in the reverse order.
        \captionbr
        See also \cref{supp-fig:min-to-drop-BH} for direct plotting of
        p-values (here, \textit{y-axis}) against estimated
        dropping-data robustness (here, \textit{color}) across genes,
        including density plots to demonstrate the portion of genes at each
        predicted robustness level.
}

We also observe an interesting asymmetry around the p-value threshold,
where Wald \textit{Fisher} testing yields more significant genes that
are susceptible to having their significance \textit{erased} by dropping a single cell
(i.e., dark red zone that is skewed \textit{above} the horizontal dotted line)
whereas Wald \textit{sandwich} testing yields more nonsignificant genes that
are susceptible to having significance \textit{bestowed} by dropping a single cell
(i.e., dark red zone that is skewed \textit{below} the horizontal dotted line).
This observation is echoed by \cref{supp-fig:min-to-drop-fisher-sandwich},
where genes that are significant under both Wald tests require dropping
fewer cells in order to \textit{erase} their \textit{Fisher} significance
or to \textit{bestow} their \textit{sandwich} significance.
We also observe an asymmetry across both standard error estimators
where genes with \textit{positive} treatment effects
(i.e., increased expression among UC cells)
are more dropping-data sensitive to flipping sign with significance
than are genes with similar effect sizes in the \textit{negative} direction.
Future work could explore these phenomenona in order to better understand the
behavior of each test for
sparse count datasets like this one.

\subsection{differential expression analysis of the ulcerative colitis dataset is sensitive to dropping a small fraction of cells} \label{sec:exp-uc-genes}

Next, we look at the dropping-data robustness of differential expression
(comparing goblet cells from subjects with ulcerative colitis
to those from healthy subjects)
from the perspective of a biologist
analyzing this dataset.

For the sake of our interpretation, we
will consider results to be
\textit{potentially dropping-data sensitive} if an outcome can be changed by
dropping less than 2\% of the data (up to 28 cells, for the UC dataset),
\textit{sensitive} if it can be changed by dropping less than 1\% (14 cells),
and
\textit{extremely sensitive} if it can be changed by dropping less than 0.5\%
(7 cells).
However, these standards are subjective\footnote{
    And, of course, subject to the same
    shortcomings
    as any analysis based on discrete cutoffs
} and should be adjusted based on the needs of each scientist
for their particular analysis---akin to balancing the consequences of
false positives and false negatives.

Among genes that were originally ruled significant (BH-corrected $p < 0.01$),
all are predicted to be
robust (up to dropping 1\% of cells)
against changes to the \emph{sign} of the treatment effect---a minimal takeaway from differential expression.
However, a handful of genes are \textit{potentially} dropping-data sensitive (up to 2\% of cells):
\textbf{four} that were ruled significant based on the Wald sandwich test,
\textbf{eight} based on the Wald Fisher test,
and \textbf{nine} based on the LR test;
0.2--0.3\% of all significant genes.
Unsurprisingly, thousands
(67--70\%)
of nonsignificant genes---whose sign is arbitrary,
for those
that truly have no underlying difference in expression between groups---are
extremely dropping-data sensitive.

On the other hand, many significant genes are nonetheless dropping-data sensitive to
whether the \emph{magnitude} of
each treatment effect is ruled as ``meaningfully large''
(based on a minimal fold-change of four\footnote{
    i.e., a decision threshold of two on a $\log_2$ scale
}).
For the Wald sandwich test, \textbf{749} genes (29\%) are extremely dropping-data sensitive,
\textbf{1063} (40\%) are sensitive, and \textbf{1392} (53\%) are potentially sensitive.\footnote{
    Where each is a superset of the preceding set of genes, and percentages are out of all genes
    ruled significant under the relevant test
}
For the Wald Fisher test, \textbf{710} (28\%) are extremely dropping-data sensitive,
\textbf{1026} (41\%) are sensitive, and \textbf{1361} (54\%) are potentially sensitive.
For the LR test, \textbf{832} (30\%) are extremely dropping-data sensitive,
\textbf{1188} (43\%) are sensitive, and \textbf{1537} (55\%) are potentially sensitive.
In fact, many significant genes (7--8\%)\footnote{
    Specifically,
    \textbf{198} for Wald sandwich,
    \textbf{185} for Wald Fisher, and
    \textbf{220} for the likelihood ratio test
} can be flipped across this threshold by dropping a single
cell---both those with effect sizes near the threshold
(vertical dotted lines in \cref{fig:min-to-drop} \textit{``flip threshold''})
and even some with effect sizes up to two-fold smaller or larger than the threshold
(i.e., one tick away on the x-axis).

Unexpectedly, we find that \emph{significance}
is dropping-data sensitive for around \textit{half of all genes} tested,
and extremely sensitive for around a third.
Specifically, by dropping up to 1\% of cells, we estimate that \textbf{57\%} of
genes can be flipped from significant to nonsignificant or vice versa
based on the Wald sandwich test,
and \textbf{48\%} of genes can be flipped based on the Wald Fisher test.
Further, \textbf{39\%} of all genes are extremely sensitive (and the vast majority---\textbf{77\%}---are
potentially sensitive) with respect to significance based on the Wald sandwich test,
and \textbf{30\%} are extremely sensitive (and \textbf{71\%}{ potentially sensitive) based on the Wald Fisher test.
In fact, we estimate that \textbf{10\%} or \textbf{6\%} of genes can have their significance status flipped
(based on the Wald sandwich or Wald Fisher test, respectively) by dropping a \textit{single} cell.

To break down these statistics further:
for the Wald \textit{sandwich} test,
we predict that \textbf{1345} genes flagged as significant
(51\% of all genes flagged) are dropping-data sensitive---i.e.,
can have their significance erased by dropping <1\% of cells.
Further, \textbf{1751} genes (67\%) are potentially sensitive,
while
\textbf{921} (35\%) are extremely sensitive
including
\textbf{213} (8\%)
where dropping a single cell would erase significance.

For the Wald \textit{Fisher} test,
we predict that \textbf{1459} genes flagged as significant
(58\% of all genes flagged) are dropping-data sensitive.
Further,
\textbf{1861} genes (74\%) are potentially sensitive,
while
\textbf{1121} (44\%) are extremely sensitive
including
\textbf{317} (13\%)
where dropping a single cell would erase significance.

We then consider the complementary set of genes: those that were not originally
flagged as significant (BH-corrected $p > 0.01$).
For the Wald \textit{sandwich} test,
we predict that \textbf{4680} of these genes
(59\% of those not flagged) are dropping-data sensitive---i.e.,
can attain significance by dropping <1\% of cells.
Further,
\textbf{6341} genes (81\%) are potentially sensitive,
while
\textbf{3176} (40\%) are extremely sensitive
including
\textbf{829} (11\%) where dropping a single cell would bestow significance.

For the Wald \textit{Fisher} test,
we predict that \textbf{3618} nonsignificant genes
(45\% of those not originally flagged as significant) are dropping-data sensitive.
Further,
\textbf{5592} genes (70\%) are potentially sensitive,
while
\textbf{2070} (26\%) are extremely sensitive
including
\textbf{344} (4\%) where dropping a single cell would bestow significance.

Notably, these nonrobust results include genes with large estimated effect sizes
(dark red points up to $\apprx$five ticks away from the x-axis origin in either
direction, representing genes with more than $2^5=32$-fold difference in
expression between treatment groups---and whose significance is estimated
to be flipped by dropping no more than a couple cells; \cref{fig:min-to-drop}
\textit{``flip significance''}).

On the other hand,
differential expression results for this dataset are near uniformly
robust to the dramatic change of
flipping
a significant finding in one direction to a significant finding for an effect
in the opposite direction;
this is predicted to be possible by dropping $<6\%$ or $<8\%$ of cells
for one gene each under Fisher or sandwich testing, respectively
(\cref{fig:min-to-drop} \textit{``flip sign w/ significance''}).

\subsection{our robustness approximation is accurate within the regimes that matter} \label{sec:exp-accuracy}

So far we have assessed
dropping-data robustness based on approximations
(since it is combinatorially complex to compute exactly).
We hypothesize that these approximations will be reasonably accurate so long as
they are based on dropping only a small fraction of cells---conveniently,
pertaining to the sensitivities
of highest concern---and that accuracy will degrade as more cells are dropped
(i.e., as weight vector $\vec{w}$ moves farther from $\vec{1}$,
where the Taylor approximation was formed).

In order to compare the fidelity of our approximation globally (across genes),
we drop the most influential $\ninfluence$ cells per gene and compare the
predicted versus actual change in the statistic-of-interest as a result of this
perturbation. In \cref{fig:pred-v-actual} (and \cref{supp-fig:pred-v-actual}),
we show that the predicted change to the statistic-of-interest
(\textit{x-axis}) is strongly correlated with the actual change
(\textit{y-axis}).
As expected, the quality of the approximation deteriorates as more cells are
dropped (from left to right across each row of \cref{fig:pred-v-actual})---i.e.,
as we move further from where the approximation was formed---though
it is still quite reasonable
(correlation >0.87, among all changes except \textit{``bestow significance''})
when as many as 2\% of cells are dropped (rightmost column).
Importantly, because we
aim to detect
nonrobustness with respect to dropping data,
our primary concern is the fidelity of the approximation at small proportions
of cells---where the approximation is, incidentally, most accurate.
In other words, we consider results to be sensitive
if the outcome can be changed by dropping a small handful
of data points---the smaller, the more concerning---so
it is not important that the approximation hold up at large proportions.

\begin{figure}[htbp!]
    \centering
    \subcaptionbox{\label{fig:pred-v-actual_flip-sign}}{\includegraphics[width=\linewidth]{%
        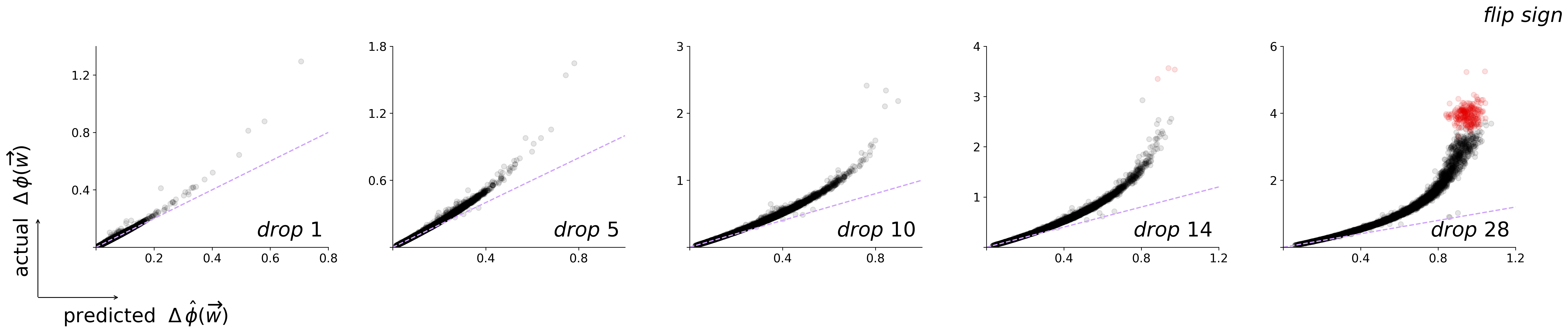}}%
    \hfill%
    \subcaptionbox{\label{fig:pred-v-actual_erase-sig-sandwich}}{\includegraphics[width=\linewidth]{%
        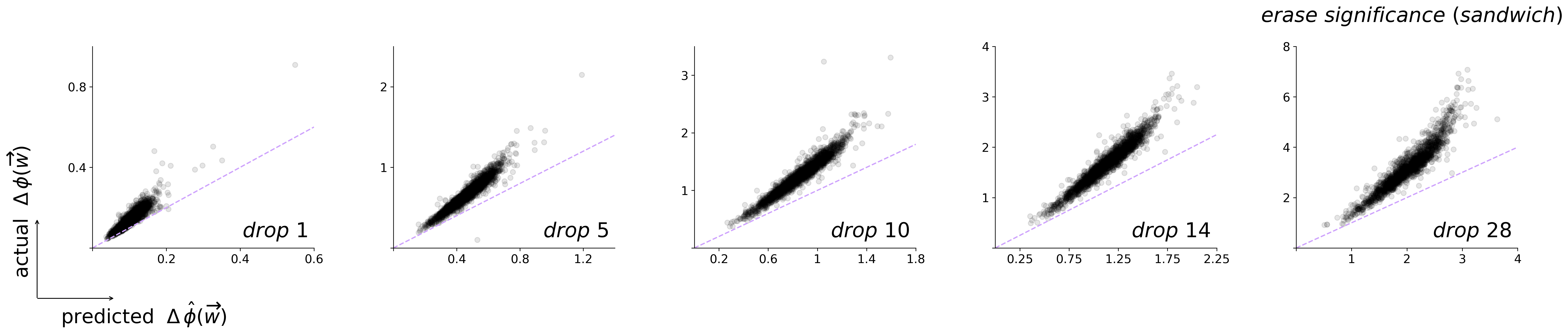}}%
    \hfill%
    \subcaptionbox{\label{fig:pred-v-actual_erase-sig-Fisher}}{\includegraphics[width=\linewidth]{%
        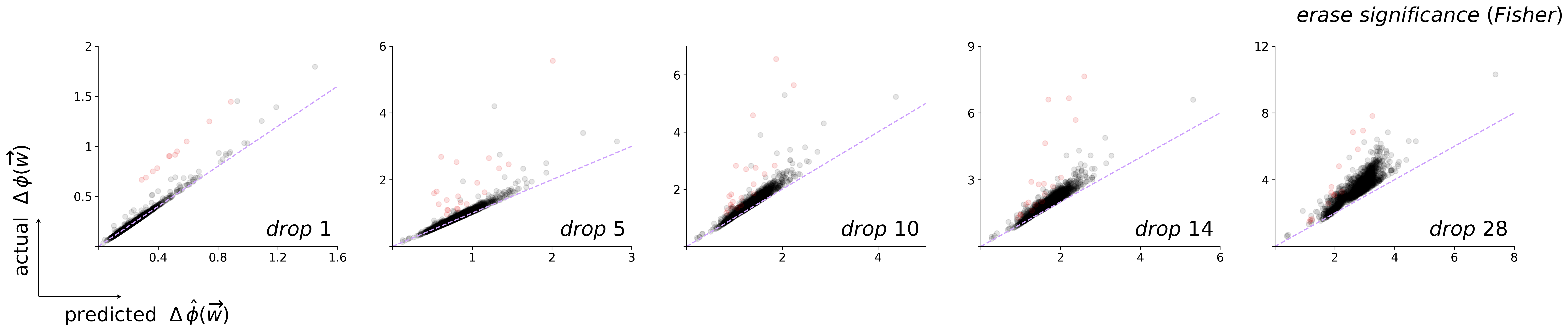}}%
    \hfill%
    \subcaptionbox{\label{fig:pred-v-actual_bestow-sig-sandwich}}{\includegraphics[width=\linewidth]{%
        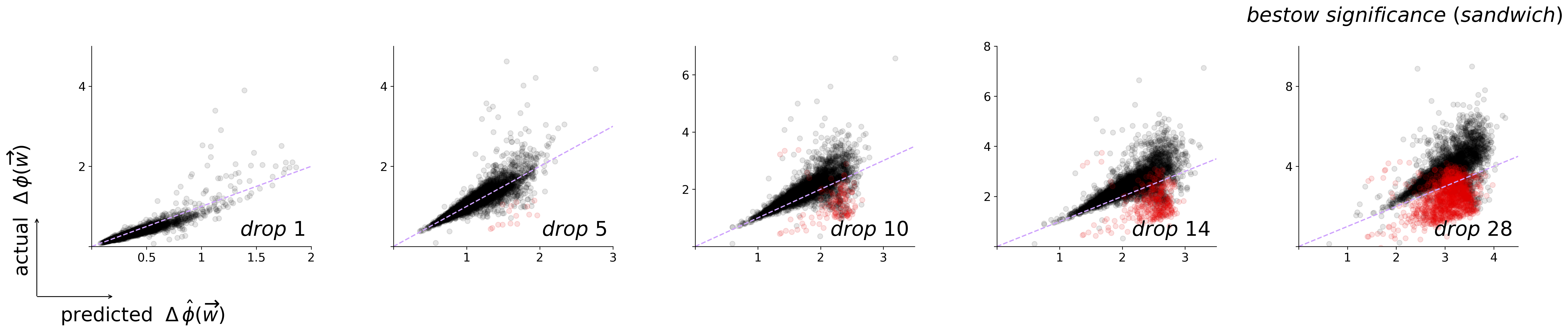}}%
    \hfill%
    \subcaptionbox{\label{fig:pred-v-actual_bestow-sig-Fisher}}{\includegraphics[width=\linewidth]{%
        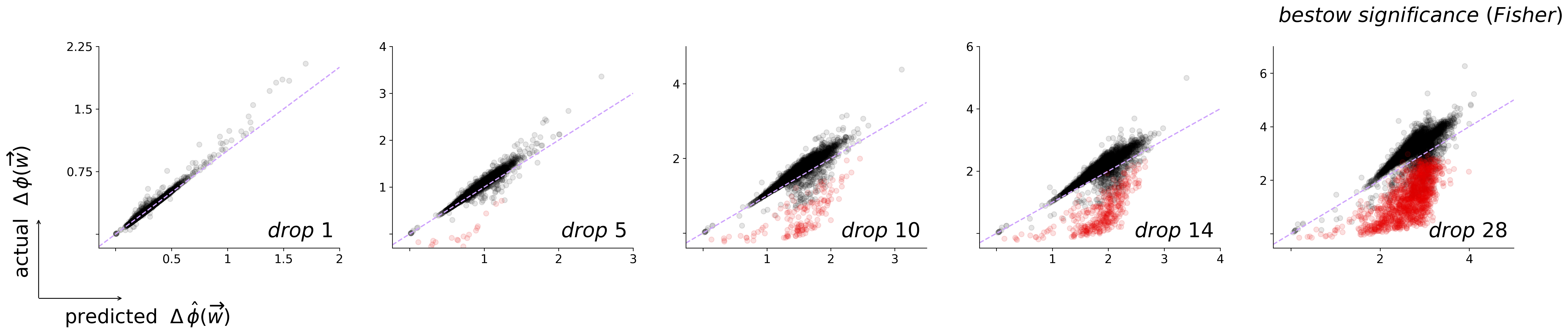}}%
    \caption{
        \textbf{Fidelity of the approximation for dropping the $\ninfluence$ most influential cells.}
        Plots are predicted (\textit{x-axis}) versus actual (\textit{y-axis})
        change to the statistic-of-interest $\stat$
        after dropping the top $\ninfluence$ most influential cells
        (up to 2\% of cells, out of 1440).
        Newly created zero-group genes (after dropping cells) are highlighted in red.
        Lilac dotted lines represent the 1-to-1 line (i.e., perfect predictions).
        \captionbr
        To avoid trivial results (like dropping all nonzero counts), and to improve
        the overall fidelity of the approximation, genes (\textit{points}) are filtered
        to those with a sufficient number of nonzero observations.
    } \label{fig:pred-v-actual}
\end{figure}

\continueCaption{%
        Specifically, we filter to genes where the maximal number of nonzero
        observations per group (treatment or control)---after dropping the
        selected cells---is at least 20.
        (See \cref{supp-fig:nnz-v-err} for details on this cutoff.)
        We also filter to relevant genes (e.g., for ``erase significance,''
        genes that are originally significant under the relevant test).
        \captionbr
        Correlations range from
        0.90--0.99 (\textbf{\subref{fig:pred-v-actual_flip-sign}});
        0.92--0.96 (\textbf{\subref{fig:pred-v-actual_erase-sig-sandwich}});
        0.87--0.97 (\textbf{\subref{fig:pred-v-actual_erase-sig-Fisher}});
        0.44--0.89 (\textbf{\subref{fig:pred-v-actual_bestow-sig-sandwich}}),
        where the low end can be raised to 0.62 by excluding newly created zero-group genes;
        and
        0.53--0.99 (\textbf{\subref{fig:pred-v-actual_bestow-sig-Fisher}}),
        where the low end can be raised to 0.75 by excluding newly created zero-group genes.
        \captionbr
        For the remaining key gene-level outcomes
        (\textit{``shrink below threshold,''
         ``increase above threshold,''
         ``flip sign w/ significance (sandwich),''} and
         \textit{``flip sign w/ significance (Fisher)''}),
         see~\cref{supp-fig:pred-v-actual}.
}

Further, when the actual change diverges from its approximation, it is biased
toward being \textit{more extreme} than predicted
(\textit{above} the dotted 1-to-1 line).\footnote{
    Recall that $\stat$ is constructed to be a decision function that
    moves toward the relevant decision boundary when \textit{increased}
}
In other words, our procedure allows us to pick out highly influential cells
whose effect on the outcome-of-interest is as dramatic as predicted, if not more so.

This is true for every key outcome with the exception of \textit{bestowing} significance
(\cref{fig:pred-v-actual_bestow-sig-sandwich,fig:pred-v-actual_bestow-sig-Fisher}),
which---like other outcomes---deteriorates for newly minted zero-group genes\footnote{
    i.e., genes that have at least one nonzero observation per group in the original dataset,
    but that become zero-group genes after dropping the $\ninfluence$ most influential cells
}
and---unlike other outcomes---is dominated by such genes,
whose
dropping-data effect tends to be \textit{less} dramatic than predicted.
While it is expected that bestowing significance will often entail creating newly zero-group
genes (thus increasing the discrepancy in expression between groups),
it is notable that our approximation breaks down for these genes
(indicating that the consequences are highly nonlinear when the observations in one group go to zero,
despite the smoothing induced by the pseudocell prior).
This observed nonlinearity in summary statistics when all counts in a group go to zero
is echoed by \cref{supp-fig:spectrum-a,supp-fig:spectrum-b,supp-fig:spectrum-c}, where
we explore the same phenomenon
on a gene-by-gene basis by interpolating across a spectrum of weights
(i.e., by gradually
dropping cells).
This phenomenon should be explored in future work, to understand how it arises and
how the approximation might be improved
(such as through a higher-order approximation for
relevant genes).

Consider our specific predictions
in \cref{sec:exp-uc-genes}
about a substantial number of genes whose
significance can be erased by dropping just one cell.
With respect to our assumed significance level (analogous to 0.01 for BH-corrected
p-values, conditioning on the original number of significant genes), \textbf{100\%} of these
predictions were accurate (i.e., all 213 genes for Wald sandwich and all 317 genes for Wald Fisher,
which were originally significant, had p-values above this fixed significance level
after dropping the most influential cell).
An additional \textbf{82} genes (Wald sandwich) or \textbf{22} genes (Wald Fisher)
were also rendered nonsignificant by dropping a single cell; this reflects our
observation that influence scores skew toward underestimates
(\cref{fig:pred-v-actual,supp-fig:pred-v-actual,supp-fig:spectrum-a,supp-fig:spectrum-b,supp-fig:spectrum-c}).

Similarly, \textbf{100\%} of predictions were borne out, for over a thousand genes,
that significance could be erased by dropping <1\% of the data (14 cells).
This was validated for 1204 genes for Wald sandwich\footnote{
    The other 141 predicted genes would have so few remaining nonzero
    counts---three or fewer in the group with the most---that we can safely
    assume nonsignificance
} and for 1394 genes for Wald Fisher.\footnote{
    The other 65 genes were, similarly, safely assumed nonsignificant
}
An additional \textbf{284} genes (Wald sandwich) or \textbf{161} genes (Wald Fisher) were
also rendered nonsignificant, with respect to a fixed threshold,
after dropping the most influential 1\% of cells. 

A more complex question is whether these genes truly lost significance with respect
to their BH-corrected p-values; this entails refitting \textit{all} genes
after dropping each most influential cell per gene-of-interest (in order to
rank p-values and properly correct them).
This procedure is more compute-intensive
(two+ minutes per gene, on average, versus four--five seconds
to refit each gene alone\footnote{
    Though times can range widely depending on factors including
    sparsity (more compute required for genes with sparser observations, such as zero-group genes)
    and the number of cells being dropped
    (more compute required to refit $\hat{\beta}(\vec{w})$
    when $\vec{w}$ is farther from that used to fit the original estimates
    $\hat{\beta}(\vec{1})$; i.e., when more cells are dropped)
})---exactly the type of analysis our
approximation seeks to avoid---so we verify a subset of our predictions.

\pagebreak

Specifically, we filter 213 genes-of-interest---where the BH-corrected
Wald sandwich test is significant at level 0.01 and dropping a single cell is
predicted to erase this finding---to 127 genes where
\begin{enum-inline}
    \item the BH-corrected likelihood ratio test is also significant, and
    \item the maximal number of nonzero counts per group (after dropping
    the influential cell) is at least 20.\footnoteV{
    ~\\
    \includegraphics[height=\vennHeightTaller]{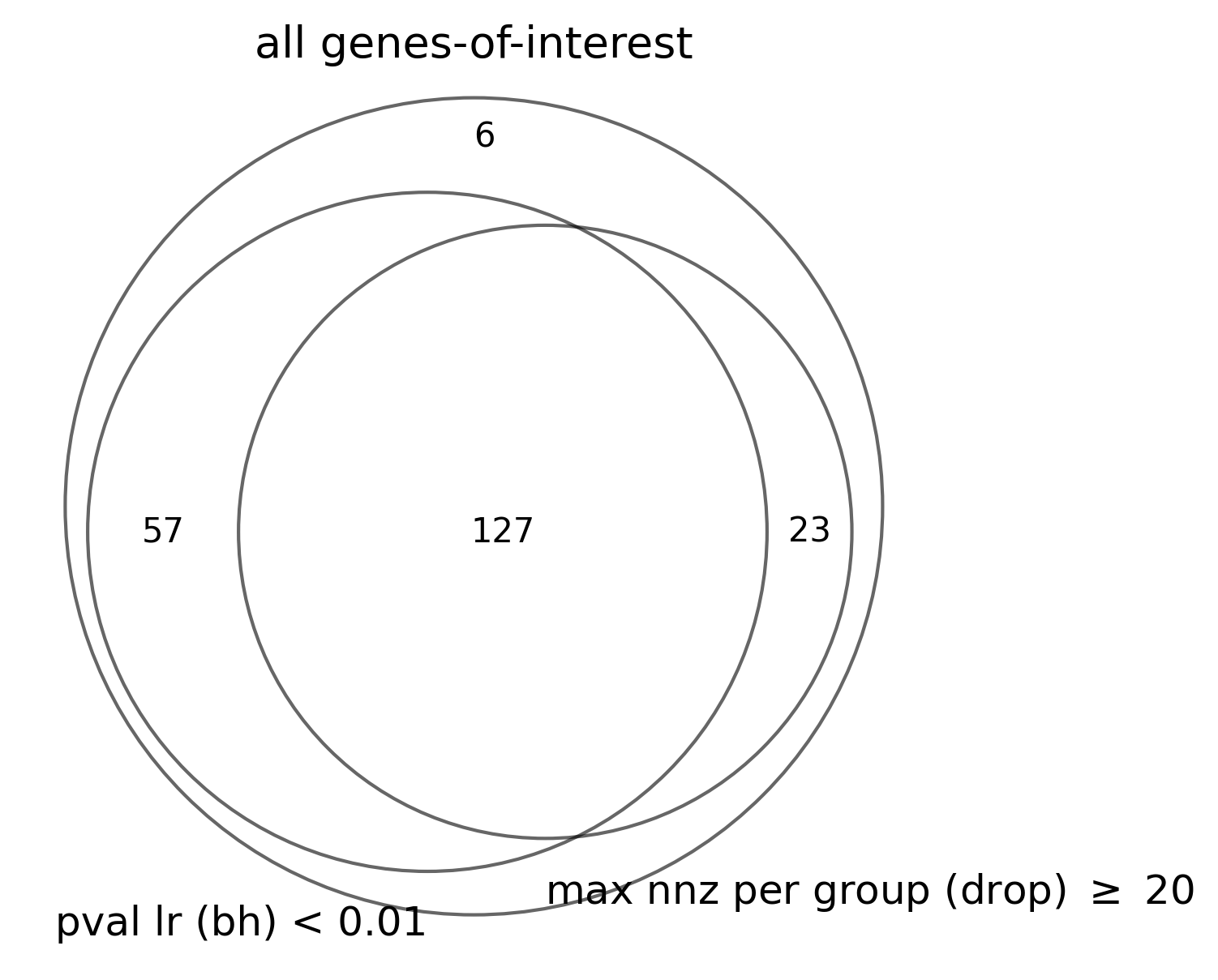}
}
\end{enum-inline}
Hypothetically, this subset of genes ought to be \textit{more} resilient
against losing significance,
since they are significant under an additional test
and are not prone to losing significance merely by
chipping away at
the handful of nonzero counts
from the group with higher levels of nonzero gene expression.
After refitting all gene regressions 127 times, with 127 weight vectors $\vec{w}$
such that the most influential cell was dropped for each of the filtered set of genes,
we compute exact BH-corrected p-values and
find that \textbf{100\%} of our predictions
are correct---i.e., dropping a single cell was indeed sufficient to eliminate
significance for all 127 genes.
Further, we find that significance under the BH-corrected likelihood ratio
test---which we did not directly target with our robustness approximation---was
also erased for \textbf{73\%} (93) of these genes.

Similarly, we consider the 86 genes whose significance (under the BH-corrected
Wald sandwich test) is predicted to be erased by dropping nearly 1\% of the data
(13 or 14 cells).
We narrow these genes to 74 by the same criteria above.\footnoteV{
    ~\\
    \includegraphics[height=\vennHeightTaller]{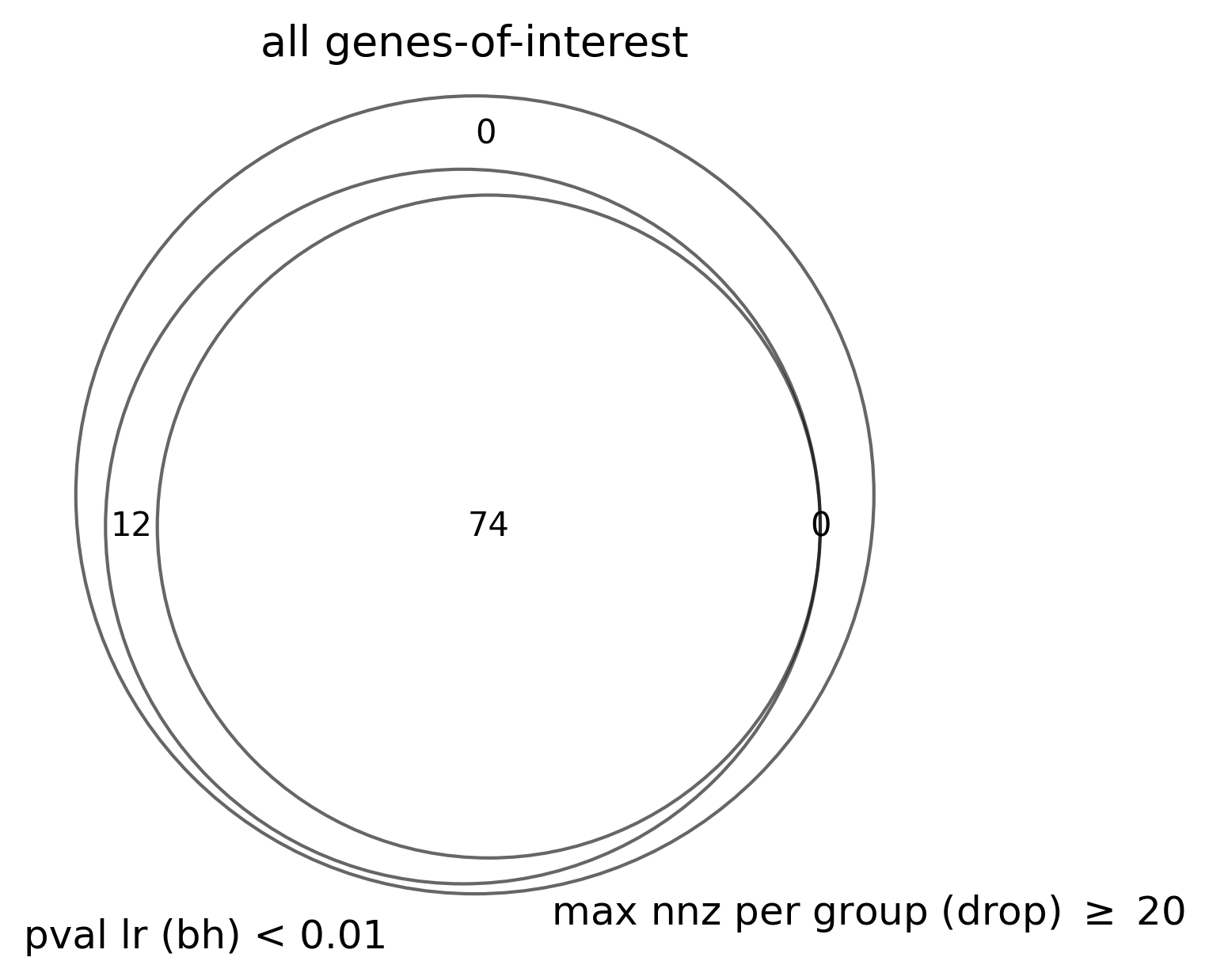}
    \\[0.12cm] 
}
After refitting all regressions
with 74 weight vectors $\vec{w}$, corresponding to dropping
the 13 or 14 most influential cells for each of these genes,\footnoteV{
    \hspace{0.5\footnotemargin}Whichever is predicted to be the minimal sufficient
}
we find that \textbf{97\%} of these genes (72 of 74) fail to retain significance,
as predicted,
under this data perturbation.
Further, \textbf{95\%} of these genes (70 of 74) additionally lose significance under
the BH-corrected likelihood ratio test.

\pagebreak 

\subsection{dropping-data sensitivity at gene level translates to sensitivity of high-level takeaways} \label{sec:exp-gsea}

These experiments validate the fidelity of our robustness approximation
for gene-level differential expression results,
and reveal widespread dropping-data sensitivity across
results for a sample dataset.
However, the ultimate outcome of such an analysis is
generally not a table of significance testing across tens of thousands of genes,
but rather a \emph{gene set enrichment analysis}
to detect
biologically meaningful
patterns
(based on a collection of predefined gene sets)
among differentially expressed genes.
Often, a biological story is spun from the analysis
based on the top 10 gene sets
(vis-\`a-vis a downstream test for enrichment).
Having demonstrated that some individual gene findings are susceptible
to a dropping-data perturbation,
we next sought to demonstrate whether high-level, biologically relevant
takeaways could be disrupted by dropping a handful of data points.

In other words, we set out to identify
a small subset of influential cells
to drop that are predicted to disrupt significance \textit{across} genes,
rather than identifying influential cells on a gene-by-gene basis.
Recall that, unlike gene-level robustness, we could not
directly extend
the original dropping-data framework~\cite{ryan-amip}
to predict how dropping cells would disrupt gene set results,
since this analysis is predicated on either a discrete subset (of all significant genes)
or a ranking (of genes, by notability of their results), neither of which is
differentiable and therefore amenable to automatic robustness.
Instead, we invent a procedure (\cref{sec:gsea-robustness})
to use the cell-by-gene influence matrix $\mat{\influence}$
to estimate the
dropping-data robustness of
a biologically meaningful summary of differential expression;
namely, the top-ranked gene sets.

Specifically,
we use hypergeometric testing to
look for enrichment of biologically coherent
gene sets
among genes ruled as differentially expressed.
We separately analyze
enrichment among upregulated and downregulated
genes (based on the sign of the effect)~\cite{gsea-up-down},
and use GO Biological Processes (\texttt{GO:BP})~\cite{go-2000,go-2020}
as our curated collection of
gene sets
(filtered to sets of size 15--500);
see \cref{app:gsea} for details.

Finally, we use cell influence scores across genes
to select a small number of cell subsets to test empirically
and, ultimately, to bound the
dropping-data robustness of the
top gene sets.
In other words, across data perturbations of
varying sizes,
we identify influential groups of cells that are predicted to be maximally disruptive
to the composition of the top enriched gene sets, and verify the validity of
these predictions (by recomputing results after dropping those cells).\footnote{
    Thus establishing an upper bound on the minimal number of cells
    that can be dropped in order to effect a given change
    (while leaving open the possibility that
    a smaller subset of cells with a similar effect may exist)---or,
    equivalently, establishing a lower bound on the maximal number of gene sets
    that can be disrupted by dropping a given number of cells
}

In \cref{fig:change-in-go-up,fig:change-in-go-down}, we highlight the gene sets
that are elevated to (\textit{red}) or demoted from (\textit{blue})
the top 10---altering the interpretation of this differential expression analysis,
regarding the most notable
functional differences in goblet cells associated with ulcerative colitis---in
response to dropping the most influential handful of cells
(from one to $\apprx$2\% of the data)
that our methods identify.
\vspace{1.1\parskip}

\begin{figure}[H]
    \centering
    \parbox{\textwidth}{
        \parbox{.67\textwidth}{%
            \subcaptionbox{\makeSubcaption{one}}{\includegraphics[height=\goHeight]{%
                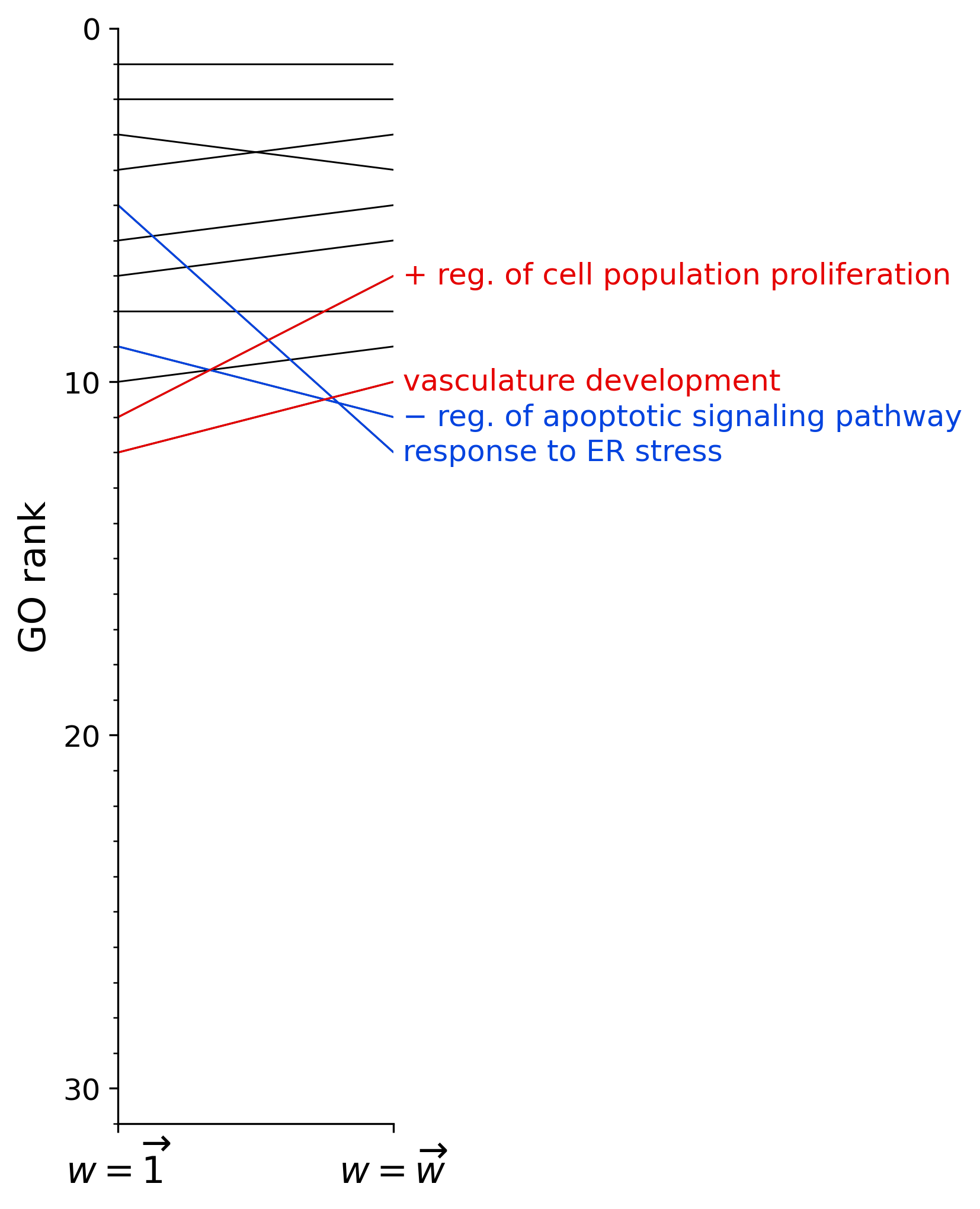}}%
            \hfill%
            \subcaptionbox{\makeSubcaption{four}}{\includegraphics[height=\goHeight]{%
                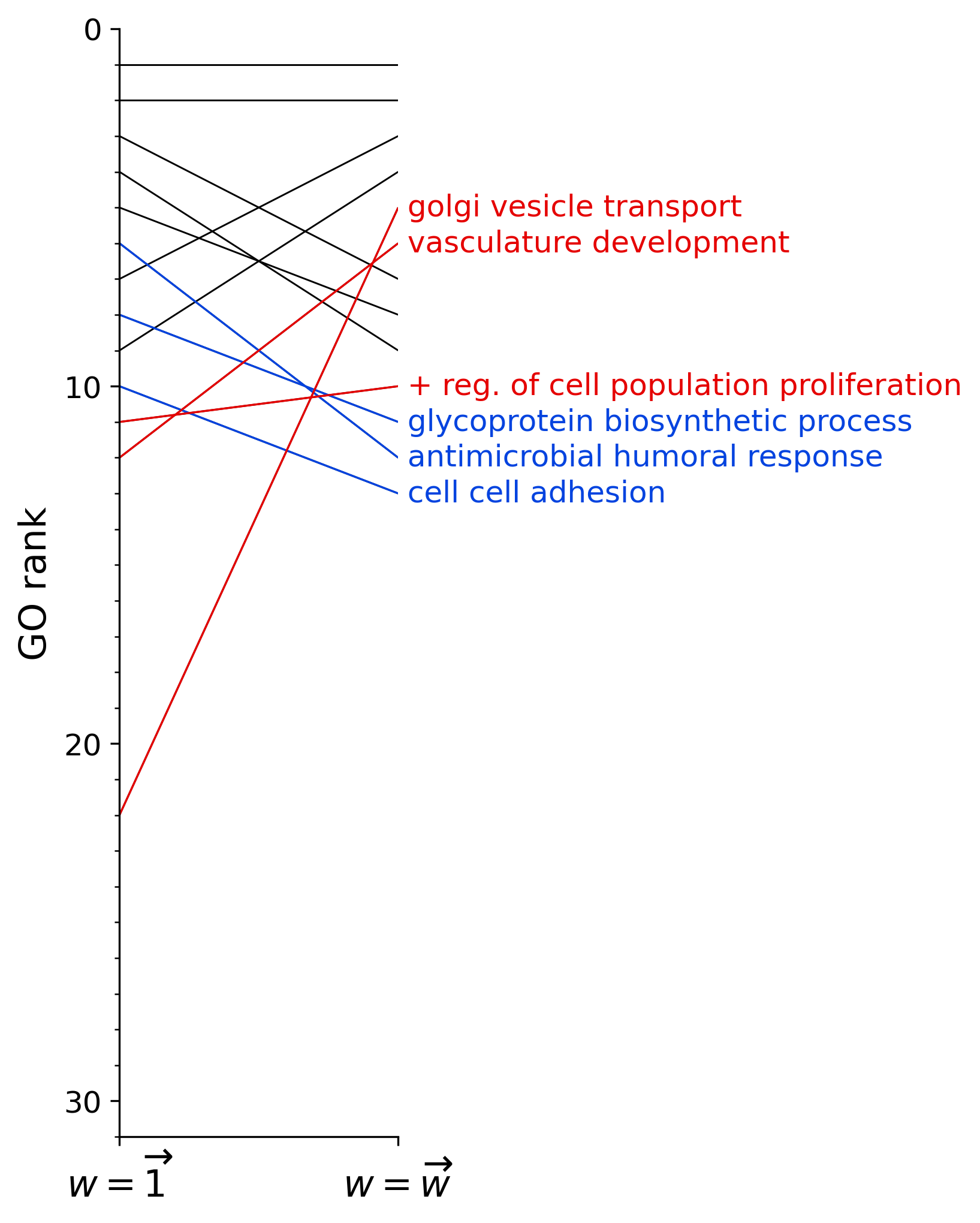}}%
            \vspace{\vspaceBwRows}
            \subcaptionbox{\makeSubcaption{seven}}{\includegraphics[height=\goHeight]{%
                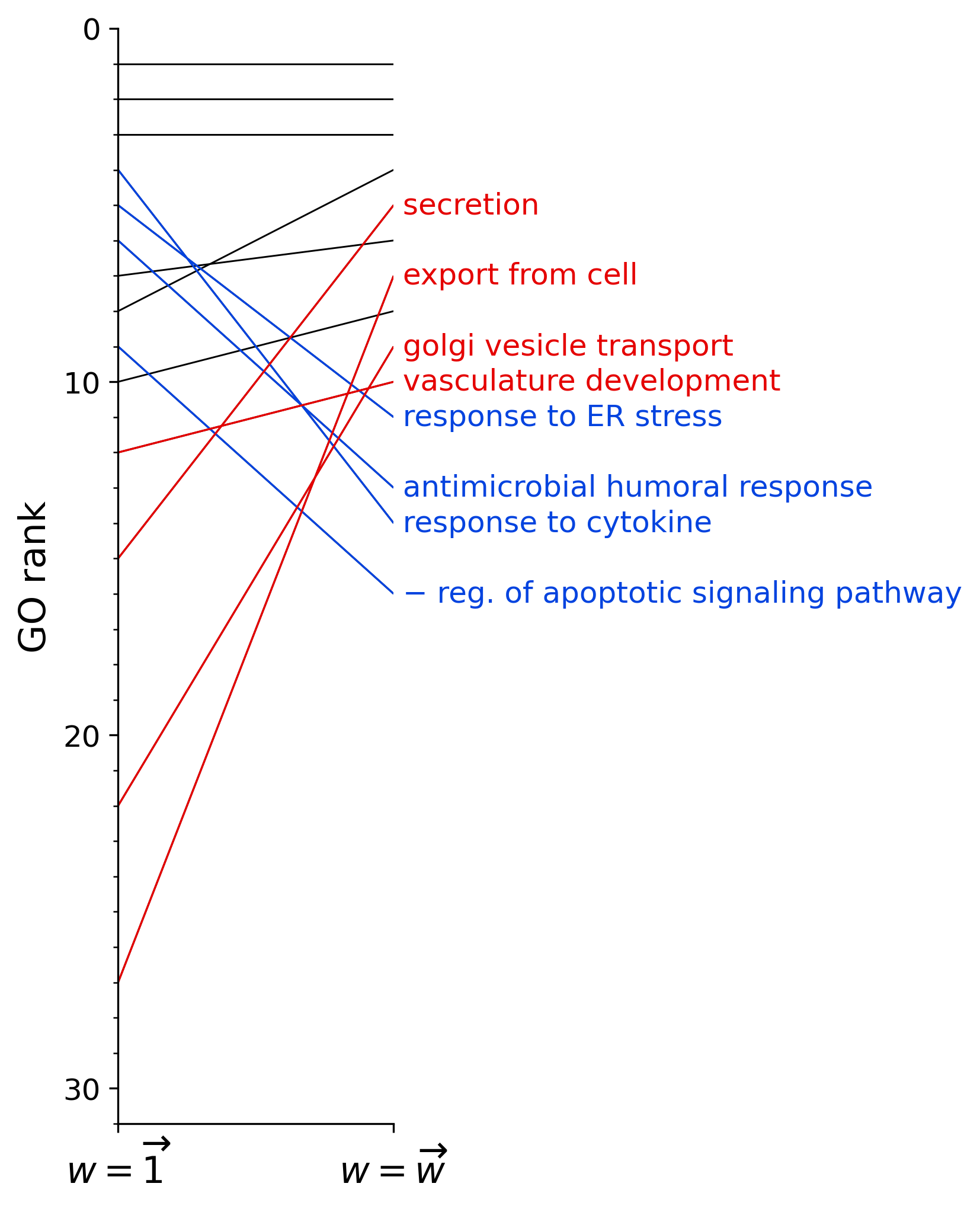}}%
            \hfill%
            \subcaptionbox{\makeSubcaption{14}}{\includegraphics[height=\goHeight]{%
                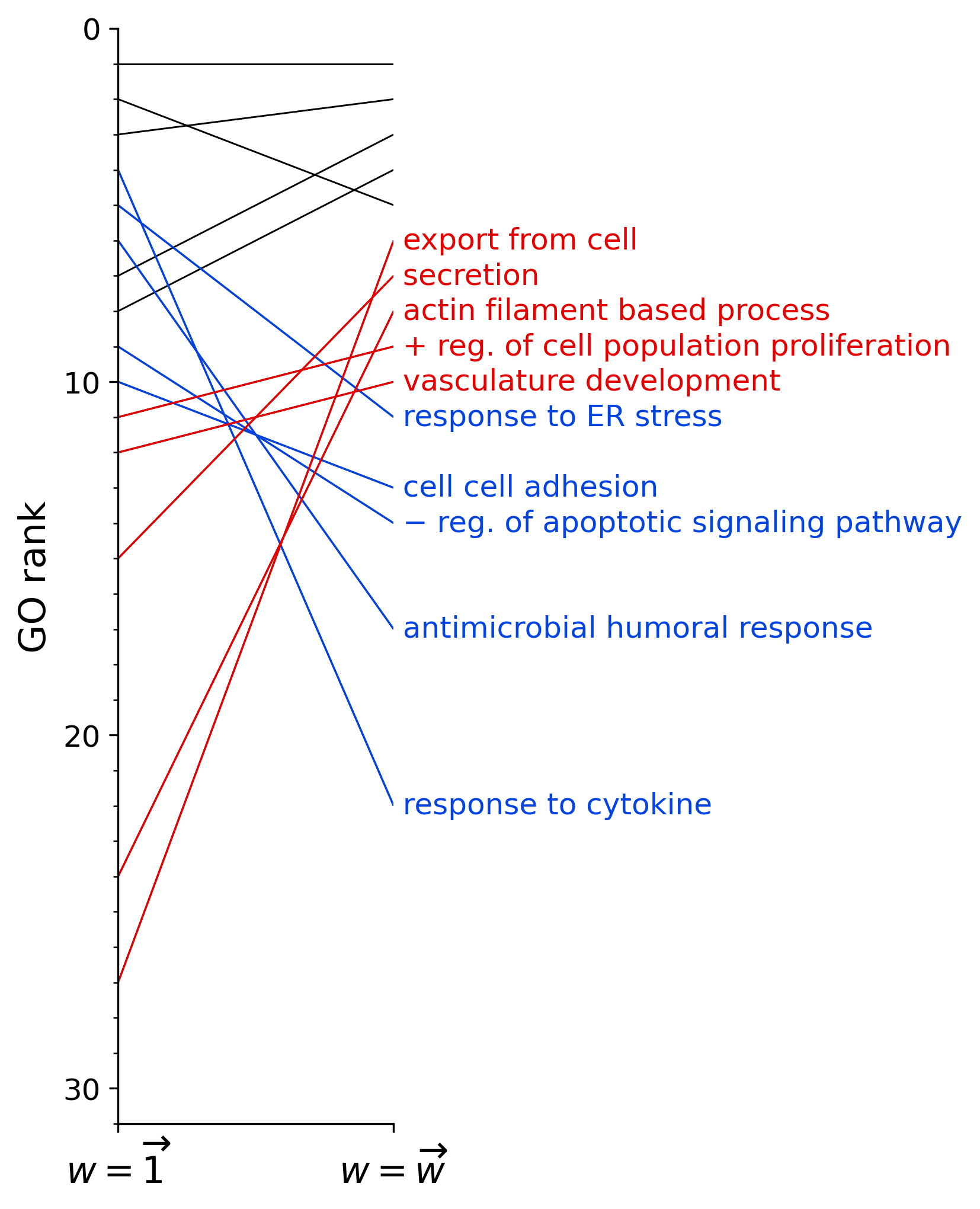}}%
        }
        \hfill
        \parbox{.33\textwidth}{%
            \subcaptionbox{\makeSubcaption{28}}{\includegraphics[height=2.1\goHeight]{%
                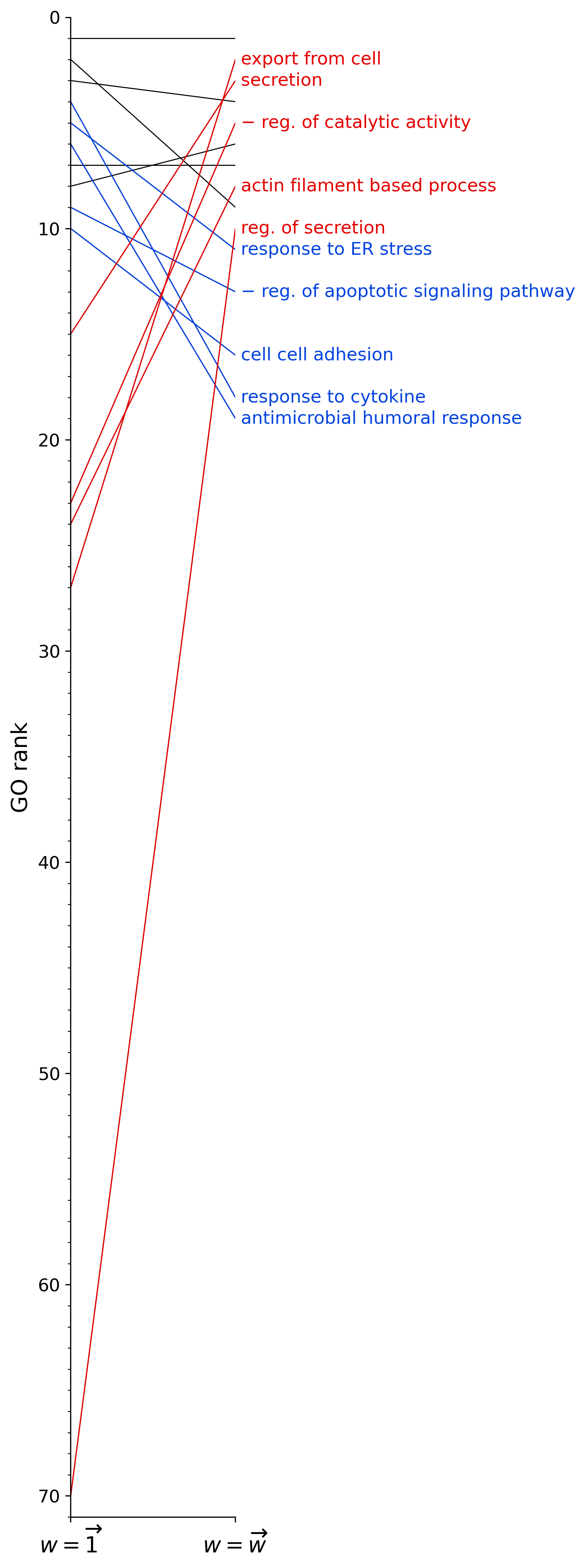}}%
        }
    }
    \caption{
        \textbf{Perturbation to top GO sets (among upregulated genes),
                by dropping a handful of influential cells.}
        Plots show changes to the top 10 ranked \texttt{GO:BP} gene sets
        when an influential cell set of
        the indicated size is dropped.
        Blue lines indicate the change in rank for gene sets that are \textit{demoted},
        red lines indicate the change in rank for those that are \textit{promoted},
        and black lines indicate the change in rank for those that \textit{remain} in the top 10.
        +/- reg.; positive/negative regulation.
        ER; endoplasmic reticulum.
        \captionbr
        See \cref{supp-fig:change-in-de-pvals-up} for the corresponding perturbations,
        actual and predicted, for DE p-values of individual genes that give
        rise to these gene-set-level changes.
    } \label{fig:change-in-go-up}
\end{figure}

Namely, among
upregulated genes (\cref{fig:change-in-go-up}), we find that
\begin{itemize}
    \item \standout{20\%} of the top 10 gene sets can be disrupted---i.e.,
        downranked below the top 10 (by hypergeometric
        p-value)
        and replaced with alternate, upranked gene sets---by
        dropping a \standout{single} cell (<0.07\% of data points),
    \item \standout{30\%} can be disrupted by dropping as few as \standout{four} cells (<0.3\%),
    \item \standout{40\%} can be disrupted by dropping as few as \standout{seven} cells (<0.5\%), and
    \item \standout{50\%} can be disrupted by dropping as few as \standout{14} cells (<1\%).
\end{itemize}
\vspace{\parskip}

By dropping more cells (up to 28, a little less than 2\% of the data),
we could push the rankings for gene sets that could be elevated into the top 10
to be more extreme---\textit{lowering} the original ranking
(down to original rank 70, even though our heuristics for clustering only
focused on the top 30\footnote{
    This gene set (``regulation of secretion'') presumably benefited from
    overlap with related gene sets in the top 30 (e.g., ``secretion''),
    such that this cell cluster was influential for the significance
    of genes involved in both gene sets.
    Recall that we \textit{do} account for such lower-ranked gene sets
    when scoring clusters (\ref{it:score-genesets}).
})
or \textit{elevating} the new ranking
(up to the top two gene sets).
However, we did not uncover a set of cells (up to 2\% of the data) capable of
perturbing more than half of the top 10 upregulated gene sets.
This finding does not preclude the existence of such a set of cells
but, rather, lower-bounds the maximal perturbation to the top gene sets
(by dropping up to 2\% of data) at 50\%.\footnote{
    In fact, ruling out the existence of such a set of cells
    (i.e., upper-bounding the maximal perturbation
    by dropping a given number of cells)
    is an active area of research; see, e.g., \cite{robustness-audit}.
}

Even more dramatically, among downregulated genes (\cref{fig:change-in-go-down}), we find that
\begin{itemize}
    \item \standout{30\%} of the top 10 gene sets can be disrupted by dropping
        a \standout{single} cell (<0.07\% of data points),
    \item \standout{40\%} can be disrupted by dropping as few as \standout{four} cells (<0.3\%), and
    \item \standout{60\%} can be disrupted by dropping as few as \standout{28} cells (<2\%).
\end{itemize}
\vspace{\parskip}

Incidentally, we find that up to \standout{70\%} of the top 10 gene sets can in fact
be disrupted by dropping an alternate set of 28 cells, identified through
a different method (\cref{app:gsea-alt}) that is otherwise inferior.
While we recommend estimating gene set robustness based on a standard
protocol (outlined in \cref{sec:gsea-robustness}) that
generalizes
across $K$,
this finding reinforces that our procedure
is heuristic rather than guaranteed optimal---and thus
(meaningfully) \textit{bounds}
the maximal disruption to top gene sets by dropping a given number of cells.

\begin{figure}[htb!]
    \centering
    \parbox{\textwidth}{
        \parbox{.67\textwidth}{%
            \subcaptionbox{\makeSubcaption{one}}{\includegraphics[height=0.8\goHeight]{%
                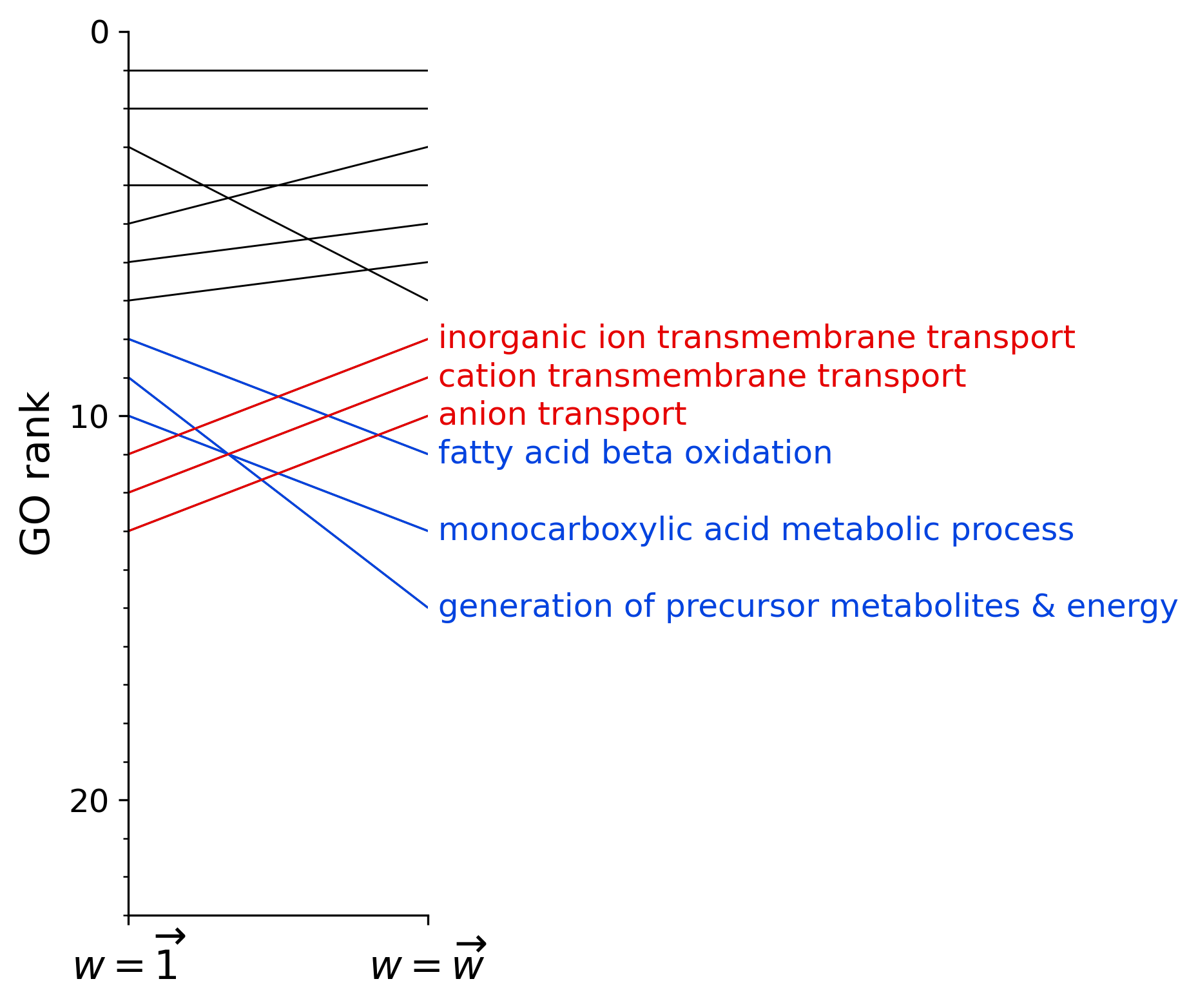}}%
            \hfill%
            \subcaptionbox{\makeSubcaption{four}}{\includegraphics[height=0.8\goHeight]{%
                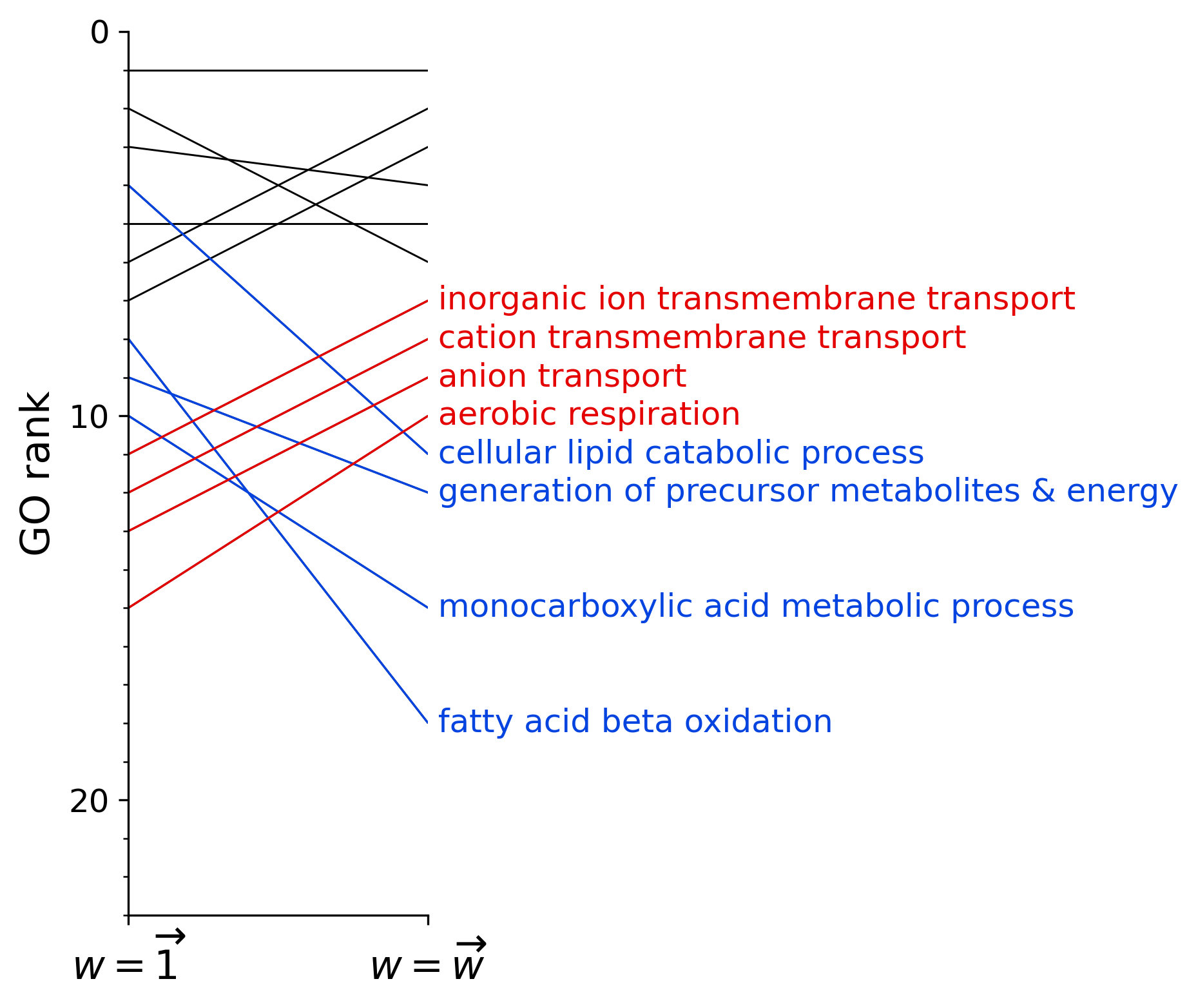}}%
            \vspace{\vspaceBwRows}
            \hfill%
            \subcaptionbox{\makeSubcaption{seven}}{\includegraphics[height=0.8\goHeight]{%
                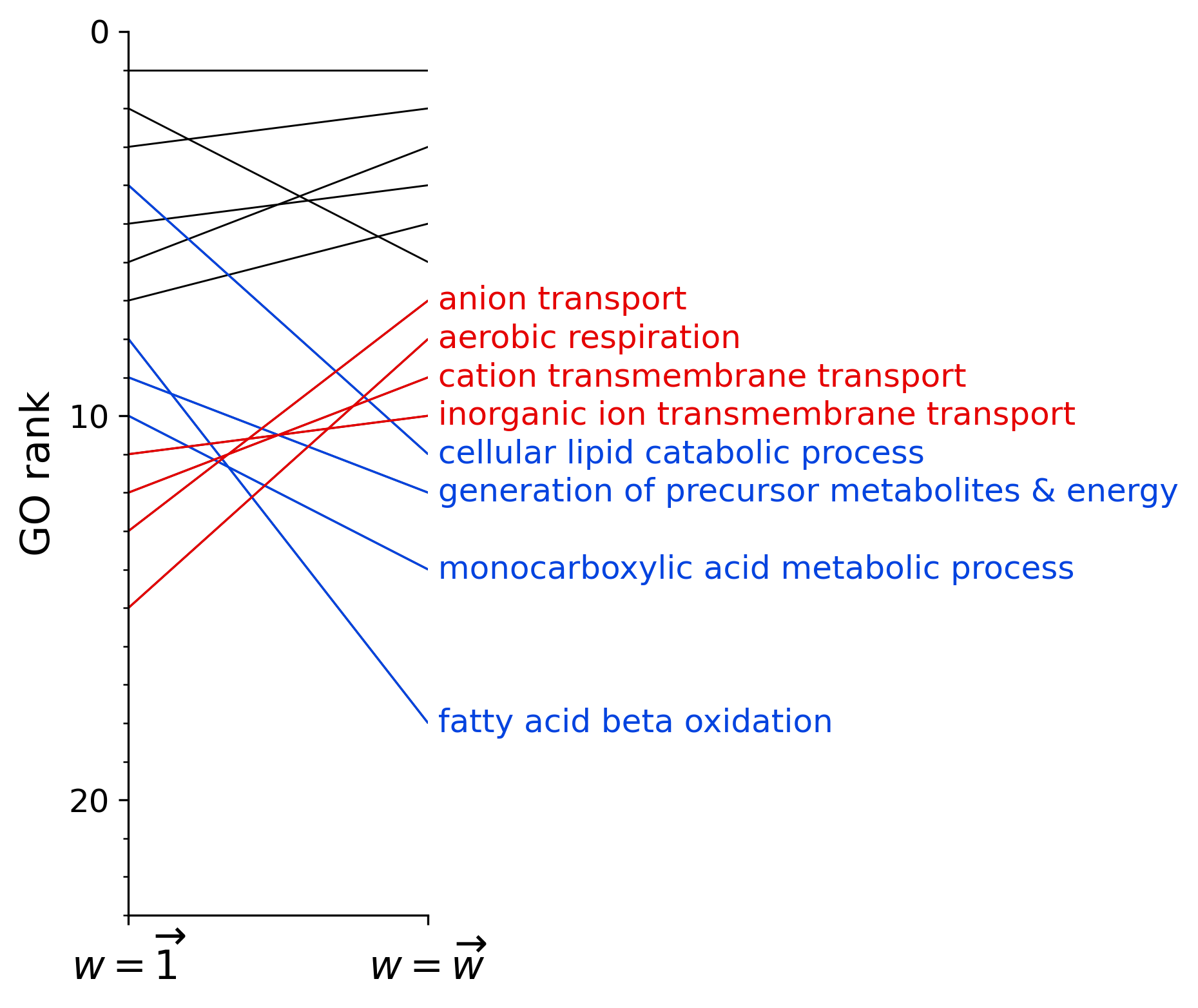}}%
            \hfill%
            \subcaptionbox{\makeSubcaption{14}}{\includegraphics[height=0.8\goHeight]{%
                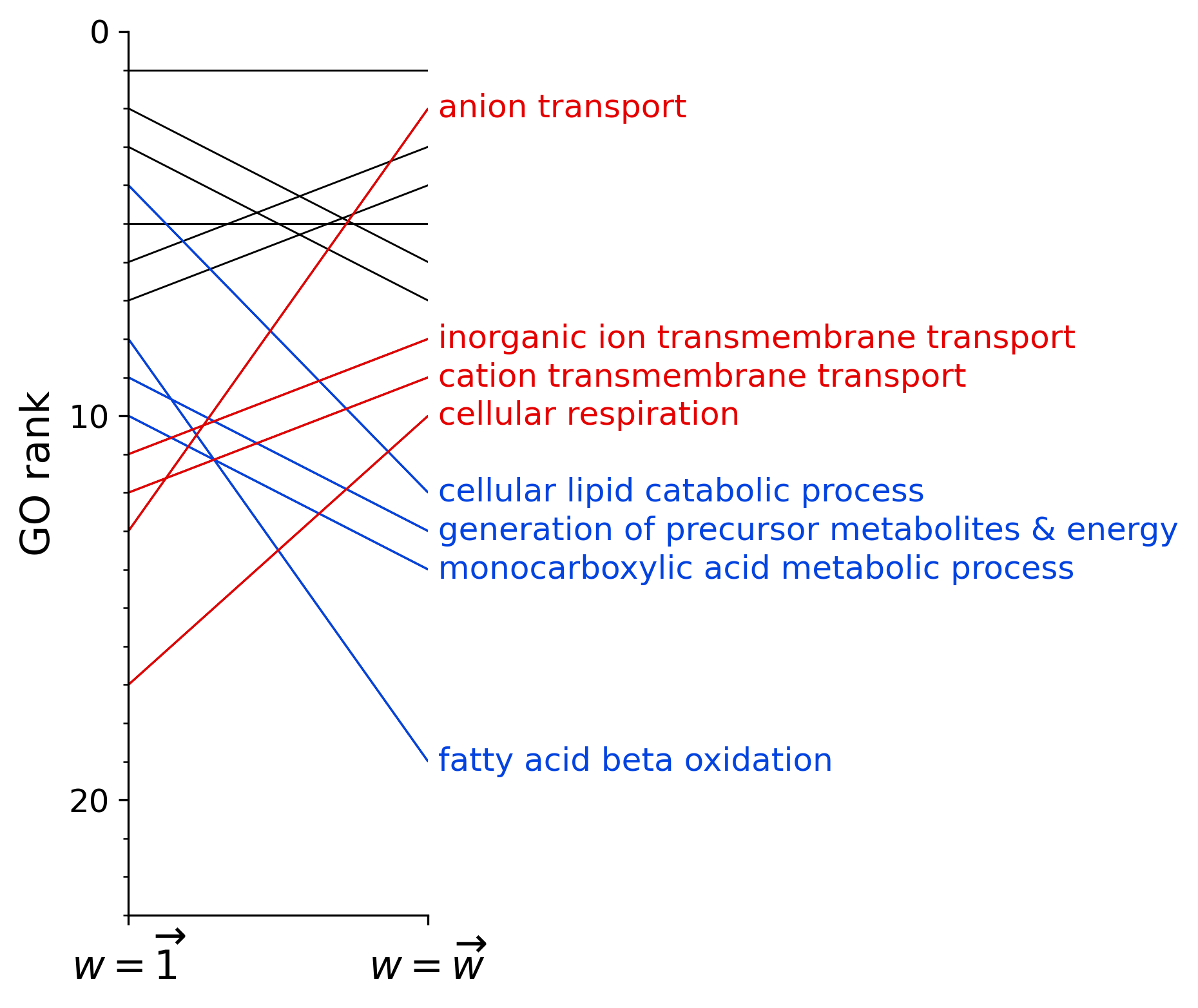}}%
            \hfill%
        }
        \hfill
        \parbox{.33\textwidth}{%
            \subcaptionbox{\makeSubcaption{28} \label{fig:change-in-go-down-drop28}}{%
                \includegraphics[height=1.9\goHeight]{%
                    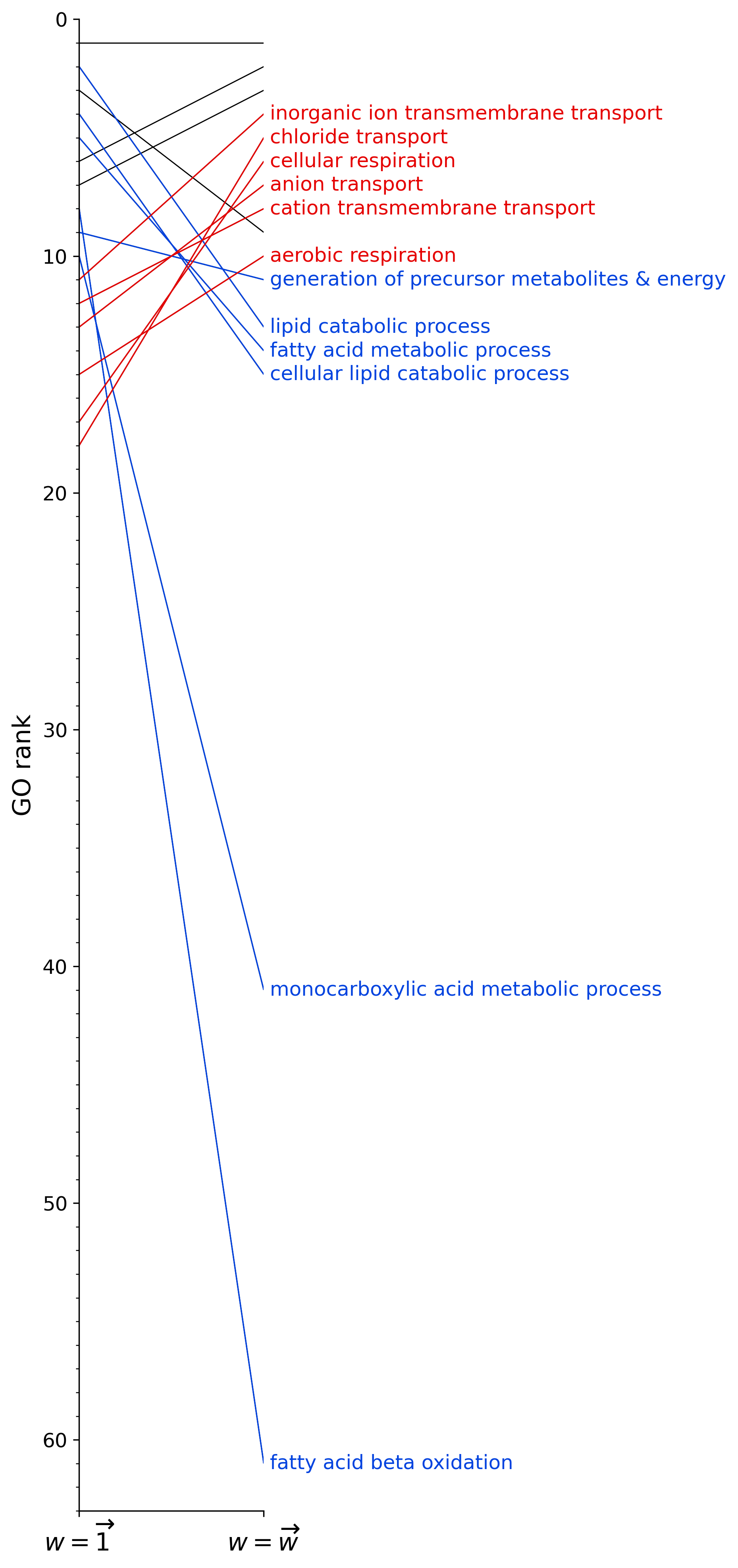}}%
        }
    }
    \caption{
        \textbf{Perturbation to top GO sets (among downregulated genes),
                by dropping a handful of influential cells.}
        Plots show changes to the top 10 ranked \texttt{GO:BP} gene sets
        when an influential cell set of
        the indicated size is dropped.
        Blue lines indicate the change in rank for gene sets that are \textit{demoted},
        red lines indicate the change in rank for those that are \textit{promoted},
        and black lines indicate the change in rank for those that \textit{remain} in the top 10.
        \captionbr
        See \cref{supp-fig:change-in-de-pvals-down} for the corresponding perturbations,
        actual and predicted, for DE p-values of individual genes that give
        rise to these gene-set-level changes.
    } \label{fig:change-in-go-down}
\end{figure}

These perturbations are not necessarily unique; in other words, for several of
these statements, we confirm that there are multiple sets of $K$ cells that
induce a similar effect when dropped (disrupting
the same number, albeit not necessarily the same ranks or entities,
of top gene sets).

For these particular sets of influential cells (whose effect is plotted in
\cref{fig:change-in-go-up,fig:change-in-go-down}), some cells are shared
across clusters of different sizes,
but each cluster contains at least one cell that is unique to that
perturbation (\cref{supp-fig:cell-overlap}).
Further, most influential clusters
(all but one with $K>1$)
are composed of
cells from both the healthy baseline group and the UC group.
While these patterns must be interpreted cautiously (given the non-uniqueness
of influential clusters), they do reflect that our methods
successfully exploit latent synergies between cells in order to
\textit{collectively} disrupt gene-level results---in such a way
that disrupts high-level patterns across differentially expressed genes,
and tailored to
each cell ``budget'' $K$.

These results confirm that robustness
of differential expression for individual genes
can be used to estimate robustness of high-level biological conclusions,
on a pathway or gene set level,
while circumventing the need to differentiate through ranking or subsetting operations
(which are intrinsic to gene set enrichment analyses).\footnote{
    Specifically, this work serves as a proof-of-concept for perturbing
    threshold-based enrichment (the hypergeometric test); future work could use
    similar tactics to identify groups of cells that are predicted to maximally
    perturb rank-based methods (like \cite{broad-gsea}).
    Similarly, with a few tweaks, we could seek cells that perturb
    hypergeometric gene set enrichment results with a minimal magnitude for
    effect size (such that knocking genes in or out of the set used to detect
    enrichment is a function of both p-value \textit{and} effect size).
}

\subsection{interpreting dropping-data robustness} \label{sec:exp-interpretation}

In these experiments, we have demonstrated that our approximate dropping-data metric identifies
widespread nonrobustness across
differential expression results for the UC dataset---including $\apprx$40\% of significant genes
that can be flipped from having a meaningfully large effect $\leftrightarrow$ not,
and $\apprx$50--60\% of all genes that can be flipped from significant $\leftrightarrow$ nonsignificant,
by dropping a handful (<1\%) of cells.
These findings
are backed by empirical experiments demonstrating
that our approximation is trustworthy
within the regimes
that we care about (dropping a small fraction of cells)---though we observe
some deterioration
for genes where dropping data creates
a newly zero-group gene; a phenomenon that should be explored in future work.
This form of nonrobustness cannot be detected through traditional tools, like p-values,
multiple-testing correction, or effect size, and
occurs
despite
multiple checks on
the robustness of differential expression results using these tools.
Further, this widespread sensitivity at the gene-level translates to high-level biological takeaways,
such that dropping a small handful of influential cells can meaningfully
alter
the top 10 gene sets enriched among up- or down-regulated genes.

Taken together, these findings
suggest that differential expression results
from this dataset, accepted
at face value, may not withstand the scrutiny of generalization.
While no result is
entirely misguided
(as we may worry if dropping a handful of cells made the difference between a
significant finding in one direction and a significant finding in the opposite direction),
we find many genes that traditional metrics would flag as ``meaningfully''
differentially expressed between groups yet are
dropping-data sensitive.
Analyses that rely on these traditional metrics to rank or subset genes,
without considering their sensitivity to dropping data,
may therefore fail to prioritize the results that are most likely to generalize
to new datasets and so best characterize the underlying biology of the system.

For such
genes---whose sign, magnitude, and/or significance would meaningfully
change if a small handful of cells were ignored---these outcomes are
seemingly a ``lucky'' (or unlucky) fluke of the particular dataset that was sampled.
If results are not stable, had a few cells not been observed,
then
we have reason to suspect that
a newly collected dataset of goblet cells from
subjects with and without UC (or even new cells from the same subjects)
may fail to corroborate these findings.
Of course,
some level
of nonrobustness is expected for any analysis that is based on discrete decision boundaries
(like
a significance threshold for p-values, or a ``meaningfully large'' threshold for effect sizes).
But, if we believe that statistical testing for differential expression is a valid approach to
detect biological
differences in expression between treatment groups,
then results that are brittle to the exclusion of just a few cells
should give us pause.
It is plausible that such a finding
(e.g., a gene that is ruled to be noteworthy for UC-associated inflammation)
may in fact be an artifact of the particular dataset that was sampled, rather
than
a testament to the underlying biology of the disease.
This suspicion is reinforced by the finding that brittleness is not limited to
a few isolated genes---superficially
affecting results while leaving high-level takeaways intact---but
rather is reflected at the functional level by a corresponding brittleness among
top enriched gene sets.

The takeaway from our dropping-data metric is not (necessarily) to discount
spurious results, but rather to pointedly highlight
where
apparent
outcomes from differential expression may be
misleading.
At minimum, we advise that dropping-data robustness be
\begin{itemize}
    \item reported (alongside the usual p-values and effect sizes)
        when sharing differential expression results, such that others can decide
        whether it meets their standards of replicability, and
    \item used as a lens through which to pointedly re-examine
        the data and chosen model/analysis.
\end{itemize}
\vspace{\parskip}

In some cases, the specified model
may be insufficient to capture the biological and technical factors underlying
measured RNA counts.
Results that are driven by a small population
of cells could point to unexpected biological heterogeneity
(a rare cell subtype or transient transcriptional state).
On the other hand, influential cells may be outliers caused by technical
problems, such as doublets or contaminating (or mislabeled) cells.
Dropping-data robustness empowers researchers with domain knowledge about their
particular system and dataset
to re-examine influential cells in light of the gene or gene set
result that their absence would disrupt.

Looking at
the ulcerative colitis dataset,
we find that gene set sensitivity may align with
previously observed
spatial and functional
diversity within goblet cells.\footnote{
    Which have recently been described as less homogenous than previously
    appreciated; e.g.,~\cite{upper-GC-mucins,senGCs,epithelial-diversity,icGCs}
}
For example, among upregulated genes,
top gene sets that are dropping-data sensitive
(specifically, those that are \textit{demoted} when
the influential cells in \cref{fig:change-in-go-up} are dropped)
primarily revolve around
response to microbial and other stressors,\footnote{
    Namely, ``response to cytokine,'' ``antimicrobial humoral response,''
    and ``response to ER stress''
} as well as cell adhesion and apoptosis.
These resemble the functions that were recently described as characteristic\footnote{
    In comparison to other goblet cells
} of intercrypt goblet cells, a particular subpopulation
located at the surface epithelium between crypts~\cite{icGCs}.
(The complementary gene sets that are \textit{promoted} when those
cells are dropped primarily involve
vesicle transport and secretion,\footnote{
    Namely, ``export from cell,'' ``secretion,'' ``regulation of secretion,''
    ``golgi vesicle transport,'' and ``actin filament based process''
} presumably of mucins.)
Among downregulated genes,
top gene sets that are dropping-data sensitive
(specifically, those that are \textit{demoted} when
the influential cells in \cref{fig:change-in-go-down} are dropped)
virtually all involve
lipid metabolism.\footnote{
    Namely, ``lipid catabolic process,'' ``cellular lipid catabolic process,''
    ``fatty acid beta oxidation,'' and ``monocarboxylic acid metabolic process''
}
In the same work, which
explored functional diversity among goblet cells,
lipid metabolism was the dominant pathway distinguishing
``non-canonical'' goblet cells from
those with a canonical maturation process and
expression profile~\cite{icGCs}.
(The complementary gene sets that are \textit{promoted} when those
cells are dropped all involve ion transport or cellular respiration.)

These associations (of distinct subpopulations within goblet cells,
with the functional impact of dropping a fraction of cells) suggest
that differential expression results could be driven by
shifts in
the population makeup
of goblet cells rather than (solely)
fluctuations in expression within the baseline population.
This distinction does not invalidate
the original differential expression analysis,
but rather suggests that, in order to understand the
etiology and impact of ulcerative colitis,
more work is needed to disentangle
change in composition of the goblet cell population
from changes in expression within
distinct
subtypes.
This is one example of how
dropping-data robustness serves as a tool to more carefully
comb through
and interpret
the results of differential expression, rather than
to nullify results outright.

\pagebreak

Another takeaway from examining dropping-data robustness is to re-examine
the analysis itself, including the chosen summary statistics.
While practitioners are surely aware that reporting
a fixed number of top gene sets
(as a shorthand to summarize complex results in a digestible way)
is susceptible to the downsides
of any somewhat arbitrary hard
cutoff,
it is nonetheless surprising that dropping such a tiny fraction of data---down
to a single cell---is sufficient to disrupt \textit{multiple}
members of the top gene sets.

Through a new lens
(of dropping-data robustness)
this observation
echoes and unifies past findings that
varying the threshold for gene significance can have major implications
for gene set results~\cite{gsea-threshold-microarrays}, as well
as the
caveats of testing gene sets that are far from independent
(due to overlapping genes)~\cite{gsea-redundancy-set-theory, gsea-redundancy-setrank, gsea-redundancy-downweighting, gsea-redundancy-specificity}.\footnote{
    See correlated disruptions to the top gene sets that leverage this overlap,
    e.g., the downranked sets (all involving fatty acid metabolism)
    and the upranked sets (involving ion transport and respiration)
    in \cref{fig:change-in-go-down}.
}
Whereas past work examined the robustness of GSEA as a consequence of analysis
decisions---assuming that results, if flawed, would at least be consistent
across future samples---our results suggest that top gene sets are
meaningfully nonrobust to
even tiny perturbations to the data itself.
It remains to be seen whether this result
is typical of scRNA-seq datasets
or confined to particular examples like this one---as well as the degree to
which pseudobulk analysis (as opposed to individual cell) addresses this
instability.

%% file: 6-conclusions.tex
We
set forth
a framework to efficiently
estimate dropping-data robustness for differential expression
analyses,
with respect to gene-level results
(building on the framework established in \cite{ryan-amip})
as well as high-level
functional
takeaways
(based on a novel approach to synthesize robustness results across regressions).
For a sample scRNA-seq dataset, we find
that many of these results can be consequentially disrupted by dropping a
handful of influential cells from the analysis
(<1--2\% \ldots
or even just a single cell).

We reiterate that
we do not suggest throwing out differential expression results that
survive the scrutiny of classical inference but not dropping-data robustness.
Rather, we suggest that these results be \textit{interpreted} differently
(with respect to their generalizability, and potential influence by
unappreciated sources of conditional structure within the data)---analogous
to how
significance testing that fails to detect an effect
is not equivalent to
``positively detecting the absence of an effect''~\cite{ryan-amip}.
For example,
dropping-data robustness can be used to prioritize
which biological hypotheses merit
further investigation (particularly under limited resources)---as well as
to diagnose unforeseen technical issues
and/or
point to
interesting biological heterogeneity
within the data.

\linerule

We close by highlighting fruitful directions for future work building on our
results:

First, while we develop a framework for dropping-data robustness
based on both individual cell and pseudobulk approaches to
single-cell measurements,
the experiments we present are based on the individual cell model.
We leave it to future work to apply our robustness framework for
the pseudobulk model
to single-cell data, and to compare the dropping-data robustness of differential expression results
(from the same dataset)
across models.
Discussions around this choice (of whether and how to aggregate single-cell measurements)
have largely been based on statistical power;\footnote{
    Namely, that the individual cell approach provides false power by treating cells from the
    same subject
    as independent samples, whereas the pseudobulk approach loses the resolution provided by single-cell
    measurements and may be under-powered
} another important, yet distinct, lens into robustness and replicability
(under the assumption that the data in hand is sampled precisely from the target population).
Understanding how these models behave under realistic data perturbations (dropping a handful of cells),
for real single-cell datasets, would provide insight into the tradeoffs of this choice
from a new angle of
generalizability (to future samples that may systematically differ from the data in hand).
It would be also interesting to explore whether
cells
are similarly
influential across models,
or if some cells
play a keystone role under only one approach.

Second, we develop an approach to dropping-data robustness of gene set enrichment analysis where
\begin{enum-inline}
    \item enrichment is based on a hypergeometric test (thresholding genes by significance)
    and
    \item robustness is measured with respect to the composition of the top 10 gene sets.
\end{enum-inline}

Future work could adapt our approach
in order to measure robustness of GSEA based on effect size as well as significance
(by clustering influential cells
based on
\textit{two} influence matrices, formed with respect to the
unsigned Wald statistic $\stat_W^+$ as well as the unsigned treatment effect $\stat_\text{LFC}^+$\footnote{
    Or, even more simply, by constructing a Wald test with respect to the minimal meaningful effect size
    in order to choose differentially expressed genes for GSEA, and directly
    applying our existing dropping-data framework to this alternate Wald statistic
    (rather than multiply filtering genes based on a Wald test with a null hypothesis
    of zero as well as a separate filter on the size of the effect)
}),
as well as GSEA based on ranking rather than thresholding genes
(a less straightforward task, since ranking genes is
non-differentiable and thus
not readily amenable to approximating influences\footnote{
    Recent developments in scalable
    relaxations
    for differentiable ranking
    (e.g.,~\cite{diffable-ranking})
    may be a promising direction
}).

Further, in addition to estimating robustness of the top 10 gene sets as a whole,
future work could estimate the individual robustness of each top gene set of interest
(i.e., the minimal number of cells that could be dropped in order to knock that gene set \textit{into}
or \textit{out of} the top 10\footnote{
    Or, alternately, the minimal or maximal rank that that gene set could achieve
    by dropping a given number of cells
}).
This, too, can be done by directly adapting our clustering approach,
but tailoring the selection of gene features for clustering and scoring
to target one gene set at a time.

Finally, we suggest that this framework (of dropping-data robustness) is a
generally useful construct for biology, which increasingly depends on large and
high-dimensional datasets and an ever-expanding array of computational methods.
For example,
genome-wide associate studies (GWAS) present an obvious candidate for dropping-data robustness,
because they involve a methodology (linear regression) that
straightforwardly lends itself to robustness\footnote{
    Once coupled with our simple approach to estimate robustness of gene p-values
    after multiple-testing correction based on rank
    (\cref{sec:de-analysis-for-sensitivity})
}
\textit{and} is widely adopted (as opposed to single-cell analyses, where methodological approaches are more splintered).
Dropping-data robustness would be a powerful tool to measure the effect of dropping a small handful of individuals
on GWAS effect sizes, or on polygenic risk scores
(synthesized from multiple GWAS).
More broadly, any methods that can be formulated as optimizing a twice
differentiable objective, such as a log-likelihood,
are directly amenable to the dropping-data approximation.\footnote{
    More precisely, so long as they yield statistics-of-interest that themselves are differentiable
    functions of those estimators
}
On the other hand, many biological analyses involve multi-step heuristic procedures---and these
may be precisely the cases where robustness
is a worry.
While such analyses require more hands-on work to develop tools for dropping-data robustness,
influences can be propagated from step to step, and these methods may still be amenable to well-behaved approximations.

Notably, the tools we develop allow us to flexibly compute dropping-data robustness across
many flavors of GLMs
(such as varying the
link and distribution of the response),
and
to modularly incorporate data sensitivity of additional parameters
(such as adjusting the
normalization scheme for cell sizes,
or expressing the overdispersion
as the solution to an additional optimization),
thanks to automatic differentiation.
While methods in computational biology may remain fractured,\footnote{
    Not that this is necessarily a bad thing;
    biological datasets
    may benefit from being exposed to a wider variety of approaches,
    each with its own biases and blind spots
} we argue that this is one good reason (of many) to express analyses,
where possible,
in the common framework of differentiable programming languages.\footnote{
    This can be as simple as writing clean
    \texttt{numpy}, and instead importing \texttt{jax.numpy} \cite{jax}
}
This would allow for the development of a toolkit of
differentiation-based metrics,
including dropping-data robustness,\footnote{
    As well as metrics for convergence based on the gradient and Hessian,
    which would have readily detected the issue with zero-group genes
    (\cref{sec:pseudocells,fig:convergence})
}
that could easily be ported across analyses in order to audit
the generalizability of
new methods and datasets.

%% file: 7-acknowledge.tex
We thank Michael Hoffman and Michael Love for helpful discussions.
This work was supported in part by an NSF CAREER Award and an ONR Early Career Grant.

%% file: _-appendix.tex
\section{negative binomial is equivalent to a gamma mix of Poissons}\label{app:gamma-poisson}

Here we'll review the well-known result that the negative binomial density can
be derived as a gamma-weighted mixture of Poissons~\cite{nb-gamma-poisson}.

Let
\begin{alignat*}{2}
    \p(\lambda \mid \gammaA, \gammaB)
        &= \frac{\gammaB^{\gammaA}}{\Gamma(\gammaA)} \,
           \lambda^{\gammaA-1} \, \exp\left\{ \minus \gammaB \lambda \right\}
        && \mathcomment{\parbox{5.5cm}{ \singlespacing
            Poisson rate distributed as gamma (parameterized by shape and rate)
        }}
    \\
    \p(y \mid \lambda)
        &= \exp\left\{ \minus\lambda \right\} \, \lambda^y \, \frac{1}{y\,!}
        && \mathcomment{observations distributed as Poisson}
\shortintertext{---then,}
    \p(y \mid \gammaA, \gammaB) &= \int \frac{\gammaB^{\gammaA}}{\Gamma(\gammaA)} \,
                                       \lambda^{\gammaA-1} \,
                                       \exp\left\{ \minus \gammaB \lambda \right\} \,
                                       \exp\left\{ \minus \lambda \right\} \,
                                       \lambda^y \,
                                       \frac{1}{y\,!} \; \d \lambda
    \\
                               &= \frac{\gammaB^{\gammaA}}{\Gamma(\gammaA)} \,
                                  \frac{\Gamma(y+\gammaA)}{(\gammaB+1)^{y+\gammaA}} \,
                                  \frac{1}{y\,!}.
\end{alignat*}

The mean of this density is
\begin{align*}
    \E_{\p(y)} [ y ]
        &= \E_{\p(\lambda)} \left[ \E_{\p(y \mid \lambda)} [ y ] \right]
        \\
        &= \E_{\p(\lambda)} [ \lambda ]
        \\
        &= \frac{\gammaA}{\gammaB}
        \eqdef \mu
\shortintertext{and the variance is}
    \var_{\p(y)} [ y ]
        &= \E_{\p(\lambda)} \left[ \var_{\p(y \mid \lambda)} [ y ] \right]
           + \var_{\p(\lambda)} \left[ \E_{\p(y \mid \lambda)} [ y ] \right]
        \\
        &= \E_{\p(\lambda)} [ \lambda ] + \var_{\p(\lambda)} [ \lambda ]
        \\
        &= \frac{\gammaA}{\gammaB} + \frac{\gammaA}{\gammaB^2}
        \eqdef V
\end{align*}

Consider what happens if we fix $\mu$
(akin to conditioning on a particular realization of the covariates in a GLM)
and parameterize the mean-variance relationship in terms of the gamma shape parameter $\gammaA$.
Since $\mu = \gammaA / \gammaB \implies \gammaB = \gammaA / \mu$,
\begin{align*}
    V(\mu) = \frac{\gammaA}{\gammaA / \mu} + \frac{\gammaA}{\left( \gammaA / \mu \right)^2}
           = \mu + \frac{1}{\gammaA} \mu^2 \, ;
\end{align*}
this is familiar as the characteristic
negative binomial mean-variance relationship
(with dispersion $\disp = 1/\gammaA$).

Rewriting the log-likelihood $\ell \defeq \log p$ in terms of $\mu$ and $\disp$,
\begin{align*}
    \p(y \mid \mu, \disp)
        &= \frac{ (1/\disp)^{1/\disp} }{ \mu^{1/\disp} \, \Gamma(1/\disp) } \,
           \frac{ \Gamma(y + 1/\disp) }{ (1 + (1/\disp)/\mu)^{y + 1/\disp} } \,
           \frac{1}{y\,!}
    \\
        &=
           \left[ \frac{\mu}{\mu + 1/\disp} \right]^y
           \left[ \frac{1/\disp}{\mu + 1/\disp} \right]^{1/\disp}
           \frac{\Gamma(y + 1/\disp)}{\Gamma(1/\disp) \, y\,!}
\\
    \implies
\\
    \ell(y \mid \mu, \disp) &=
          y \log \frac{\mu}{\mu + 1/\disp}
        + 1/\disp \log \frac{1/\disp}{\mu + 1/\disp}
        + \log \frac{\Gamma(y + 1/\disp)}{\Gamma(1/\disp) \, y\,!} \, ,
\end{align*}
recovering the negative binomial log-likelihood.

\section{likelihoods and quasi-likelihoods}\label{app:l-v-ql}

The \texttt{DESeq2} library uses a typical GLM likelihood,
whereas \texttt{glmGamPoi} uses a \textit{quasi}-likelihood.
We explain this distinction here
(in order to later justify our decisions in
designing a general framework for robustness of differential expression,
which readily extends to both settings);
a more expansive explanation is given in \cref{app:ql}.

We begin by situating differential expression GLMs within the framework of
\emph{natural exponential family} models.
Namely, as negative binomials are in the exponential family
(reviewed in more detail in \cref{app:exponential-fam}),
their likelihood can be expressed as
\[
    \log \, \p(y \mid \eta ) = \eta \, y - A(\eta)
\]
for natural parameter $\eta$ and log-normalizer $A$,
where $T(y) = y$ serves as the sufficient statistic.
A handy property of
this definition
is that differentiating $A$ yields the moments of our distribution:
\begin{align}
    \dd{A}{\eta} = \E[Y \mid \eta]
                 &\eqdef \mu \label{eq:A-mu}
\shortintertext{and}
    \dd[2]{A}{\eta}
                    = \dd{\mu}{\eta}
                    = \var[Y \mid \eta]
                    &\eqdef V(\mu). \label{eq:A-var}
\end{align}
For a generalized \textit{linear} model, the mean $\mu$ is, in turn, given
as a linear function of the covariates.\footnote{
    In particular,
    $\mu = \E[ \vec{\mu} ]$
    where
$
        \vec{\mu}
                = \link\inv(\mat{X} \, \vec{\beta} \,\oplus\, \vec{o})
    $
    for some offset vector $\vec{o}\ofSize{\ncells,1}$.
    For \texttt{DESeq2} and \texttt{glmGamPoi}, $\link=\log$ (the canonical link)
    and $\vec{o} = \log \vec{\size}$.
}
The variance function $V$ casts the variance as a function of $\mu$;
each distribution in the natural exponential family has a characteristic
such
mean-variance relationship.

For our parameterization of the negative binomial, this relationship is
\begin{align}
    V(\mu) = \mu + \disp \mu^2 \label{eq:nb-mean-var}
\end{align}
for dispersion $\disp$ ($\geq 0$).

On the other hand, the \emph{quasi-likelihood} framework only requires defining the first
two derivatives of $A(\eta)$---corresponding to the first and second
moments of the distribution (\cref{eq:A-mu,eq:A-var})---without needing to
explicitly define $A$ itself and ensure that it's a proper normalizer.
Under this framework, the score used for maximum-likelihood optimization
(a function of $\mu$ and $V(\mu)$, a characteristic of the distribution,
given by differentiating the log-likelihood)
can instead be replaced by a quasi-score
(a function of the chosen $\mu$ and $V(\mu)$, regardless of whether
they correspond to a viable log-likelihood)
and used to estimate $\mu$ (and thus $\vec{\beta}$)~\cite{ql}.
Specifically, it turns out that the Newton-Raphson update to estimate $\mu$
is invariant to scaling of $V(\mu)$:
a familiar result from the quasi-likelihood literature that we review for the
particular case of \texttt{glmGamPoi}'s implementation in \cref{app:ggp-newton}.
In other words,
though this definition of the variance
corresponds to no viable
generative model,
it is nonetheless sufficient to
estimate $\vec{\beta}$.

So, for the quasi-likelihood posited by \texttt{glmGamPoi}
(``quasi'' because it
doesn't
necessitate
the existence of a congruous
probability density~\cite{ql}),
the mean-variance relationship is
\begin{align}
    V(\mu) = \QLdisp \times (\mu + \NBdisp \, \mu^2) \label{eq:ql-mean-var}
\end{align}
for quasi-likelihood dispersion $\QLdisp$
($\geq 1$).

Since both assumptions of mean-variance relationship
(\cref{eq:nb-mean-var,eq:ql-mean-var}) hold,
\begin{align}
    V(\mu) &\defeq \mu + \disp \mu^2 \nonumber
    \\
           &\defeq \QLdisp \times (\mu + \NBdisp \, \mu^2) \nonumber
    \\
    \implies \nonumber
    \\
    \disp &= \frac{\QLdisp - 1}{\mu} + \QLdisp \, \NBdisp. \label{eq:disp}
\end{align}

\texttt{DESeq2}---which postulates a standard likelihood---uses heuristics to
determine $\disp$ for each gene,
then estimates $\hat{\vec{\beta}} \mid \disp$.
On the other hand, \texttt{glmGamPoi}---which postulates a quasi-likelihood
framework---uses heuristics to
determine $\NBdisp$ ($< \disp$)
and $\QLdisp$ for each gene,\footnote{
    In fact, \texttt{glmGamPoi}'s heuristic procedure involves
    first fitting a rough estimate of $\disp$,
    and regressing $\disp$ against $\mu$ across genes,
    in order to estimate
    $\NBdisp$ and $\QLdisp$ (via \cref{eq:disp}).
}
then estimates $\hat{\vec{\beta}} \mid \NBdisp$
and modulates its statistical test with $\QLdisp$.

\section{negative binomial as an exponential family model}\label{app:exponential-fam}

The negative binomial density---conditioned on a fixed dispersion---is in the form of
an exponential family~\cite{exponential-fam} where
$$
\begin{aligned}
    \eta(\theta) &= \eta(\mu) = \log \frac{\mu}{\mu + 1/\disp}
\\
    T(y) &= y
\\
    A(\theta) &= A(\mu) = -1/\disp \log \frac{1/\disp}{\mu + 1/\disp}
        \implies A(\eta) = -1/\disp \log \left[ 1 - \exp(\eta) \right]
\\
    h(y) &= \frac{\Gamma(y + 1/\disp)}{\Gamma(1/\disp) \, y\,!} =
            \frac{\Gamma(y + 1/\disp)}{\Gamma(1/\disp) \, \Gamma(y - 1)}
\end{aligned}
$$
with the typical exponential family log-density
\[
    \ll(y \mid \theta) = \eta(\theta) \,\, T(y) - A(\eta) + \log h(y)
\]
for parameter $\theta$, natural parameter $\eta$, sufficient statistic $T$,
log-partition $A$, and base measure $h$.

Then, by the properties of a natural exponential family,
$$
\begin{aligned}
    \E[y] &= \dd{A}{\eta} = \frac{1/\disp \exp \eta}{1 - \exp \eta}
        \eqdef \mu
\\
    \var[y] &= \dd[2]{A}{\eta} = \frac{1/\disp \exp \eta}{(1 - \exp \eta)^2}
        = \mu + \disp \mu^2 \eqdef V(\mu)
\end{aligned}
$$
and we recover the characteristic mean-variance relationship of a negative binomial.

\section{the quasi-likelihood framework of \texttt{glmGamPoi}}\label{app:ql}

\texttt{glmGamPoi}, following \texttt{edgeR} and its predecessors~\cite{ql-edgeR,ql-tjur},
adopts a quasi-likelihood framework atop
the usual exponential family GLM.

The central change is to redefine the model variance as
$V'(\mu) \defeq \QLdisp \, V(\mu)$, where $V(\mu)$
is the characteristic mean-variance relationship of a negative binomial
(\cref{eq:nb-mean-var}) and $\QLdisp$ is a positive constant (per gene).
Ostensibly, this change
(``over''-overdispersion)
serves to better calibrate p-values
by injecting additional uncertainty into the model, to
address the flaw of conditioning on $\disp$ in order to fit coefficients
when in reality the value of $\disp$ is uncertain.
However, we can no longer evaluate or sample from likelihoods under this model
(since it's not tied to a defined probability density
that sums to 1); hence the term quasi-likelihood.

Nonetheless, this definition is sufficient to form an estimator of $\mu$
(and thus $\vec{\beta}$),
by rewriting the score (a function of $\mu$ and $V(\mu)$, itself dictated
by the chosen distribution)
as a quasi-score (a function of $\mu$ and $V(\mu)$, selected at will
with no guarantee that it corresponds to a viable likelihood).
Specifically, it turns out that estimation under the typical
negative binomial log-likelihood objective
is invariant to scaling by $\QLdisp$ (\cref{app:ggp-newton}).

Completing the framework,
\texttt{glmGamPoi} places a scaled inverse $\chi^2$ prior
over the quasi-likelihood dispersion (per gene)
\[
    \QLdisp \sim ( \tau^2 \nu ) \times 1 / \chi^2_{\nu}
\]
where hyperparameters $\tau, \nu$ are set
empirically, using data across genes.

Under the assumption that observations are roughly normal,
this prior would be conjugate
and its posterior would have a closed form (as $\chi^2$-distributed).
This is
generally not the case, but
Tjur (the basis for \texttt{edgeR} and, transitively, for \texttt{glmGamPoi})
posits that
``common sense suggests that it is better to perform this correction for
randomness...than not to perform any correction at all.''~\cite{ql-tjur}
So, given maximum likelihood estimate $\hat{\QLdisp}$ with
$(\ncells - \nbetas)$ degrees of freedom,
the final quasi-likelihood overdispersion estimator is calculated as
\begin{equation} \label{eq:ql-disp}
    \hat{\QLdisp} = \frac{\nu\tau^2 + (\ncells - \nbetas)\tilde{\QLdisp} }
                         {\nu + (\ncells - \nbetas)}.
\end{equation}

The likelihood ratio test statistic \text{LR}---where likelihoods are
evaluated under the original model likelihood, ignoring ``quasi-''
amendments---is asymptotically (in $\ncells$) distributed as $\chi^2$
(with $\df_{\text{LR}}$ degrees of freedom; generally 1) under the null
(Wilks' theorem).
To incorporate additional uncertainty through ``over''-overdispersion,
\texttt{glmGamPoi} then scales this statistic by the quasi-likelihood
dispersion estimate $\hat{\QLdisp}$, yielding the test statistic
\[
    F \defeq \frac{\text{LR} / \df_{\text{LR}}}{ \hat{\QLdisp} }.
\]
The null distribution of $F$ is assumed to follow
an F-distribution with
($\df_{\text{LR}} \defeq \df_{\M} - \df_{\M\reduced} = \nbetas - \nbetas\reduced$)
and
($\df_{\QLdisp} \defeq \nu + \ncells - \nbetas$)
degrees of freedom.\footnote{
    Recall that $\M$ is the ``full'' model (with all covariates),
    and $\M\reduced$ is the ``reduced'' model (e.g., excluding $\betaTreated$).
    So, $\df_{\text{LR}}=1$ for the most common comparison-of-interest
    in differential expression.
}

The adoption of a quasi-likelihood by \texttt{glmGamPoi} implies the belief that the mean structure
of the GLM is well-specified, but the variance is overly conservative---though
the form is correct, up to a scalar multiplier ($\QLdisp > 1$).
The stated justification is that uncertainties are miscalibrated
(i.e., confidence intervals are too tight)
when the coefficients $\vec{\beta}$ are estimated by conditioning on a fixed
dispersion, since the dispersion itself ought to be a random variable with
uncertainty~\cite{ggp}.
Rather than directly treating the dispersion as a random variable and fitting
the GLM with respect to multiple parameters ($\vec{\beta}, \disp$),
the quasi-likelihood framework ostensibly provides an alternate mechanism
to inflate ``overly confident'' p-values
(by fitting $\vec{\beta}$ conditional on $\NBdisp < \disp$
and altering the test statistic and its null sampling distribution,
as described above).

\section{the \texttt{glmGamPoi} inference algorithm recovers standard Newton-Raphson}\label{app:ggp-newton}

Here we'll verify that the \texttt{glmGamPoi} inference algorithm,
which is motivated by minimizing deviance based on iteratively reweighted least
squares (IRLS)---a historically popular algorithm in the GLM literature because
of its connection to optimizing pure linear models---is numerically equivalent
to typical Newton-Raphson optimization of a GLM log-likelihood objective
(as expected).
Additionally, we'll show that it optimizes the
original objective we describe in \cref{sec:de-objective}
and that---conditional on the negative binomial dispersion parameter---it is
independent of \texttt{glmGamPoi}'s quasi-likelihood framework.
This analysis validates
these general familiar
results~\cite{hardin-glm,ql}
for the particular case of the \texttt{glmGamPoi} algorithm.

Assume that all parameters except $\vec{\beta}$ are fixed.
For example, the gene-specific dispersion $\disp$ is empirically
estimated up front and henceforth considered constant.
Then, inference proceeds by
iteratively optimizing the
log-likelihood objective,
by updating
\[
    \hat{\vec{\beta}}_{(\step+1)} = \hat{\vec{\beta}}_{(\step)} + \mathop{\Delta}\left( \hat{\vec{\beta}}_{(\step)} \right)
\]
until some convergence criteria are met,
where step function $\Delta$ is some scaling of the gradient at the
current estimate $\hat{\vec{\beta}}_{(\step)}$.
The output is the maximum likelihood estimate of the coefficients, $\hat{\vec{\beta}}$.

Consider the basic GLM log-likelihood objective.
Take the $\icell\th$ data point $(\vec{x}_{\icell}, y_{\icell})$---where
$\vec{x}_{\icell}$ is the column vector formed by transposing the $\icell\th$ row of $\mat{X}$,
and $y_{\icell}$ is the scalar formed by selecting the $\icell\th$ component of $\vec{y}$.
Under the exponential family framework (\cref{app:exponential-fam}),
the gradient for this point is
$$
\begin{aligned}
    \nabla_{\icell} \defeq \dd{\log p}{\vec{\beta}}
    &= \dd{\eta}{\vec{\beta}} \, y_{\icell} - \dd{A}{\vec{\beta}}
\\
    &= \dd{\eta(\mu_{\icell})}{\mu_{\icell}} \, \dd{\mu}{\vec{\beta}} \, y_{\icell}
        - \dd{A(\eta_{\icell})}{\eta_{\icell}} \, \dd{\eta(\mu_{\icell})}{\mu_{\icell}} \, \dd{\mu}{\vec{\beta}}
\\
    &= \frac{y_{\icell} - \mu_{\icell}}{V(\mu_{\icell})} \, \dd{\mu}{\vec{\beta}}.
\end{aligned}
$$

Plugging in the negative binomial mean-variance function,
and noting that here
\[
    \dd{\mu}{\vec{\beta}} = \size_{\icell} \, \exp\{ \xbn \} \, \vec{x}_{\icell}
                          = \mu_{\icell} \, \vec{x}_{\icell},
\]
this simplifies to
\[
    \nabla_{\icell} = \frac{y_{\icell} - \mu_{\icell}}{\mu_{\icell} + \disp \mu_{\icell}^2} \, \mu_{\icell} \, \vec{x}_{\icell}
                  = \boxed{ \frac{y_{\icell} - \mu_{\icell}}{1 + \disp \mu_{\icell}} \, \vec{x}_{\icell} } \;.
\]

Then consider the corresponding Hessian:
$$
\begin{aligned}
    \mat{\nabla}^2_{\icell} \defeq \ddT[2]{\log p}{\vec{\beta}}
    &= \mathop{\dd{}{\vec{\beta}}} \left[ \frac{y_{\icell} - \mu_{\icell}}{V(\mu_{\icell})} \, \dd{\mu}{\vec{\beta}} \right]\T
\\
    &= \mathop{\dd{}{\vec{\beta}}} \left[ \frac{y_{\icell} - \mu_{\icell}}{V(\mu_{\icell})} \right] \cdot \left( \dd{\mu}{\vec{\beta}} \right)\T
       + \frac{y_{\icell} - \mu_{\icell}}{V(\mu_{\icell})} \, \ddT[2]{\mu}{\vec{\beta}}
\\
    &= \frac{ -\dd{\mu}{\vec{\beta}} \, V(\mu_{\icell}) - (y_{\icell}-\mu_{\icell}) \, \dd{V(\mu_{\icell})}{\mu_{\icell}} \, \dd{\mu}{\vec{\beta}} }{V(\mu_{\icell})^2}
         \cdot \left( \dd{\mu}{\vec{\beta}} \right)\T
       + \frac{y_{\icell} - \mu_{\icell}}{V(\mu_{\icell})} \, \ddT[2]{\mu}{\vec{\beta}}
\\
\\
    &= - \frac{ V(\mu_{\icell}) + (y_{\icell}-\mu_{\icell}) \, \dd{V(\mu_{\icell})}{\mu_{\icell}} }{V(\mu_{\icell})^2}
         \, \dd{\mu}{\vec{\beta}} \cdot \left( \dd{\mu}{\vec{\beta}} \right)\T
       + \frac{y_{\icell} - \mu_{\icell}}{V(\mu_{\icell})} \, \ddT[2]{\mu}{\vec{\beta}}
\end{aligned}
$$

Again plugging in identities for the \texttt{glmGamPoi} model, and differentiating $\mu(\vec{\beta})$,
\begin{equation} \label{eq:hessian}
    \vec{\nabla}^2_{\icell} = \boxed{
        - \vec{x}_{\icell} \,
            \frac{ \mu_{\icell} \, (1 + \disp \, y_{\icell}) }
                 { (1 + \disp \mu_{\icell})^2 }
                \, \vec{x}_{\icell}\T
    } \; .
\end{equation}

Then, the \emph{Newton-Raphson} step would be
$$
\begin{aligned}
    \mat{\Delta}_{\mathrm{NR}}
    &= \minus\left( \sum\overCells \mat{\nabla}^2_{\icell} \right)\inv \,\cdot\, \sum\overCells \vec{\nabla}_{\icell}
\\
    &\defeq \minus\left( \mat{\nabla}^2 \right)\inv \, \mat{\nabla}
\\
    &= \left[ \mat{X}\T \, \left( \frac{ \vec{\mu} \odot (1 + \disp \, \vec{y}) }
                                 { (1 + \disp \vec{\mu})^2 } \odot \mat{X} \right)
       \right]\inv
            \mat{X}\T \, \frac{\vec{y} - \vec{\mu}}{1 + \disp \vec{\mu}}
\end{aligned}
$$
where $\minus\mat{\nabla}^2$
(the ``observed'' Fisher information)
is computed
by taking the sample estimate (i.e., numerically evaluating the Hessian at each data point).

The corresponding \emph{Fisher scoring} step takes the analytical expectation of the negative
Hessian (the ``expected'' Fisher information) rather than
averaging
empirically.
By construction, $\E[\vec{y}] = \vec{\mu}$, so:
$$
\begin{aligned}
    \mat{\Delta}_{\mathrm{FS}}
    &= -\left( \E[ \mat{\nabla}^2 ] \right)\inv \, \mat{\nabla}
\\
    &= \left[ \mat{X}\T \, \left( \frac{ \vec{\mu} \odot (1 + \disp \, \vec{\mu}) }
                                 { (1 + \disp \vec{\mu})^2 } \odot \mat{X} \right)
       \right]\inv
            \mat{X}\T \, \frac{\vec{y} - \vec{\mu}}{1 + \disp \vec{\mu}}
\\
    &= \left[ \mat{X}\T \, \left( \frac{ \vec{\mu} }
                                 { 1 + \disp \vec{\mu} } \odot \mat{X} \right)
       \right]\inv
            \mat{X}\T \, \frac{\vec{y} - \vec{\mu}}{1 + \disp \vec{\mu}}.
\end{aligned}
$$

When the model is an exponential family GLM with canonical link, Newton-Raphson
and Fisher scoring are equivalent.

Finally, observe what happens if the mean-variance relationship is redefined as
$V'(\vec{\mu}) = \QLdisp \, V(\vec{\mu})$.
Then,
\begin{align*}
    \nabla_{\icell}'
    &= \frac{y_{\icell} - \mu_{\icell}}{\QLdisp \, V(\mu_{\icell})} \, \dd{\mu}{\vec{\beta}}
\\
    &= \frac{1}{\QLdisp} \, \nabla_{\icell}
\\
\shortintertext{and}
\\
    \left(\mat{\nabla}^2_{\icell}\right)'
    &= - \frac{ \QLdisp \, V(\mu_{\icell}) + (y_{\icell}-\mu_{\icell}) \, \dd{}{\mu_{\icell}}\left[ \QLdisp \, V(\mu_{\icell}) \right] }{\left( \QLdisp \, V(\mu_{\icell}) \right)^2}
         \, \dd{\mu}{\vec{\beta}} \cdot \left( \dd{\mu}{\vec{\beta}} \right)\T
       + \frac{y_{\icell} - \mu_{\icell}}{\QLdisp \, V(\mu_{\icell})} \, \ddT[2]{\mu}{\vec{\beta}}
\\
    &= - \frac{ \QLdisp \, V(\mu_{\icell}) + (y_{\icell}-\mu_{\icell}) \, \QLdisp \, \dd{V(\mu_{\icell})}{\mu_{\icell}} }{\QLdisp^2 \, V(\mu_{\icell})^2}
         \, \dd{\mu}{\vec{\beta}} \cdot \left( \dd{\mu}{\vec{\beta}} \right)\T
       + \frac{y_{\icell} - \mu_{\icell}}{\QLdisp \, V(\mu_{\icell})} \, \ddT[2]{\mu}{\vec{\beta}}
\\
    &= \frac{1}{\QLdisp} \, \mat{\nabla}^2_{\icell}.
\end{align*}

Between the score and the inverse Hessian, the $1 / \QLdisp$ factors cancel
and the optimization steps are exactly the same as before (and so the optimal
$\hat{\vec{\beta}}$ also remains unchanged).
In other words, GLM optimization is invariant to the quasi-likelihood overdispersion $\QLdisp$,
and we can ignore this parameter when fitting the coefficients
or deriving a Z-estimator for sensitivity analysis.

Now we'll walk through the three methods implemented by \texttt{glmGamPoi} to
calculate an IRLS optimization step, building up in complexity, and
show that each is equivalent to a form of Fisher scoring.

\subsection{w/o prior (diagonal)}

First we can explain \texttt{fisher\_scoring\_diagonal\_step}\footnote{
    \url{https://github.com/const-ae/glmGamPoi/blob/6c5c93118f21ca9f663d233ab96404b27dfd5f59/inst/include/fisher_scoring_steps.h\#L76-L86}
} (no prior / ridge penalty).
Consider the implementation by \texttt{glmGamPoi}, where $\W$ is the IRLS weight vector:
$$
\begin{aligned}
    \W &\defeq \frac{\vec{\mu}}{1 + \disp \vec{\mu}}
\\
    \texttt{score\_Sec} &\defeq (\W \odot \mat{X})\T \; \frac{\vec{y}-\vec{\mu}}{\vec{\mu}}
                       = \mat{X}\T \, \frac{\vec{\mu} \odot (\vec{y} - \vec{\mu})}{(1 + \disp \vec{\mu}) \odot \vec{\mu}}
                       = \vec{\nabla}
\\
    \texttt{info\_vec} &\defeq \diag\left\{ \mat{X}\T (\W \odot \mat{X}) \right\}
                      = \diag \left\{ \mat{X}\T \left( \frac{\vec{\mu}}{1 + \disp \vec{\mu}} \odot \mat{X} \right) \right\}
                      = \diag( \, -\E[\mat{\nabla}^2] \, )
\\
    \texttt{step} &\defeq \texttt{score\_vec} / \texttt{info\_vec}
\end{aligned}
$$

This approximates the Fisher scoring step under the simple GLM objective
by exactly computing only the diagonal of the Hessian
(and so requiring just a reciprocal rather than a full matrix inversion).

\subsection{w/o prior}

The \texttt{fisher\_scoring\_qr\_step}\footnote{
    \url{https://github.com/const-ae/glmGamPoi/blob/6c5c93118f21ca9f663d233ab96404b27dfd5f59/inst/include/fisher_scoring_steps.h\#L8-L23}
} method implements the more
computationally-intensive---but presumably better conditioned---step
with full inversion of the Hessian (via QR decomposition).
Consider the implementation by \texttt{glmGamPoi}:
\begingroup
\allowdisplaybreaks
\begin{align*}
    \W \defeq& \frac{\vec{\mu}}{1 + \disp \vec{\mu}}
\\
    \mat{Q}\mat{R} \stackrel{\dagger}{=}& \rootW \odot \mat{X}
\\
    \texttt{``score\_vec''} \defeq& (\rootW \odot \mat{Q})\T \, \frac{\vec{y}-\vec{\mu}}{\vec{\mu}}
\\
    \texttt{step} \defeq& \texttt{solve}(\mat{R}, \, \texttt{``score\_vec''})
\\
    =& \left( \mat{R}\inv \mat{Q}\T \right) \left( \rootW \odot \frac{\vec{y}-\vec{\mu}}{\vec{\mu}} \right)
\\
    =& \left( \rootW \odot \mat{X} \right)\inv \left( {\rootW} \odot \frac{\vec{y}-\vec{\mu}}{\vec{\mu}} \right)
\\
    \stackrel{\star}{=}& \left[ \left(\rootW \odot \mat{X}\right)\invT \, \left( -\E[\mat{\nabla}^2] \right) \right]\inv
        \left[ \left(\rootW \odot \mat{X} \right)\invT \, \nabla \right]
\\
    =& \left( -\E[\mat{\nabla}^2] \right)\inv \, \left(\rootW \odot \mat{X} \right)\T \left(\rootW \odot \mat{X} \right)\invT \, \vec{\nabla}
\\
    =& \left( -\E[\mat{\nabla}^2] \right)\inv \, \vec{\nabla}
\end{align*}
\endgroup
where $\dagger$ is the QR decomposition, and $\;\star\;$ involves recognizing that
\begin{align*}
    \phantom{-\E[}\vec{\nabla}\phantom{^2]} &= (\W \odot \mat{X})\T \, \frac{\vec{y}-\vec{\mu}}{\vec{\mu}}
        = (\mat{X}\T \odot {\rootW}\T) \left(\rootW \odot \frac{\vec{y}-\vec{\mu}}{\vec{\mu}}\right)
\shortintertext{and}
    -\E[\mat{\nabla}^2] &= \mat{X}\T ( \W \odot \mat{X} )
                        = \left( \mat{X}\T \odot {\rootW}\T \right)
                          \left( \rootW \odot \mat{X} \right).
\end{align*}

\subsection{w/ prior}\label{app:ggp-newton-prior}

The \texttt{fisher\_scoring\_qr\_ridge\_step}\footnote{
    \url{https://github.com/const-ae/glmGamPoi/blob/6c5c93118f21ca9f663d233ab96404b27dfd5f59/inst/include/fisher_scoring_steps.h\#L47-L72}
} method adds a
Gaussian prior over coefficients $\vec{\beta}$ (i.e., ridge penalty).
This is the method that is ultimately used to optimize the coefficients
in \texttt{glmGamPoi} as called by \texttt{DESeq2}
(with a very wide prior; ``\texttt{ridge\_penalty = 0, which is internally replaced with a small positive number for numerical stability}''\footnote{
    \url{https://github.com/const-ae/glmGamPoi/blob/1702d70a8f57a5569baea195acf9418d2681b8a5/R/glm_gp.R\#L73}
}).

To incorporate the Gaussian / ridge penalty,
we update the gradient and Hessian accordingly:
$$
\begin{aligned}
    \vec{\nabla}^{\phantom{2}} &\mathrel{+}= - \frac{\vec{\beta}}{\vec{\sigma}^2} \;;
\\
    \mat{\nabla}^2 &\mathrel{+}= - \frac{1}{\vec{\sigma}^2}.
\end{aligned}
$$

The implementation by \texttt{glmGamPoi} updates the previous implementation
as follows. First, let
$$
\begin{aligned}
    \mat{X}' &\defeq \begin{pmatrix}
                        \mat{X} \\ \sqrt{\ncells} \, \vec{\lambda}\T
                  \end{pmatrix}
\\[6pt]
    \W' &\defeq \begin{pmatrix} \W \\ 1 \end{pmatrix}
\\[6pt]
    \texttt{residuals}' &\defeq \begin{pmatrix}
                                        \dfrac{\vec{y} - \vec{\mu}}{\vec{\mu}} \\
                                        -\sqrt{\ncells} \, \vec{\lambda}\T \odot \vec{\beta}\T
                                   \end{pmatrix}
\end{aligned}
$$
for ridge penalty $\vec{\lambda}$.
Then, replace these augmented matrices in the equations above:
$$
\begin{aligned}
    \mat{Q}\mat{R} \stackrel{\dagger}{=}& {\rootW}' \odot \mat{X}'
\\
    \texttt{``score\_vec''} \defeq& \left( {\rootW}' \odot \mat{Q} \right)\T
                                            \cdot \texttt{residuals}'
\end{aligned}
$$

When $\vec{\lambda} \defeq 1 / \vec{\sigma}$,
this has the effect of updating $\vec{\nabla}$ and $\mat{\nabla}^2$ as desired
(by adding the prior term as a sort of pseudo-data-point)---except
that the prior contribution is scaled by the number of data points $\ncells$.
As a result, the effective prior variance is actually
$\vec{\sigma}^2 / \ncells$.

Since \texttt{glmGamPoi} sets each component of $\vec{\lambda}$ to
$10^{\minus10} / \ncells$
by default, the effective prior variance over each component of $\vec{\beta}$ is
$\ncells \times 10^{20}$.

\subsection{convergence}

Optimization is reported as converged at step $\step+1$ when
\[
    \frac{|d_{(\step+1)} - d_{(\step)}|}{|d_{(\step)}| + 0.1} \leq 10^{\minus8}
\]
for deviance
$d \defeq -2 \, \left[ \ll(\vec{y}, \hat{\vec{\mu}}) - \ll(\vec{y}, \vec{y}) \right]$,
with $\ll(\vec{y}, \hat{\vec{\mu}}) \defeq \log \p(\vec{y} \mid \vec{\mu} = \hat{\vec{\mu}}, \ldots)$,\footnote{
    Where \texttt{glmGamPoi} calculates deviance based on
    the standard negative binomial log-likelihood $\ll$
    (with dispersion $\NBdisp$)
    rather than the quasi-likelihood
    (which has no corresponding likelihood function to evaluate)
}
i.e. twice the difference in log-likelihoods between the fitted and ``saturated''
models. Presumably the $0.1$ is there to avoid numerical instability for small deviances.

\section{Z-estimators for GLMs}\label{app:z-estimators}

In this section, we'll review and synthesize familiar results to
derive the estimating equations that give rise to the coefficients
for GLMs with Gaussian (as a base case) or negative binomial observations.

The equation that defines the Z-estimator for a parameter-of-interest $\vec{\beta}$
is given by the gradient of the log-likelihood,
$\nabla \ll(\beta) \defeq \frac{\partial}{\partial \vec{\beta}} \, \ll(\vec{\beta}, \cdots)$,
since this function goes to zero at the optimal
solution.

Consider fitting data $\left[ \ldots, \, (\vec{x}_{\icell}, y_{\icell}), \, \ldots \right]$
with a generalized linear model of the form
\[
    \linked \defeq \link(\mu) = \xb
\]
where mean $\mu$
parameterizes the distribution-of-choice over
outcomes $\vec{y}$
for some (monotonic, increasing, differentiable) link $\link$.

The Z-estimator will be the solution $\hat{\vec{\beta}}$ such that
\begin{equation} \label{eq:G}
    \estimating_0(\hat{\vec{\beta}}) + \sum\overCells
                                        \estimating_{\icell}(\hat{\vec{\beta}})
    = \vec{0}\ofSize{\nbetas,1}
\end{equation}
for estimating equation
$\estimating_{\icell} \defeq \nabla \ll\left(\vec{\beta}; \vec{x}_{\icell}, y_{\icell}\right)$
and (optional) regularization $\estimating_0$.

\subsection{ordinary least squares (OLS)}

We recover OLS when $\link = \operatorname{identity}$ and the response is
Gaussian, i.e.,
\[
    y \sim \normal(\xb, \, \sigma^2).
\]

The log-likelihood is
\[
    \ll = \minus\frac{1}{2\sigma^2} (y - \xb)\T (y - \xb)
           + \xi
\]
(with $\xi$ soaking up terms that don't depend on $\vec{\beta}$), so the gradient of a single point is
\[
    \nabla\ll(\vec{\beta}) = \frac{1}{\sigma^2} \vec{x} (y - \xb)\T.
\]

Then, the Z-estimator $\hat{\beta}$ is the solution to \cref{eq:G} when
$\boxed{ \estimating_{\icell}(\vec{\beta}) = \vec{x}_{\icell} (y_{\icell} - \xb) }\;.$

\subsection{negative binomial}
\enlargethispage{\baselineskip}

Now let the response be negative binomial, i.e.,
\[
    y \sim \distNamed{NB}\left( \mu, \disp \right)
\]
with dispersion $\disp$ and (canonically) $\link = \log$
to link the constrained mean parameter to the unconstrained regression.
Assume $\disp$ is fixed and known.

The gradient of each data point is
\begin{equation} \label{eq:nb-gradient}
    \nabla\ll(\vec{\beta})
    = \underbrace{
        \dd*{\ll(\mu; \vec{x}, y)}{\mu}
      }_{\text{NB}}
      \,
      \underbrace{
        \dd*{\mu}{\vec{\beta}}
      }_{\text{via\ } \, \link\inv(\xb)}
\end{equation}

The log-likelihood of a negative binomial parameterized in this way is
\[
    \ll(\mu) =
    \log \, \Gamma(y + 1/\disp) - \log{y\,! \; \Gamma(1/\disp)}
    + y \, \left[ \log{\disp \mu} - \log(1 + \disp \mu) \right]
    - 1/\disp \, \log(1 + \disp\mu)
\]
so the gradient (w.r.t. $\mu$) is
\begin{align*}
    \nabla \ll(\mu) =
    \left( \frac{y \, \disp}{\disp\,\mu} \right)
    - \frac{y \, \disp}{1 + \disp\,\mu}
    - \frac{\disp}{\disp \, (1+\disp\,\mu) }
    &= \frac{y}{\mu} - \frac{y\,\disp - 1}{1 + \disp\,\mu}
\\
    &= \frac{y-\mu}{\mu \, (1 + \disp\,\mu)}.
\end{align*}

The gradient of $\mu$ (w.r.t. $\vec{\beta}$) is
\[
    (\link\inv)'(\xb) = \exp\left\{ \xb \right\} \, \vec{x}
\]
for the canonical log link.

So, the gradient in \cref{eq:nb-gradient} is
\[
    \nabla \ll(\vec{\beta}) =
       \frac{y-\exp\left\{\xb\right\}}{\exp\left\{\xb\right\} (1 + \disp \exp\left\{\xb\right\})}
        \, \vec{x} \exp\left\{\xb\right\}
     = \frac{y-\exp\left\{\xb\right\}}{1 + \disp \exp\left\{\xb\right\}}
        \, \vec{x}
\]
and the Z-estimator $\hat{\vec{\beta}}$ is the solution to \cref{eq:G} when
\begin{equation} \label{eq:G-nb}
    \boxed{
    \estimating_{\icell}(\vec{\beta}) =
          \frac{ y_{\icell} - \exp\left\{\xbn\right\} }
               { 1 + \disp \exp\left\{\xbn\right\} } \, \vec{x}_{\icell}
    }\;.
\end{equation}

Unlike OLS, there is no closed form solution---but $\hat{\vec{\beta}}$ can be
estimated by gradient descent.

In \texttt{DESeq2} and \texttt{glmGamPoi}, the mean is additionally scaled
by a (fixed) scaling factor, i.e.
$\mu = \size \, \link\inv(\linked)$.
In this case, we slightly modify \cref{eq:G-nb} to be
\[
    \boxed{
    \estimating_{\icell}(\vec{\beta}) =
          \frac{ y_{\icell} - \size_{\icell} \exp\left\{\xbn\right\} }
               { 1 + \disp \size_{\icell} \exp\left\{\xbn\right\} } \, \vec{x}_{\icell}
    }\;.
\]

\subsection{negative binomial with prior}

In \texttt{DESeq2} and \texttt{glmGamPoi}, coefficients $\vec{\beta}$
are estimated after placing a zero mean
Gaussian prior---i.e., MAP estimation rather than ML---albeit a very wide one.
The prior width can be determined by an empirical Bayes procedure
involving additional optimization steps, but by default it is set to
$10^6$ (in \texttt{DESeq2}) or $\ncells \times 10^{20}$ (in \texttt{glmGamPoi}).

Let the prior over each coefficient be
\[
    \beta_{\ibeta} \sim \normal(0, \, \sigma_{\ibeta}^2),
\]
where the prior over the intercept term is always the (very wide) default width.

Then, the posterior log-likelihood is the log-likelihood from the previous section
with an offset for the prior, i.e.,
\[
    \sum\overCells* \bigg[ \ll_{\distNamed{NB}}(\vec{\beta} \mid \vec{x}_{\icell}, y_{\icell}, \disp) \bigg] +
    \left[ -\frac{\vec{\beta}}{2\vec{\sigma}^2} + \xi \right]
\]
---with $\xi$ again soaking up irrelevant terms from the Gaussian likelihood---so
the gradient is
\[
    \nabla\ll(\beta) =
       \sum\overCells* \frac{y_{\icell} - \exp\left\{\xbn\right\}}{1 + \disp \exp\left\{\xbn\right\}}
        \, \vec{x}_{\icell}
        -
        \frac{\vec{\beta}}{\vec{\sigma}^2}
\]
(assuming the canonical log link, as above).

The Z-estimator $\hat{\vec{\beta}}$ is the solution to \cref{eq:G} when
$\estimating$ is defined as above (\cref{eq:G-nb}) and
\[
    \boxed{
    \estimating_0(\vec{\beta}) =
    - \frac{\vec{\beta}}{\vec{\sigma}^2}
    }\;.
\]

\section{differentiating the estimator with respect to data weights}\label{app:dbeta-dw}

The estimator $\hat{\vec{\beta}}$ is implicitly defined as a function of the data weights,
$\hat{\beta}(\vec{w})$, as the solution to the weighted estimating equation
(\cref{eq:w-estimating})~\cite{ryan-amip,implicit-fn-thm}.
Following \cite{ryan-amip,influence-derivative},
so long as \cref{eq:w-estimating} is continuously differentiable with
respect to $\vec{w}$---and the Jacobian matrix is full-rank
(and therefore invertible)---the derivative
$\ddd*{}{w_{\icell}}\hat{\beta}(\vec{w})$
exists and can be calculated as follows:
\begin{align*}
    \vec{0}
     &= \dd*{}{\vec{w}\T} \left[
         \estimating_0\ofBetaHatW* + \sum\overCells*
                                     w_{\icell} \, \estimating_{\icell}\ofBetaHatW
        \right]
        \bigg|_{\vec{w}}
    \\
     &= \dd*{}{\vec{\beta}\T} \left[
         \estimating_0\ofBeta + \sum\overCells
                                w_{\icell} \, \estimating_{\icell}\ofBetaW
        \right]
        \Bigg|_{\hat{\beta}(\vec{w}), \vec{w}}
        \cdot \,
        \dd{\hat{\beta}(\vec{w})}{\vec{w}\T}
        \Bigg|_{\vec{w}}
        + \,
        \dd*{}{\vec{w}\T}\left[
            \sum\overCells
            w_{\icell} \, \estimating_{\icell}\ofBetaW
        \right]
        \Bigg|_{\hat{\beta}(\vec{w}), \vec{w}}
    \\
    \implies
    \\
     \dd{\hat{\beta}(\vec{w})}{\vec{w}\T}
     \bigg|_{\vec{w}}
     &= \, -\left(
        \dd*{\estimating_0\ofBeta}{\vec{\beta}\T}
        \bigg|_{\hat{\beta}(\vec{w})}
        + \;
        \sum\overCells
            w_{\icell} \,
            \dd*{\estimating_{\icell}\ofBetaW}{\vec{\beta}\T}
            \bigg|_{\hat{\beta}(\vec{w}), \vec{w}}
        \right)\inv
        \cdot \;
        \left(
            \sum\overCells
            \dd*{\left[ w_{\icell} \estimating_{\icell}\ofBetaW \right]}{\vec{w}\T}
            \bigg|_{\hat{\beta}(\vec{w}), \vec{w}}
        \right)
    \\
     &= \, -\Bigg(
        \dd*{\estimating_0\ofBeta}{\vec{\beta}\T}
        \bigg|_{\hat{\beta}(\vec{w})}
        + \;
        \sum\overCells
            w_{\icell} \,
            \dd*{\estimating_{\icell}\ofBetaW}{\vec{\beta}\T}
            \bigg|_{\hat{\beta}(\vec{w}), \vec{w}}
        \Bigg)\inv
    \\
     &\hphantom{= \;\;} \cdot \;
        \Bigg(
            \underbrace{
                \strut
                \sum\overCells
                w_{\icell} \dd*{\estimating_{\icell}\ofBetaW}{\vec{w}\T}
                \bigg|_{\hat{\beta}(\vec{w}), \vec{w}}
            }_{\star}
        + \;
            \bigg[
                \estimating_1\ofBetaHatW,
                \,\ldots\, ,
                \estimating_{\ncells}\ofBetaHatW
            \bigg]
        \Bigg)
\end{align*}

There are two elements that differ from the original derivation:
the $\estimating_0$ term is present because we've included regularization
(lacking from the original Z-estimator), and
$\;\star\;$ is present because we've relaxed the assumption
that $\estimating_{\icell}$ depends on $\vec{w}$ \textit{only}
through its dependence on $\hat{\beta}(\vec{w})$
(assumed throughout \cite{ryan-amip}).

\section{the likelihood ratio test is not amenable to first-order sensitivity approximations}\label{app:lrt-sensitivity}

Here, we outline why the likelihood ratio test is not suitable for the
original
first-order
approach~\cite{ryan-amip} to estimating sensitivity to dropping data
(reviewed in \cref{sec:amip}).

Recall (\cref{sec:testing}) that the likelihood ratio test
statistic is---as it says on the tin---a log ratio of two likelihoods; namely,
\begin{align*}
    LR \defeq -2 \log \frac
        { \p(\vec{y}, \mat{X}; \hat{\vec{\beta}}\reduced, \cdots) }
        { \p(\vec{y}, \mat{X}; \hat{\vec{\beta}}, \cdots) }
       = -2 \left[ \LL(\hat{\vec{\beta}}\reduced) - \LL(\hat{\vec{\beta}}) \right].
\end{align*}
In brief, the likelihood in the numerator is that of the reduced model
$\M\reduced$ (where, in the context of differential expression, there is no
coefficient in the GLM for the treatment effect)
while the likelihood in the denominator is that of the full model
$\M$ (i.e., the GLM we fit in order to estimate the treatment effect $\hatBetaTreated$).
In practice, \texttt{glmGamPoi} uses a slightly modified test statistic
where $LR$ is scaled by two scalar estimates (\cref{sec:testing})
and $\M, \M\reduced$ are quasi-likelihood models (\cref{sec:model}).

To form this statistic as an (implicit) function of data weights,
we will, equivalently, write it as the function $\stat_{LR}\ofBetaBetaRed$,
where $\vec{\beta} = \hat{\beta}(\vec{w})$ and
$\vec{\beta}\reduced = \hat{\beta}\reduced(\vec{w})$.

When using the likelihood ratio test to assess differential expression,
the key statistics-of-interest revolving around significance\footnote{
    i.e., $\stat_{\text{erase significance}},\;\stat_{\text{bestow significance}},\;\stat_{\text{flip sign w/ significance}}$
    as defined for the Wald test in \cref{sec:stats-of-interest}
} will all be functions of $\stat_{LR}\ofBetaBetaRed$.
Then, to assess sensitivity of these statistics with respect to dropping data points,
we will ultimately need to differentiate $\stat_{LR}$ with respect to each data weight $w_{\icell}$
in order to compute a first-order Taylor approximation of
$\stat_{LR}$ at arbitrary data weights $\vec{w}$
(\cref{sec:amip}, particularly \cref{eq:taylor-expansion,eq:influence}).

Adapting the influence computation
(for the fact that $\stat_{LR}$ depends on the outcome of not one, but two optimizations),
\cref{eq:influence-calc} becomes
\begin{align*}
    \dd*{\stat_{LR}\ofBetaHatBetaRedHatW}{w_{\icell}} \bigg|_{\vec{w}} =\;
    &\underbrace{
        \strut
        \dd*{\stat_{LR}\ofBetaBetaRedW}{\vec{\beta}\T}
            \bigg|_{\hat{\beta}(\vec{w}),\, \hat{\beta}\reduced(\vec{w}),\, \vec{w}}
    }_{\star}
    \, \cdot \,
        \dd*{\hat{\beta}(\vec{w})}{w_{\icell}} \bigg|_{\vec{w}\vphantom{\hat{\beta}\reduced(\vec{w})}}
     \\
     +\;
    &\underbrace{
        \strut
        \dd*{\stat_{LR}\ofBetaBetaRedW}{\vec{\beta}^{\ddagger \intercal}}
            \bigg|_{\hat{\beta}(\vec{w}),\, \hat{\beta}\reduced(\vec{w}),\, \vec{w}}
    }_{\star}
    \, \cdot \,
        \dd*{\hat{\beta}\reduced(\vec{w})}{w_{\icell}} \bigg|_{\vec{w}\vphantom{\hat{\beta}\reduced(\vec{w})}}
     \\
     +\;
        &\dd*{\stat_{LR}\ofBetaBetaRedW}{w_{\icell}}
            \bigg|_{\hat{\beta}(\vec{w}),\, \hat{\beta}\reduced(\vec{w}),\, \vec{w}}
    \,.
\end{align*}
The $\;\star\;$ terms are the root of the issue with approximating sensitivity of the likelihood ratio test.
Since $\stat_{LR}$
is a function of the objectives themselves,
the gradient of these terms (with respect to the parameter estimates
$\hat{\vec{\beta}}$ and $\hat{\vec{\beta}}\reduced$, respectively)
is---by definition---zero.
Therefore, the first derivative of the likelihood ratio test statistic does \textit{not}
provide useful information to approximate the Taylor expansion around
$\stat_{LR} \big( \hat{\vec{\beta}}, \hat{\vec{\beta}}\reduced, \vec{1} \big)$.
For the purposes of this work, we focus only on statistics-of-interest for differential expression
that are amenable to a first-order dropping-data robustness approximation.

\section{Fisher and sandwich covariance estimators}\label{app:cov-estimators}

There are two standard statistical estimators for the covariance of the sampling distribution
of a fitted parameter.
Either could be appropriate to estimate the standard error of $\hatBetaTreated$
under the differential expression objective, under different assumptions.
As we'll show below, the \emph{Fisher} estimator reflects the assumption that the model is correctly specified,
whereas the \emph{sandwich} estimator is valid regardless of model
specification~\cite{se-geyer,se-m-estimators}.

Recall that $\hat{\vec{\beta}}$ is our solution to the estimating equation formed by the
gradient of the log-likelihood; i.e., it solves
\[
    \nabla \LL\ofBeta \defeq \sum\overCells
        \nabla \ll\left( \vec{\beta}; \vec{x}_{\icell}, y_{\icell} \right)
                      \defeq \sum\overCells \nabla \ll_{\icell}\ofBeta,
\]
potentially with an additional term for the prior.

Define two useful quantities:
\begin{align}
    \mat{H} &\defeq -\E\big[ \nabla^2 \ll_{\icell}\ofBetaTrue \big]
        && \mathcomment{negative Hessian} \label{eq:H}
\shortintertext{and}
    \mat{S} &\defeq \E\big[ \cov \left[ \nabla \ll_{\icell}\ofBetaTrue \right] \big]
        && \mathcomment{variance of the score} \label{eq:S}
\end{align}
where $\betaTrue$ is the true solution to our optimization problem.

By the asymptotic properties of a smooth estimator and the central limit theorem,
\begin{align*}
    \hat{\vec{\beta}} - \betaTrue
    &\xrightarrow{d}
    \frac{1}{\sqrt{\ncells}} \, \mat{H}\inv \, \normal\left( 0, \, \mat{S} \right)
    \\
    &=
    \normal\big( 0, \; \underbrace{
        \strut
        \frac{1}{\ncells} \, \mat{H}\inv \mat{S} \mat{H}\inv
    }_{ \mat{\Sigma} }
    \big).
\end{align*}
$\mat{\Sigma}$ is the theoretical covariance-of-interest in order to compute
the standard error of any $\hat{\vec{\beta}}_{\ibeta}$
(to compute a Wald statistic, for example).

How to estimate the expectations in \Cref{eq:H,eq:S}? Empirically,
\begin{align*}
    \widehat{\mat{H}} &\defeq -\frac{1}{\ncells} \sum\overCells
            \nabla^2 \ll_{\icell}\ofBetaHat
\shortintertext{and}
    \widehat{\mat{S}} &\defeq \frac{1}{\ncells} \sum\overCells
        \left[ \nabla \ll_{\icell}\ofBetaHat \right] \left[ \nabla \ll_{\icell}\ofBetaHat \right]\T
\end{align*}
where $\widehat{\mat{S}}$ corresponds to the (sample) variance of the score because
the usual centering factor---the expectation of the observed scores---is zero
(by definition of the optimization problem).

Note that we have so far made no assumptions about the sampling distribution
of the data, other than the independence of each
$\left( \vec{x}_{\icell}, y_{\icell} \right)$
for the central limit theorem.

The sandwich (or ``robust'') estimator, then, is
\[
    \boxed{
        \widehat{\mat{\Sigma}}_{\text{sandwich}} \defeq
        \frac{1}{\ncells} \, \widehat{\mat{H}}\inv \widehat{\mat{S}} \widehat{\mat{H}}\inv
    }\;.
\]

Now assume that the model is perfectly specified, meaning that data
$\left( \vec{x}_{\icell}, y_{\icell} \right)$ is drawn i.i.d.
according to the proposed log-likelihood.
Then, \Cref{eq:H,eq:S} are equivalently two ways
to calculate the Fisher information; $\mat{H} \eqdef \Info \defeq \mat{S}$.
The theoretical covariance-of-interest becomes
\[
    \mat{\Sigma} = \frac{1}{\ncells} \Info\inv \Info \Info\inv = \frac{1}{\ncells} \Info\inv.
\]

A handy way to calculate this empirically (avoiding expensive integration of the expectation)
is to compute the ``observed'' Fisher information,
\[
    \widehat{\Info} \defeq -\frac{1}{\ncells} \sum\overCells
            \nabla^2 \ll_{\icell}\ofBetaHat.
\]

So, the Fisher estimator of the covariance-of-interest is
\[
    \boxed{
        \widehat{\mat{\Sigma}}_{\text{Fisher}} \defeq
        \frac{1}{\ncells} \, \widehat{\Info}\inv
    }\;.
\]

Each estimator $\widehat{\Sigma}(\cdot)$ can be viewed as a function of $\vec{\beta}$
(under the original objective) or of $\ofBetaW$ (under the weighted objective).

On the other hand,
\texttt{DESeq2} uses an IRLS formula to calculate the covariance for its Wald test,
based on the IRLS weights used to reweight the final optimization step.
This formula---which is specific to a negative binomial GLM
with Gaussian $\vec{\beta}$ prior---is equivalent to the generic Fisher estimator
so long as
the GLM is parameterized
by its canonical (log) link.

\section{quasi-likelihood statistics-of-interest}\label{app:ql-stats-of-interest}

To compute the differential expression statistics-of-interest (and their associated sensitivities)
under the quasi-likelihood framework assumed by \texttt{glmGamPoi},\footnote{
    Note that we describe how to evaluate sensitivity of the quasi-likelihood \textit{Wald} statistic
    rather than the quasi-likelihood \textit{likelihood ratio} statistic that \texttt{glmGamPoi} uses
    (which is not amenable to our first-order sensitivity approximation; \cref{app:lrt-sensitivity})---though
    these tests are asymptotic in $\ncells$ (\cref{sec:testing}).
}
we would modify our approach in a few ways.

First, we would estimate the coefficients $\hat{\vec{\beta}}$ under the modified negative
binomial GLM outlined in \cref{sec:de-analysis-for-sensitivity}---but with
dispersion $\NBdisp$ rather than $\disp$ (cf. \cref{eq:disp}).

We would then modify one of the building blocks used to compute key
statistics-of-interest (\cref{sec:stats-of-interest}) in order to account for the
quasi-likelihood framework. Namely, in lieu of $\stat_W^+$, we would instead compute
$$
    \stat_{W'}^+\ofBetaW =
        \frac{(\vec{\contrast}\T \, \vec{\beta})^2}
             {\left[ \vec{c}\T \cdot \widehat{\Sigma}\ofBetaW \cdot \vec{c} \right] \times \df \times \hat{\QLdisp}}
    \mathcomment{\;\;\;unsigned quasi-likelihood Wald statistic}
$$
where $\df$ is the degrees-of-freedom of the contrast being estimated (generally 1)
and $\hat{\QLdisp}$ is the estimated quasi-likelihood dispersion (\cref{eq:disp}).
In other words, $\;\stat^+_{W'} \defeq (\stat_W^+)^2 \,/\, \df \,/\,  \hat{\QLdisp}$.

Finally, we would alter the statistics-of-interest involving the test statistic as follows:
\begin{align*}
    \stat_{\text{erase significance}}\ofBetaW \,&=\,
        -\left[ \stat_{W'}^+\ofBetaW - \CIwidth \right]
    && \mathcomment{-CI lower bound}
    \\
    \stat_{\text{bestow significance}}\ofBetaW \,&=\,
        +\left[ \stat_{W'}^+\ofBetaW - \CIwidth \right]
    && \mathcomment{\hphantom{-}CI lower bound}
    \\
    \stat_{\text{flip sign w/ significance}}\ofBetaW \,&=\,
        -\left[ \stat_{W'}^+\ofBetaW + \CIwidth \right]
    && \mathcomment{-CI upper bound}
\end{align*}
where $\CIwidth$ is the one-sided width of a confidence interval (CI)
at the chosen significance level.
Unlike before, this width would now be
based on an F-distributed null
with $(\df, \df_{\QLdisp})$ degrees of freedom
(as estimated by \texttt{glmGamPoi}; \cref{sec:testing,app:ql}).

We could then compute sensitivities of these key quasi-likelihood outcomes as previously described
(\cref{sec:de-robustness})---either
conditioning on $\hat{\QLdisp}$ as a constant (\cref{sec:not-accounted-for}),
or (with more work, if its estimation procedure permits)
differentiating through its dependence on each data weight,
$\ddd*{\QLdisp}{w_{\icell}}$,
when computing term \circled{3} in \cref{eq:influence-calc}.

\section{a sample scRNA-seq dataset}\label{app:data}

Throughout, we focus on single-cell RNA-seq data from a study of ulcerative colitis (UC)~\cite{data-uc}.
In this dataset, \emph{treatment} is the natural biological ``perturbation'' of
disease---i.e., cells from subjects with UC.
Specifically, we compare expression within goblet cells
(based on the original authors' cell type annotations)
for cells that are ``healthy'' versus ``inflamed.''\footnote{
    Ignoring the third health status ``non-inflamed''
}
This slice of the data comprises $\ncells=1440$ cells and $\ngenes=15,516$
genes with at least one nonzero observation (reduced from 20,028 genes measured).
Cells are sampled from 12 healthy subjects and 14 subjects with UC.

\section{gene set enrichment}\label{app:gsea}

We consider the simplest, and an extremely common, method for gene set
enrichment analysis; namely, the hypergeometric test.
Under this procedure, we first identify significant genes
(based on some significance cutoff applied to multiple-testing corrected p-values).
We then segment this set of significant genes into two groups:
significant genes that are upregulated among treated cells (``targets up''), and
those that are downregulated among treated cells (``targets down'')~\cite{gsea-up-down}.
A third entity, the ``gene universe,'' is defined as the set of all genes that
were tested for differential expression.

We use GO Biological Processes (\texttt{GO:BP})~\cite{go-2000,go-2020}
as our curated collection of gene sets (also termed ``pathways''\footnote{
    Though each set, while related in biological function,
    does not necessarily constitute a true pathway
}).
Specifically, we access
this collection through the Molecular Signatures Database (MSigDB)~\cite{broad-gsea,msigdb}
via the R command
\texttt{msigdbr::msigdbr(category=`C5', subcategory=`GO:BP', species=\$SPECIES)}~\cite{msigdbr},
where \texttt{\$SPECIES} is set to concord with the organism whose gene expression was measured
(e.g., ``human'' or ``mouse'').
To eliminate gene sets that are trivial or overly broad, we limit this
collection to those with a minimum size of 15 and a maximum size of 500 genes
(after overlapping with the ``gene universe'').

We convert gene names measured in the experiment to their corresponding gene symbols
(to concord with GO gene sets)
via the R command
\texttt{limma::alias2SymbolUsingNCBI}~\cite{limma}
and the NCBI ``gene info'' mapping for the species-of-interest.\footnote{
    Downloaded from \url{ftp://ftp.ncbi.nlm.nih.gov/gene/DATA/GENE_INFO}
}
Duplicate symbol conversions (multiple genes mapping to the same symbol)
are resolved by selecting
\begin{enum-inline}
    \item the gene whose name matches the symbol exactly,
    \item the gene whose name starts or ends with the symbol,
    or
    \item as a final fallback, the gene that is first alphabetically.
\end{enum-inline}

We then test for gene set enrichment
(of differentially expressed genes, among each \texttt{GO:BP} gene set)
using a hypergeometric test.
Namely, for each list of genes (``targets up'' or ``targets down''),
we test each gene set for overrepresentation of target genes
via \texttt{scipy.stats.hypergeom.sf(ts$\minus$1, U, us, T)},\footnote{
    The \ul{s}urvival \ul{f}unction (\texttt{sf}) corresponds to
    1 - the relevant CDF (of \texttt{ts-1}),
    so the output of this function is a vector of
    one-tailed
    probabilities
    that gene overlap would be at least as large as
    observed ($\geq \texttt{ts}$), under the expected null overlap for a sample
    of size \texttt{T} drawn uniformly at random from the universe as a whole.
}
where \texttt{ts} is the size of the overlap between targets
and each gene set, \texttt{U} is the size of the gene universe
(i.e., the total number of genes tested), \texttt{us} is the size
of the overlap between the universe and each gene set, and
\texttt{T} is the size of the target list.
We confirm that our implementation yields identical results to
the R command \texttt{fgsea::fora}
(\ul{f}ast \ul{o}ver-\ul{r}epresentation \ul{a}nalysis)~\cite{fgsea}.
Finally, we identify the top gene sets---representing the ``most notable''
biological processes that are up- or down-regulated among treated cells---by
ranking results by their hypergeometric p-values.\footnote{
    We also correct p-values by the Benjamini-Hochberg procedure~\cite{bh}
    to control the false discovery rate, as does \texttt{fgsea::fora},
    but this rank-based correction does not affect the ranking.
}

For convenience,
all R commands above are wrapped within a Python pipeline using
\texttt{rpy2}.\footnote{
    \url{https://github.com/rpy2/rpy2}
}

\section{another approach to perturbing gene set enrichment}\label{app:gsea-alt}

Our method for identifying groups of $K$ cells that, when dropped,
will maximally disrupt the top gene sets (\cref{sec:gsea-robustness})
is designed to work well across a range of $K$s.
However, it is heuristic and, ultimately, provides a lower bound on the maximal
disruption to the top 10 gene sets for a given $K$.
In fact, in the process of developing this procedure, we occasionally observe
that a tweaked procedure
for clustering cell influences
gives rise to better results
(i.e., more disruption to the composition of the top gene sets)
in one particular setting.

For example, when analyzing the dataset in \cref{app:data},
we identify a set of 28 cells whose removal from the dataset
disrupts \standout{70\%} of the top 10 gene sets
enriched among downregulated genes
(versus \standout{60\%} of the top 10 when choosing a set of cells by our
overall best method; \cref{fig:change-in-go-down-drop28}).

\begin{SCfigure}[][!hbt]
    \centering
    \includegraphics[height=2.5\goHeight]{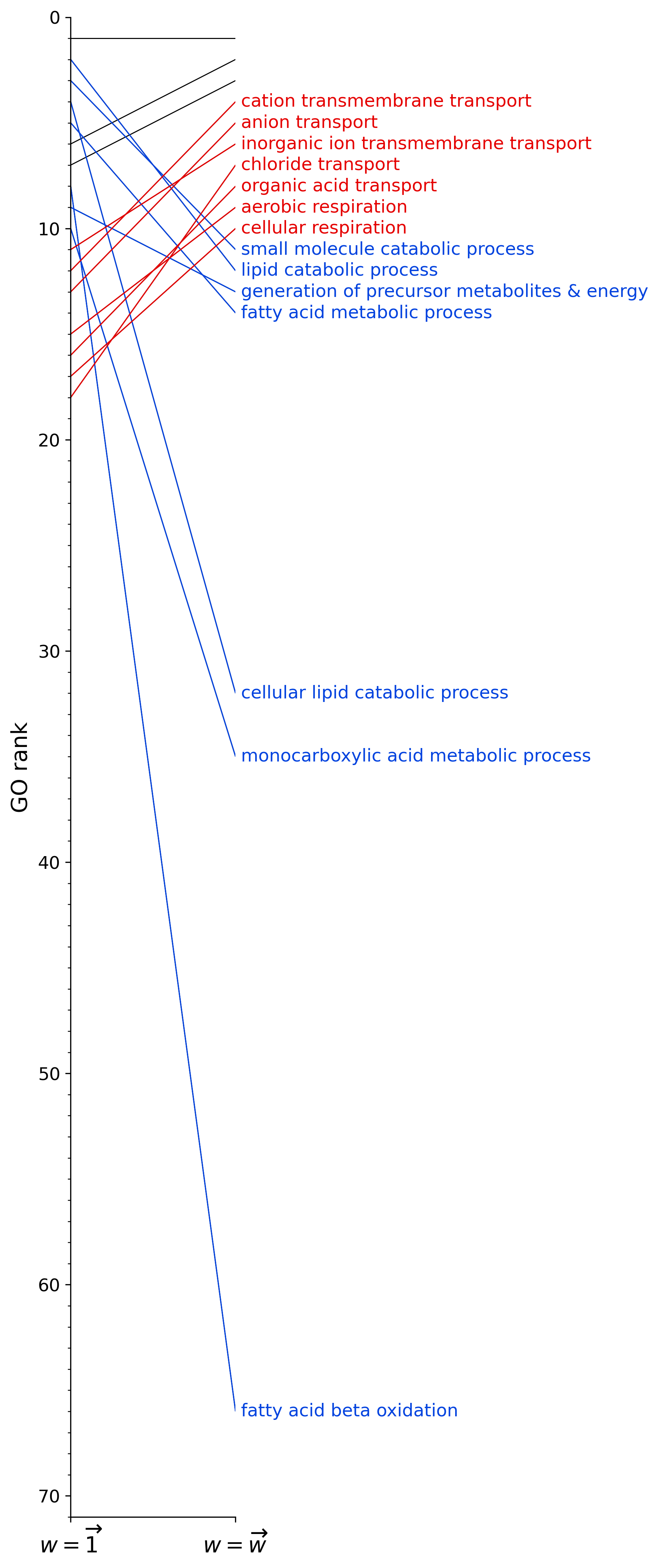}%
    \caption{
        \textbf{Perturbation to top GO sets (among downregulated genes),
                by dropping a handful of influential cells.}
        Changes to the top 10 ranked \texttt{GO:BP} gene sets
        when an influential group of 28 cells (<2\%) is dropped.
        Blue lines indicate the change in rank for gene sets that are \textit{demoted},
        red lines indicate the change in rank for those that are \textit{promoted},
        and black lines indicate the change in rank for those that \textit{remain} in the top 10.
        \captionbr
        After clustering cells via the alternate method described in this section,
        this particular group of cells
        (whose effect, when dropped, is plotted at right)
        represents the 8\th-ranked cluster (scored as per \cref{sec:gsea-scoring}).
    } \label{supp-fig:drop-28-alt}
\end{SCfigure}

We identified this particular set of cells via iterative greedy clustering,
using each cell as a seed, but with a different objective (cf.~\cref{eq:greedy-cluster}).
Specifically, let $\mathbb{\ncells}$ be the set of all cells and let
$\mathbb{K}$ be the set of all cells in the cluster so far.
Then, the next cell we'd add to the cluster is
\[
\argmax\limits_{\icell \,\in\, \mathbb{\ncells} \setminus \mathbb{K}}
    \; \sum\limits_{\substack{
            \setlength{\jot}{-0.95\baselineskip}
            \begin{aligned}
                \igene \,\in\, &\mathbb{\ngenes}^K_{\text{promote}} \,\cup
                \\
                &\mathbb{\ngenes}^K_{\text{demote}}
            \end{aligned}
        }}
        \sign\left[ \, \sum\limits_{k \,\in\, \mathbb{K}} \vec{\influence}^{(\igene)}_k
             \right]
        \times \,
        \vec{\influence}^{(\igene)}_{\icell}.
\]
In other words, in order to group cells with synergistic effects,
we greedily add cells to the cluster that maximize the total
influence along the gene direction vector defined by the cumulative influence.

While this method happened to identify this particularly disruptive set of cells
(at $K=28$, for gene set enrichment among downregulated genes, for this particular dataset),
it otherwise yielded subpar results in other settings.

\FloatBarrier
\pagebreak
\section{supplementary methods figures}

\begin{figure}[H]
    \centering
    \includegraphics[width=\linewidth]{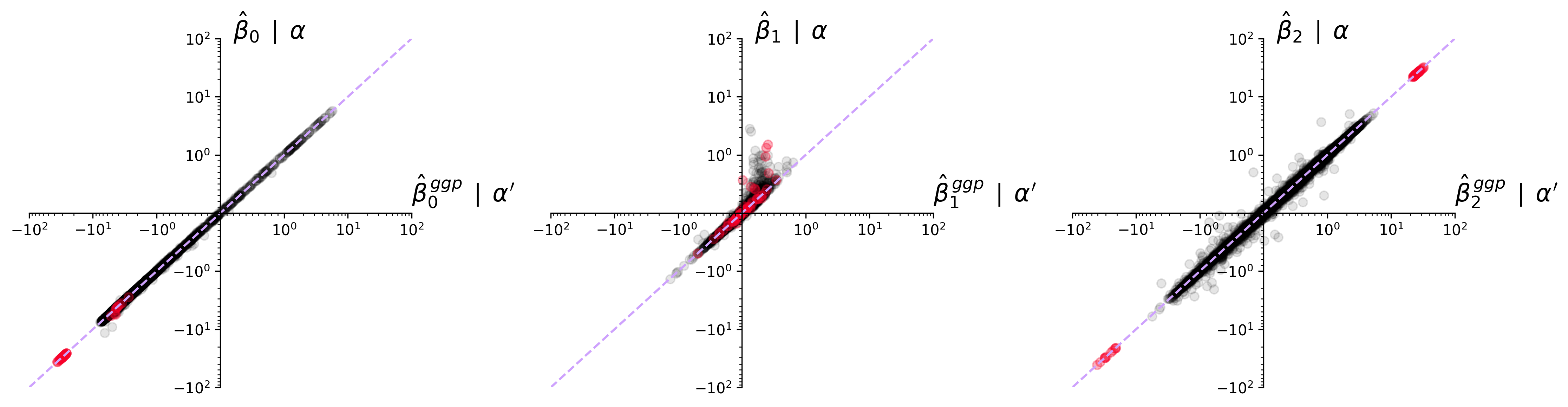}%
    \caption{
        \textbf{Changes to fitted coefficients under NB vs. quasi-likelihood NB dispersion.}
        The estimated coefficients $\hat{\vec{\beta}}$ across genes (\textit{points}),
        under the quasi-likelihood model fitted by \texttt{glmGamPoi} (\textit{x-axis}),
        where the dispersion $\alpha'$ used to fit the negative binomial model
        is $\Lt$ the overall estimated dispersion, versus the simpler classical
        negative binomial model that we fit (\textit{y-axis}) with a single overall
        dispersion $\disp$.
        Each column reflects a different GLM coefficient, where $\beta_2$ is
        the treatment effect.
        Zero-group genes are highlighted in red.
    } \label{supp-fig:betas-alpha}
\end{figure}

\begin{figure}[!htb]
    \centering
    \includegraphics[width=\linewidth]{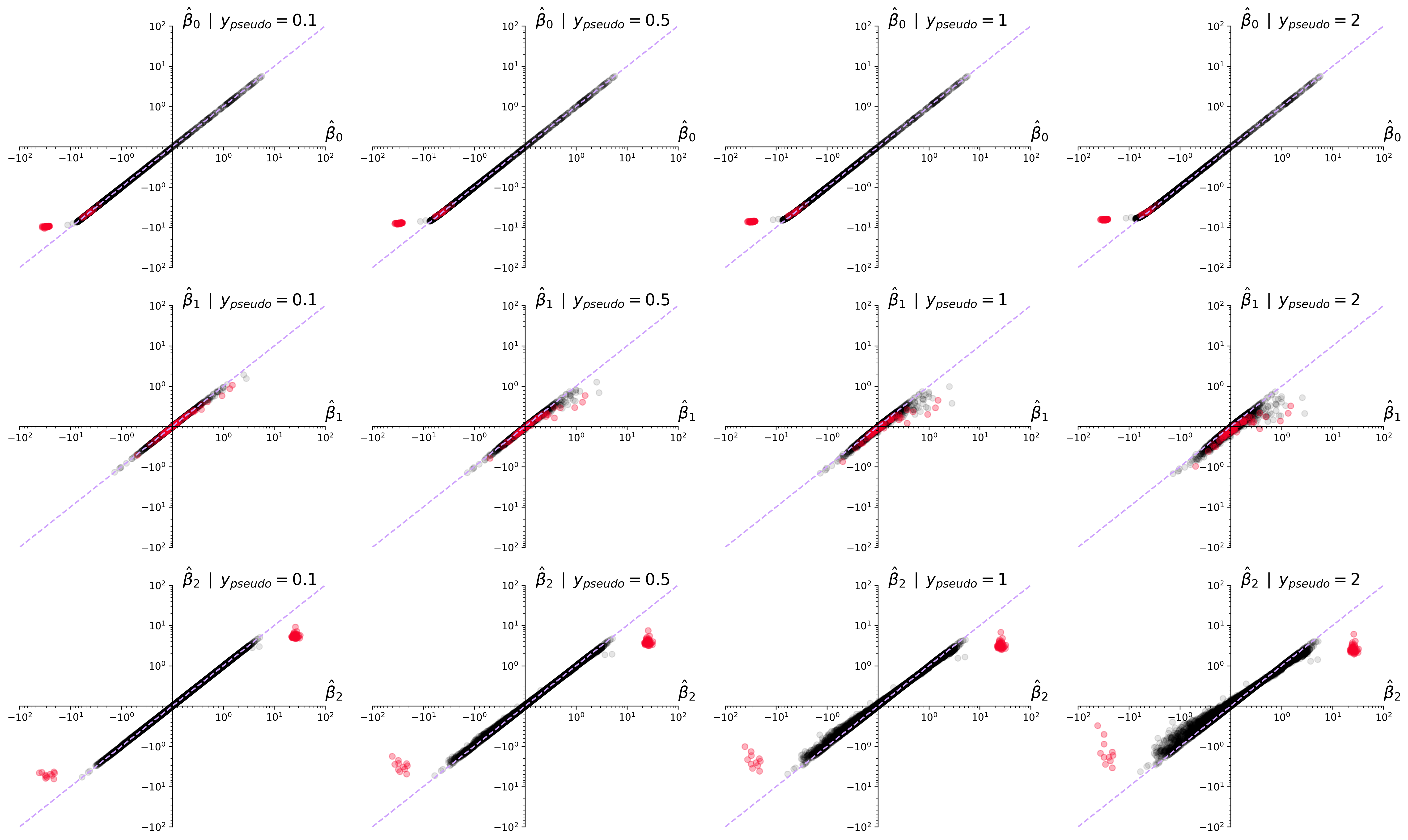}%
    \caption{
        \textbf{Changes to fitted coefficients under pseudocell prior.}
        The estimated coefficients $\hat{\vec{\beta}}$ across genes (\textit{points}),
        with (\textit{y-axis}) and without (\textit{x-axis}) a pseudocell prior.
        Each row reflects a different GLM coefficient, where $\beta_2$ is the treatment effect.
        The strength of the pseudocell prior
        (i.e., size of the pseudocell observation $\ypseudo$)
        increases from left to right across columns.
        Zero-group genes are highlighted in red.
    } \label{supp-fig:betas-pseudo}
\end{figure}

\begin{figure}[!htb]
    \centering
    \subcaptionbox{\;\textit{Wald Fisher vs. Wald sandwich}}{\includegraphics[width=\linewidth]{%
        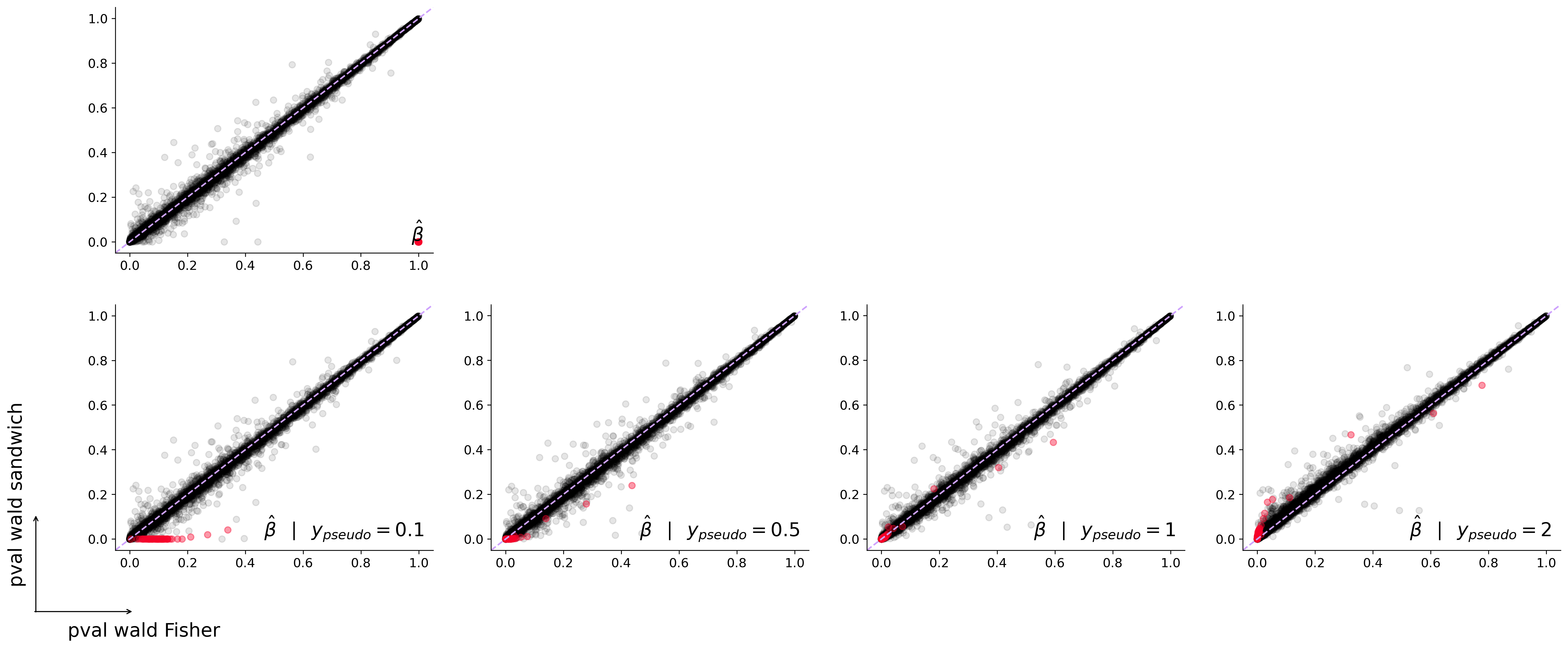}}
    \subcaptionbox{\;\textit{likelihood ratio vs. Wald sandwich}}{\includegraphics[width=\linewidth]{%
        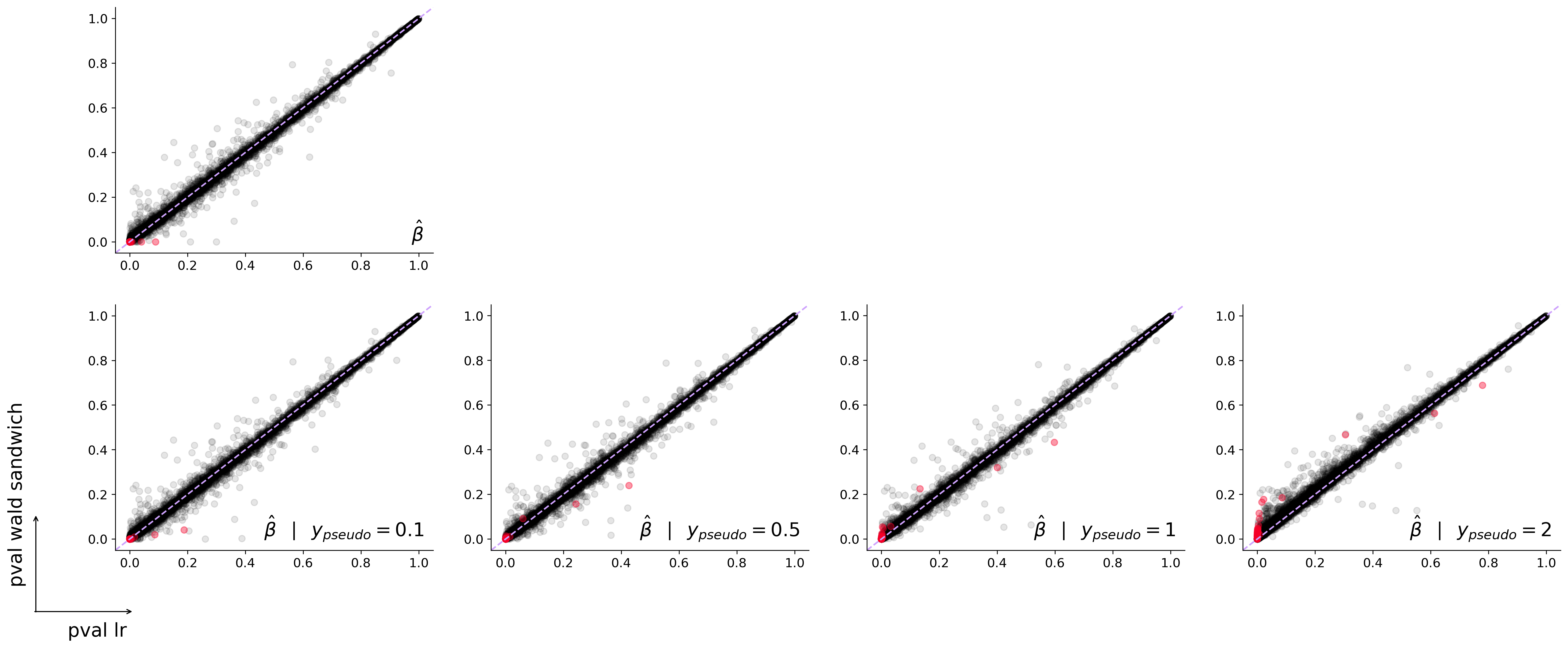}}
    \subcaptionbox{\;\textit{likelihood ratio vs. Wald Fisher}}{\includegraphics[width=\linewidth]{%
        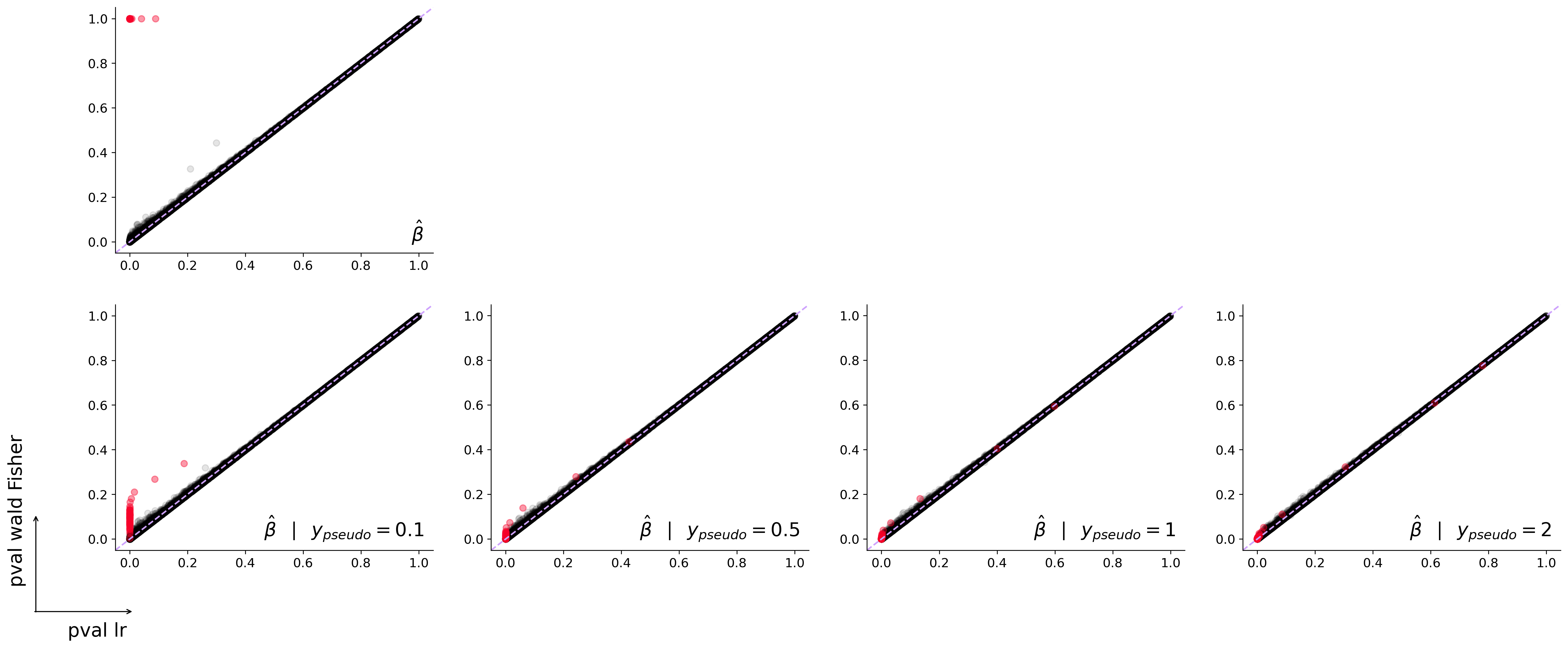}}
    \caption{
        \textbf{Relationship between test p-values under a pseudocell prior of varying strength.}
        \textit{Top} (within each subfigure),
        the relationship between
        tests, across gene p-values,
        when no pseudocell prior is enforced.
        \textit{Bottom},
        the relationship under a pseudocell prior as strength
        (size of the observed count $\ypseudo$) increases from left to right.
        Zero-group genes are highlighted in red.
        At the pseudocell prior that we choose for further analysis ($\ypseudo=0.5$),
        correlation between p-values across all pairs of tests is >0.99.
        In contrast, for the GLM with no pseudocell prior, correlation between
        the Wald Fisher test and either other test is $\apprx$0.96.
    } \label{supp-fig:pvals-pseudo}
\end{figure}

\FloatBarrier
\pagebreak
\enlargethispage{\baselineskip}
\section{supplementary experimental figures}

\begin{figure}[H]
    \centering
    \includegraphics[width=\volcanoWidth]{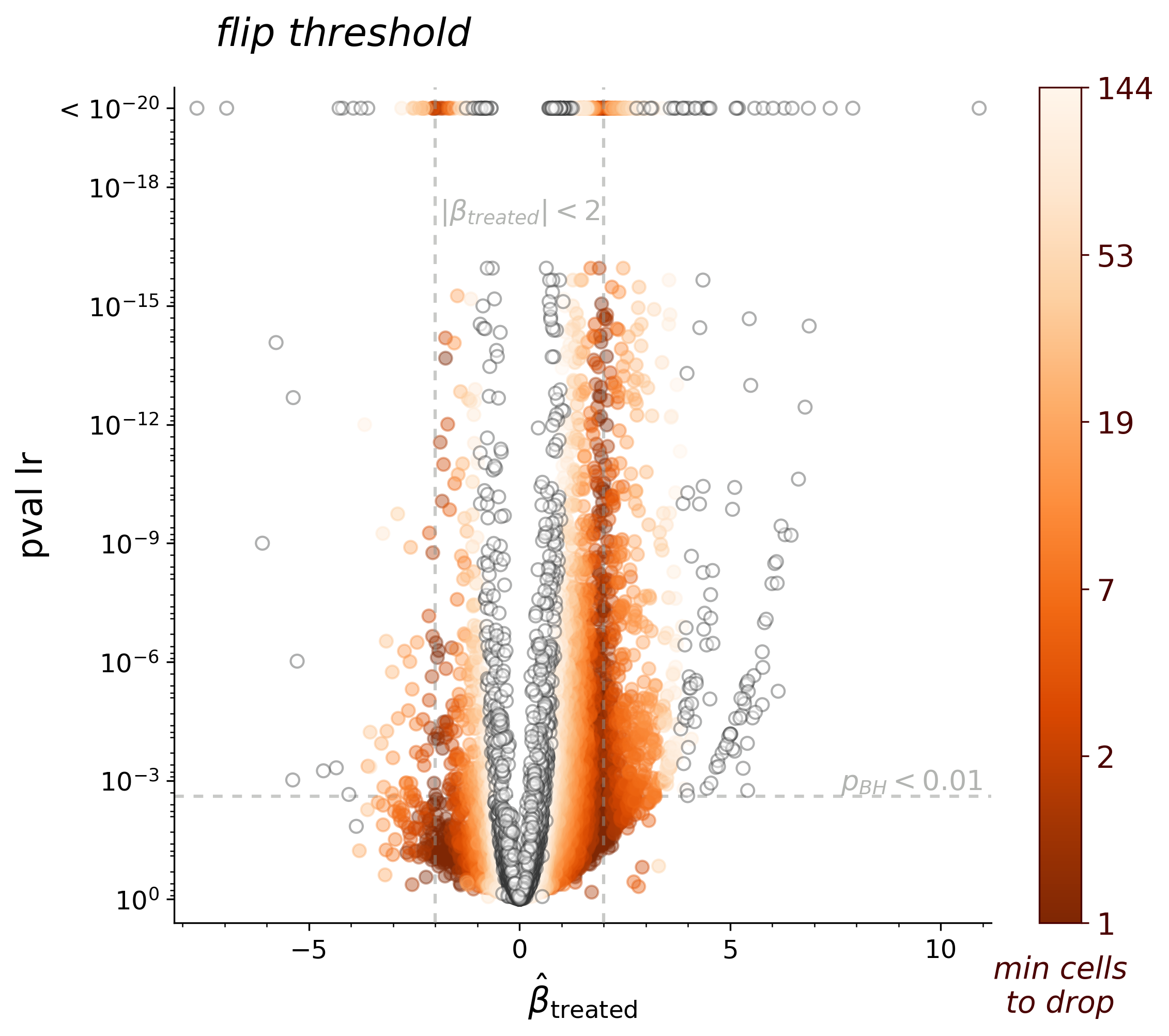}%
    \hfill%
    \includegraphics[width=\volcanoWidth]{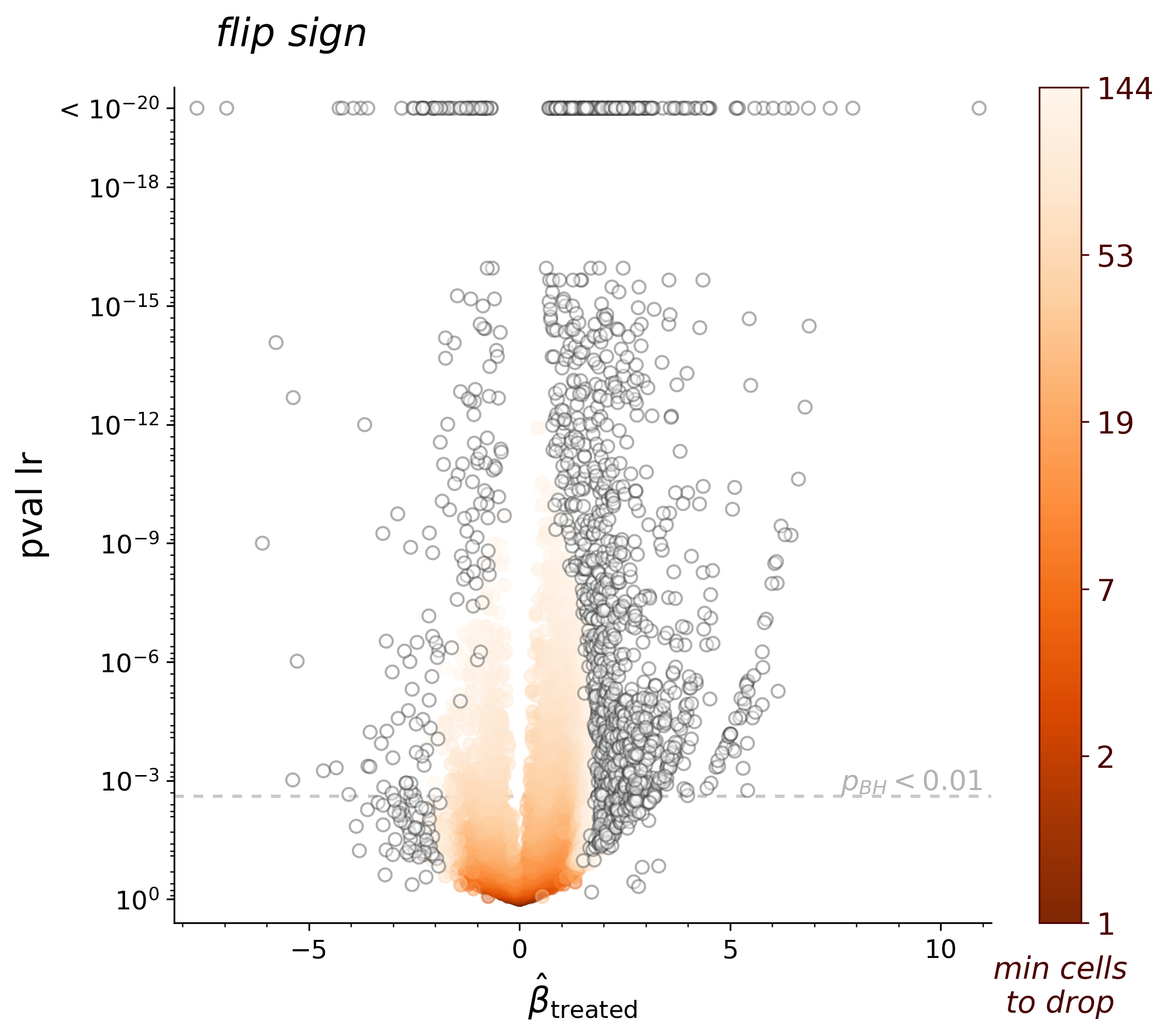}%
    \vspace{\vspaceBwRows}
    \includegraphics[width=\volcanoWidth]{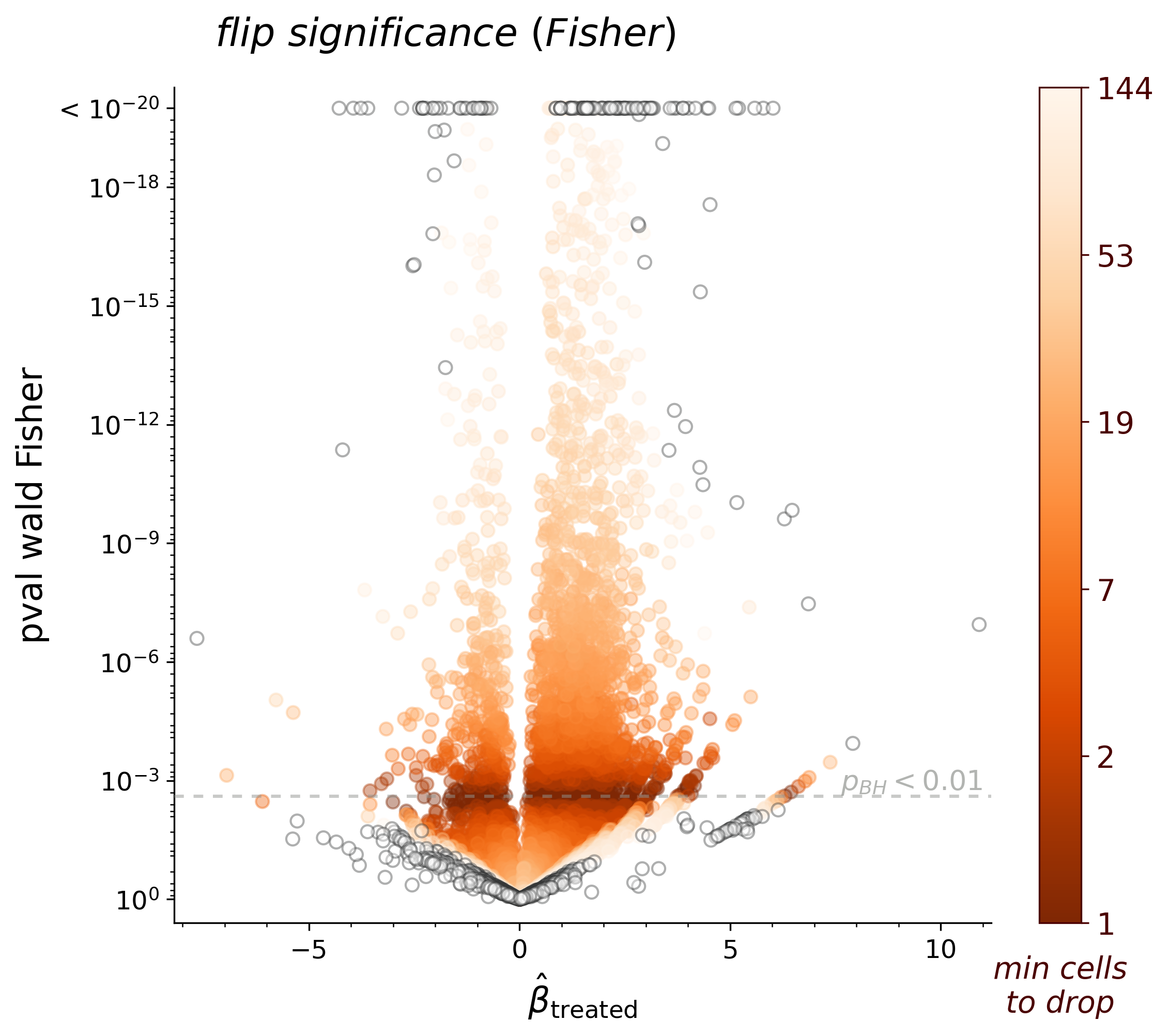}%
    \hfill%
    \includegraphics[width=\volcanoWidth]{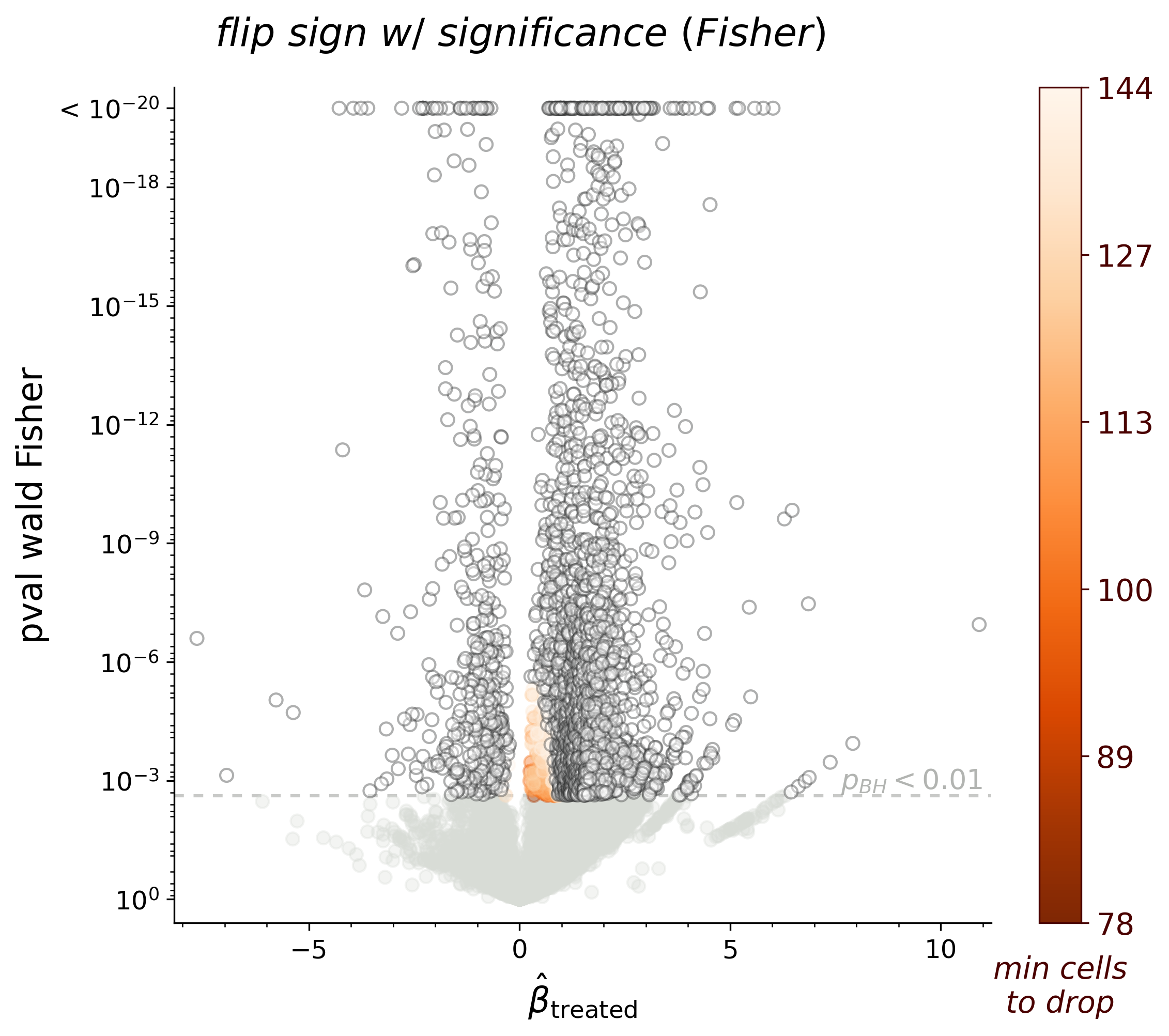}%
    \vspace{\vspaceBwRows}
    \includegraphics[width=\volcanoWidth]{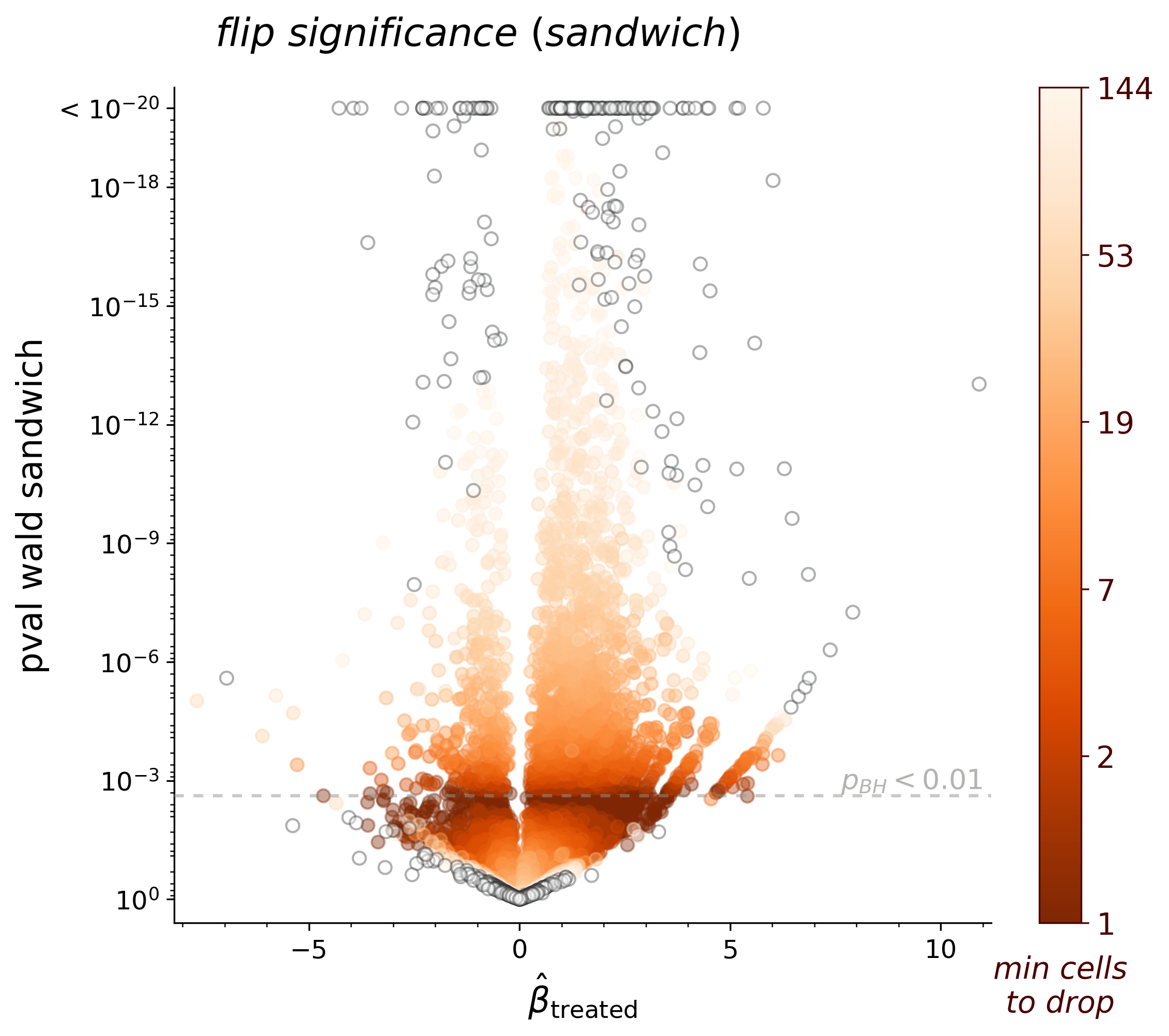}%
    \hfill%
    \includegraphics[width=\volcanoWidth]{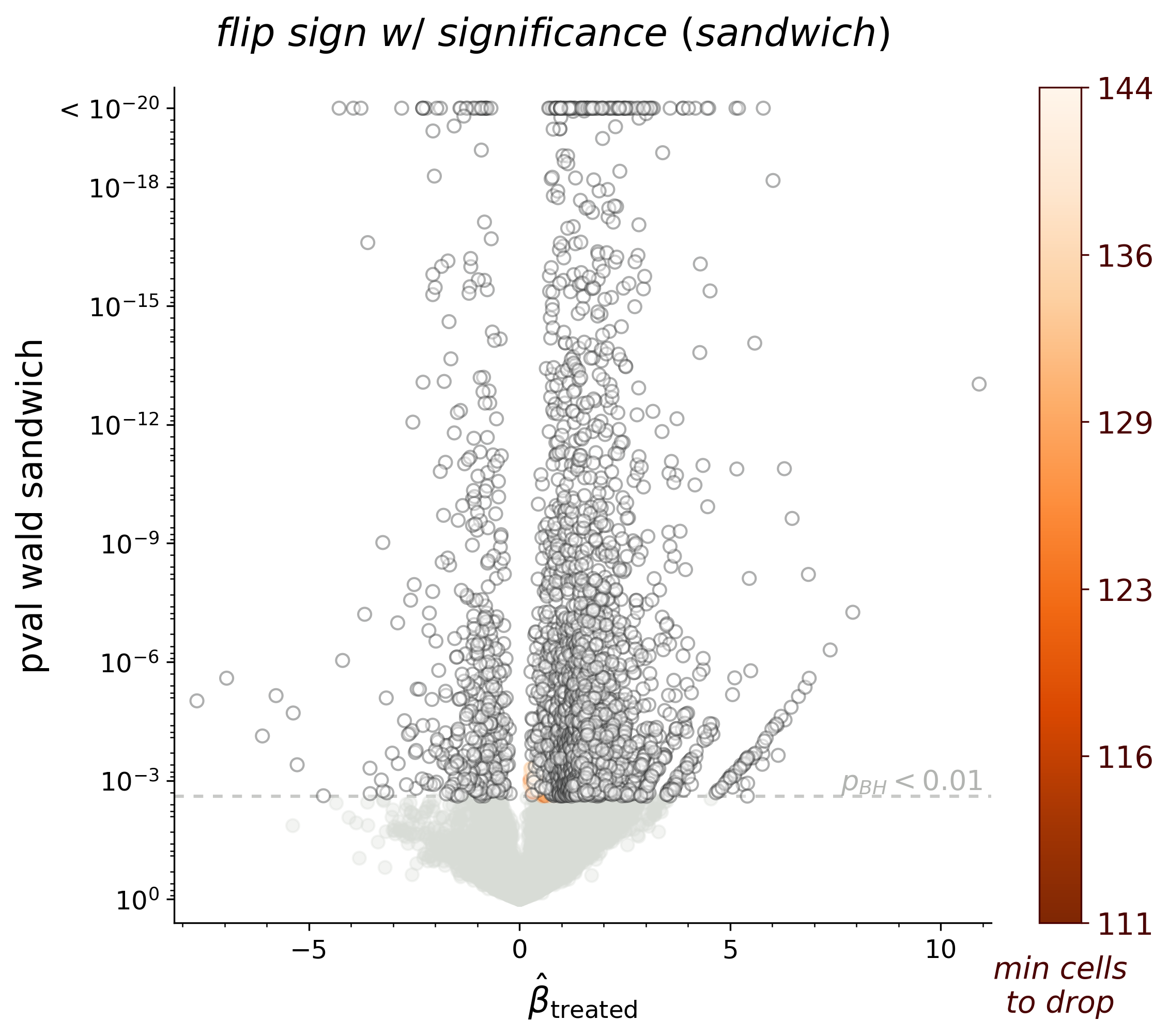}%
\end{figure}

\pagebreak

\continueCaption*{%
        \textbf{Minimal number of cells to drop to enact the change-of-interest, across genes.}
        Volcano plots of effect size
        (on a $\log_2$ scale) versus p-value
        (for the test indicated on the y-axis).
        Genes (\textit{points}) are colored by the size of the minimal cell
        subset---up to 10\% of cells ($N=1440$)---that, when dropped, are predicted to
        effect the change-of-interest (\textit{title}).
        The key (and sole) difference from \cref{fig:min-to-drop} is that genes
        are plotted in reverse order; i.e., from least to most robust.
        \label{supp-fig:min-to-drop-reversed}
        \vspace{3\baselineskip}
}

\begin{figure}[!htb]
    \centering
    \includegraphics[width=\textwidth]{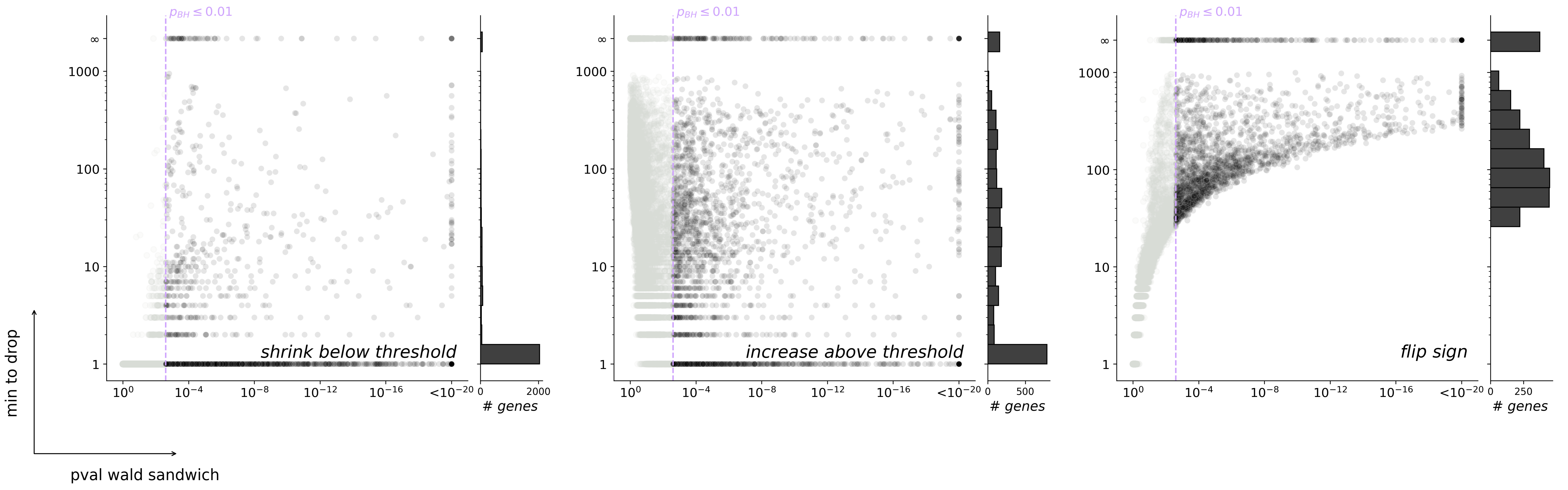}%
    \vspace{\vspaceBwRows}
    \includegraphics[width=\textwidth]{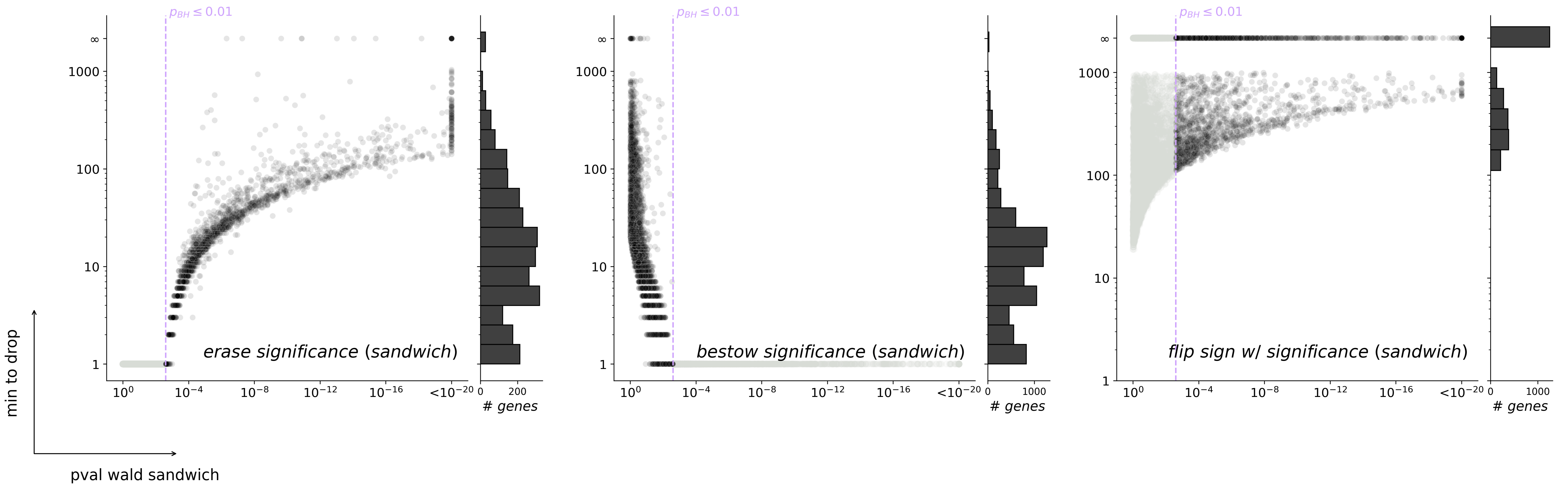}%
    \vspace{\vspaceBwRows}
    \includegraphics[width=\textwidth]{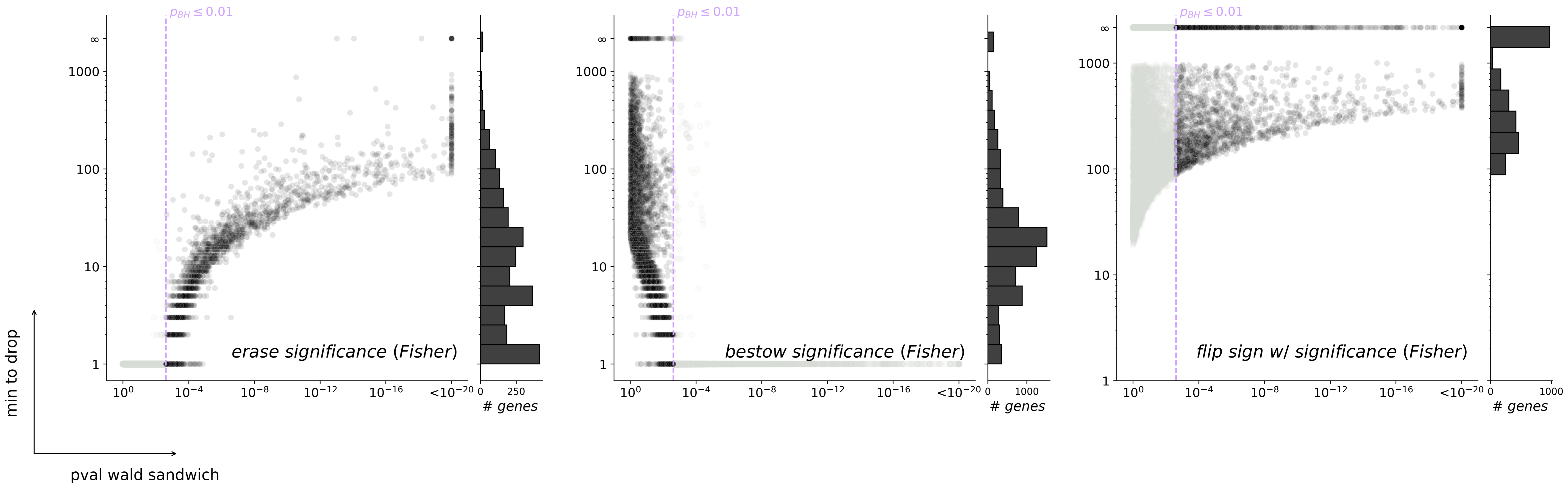}%
    \caption{
        \textbf{Minimum cells to drop to enact the change-of-interest, across genes.}
        Plots are raw p-values (for the test indicated on the x-axis)
        versus predicted minimal number of cells to drop (out of 1440).
        Black points highlight genes that are germane to the change-of-interest, based on
        significance level 0.01 for BH-corrected p-values.
    } \label{supp-fig:min-to-drop-BH}
\end{figure}

\begin{figure}[!htb]
    \centering
    \includegraphics[width=\linewidth]{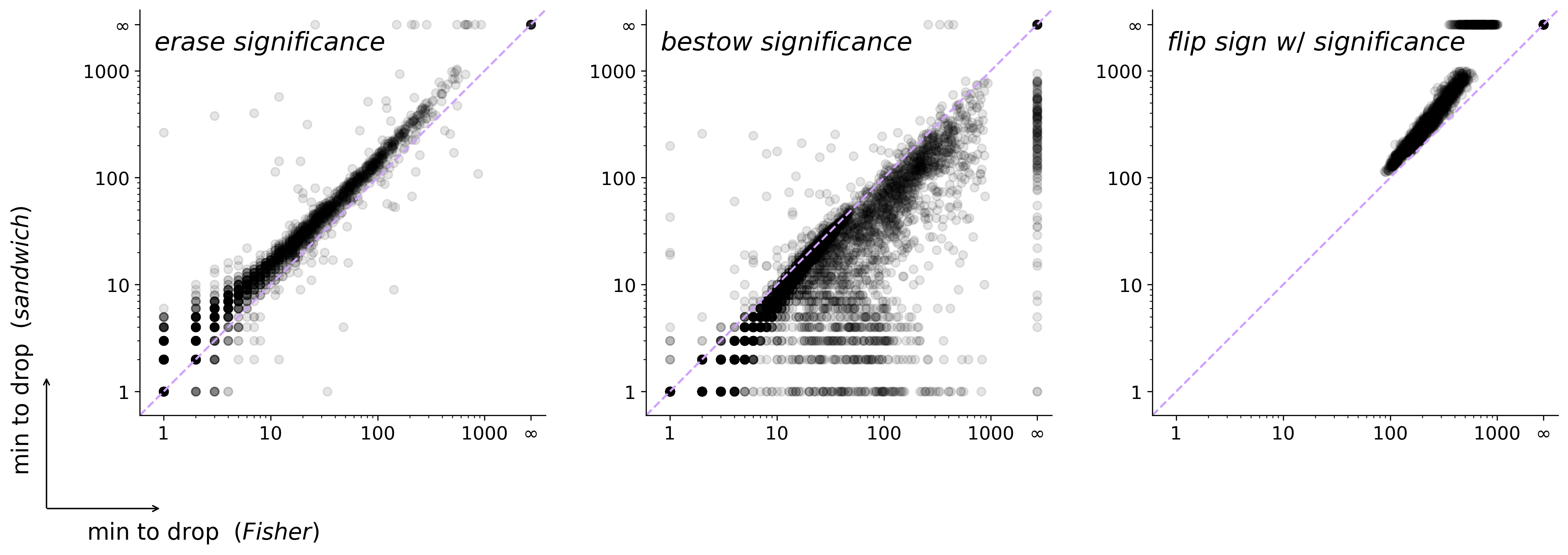}%
    \caption{
        \textbf{Minimum cells to drop in order to enact the change-of-interest involving a (Fisher vs. sandwich) Wald statistic.}
        Plots are the predicted minimal number of cells to drop (out of 1440)
        to enact the change-of-interest if the test is the Wald with Fisher estimator
        versus Wald with sandwich estimator.
        If this change is never predicted, the value is denoted as $\infty$
        (and plotted here on ``broken'' axes).
        Genes (\textit{points}) are filtered to those that are relevant across both
        standard error estimators for the change-of-interest (e.g., for
        ``erase significance,''
        genes are filtered to those that are
        significant at level 0.01 for BH-corrected p-values with respect to
        both Wald tests).
    } \label{supp-fig:min-to-drop-fisher-sandwich}
\end{figure}

\begin{figure}[!htb]
    \centering
    \subcaptionbox{\label{supp-fig:pred-v-actual_shrink-below-threshold}}{\includegraphics[width=\linewidth]{%
        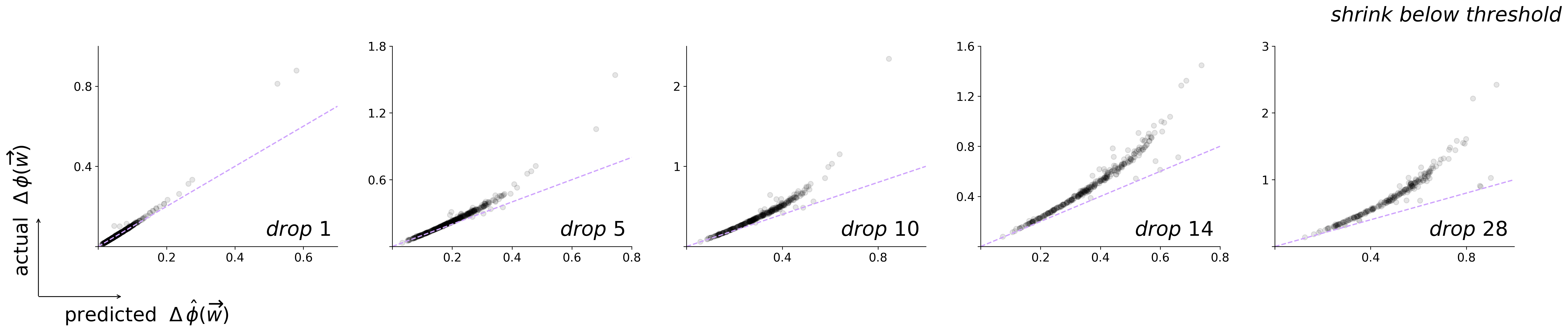}}%
    \hfill%
    \subcaptionbox{\label{supp-fig:pred-v-actual_increase-above-threshold}}{\includegraphics[width=\linewidth]{%
        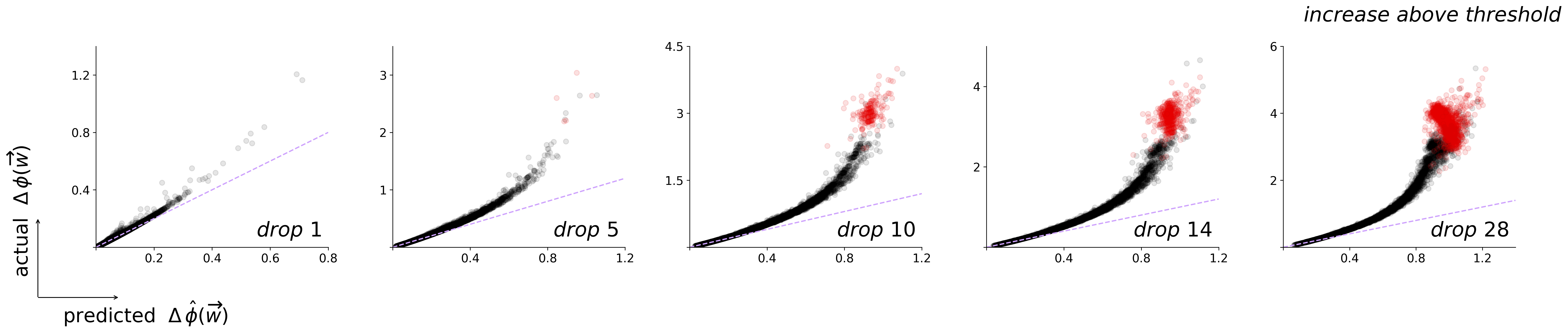}}%
    \hfill%
    \subcaptionbox{\label{supp-fig:pred-v-actual_flip-sign-w-sig-sandwich}}{\includegraphics[width=\linewidth]{%
        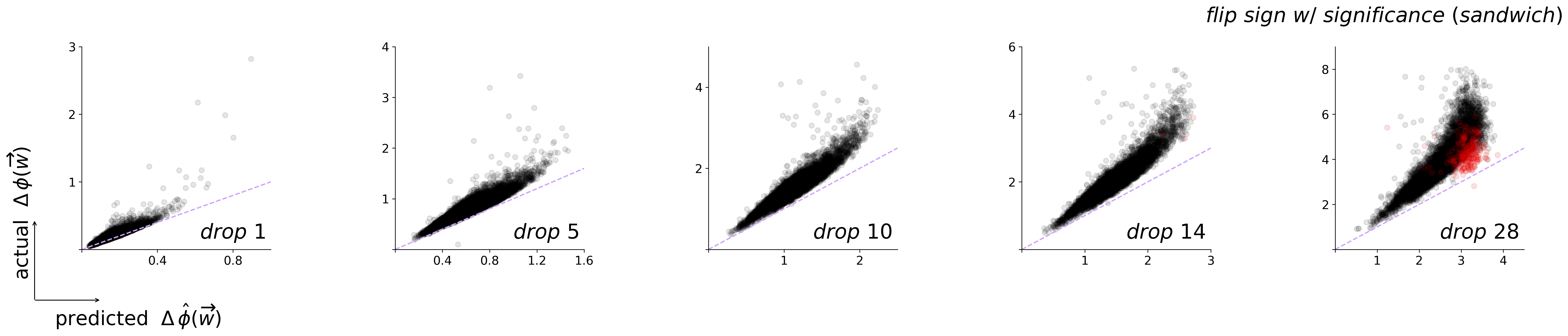}}%
    \hfill%
    \subcaptionbox{\label{supp-fig:pred-v-actual_flip-sign-w-sig-fisher}}{\includegraphics[width=\linewidth]{%
        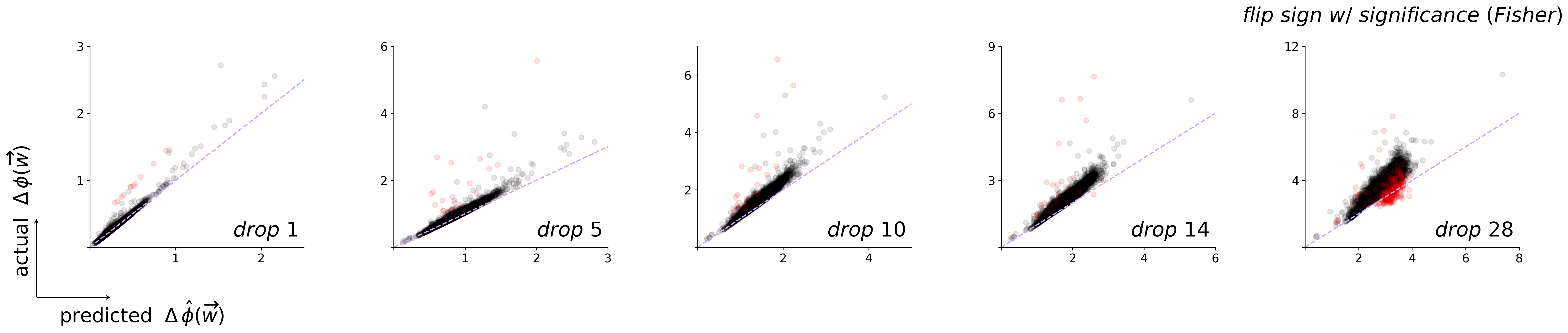}}%
    \caption{
        \textbf{Fidelity of the approximation for dropping the $\ninfluence$ most influential cells.}
        Plots are predicted (\textit{x-axis}) versus actual (\textit{y-axis})
        change to the statistic-of-interest $\stat$
        after dropping the top $\ninfluence$ most influential cells
        (up to 2\% of cells, out of 1440).
        Newly created zero-group genes (after dropping cells) are highlighted in red.
        Lilac dotted lines represent the 1-to-1 line (i.e., perfect predictions).
        \captionbr
        To avoid trivial results (like dropping all nonzero counts), and to improve
        the overall fidelity of the approximation, genes (\textit{points}) are filtered
        to those with a sufficient number of nonzero observations.
        Specifically, we filter to genes where the maximal number of nonzero
        observations per group (treatment or control)---after dropping the
        selected cells---is at least 20.
        (See \cref{supp-fig:nnz-v-err} for details on this cutoff.)
        We also filter to relevant genes (i.e., for ``erase significance,''
        genes that are originally significant under the relevant test).
        \captionbr
        Correlations range from
        0.94--0.98 (\textbf{\subref{supp-fig:pred-v-actual_shrink-below-threshold}}),
        0.93--0.99 (\textbf{\subref{supp-fig:pred-v-actual_increase-above-threshold}}),
        0.86--0.93 (\textbf{\subref{supp-fig:pred-v-actual_flip-sign-w-sig-sandwich}}), and
        and
        0.87--0.98 (\textbf{\subref{supp-fig:pred-v-actual_flip-sign-w-sig-fisher}}).
    } \label{supp-fig:pred-v-actual}
\end{figure}

\newlength{\spectrumWidth}
\setlength{\spectrumWidth}{\linewidth}

\begin{figure}[!htb]
    \centering
    \includegraphics[width=\spectrumWidth]{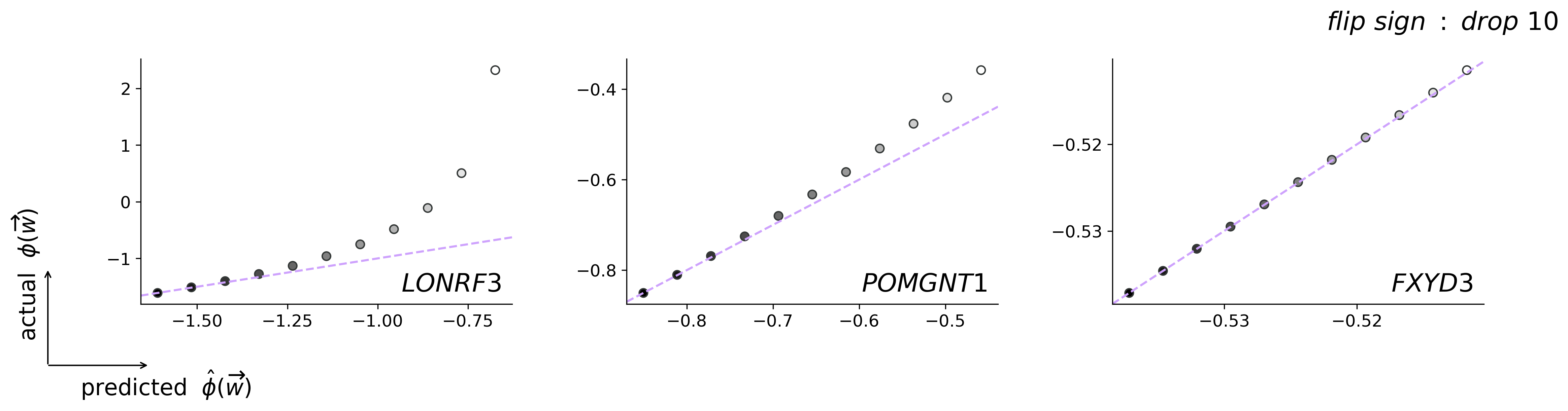}%
    \vspace{\vspaceBwRows}
    \includegraphics[width=\spectrumWidth]{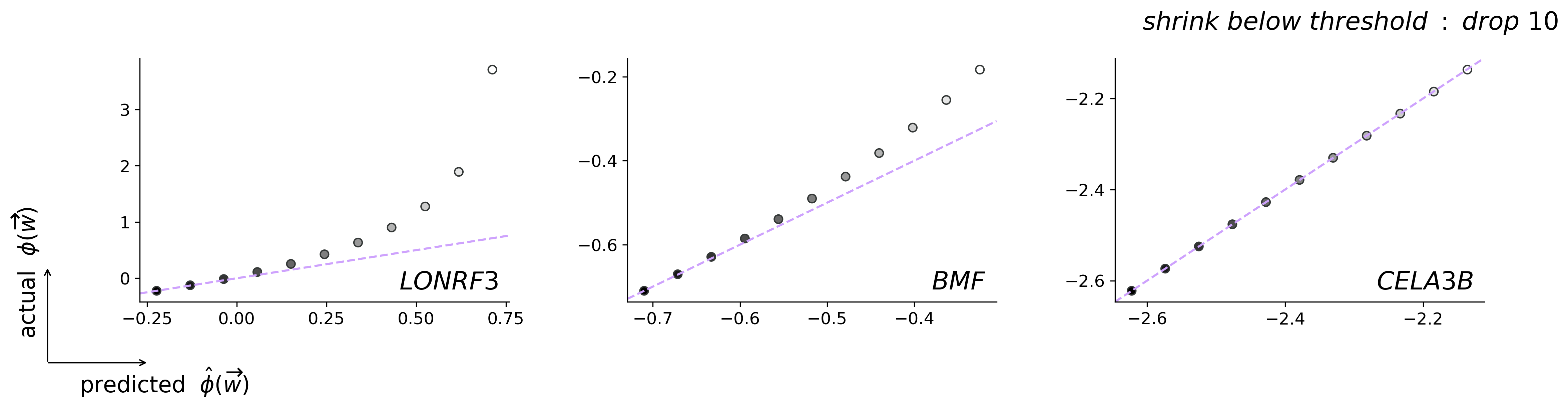}%
    \vspace{\vspaceBwRows}
    \includegraphics[width=\spectrumWidth]{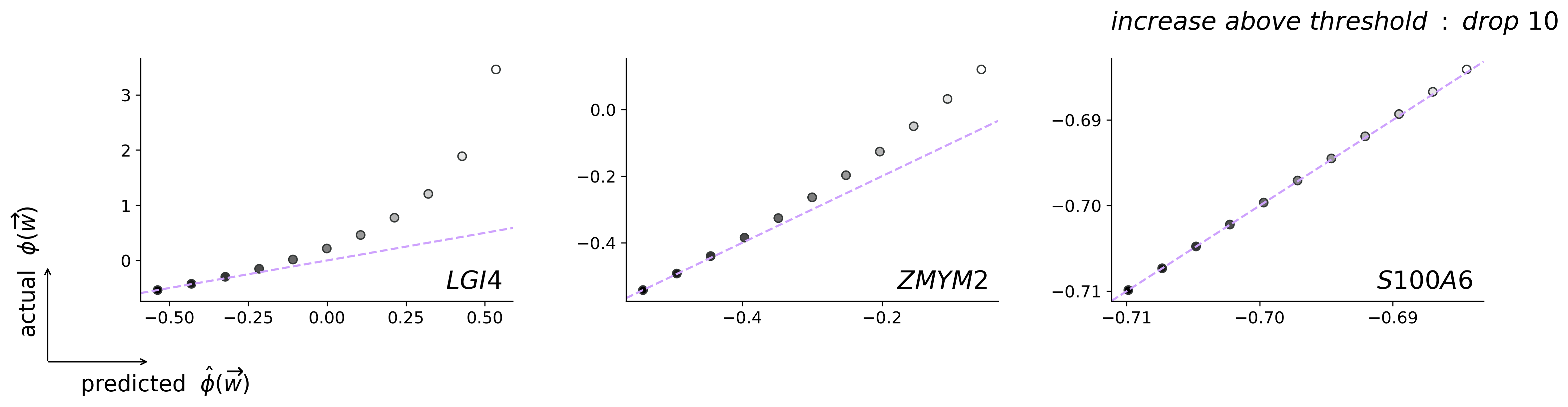}%
    \vspace{0.1cm}
    \caption{
        \textbf{Fidelity of the approximation (regarding effect size)
                across linearly interpolated weights,
                for the ``worst,'' ``median,'' and ``best'' gene predictions.}
        Each plot is the predicted versus actual value
        of a statistic-of-interest $\stat$
        (\textit{upper-right corner per row})
        for a given gene (\textit{lower-right corner}),
        evaluated across a spectrum of weights.
        Specifically, we identify the top 10 most influential cells for the
        statistic-of-interest, and evaluate the fidelity of the approximation
        as we move further from the place where the Taylor approximation was formed
        ($\vec{w}=\vec{1}$) by linearly modulating the weights for these cells
        (\textit{darkness of the points}) from 1 (\textit{black}) to 0 (\textit{white}).
        The one-to-one line (perfect concordance) is drawn in dashed lilac.
        Genes are selected to represent the ``worst,'' ``median,'' and ``best''
        approximations (\textit{left to right across each row}), based on
        the prediction error $| \hat{\stat}(\vec{w}) - \stat(\vec{w}) |$
        when the top 10 cells are fully dropped.
        \captionbr
        For genes with poor fidelity, we see as expected that the approximation
        itself is reasonable but the actual change in the statistic is too
        nonlinear to be captured by a first-order method.
        We also see that, when the actual statistic diverges from the approximation,
        it tends to change even more dramatically than predicted
        (recall that $\stat$ is constructed to be a decision function that
        moves toward the relevant decision boundary when \textit{increased}).
    } \label{supp-fig:spectrum-a}
\end{figure}

\begin{figure}[!htb]
    \centering
    \includegraphics[width=\spectrumWidth]{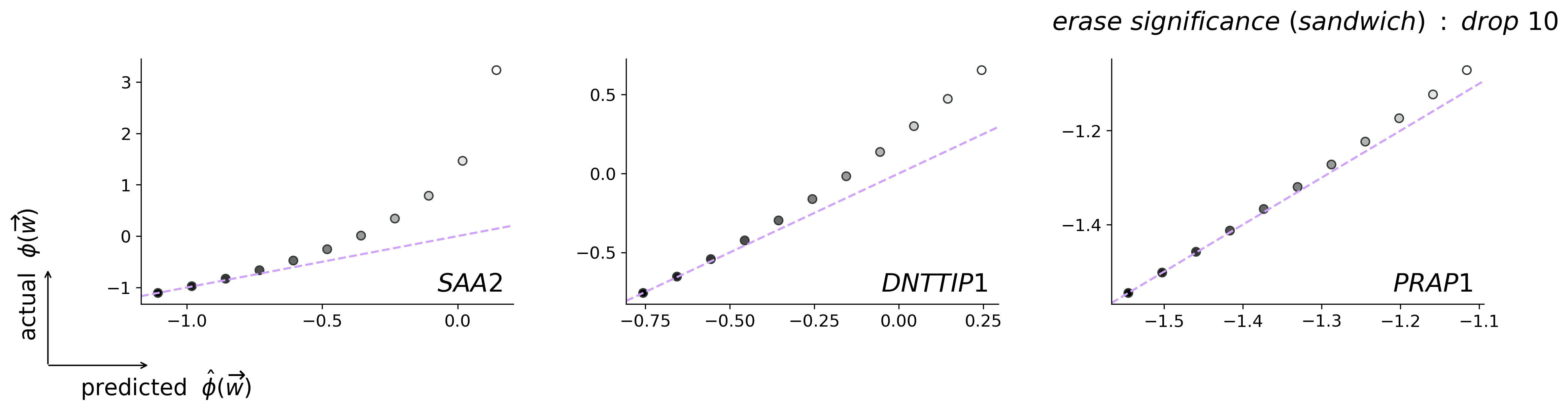}%
    \vspace{\vspaceBwRows}
    \includegraphics[width=\spectrumWidth]{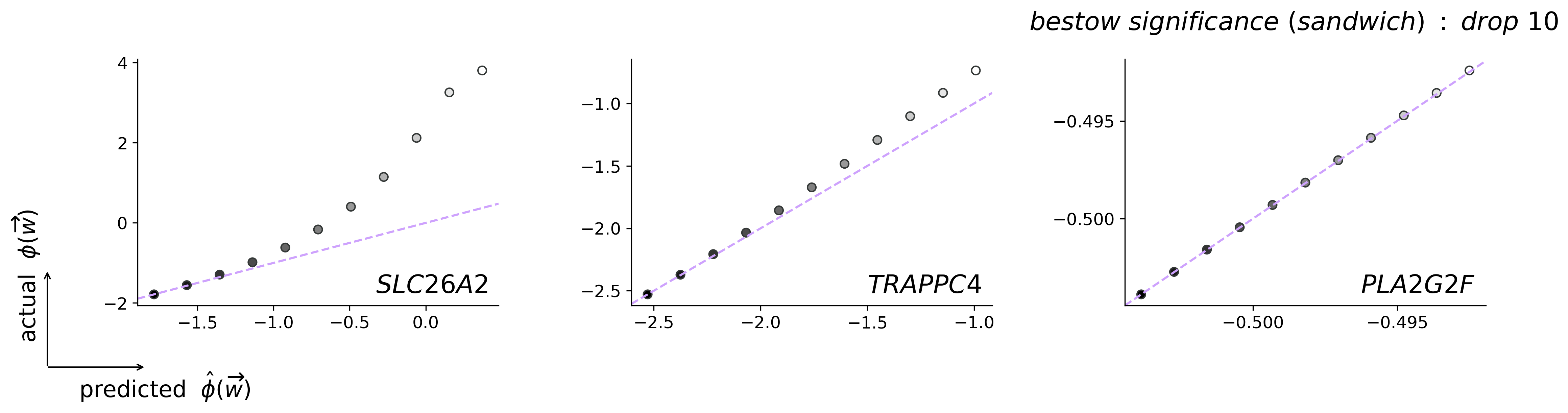}%
    \vspace{\vspaceBwRows}
    \includegraphics[width=\spectrumWidth]{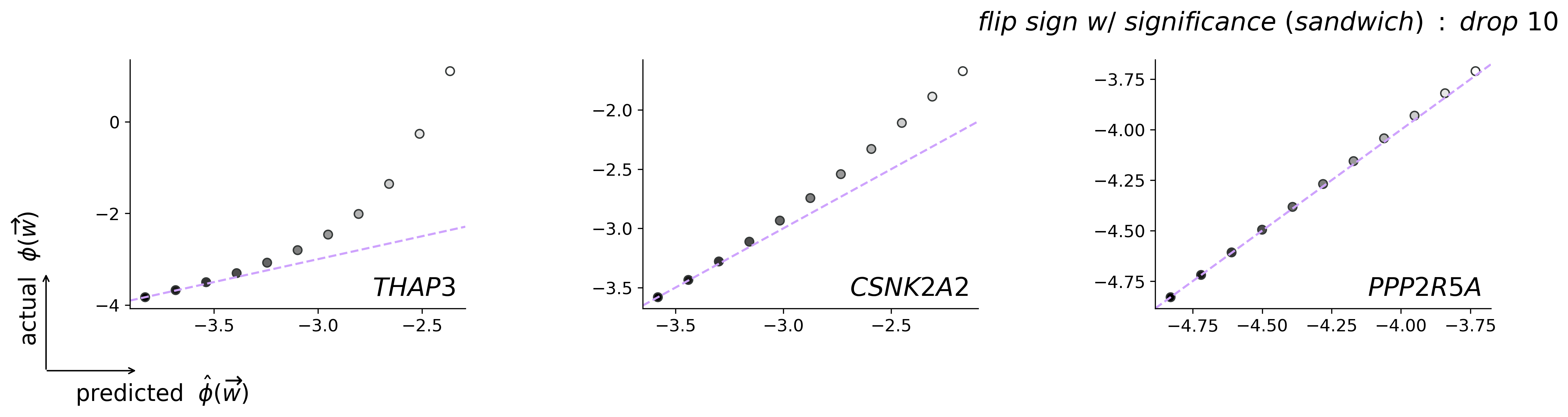}%
    \vspace{0.1cm}
    \caption{
        \textbf{Fidelity of the approximation
                (regarding Wald sandwich significance)
                across linearly interpolated weights,
                for the ``worst,'' ``median,'' and ``best'' gene predictions.}
        Each plot is the predicted versus actual value
        of a statistic-of-interest $\stat$
        (\textit{upper-right corner per row})
        for a given gene (\textit{lower-right corner}),
        evaluated across a spectrum of weights.
        Specifically, we identify the top 10 most influential cells for the
        statistic-of-interest, and evaluate the fidelity of the approximation
        as we move further from the place where the Taylor approximation was formed
        ($\vec{w}=\vec{1}$) by linearly modulating the weights for these cells
        (\textit{darkness of the points}) from 1 (\textit{black}) to 0 (\textit{white}).
        The one-to-one line (perfect concordance) is drawn in dashed lilac.
        Genes are selected to represent the ``worst,'' ``median,'' and ``best''
        approximations (\textit{left to right across each row}), based on
        the prediction error $| \hat{\stat}(\vec{w}) - \stat(\vec{w}) |$
        when the top 10 cells are fully dropped.
        \captionbr
        For genes with poor fidelity, we see as expected that the approximation
        itself is reasonable but the actual change in the statistic is too
        nonlinear to be captured by a first-order method.
        We also see that, when the actual statistic diverges from the approximation,
        it tends to change even more dramatically than predicted
        (recall that $\stat$ is constructed to be a decision function that
        moves toward the relevant decision boundary when \textit{increased}).
    } \label{supp-fig:spectrum-b}
\end{figure}

\begin{figure}[!htb]
    \centering
    \includegraphics[width=\spectrumWidth]{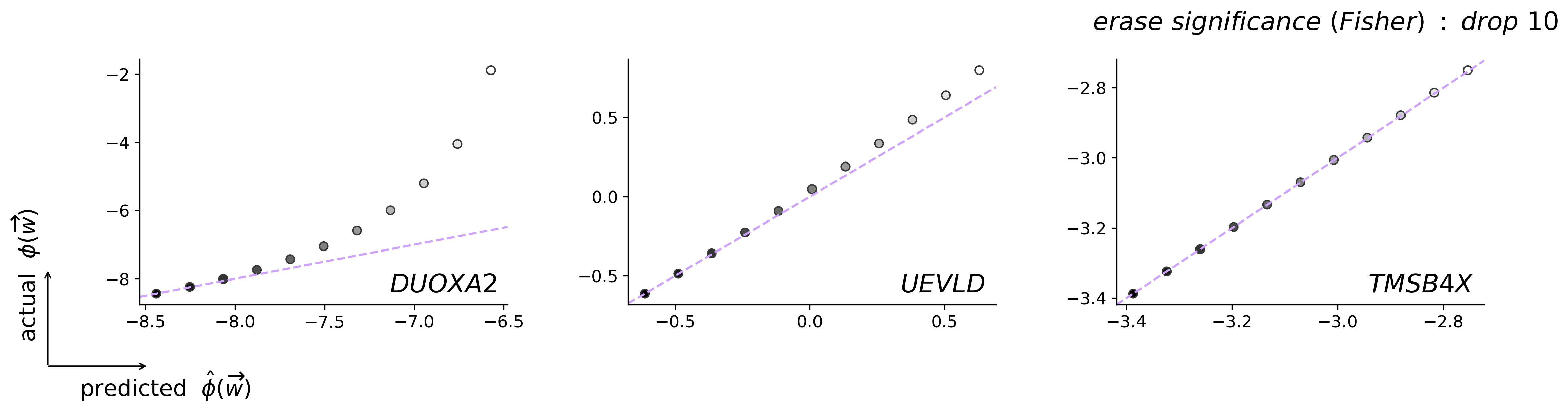}%
    \vspace{\vspaceBwRows}
    \includegraphics[width=\spectrumWidth]{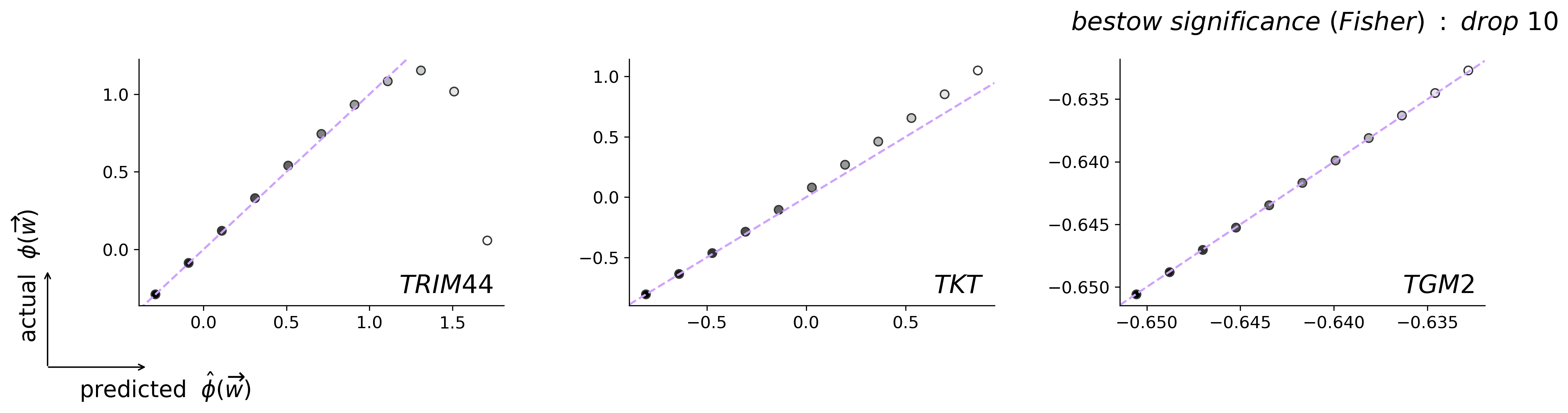}%
    \vspace{\vspaceBwRows}
    \includegraphics[width=\spectrumWidth]{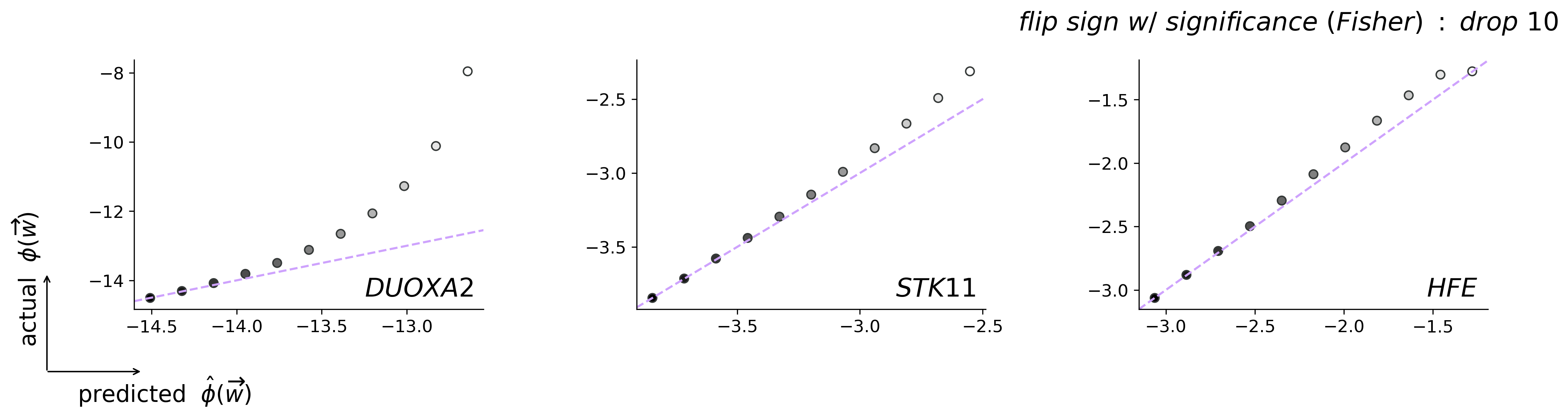}%
    \vspace{0.1cm}
    \caption{
        \textbf{Fidelity of the approximation
                (regarding Wald Fisher significance)
                across linearly interpolated weights,
                for the ``worst,'' ``median,'' and ``best'' gene predictions.}
        Each plot is the predicted versus actual value
        of a statistic-of-interest $\stat$
        (\textit{upper-right corner per row})
        for a given gene (\textit{lower-right corner}),
        evaluated across a spectrum of weights.
        Specifically, we identify the top 10 most influential cells for the
        statistic-of-interest, and evaluate the fidelity of the approximation
        as we move further from the place where the Taylor approximation was formed
        ($\vec{w}=\vec{1}$) by linearly modulating the weights for these cells
        (\textit{darkness of the points}) from 1 (\textit{black}) to 0 (\textit{white}).
        The one-to-one line (perfect concordance) is drawn in dashed lilac.
        Genes are selected to represent the ``worst,'' ``median,'' and ``best''
        approximations (\textit{left to right across each row}), based on
        the prediction error $| \hat{\stat}(\vec{w}) - \stat(\vec{w}) |$
        when the top 10 cells are fully dropped.
        \captionbr
        For genes with poor fidelity, we see as expected that the approximation
        itself is reasonable but the actual change in the statistic is too
        nonlinear to be captured by a first-order method.
        We also see that, when the actual statistic diverges from the approximation,
        it tends to change even more dramatically than predicted
        (recall that $\stat$ is constructed to be a decision function that
        moves toward the relevant decision boundary when \textit{increased}).
    } \label{supp-fig:spectrum-c}
\end{figure}

\newcommand{\nnzVErrWidth}{\linewidth}
\begin{figure}[!htb]
    \centering
    \subcaptionbox{\;drop \textit{one}\label{fig:nnz-first}}{\includegraphics[width=\nnzVErrWidth]{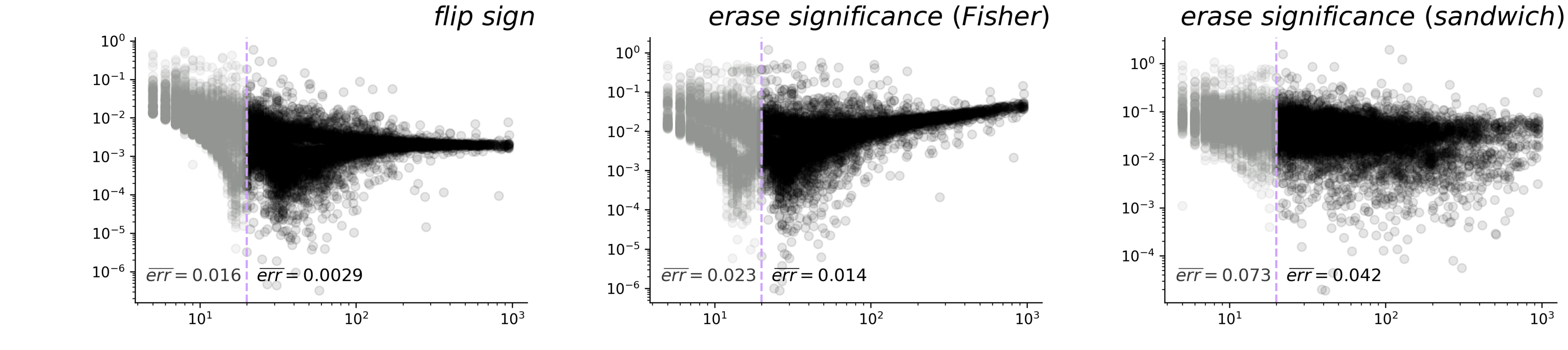}}
    \subcaptionbox{\;drop \textit{five}}{\includegraphics[width=\nnzVErrWidth]{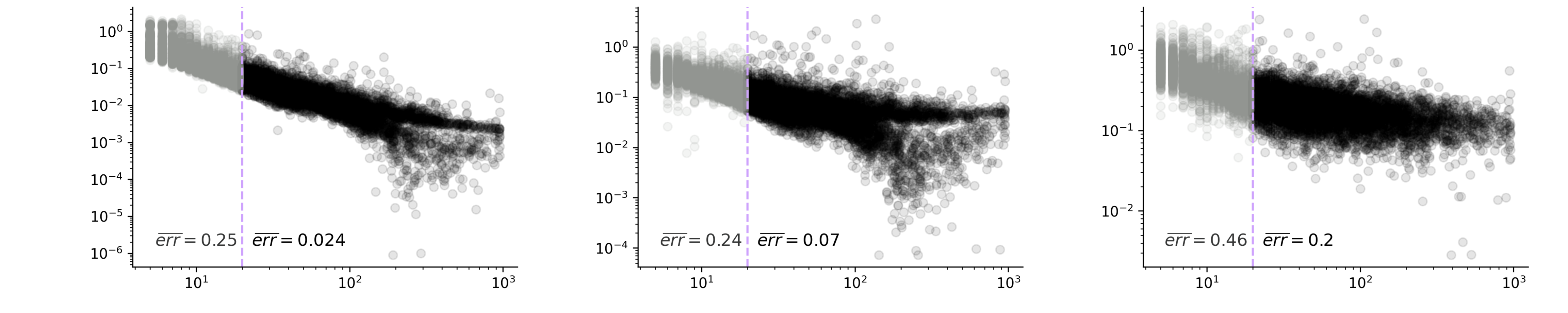}}
    \subcaptionbox{\;drop \textit{10}}{\includegraphics[width=\nnzVErrWidth]{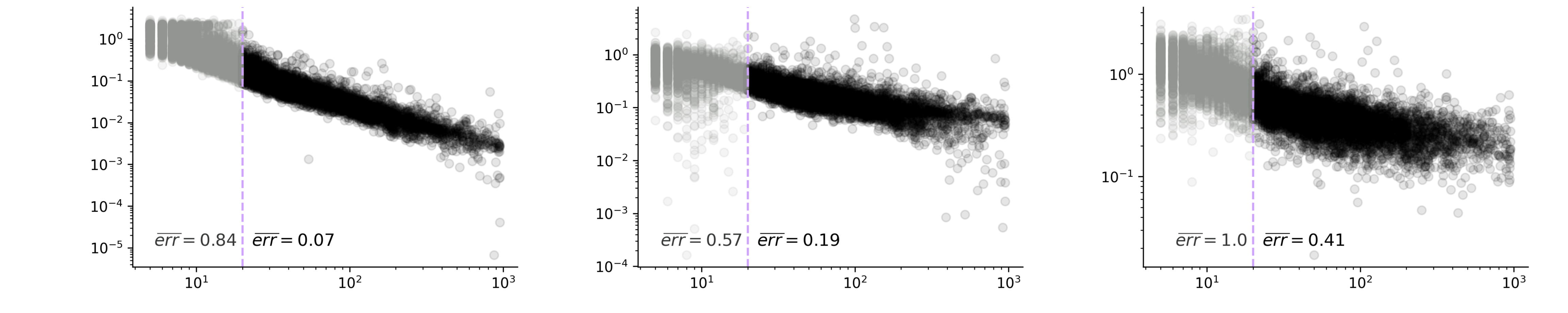}}
    \subcaptionbox{\;drop \textit{14}}{\includegraphics[width=\nnzVErrWidth]{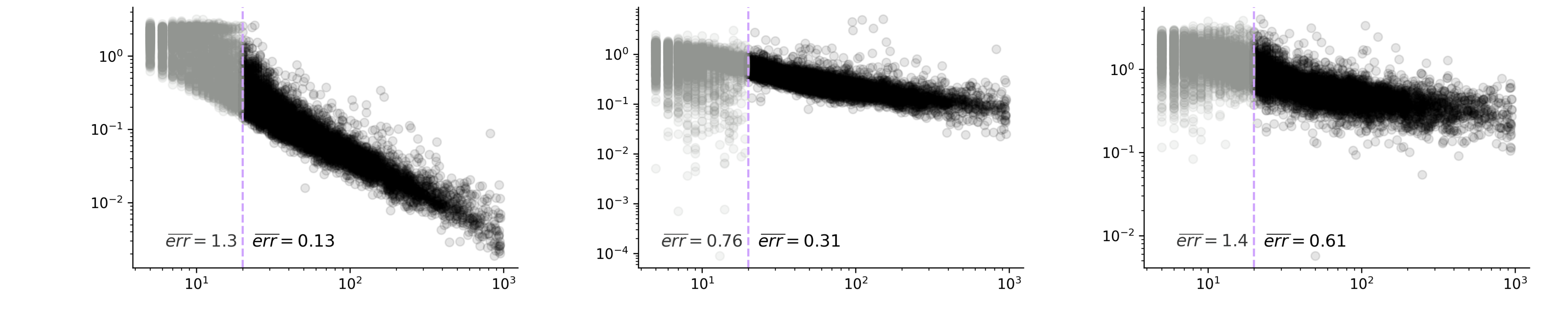}}
    \subcaptionbox{\;drop \textit{28}\label{fig:nnz-last}}{\includegraphics[width=\nnzVErrWidth]{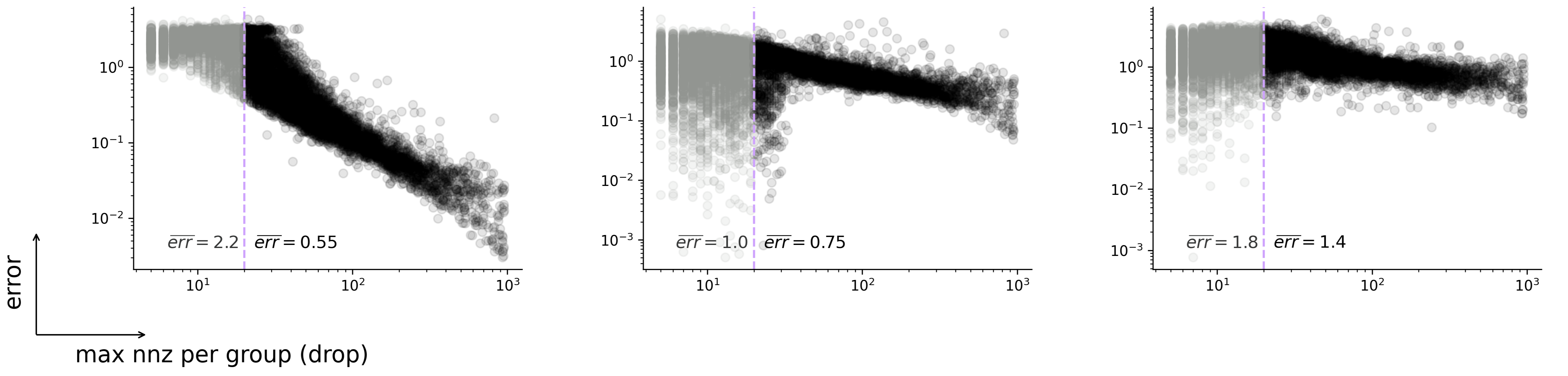}}
    \caption{
        \textbf{Maximal nonzero counts per group vs. approximation error.}
        Here, the maximal number of nonzero counts (\textit{nnz}) per group
        is plotted against the approximation error
        ($\mkern1mu| \hat{\stat}(\vec{w}) - \stat(\vec{w}) |\mkern1mu$).
        Each row plots the
        approximation error
        when the top $\ninfluence$ cells are dropped
        (\textit{increasing from \textbf{\subref{fig:nnz-first}} $\rightarrow$ \textbf{\subref{fig:nnz-last}}}),
        where each column corresponds to
        the statistic-of-interest (\textit{top right per column}).
        We find that the quality of the approximation is
        correlated with the sparsity of the \textit{least sparse} group
        (treatment or control). As genes with few nonzero counts in either
        group should not show up as significant, it is reasonable to
        exclude them from the analysis.
        The mean error on either side of our chosen cutoff ($nnz \geq 20$) is annotated.
    } \label{supp-fig:nnz-v-err}
\end{figure}

\renewcommand{\makeSubcaption}[3]{
    \;\;drop \textit{#1} cells $\implies$\\
    displaces \textit{#2} of the top 10 GO sets\\
    (enriched among \ul{#3}regulated genes)
}
\WithSuffix\newcommand{\makeSubcaption}*[3]{
    \;\;drop \textit{#1} cell $\implies$\\
    displaces \textit{#2} of the top 10 GO sets\\
    (enriched among \ul{#3}regulated genes)
}

\begin{figure}[!htb]
    \centering
    \captionsetup[subfigure]{justification=centering}
    \parbox{\textwidth}{
        \parbox{0.5\textwidth}{%
            \subcaptionbox{\makeSubcaption*{one}{two}{up}}{%
                \includegraphics[width=\linewidth]{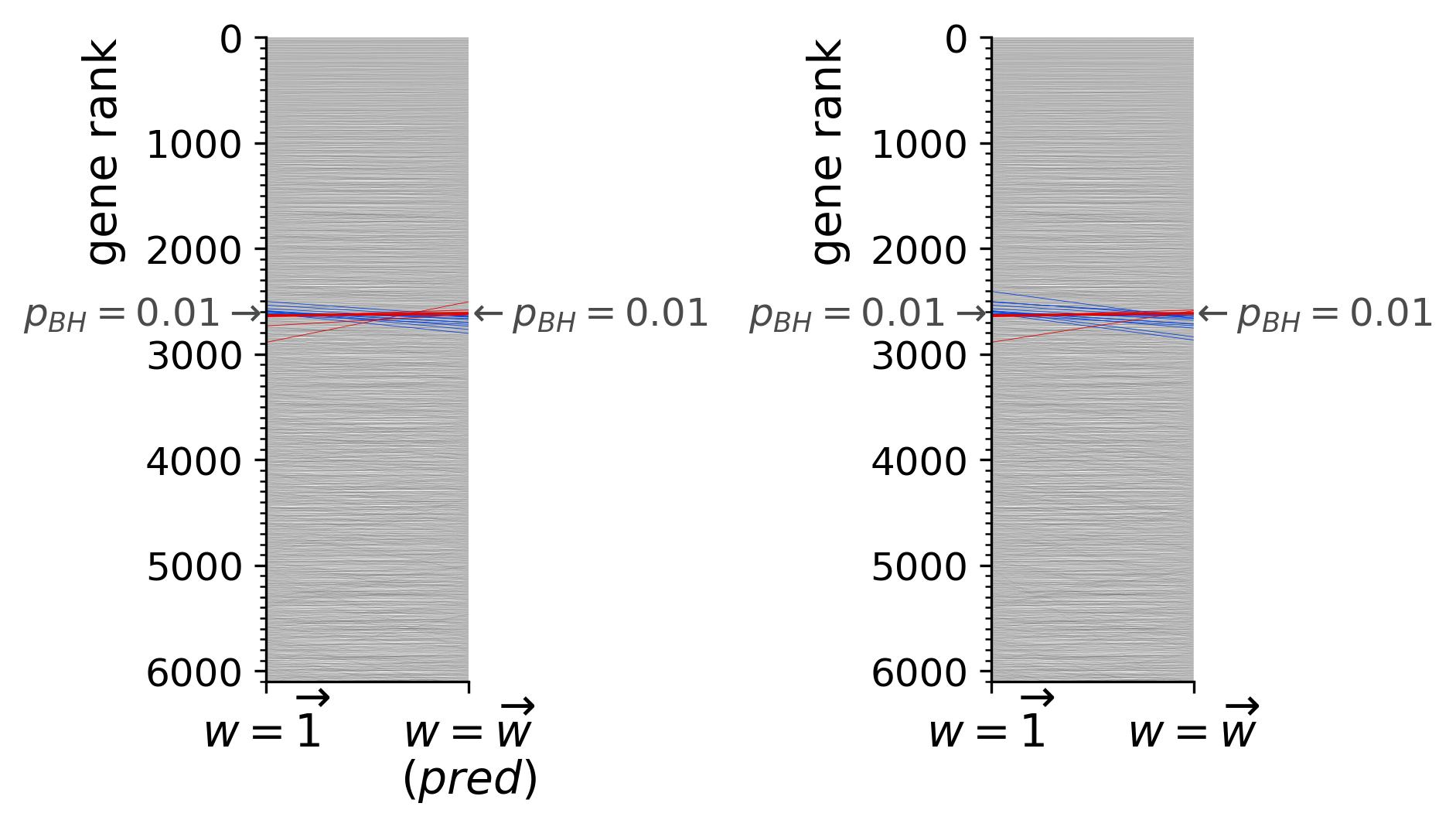}}%
            \vspace{1.5\vspaceBwRows}
            \subcaptionbox{\makeSubcaption{four}{three}{up}}{%
                \includegraphics[width=\linewidth]{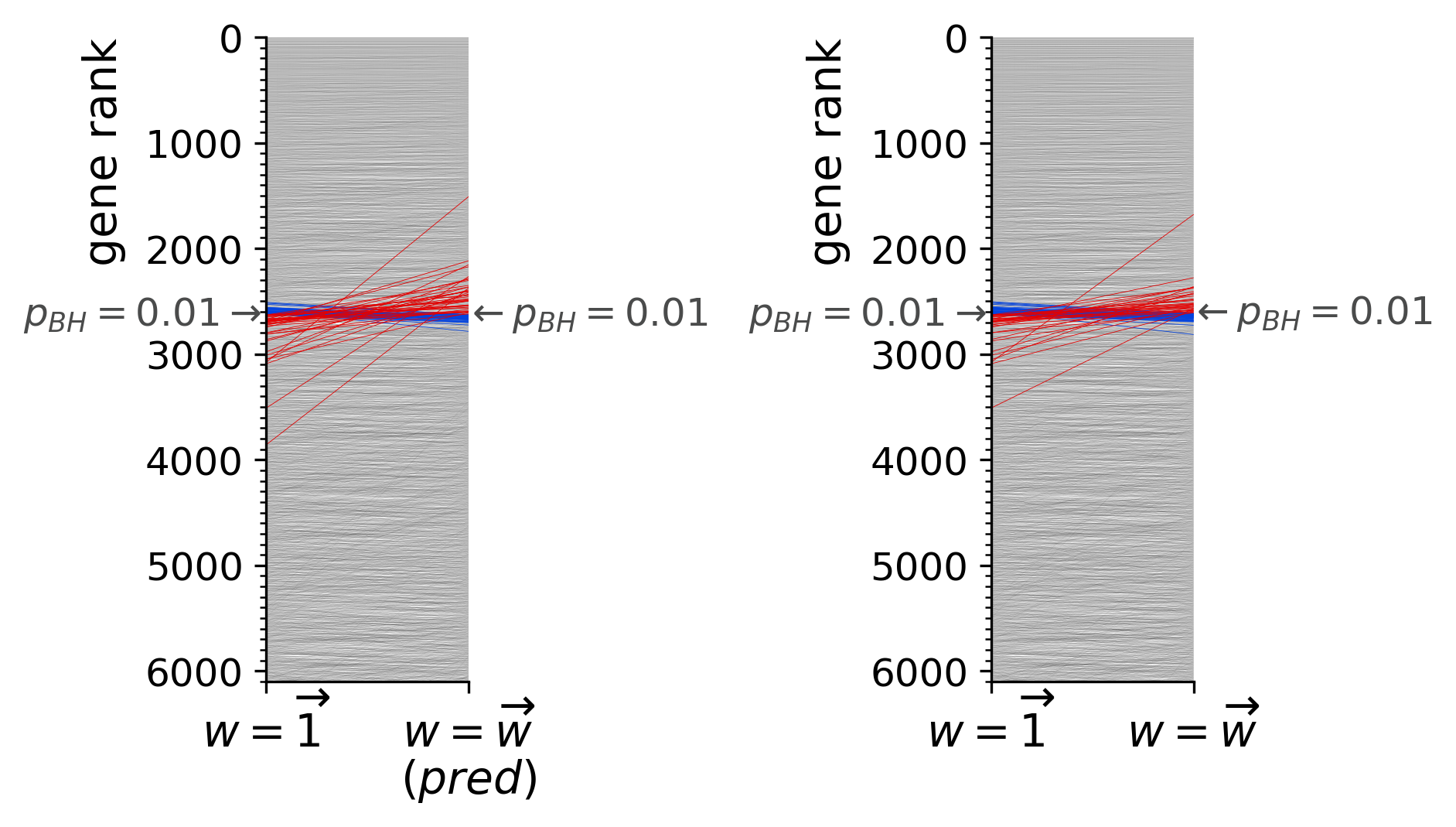}}%
            \vspace{1.5\vspaceBwRows}
            \subcaptionbox{\makeSubcaption{seven}{four}{up}}{%
                \includegraphics[width=\linewidth]{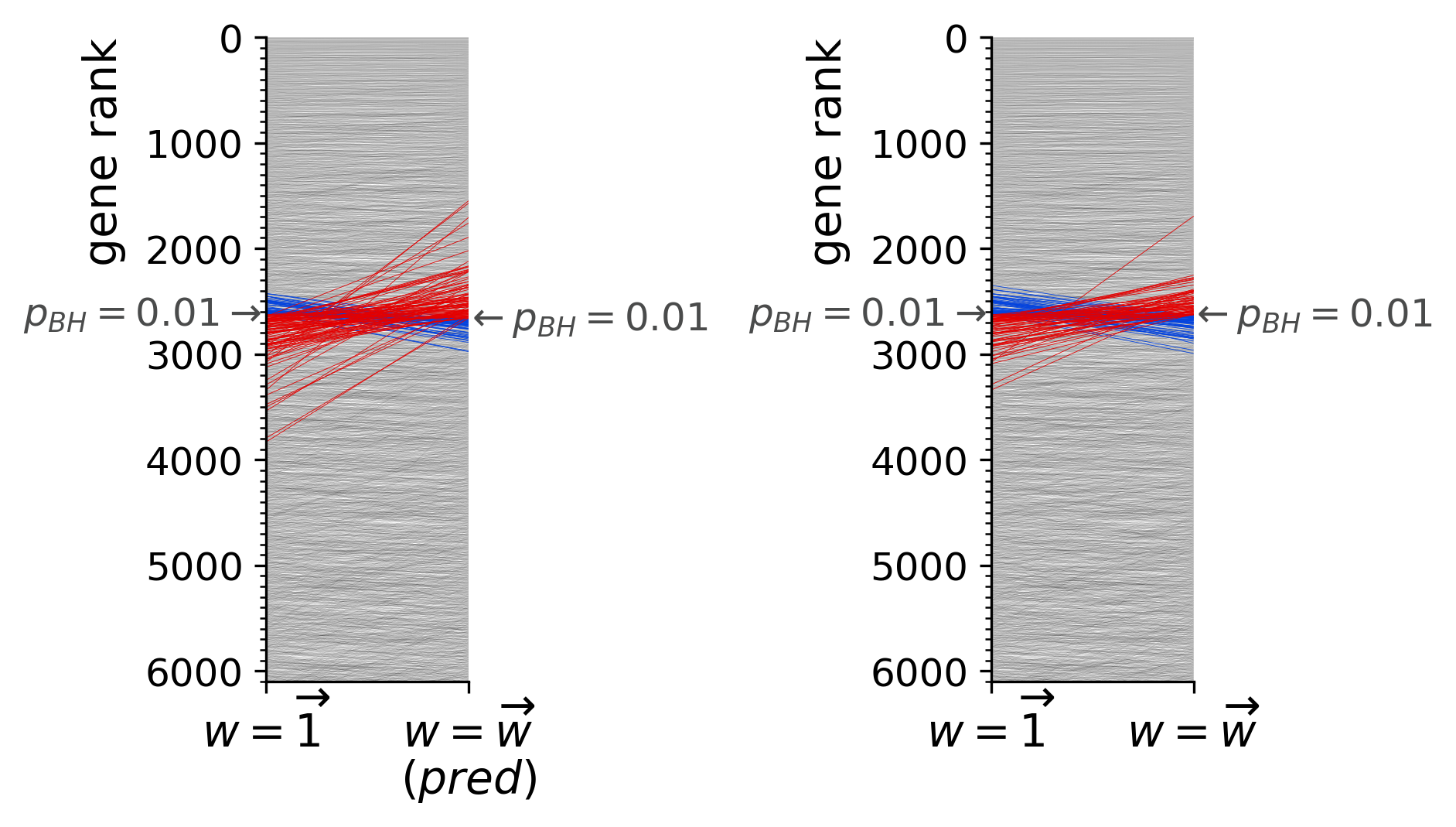}}%
        }
        \hfill
        \parbox{0.5\textwidth}{%
            \subcaptionbox{\makeSubcaption{14}{five}{up}}{%
                \includegraphics[width=\linewidth]{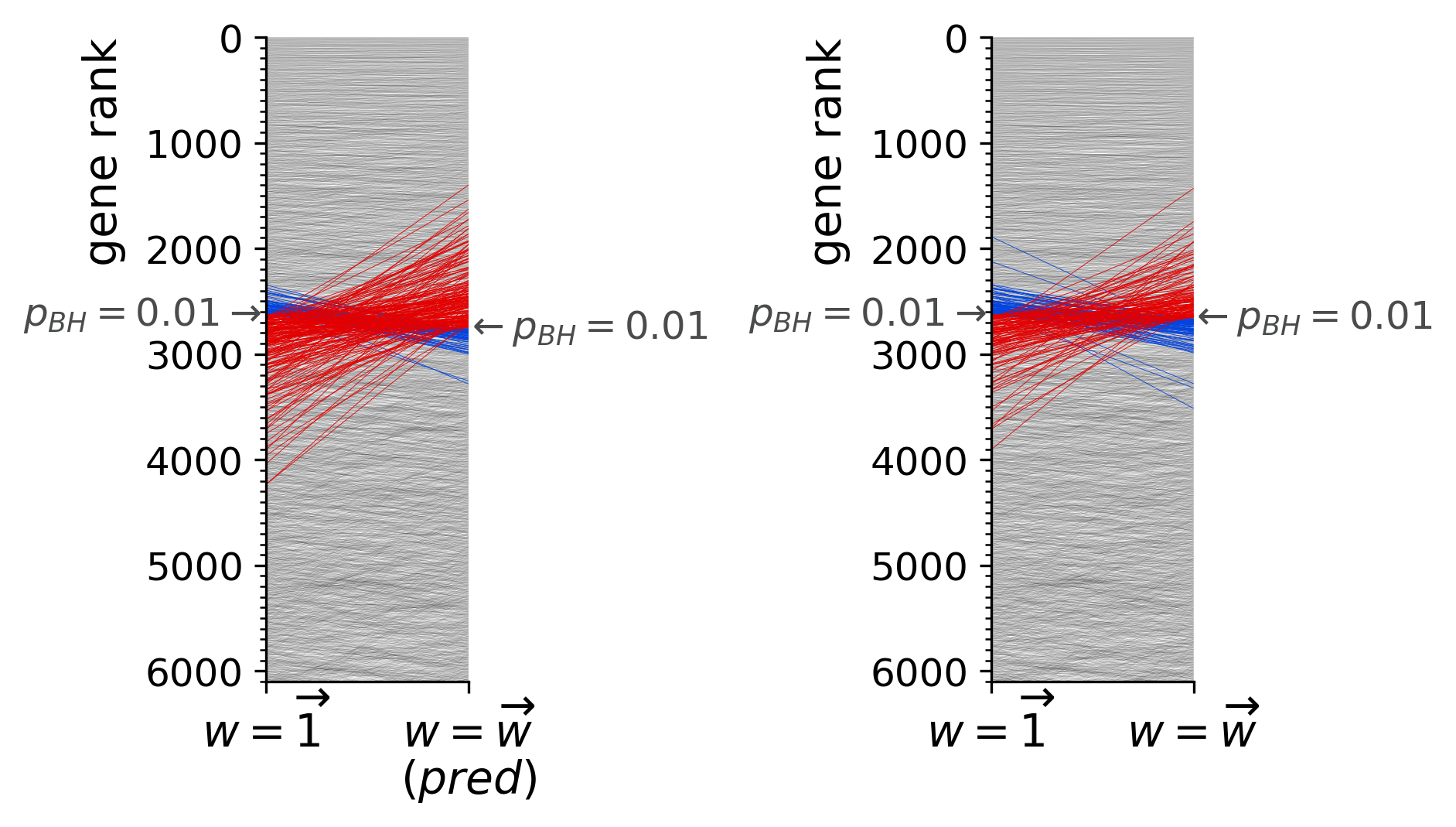}}%
            \vspace{1.5\vspaceBwRows}
            \subcaptionbox{\makeSubcaption{28}{five}{up}}{%
                \includegraphics[width=\linewidth]{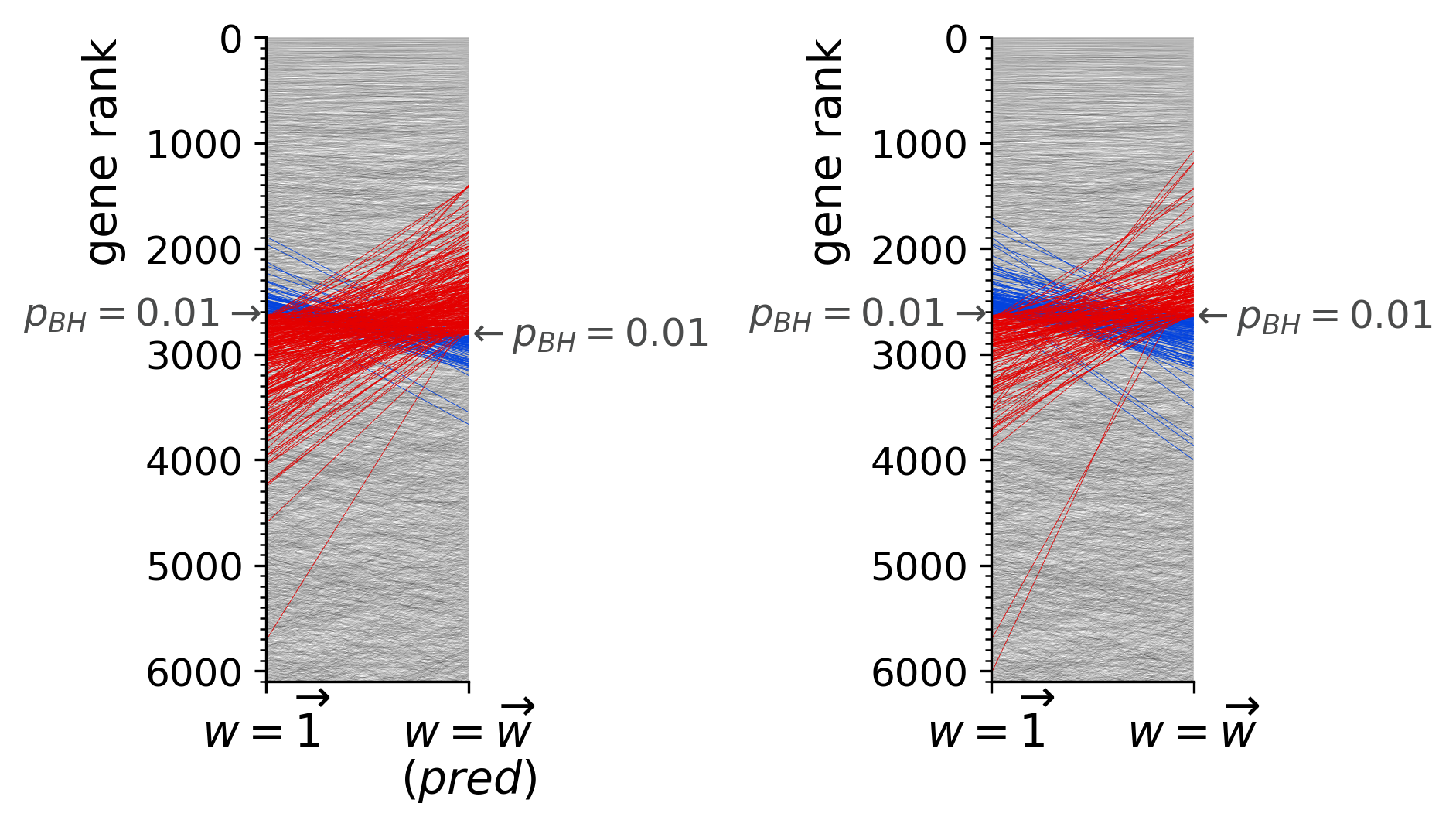}}%
        }
    }
    \caption{
        \textbf{Predicted vs. actual perturbation to DE p-values when a handful
                of influential cells (with respect to upregulated genes) are dropped.}
        Plots show the predicted (\textit{left}) and actual (\textit{right}) changes to ranked p-values
        for differential expression based on the Wald sandwich test.
        Annotated arrows indicate the ranking cutoff for BH-corrected p-values
        at level $0.01$.
        Blue lines indicate the change in ranking for genes that are \textit{demoted}
        from the significant set,
        red lines indicate the change in ranking for those that are \textit{promoted},
        and black lines indicate the change in ranking for those that \textit{retain} their
        significance status. Rankings are truncated; over 10,000 genes are tested.
    } \label{supp-fig:change-in-de-pvals-up}
\end{figure}

\begin{figure}[!htb]
    \centering
    \captionsetup[subfigure]{justification=centering}
    \parbox{\textwidth}{
        \parbox{0.5\textwidth}{%
            \subcaptionbox{\makeSubcaption*{one}{two}{down}}{%
                \includegraphics[width=\linewidth]{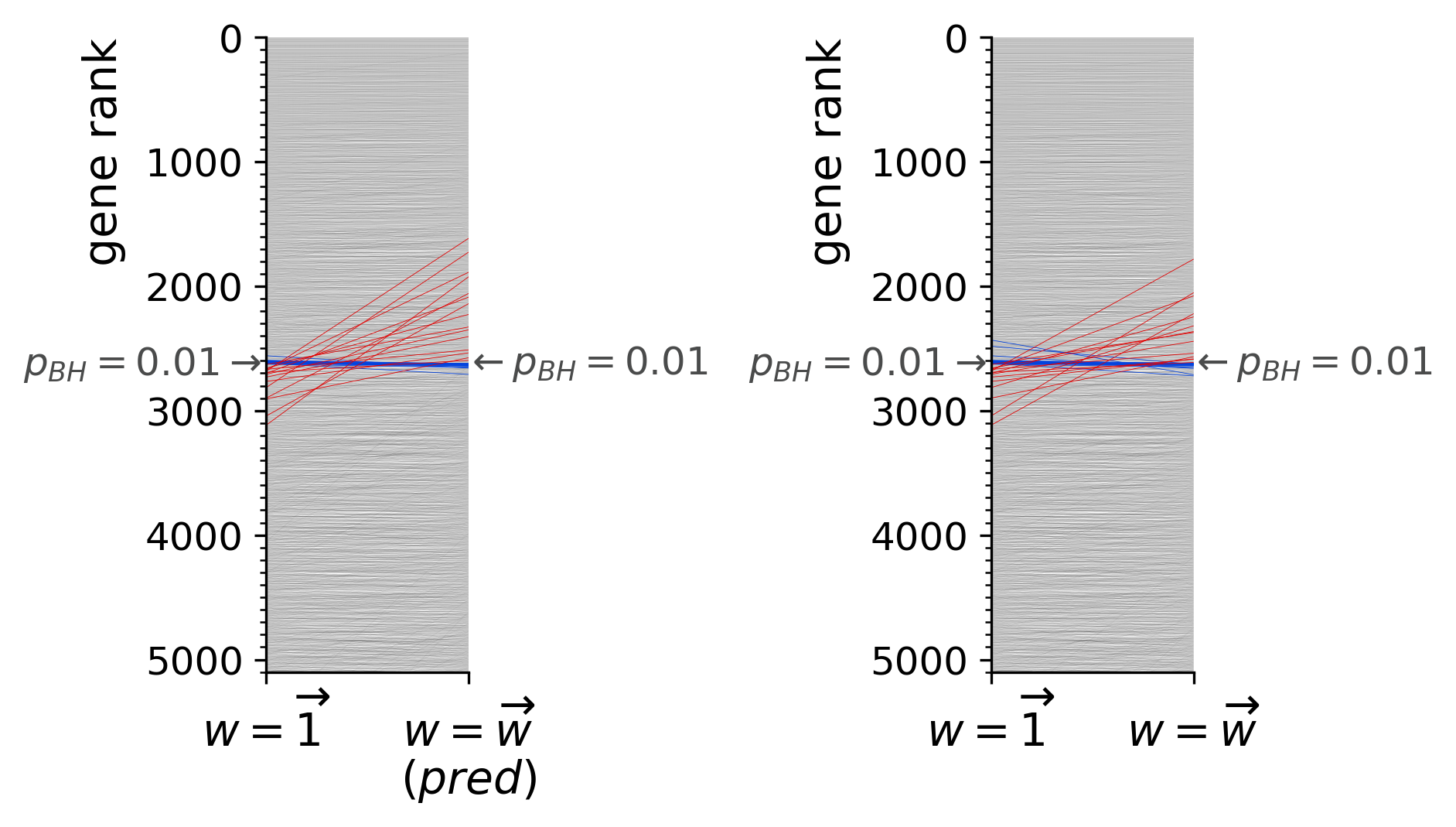}}%
            \vspace{1.5\vspaceBwRows}
            \subcaptionbox{\makeSubcaption{four}{three}{down}}{%
                \includegraphics[width=\linewidth]{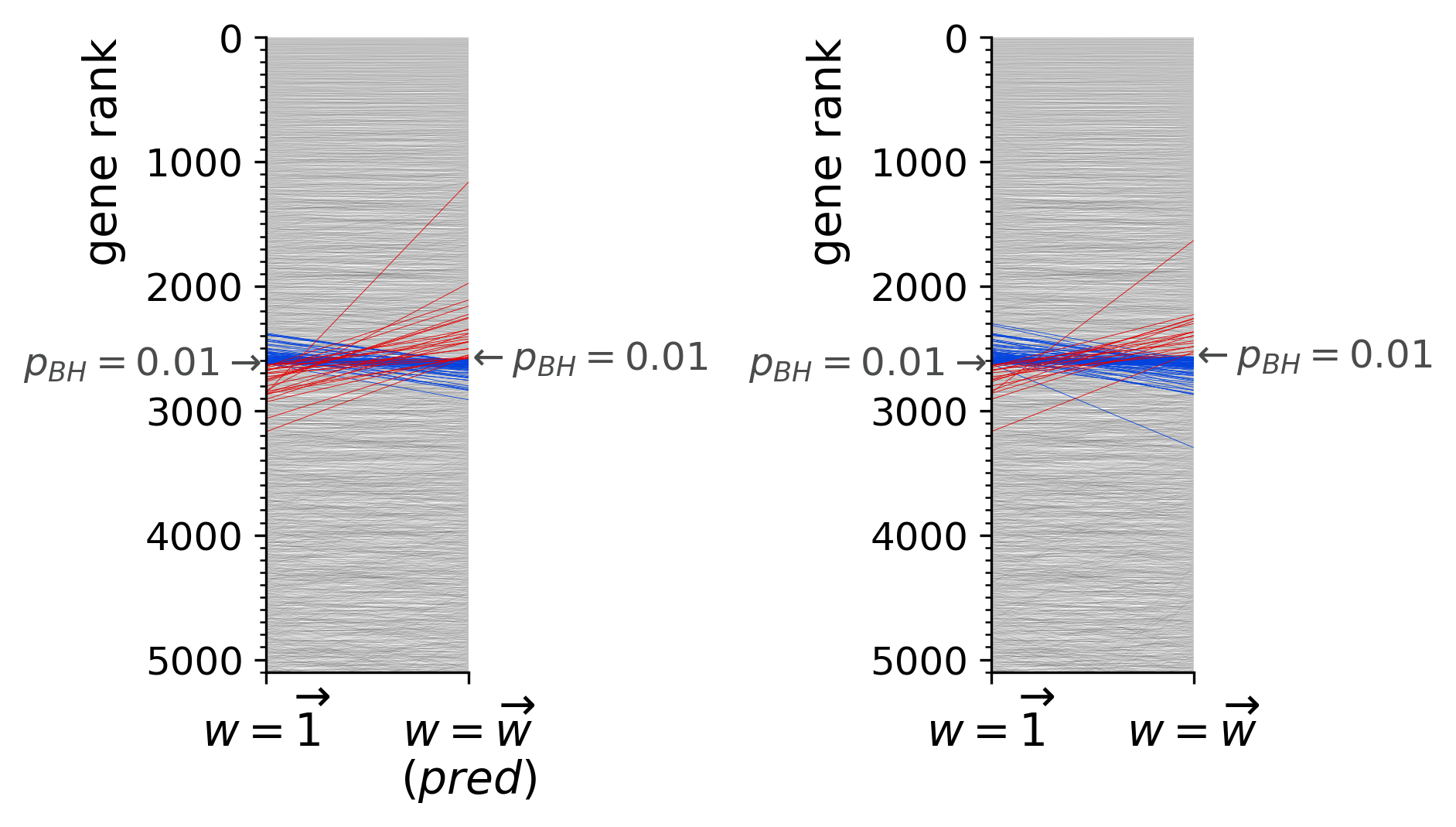}}%
            \vspace{1.5\vspaceBwRows}
            \subcaptionbox{\makeSubcaption{seven}{four}{down}}{%
                \includegraphics[width=\linewidth]{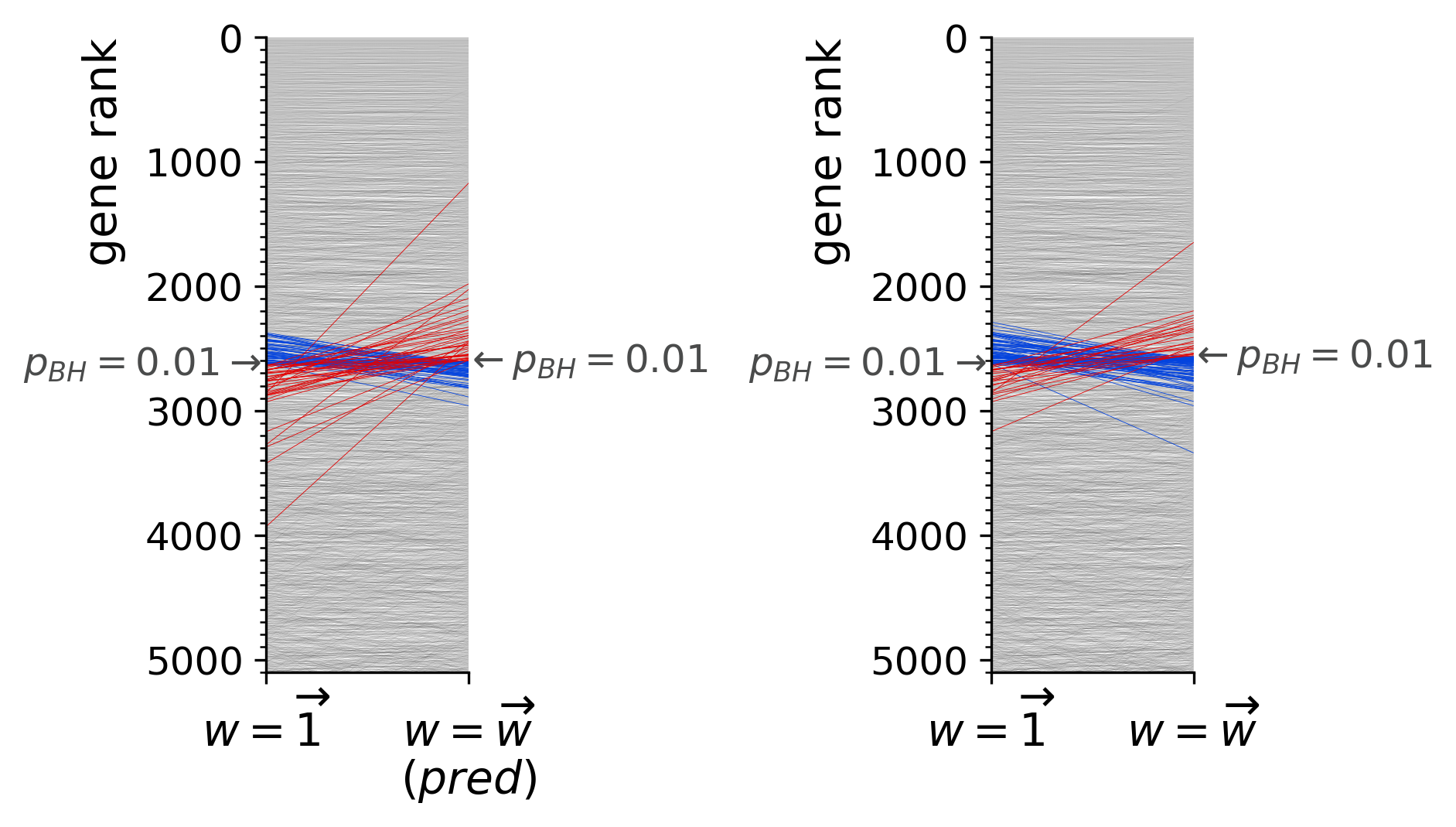}}%
        }
        \parbox{0.5\textwidth}{%
            \subcaptionbox{\makeSubcaption{14}{five}{down}}{%
                \includegraphics[width=\linewidth]{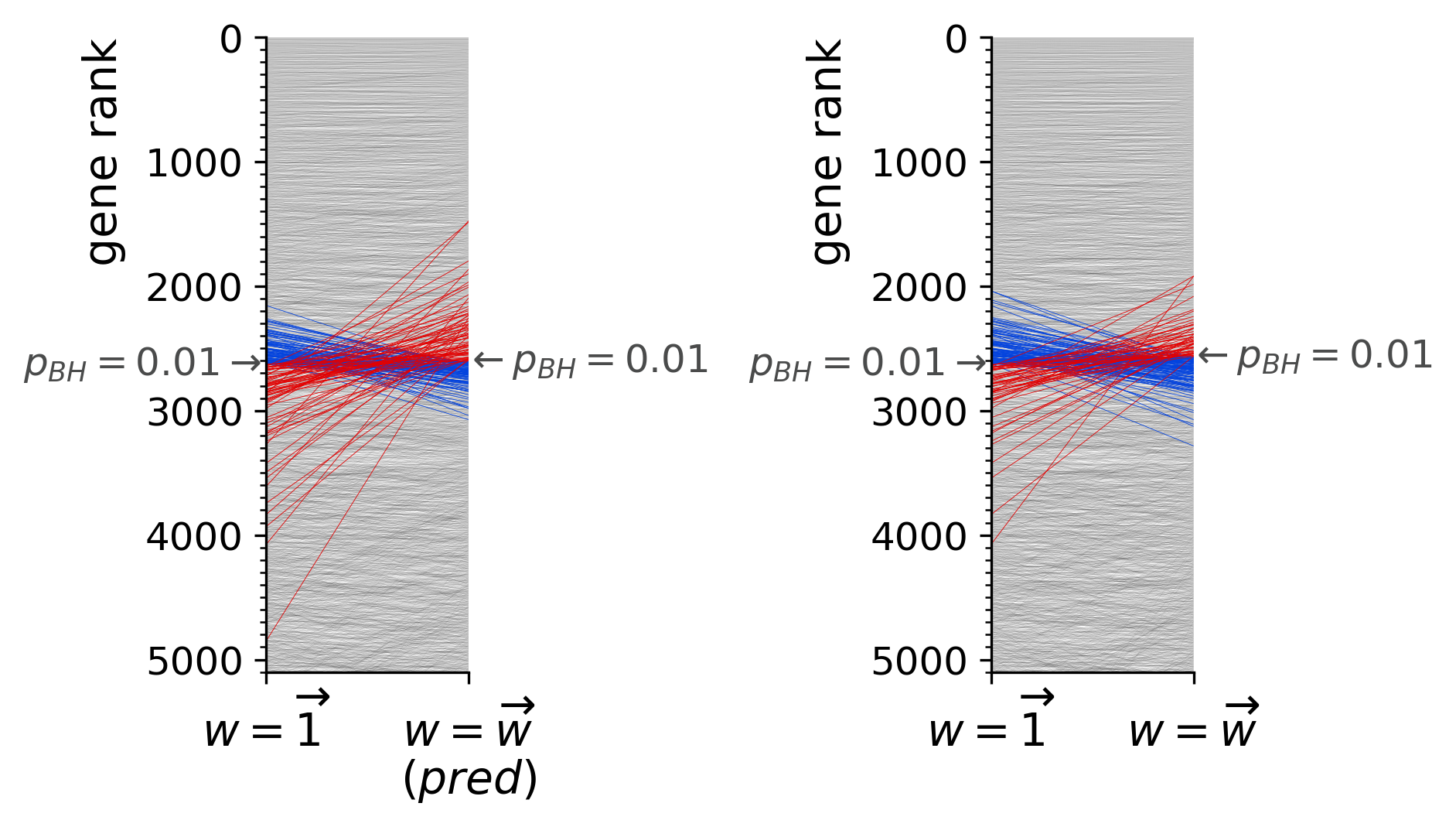}}%
            \vspace{1.5\vspaceBwRows}
            \subcaptionbox{\makeSubcaption{28}{five}{down}}{%
                \includegraphics[width=\linewidth]{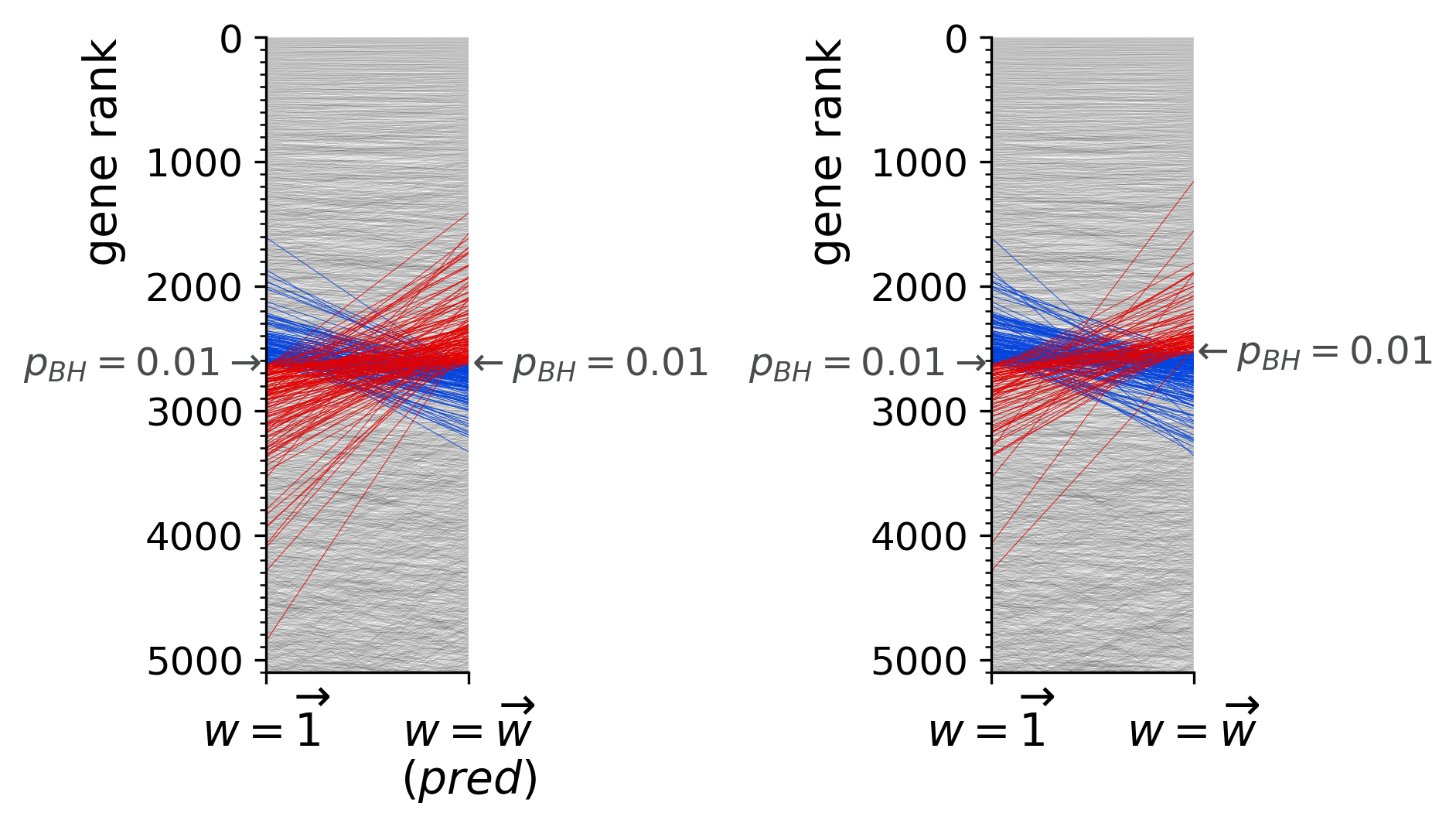}}%
        }
    }
    \caption{
        \textbf{Predicted vs. actual perturbation to DE p-values when a handful
                of influential cells (with respect to downregulated genes) are dropped.}
        Plots show the predicted (\textit{left}) and actual (\textit{right}) changes to ranked p-values
        for differential expression based on the Wald sandwich test.
        Annotated arrows indicate the ranking cutoff for BH-corrected p-values
        at level $0.01$.
        Blue lines indicate the change in ranking for genes that are \textit{demoted}
        from the significant set,
        red lines indicate the change in ranking for those that are \textit{promoted},
        and black lines indicate the change in ranking for those that \textit{retain} their
        significance status. Rankings are truncated; over 10,000 genes are tested.
    } \label{supp-fig:change-in-de-pvals-down}
\end{figure}

\begin{figure}[!htb]
    \centering
    \subcaptionbox{\label{supp-fig:cell-overlap-up}}{%
        \includegraphics[width=\linewidth]{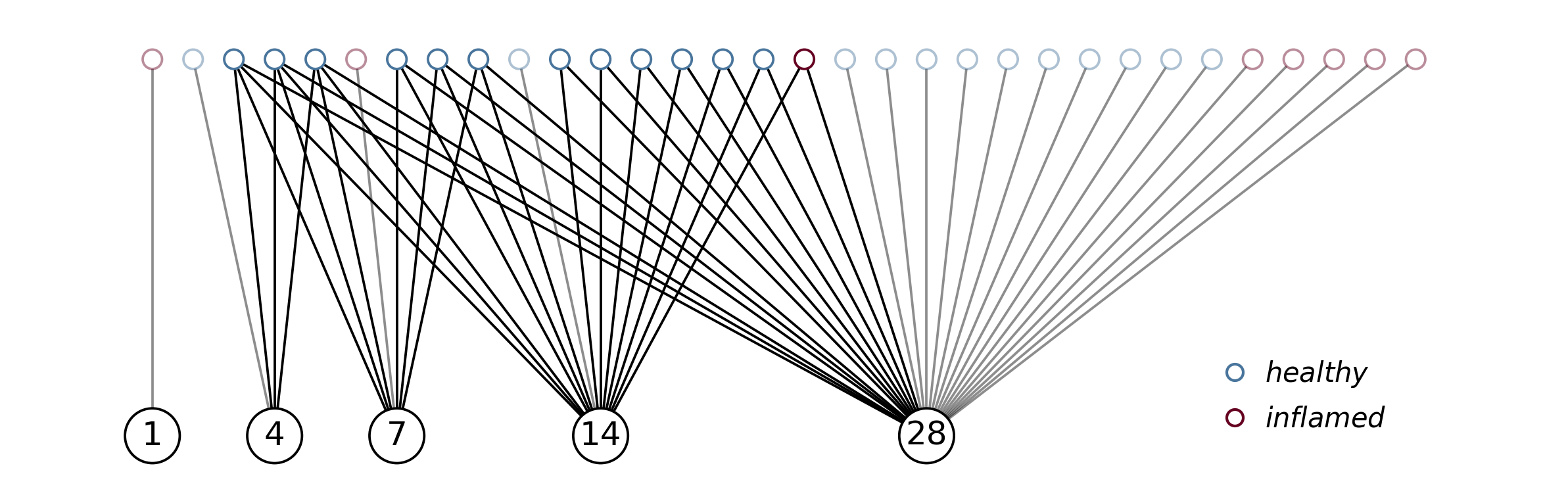}}%
    \vspace{\vspaceBwRows}
    \subcaptionbox{\label{supp-fig:cell-overlap-down}}{%
        \includegraphics[width=\linewidth]{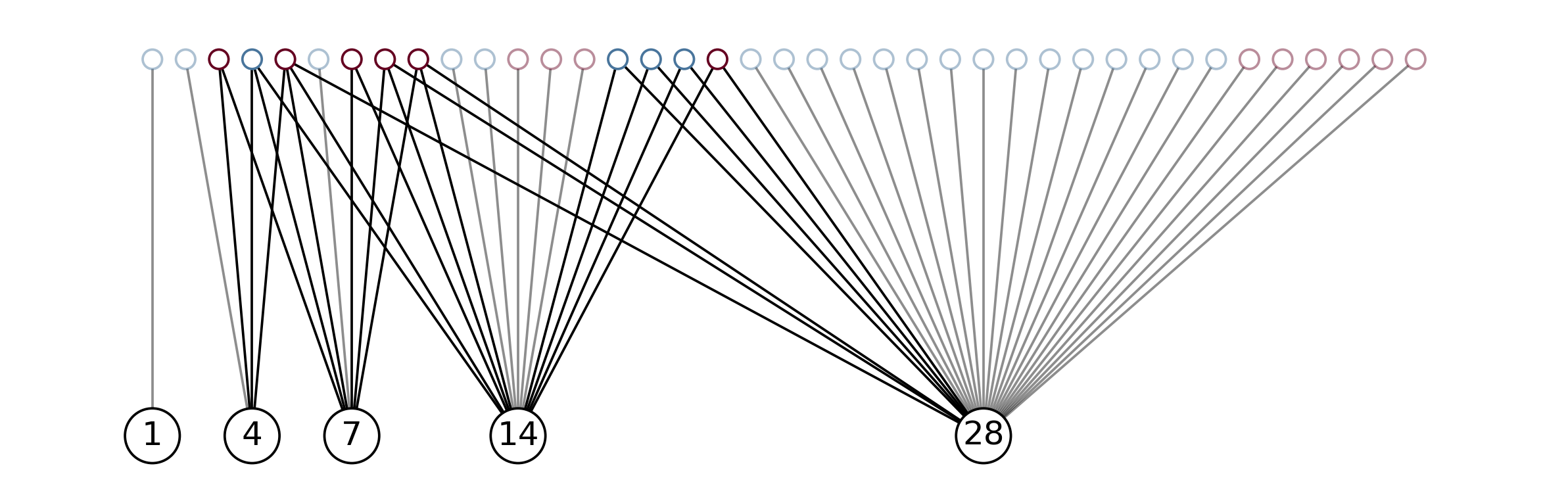}}%
    \caption{
        \textbf{Cell overlap between influential clusters of varying size.}
        Plots show the overlap of cells
        (\textit{small colored circles}; colored according to \texttt{Health},
        the grouping-of-interest for differential expression)
        across the most influential cluster (with respect to gene set enrichment)
        that we identify at each size
        (\textit{large black circles}; labeled according to the size of the cluster, $K$).
        After clustering cells at a given $K$, we use heuristics to select the
        most influential cluster based on the maximal disruption to the top 10
        gene sets---enriched among differentially upregulated
        (\textbf{\subref{supp-fig:cell-overlap-up}}) or downregulated
        (\textbf{\subref{supp-fig:cell-overlap-down}}) genes---when those cells are dropped.
        \captionbr
        See \cref{fig:change-in-go-up,fig:change-in-go-down} for the corresponding
        effect on top gene sets when each of these clusters is dropped.
    } \label{supp-fig:cell-overlap}
\end{figure}